%% file: Thesis.tex
\documentclass[12pt]{report}
\usepackage{harvard}
\citationstyle{dcu} \textwidth=14cm \textheight=23cm
\usepackage{amsfonts}
\usepackage{epsfig}
\usepackage{lscape}
\renewcommand{\baselinestretch}{2}

\begin{document}
\pagenumbering{roman} \setcounter{page}{1}
\begin{abstract}
In this thesis we investigate the stability of few body
symmetrical dynamical systems which include four and five body
symmetrical dynamical systems.

We divide this thesis into three parts. In the first part we
determine some special analytical solutions for restricted,
coplanar, four body problem both with equal masses
\cite{RoyandSteves1998} and with two pairs of equal masses. The
Lagrange solutions $L_1$ and $L_4$ are obtained in the two
triangular solutions. The equilateral triangle and the isosceles
triangle solutions are also obtained.  We also provide a
comprehensive literature review on the four and five body problems
to put our research on these problems in the wider context.

In the second part  we investigate more complicated and general
four body problems. We analyze numerically the stability of the
phase space of the Caledonian Symmetric Four Body Problem (CSFBP),
a symmetrically restricted four body configuration first
introduced by \citeasnoun{RoyandSteves1998}, by perturbing the
position of one of the bodies and using the general four body
equations. We show that the CSFBP is stable towards small
perturbations and there is no significant change in the symmetry
before and after the perturbations.

In the third part we introduce a stationary mass to the centre of
mass of the CSFBP, to derive analytical stability criterion for
this five body system and to use it to discover the effect on the
stability of the whole system by adding a central body. To do so
we define a five body system in a similar fashion to the CSFBP
which we call the Caledonian Symmetric Five body Problem (CS5BP).
We determine the maximum value of the Szebehely constant,
$C_0=0.659$, for which the CS5BP system is hierarchically stable
for all mass ratios. The CS5BP system has direct applications in
Celestial Mechanics. The analytical stability criterion tells us
for what value of $C_0$ the system will be hierarchically stable
but we would like to know what happens before this point. To
understand this, we determine a numerical stability criterion for
the CS5BP system which compares well with the analytical stability
criterion derived earlier. We conclude this thesis with the
generalization of the above analytical stability criterion to the
$n$ body symmetrical systems. This new system which we call the
Caledonian Symmetric N Body Problem (CSNBP) has direct
applications in Celestial Mechanics and Galactic dynamics.

Research presented in this thesis includes the following original
investigations: determination of some analytical solutions of the
four body problem; stability analysis of the near symmetric
coplanar CSFBP ; derivation of the analytical stability criterion
valid for all time  for a special symmetric configuration of the
general five-body problem, the CS5BP, which exhibits many of the
salient characteristics of the general five body problem;
numerical investigation of the hierarchical stability of the CS5BP
and derivation of the stability criterion for the CSNBP.

\end{abstract}
\renewcommand{\contentsname}{Table of Contents}
\tableofcontents \listoffigures \listoftables
\chapter{Introduction}
\pagenumbering{arabic} \setcounter{page}{1} The motion of systems
of three or more bodies under their mutual gravitational
attraction is a fascinating topic that dates back to the studies
of Isaac Newton. Because of the complicated nature of the
solutions, few-body orbits in the most general cases could not be
determined before the age of computers and the development of
appropriate numerical tools. Today the few body problem is
recognized as a standard tool in astronomy and astrophysics, from
solar system dynamics to galactic dynamics \cite{CarlMurray}.

During the past century, Celestial Mechanics has principally been
devoted to the study of the three body problem. Due to the
difficulty in handling the additional parameters in the four and
five body problems very little analytical work has been completed
for greater than three bodies.

In our galaxy it is estimated that roughly two-thirds of all stars
exist in binary systems. Furthermore it is estimated that about
one-fifth of these systems actually exist in triple systems, while
a further one-fifth of these triple systems are believed to exist
in quadruple or larger multiple systems. Therefore, out of the
$10^{11}$ stars in the Galaxy, of the order of $2 \times 10^9$ of
them exist in quadruple stellar systems \cite{BonnieRoy1998}. It
is therefore of important to study the dynamical behavior of such
systems.

In this thesis we investigate the dynamics of small clusters of
stars and planetary systems using  $n$-body symmetrical dynamical
systems where $n \ge 4$. We have generalized the investigations of
Steves and Roy on symmetrical four body problems to symmetrical
N-Body problems which gives us insight into the stability of
symmetrical stellar clusters with planetary systems.

The main thesis of research is divided into two parts. In the
first section (chapters 3), we discuss the equilibrium
configurations of four body problems as particular solutions of
four body problems. In the second section (chapter 4 to 7), the
more complicated four and five body restricted symmetrical
problems are studied. We particularly focus on the Caledonian
Symmetric Four Body Problem (CSFBP) Steves and Roy (1998, 2000,
2001) and the Caledonian Symmetric 5 Body problem (CS5BP).

The current state of research on the stability of the four and
five body problems is reviewed in chapter 2. Because of the
greater complexity of having a higher number of bodies, the main
focus of the literature on the four and five body problems has
been on the analytical study of their Equilibrium Configurations .
\emph{An equilibrium configuration of four-bodies is a geometric
configuration of four bodies in which the gravitational forces are
balanced by the centrifugal force so that the configuration is
maintained for all time.} We also review the few existing papers
which address the analytical stability of four body problems in
general: In particular the papers of \citeasnoun{loks1985} and
\citeasnoun{loks1987} are considered. Publications on the
equilibrium solutions and the stability of the Caledonian
Symmetric Four Body Problem ( Steves and Roy (1998, 2000, 2001),
Roy and Steves (1998) and \citeasnoun{Andras1}) form the main
source of knowledge relevant to the research of this thesis. These
papers are reviewed
 separately in detail at different places of the thesis.

In chapter 3, we review the equilibrium solutions of symmetric
four body problems given by \citeasnoun{RoyandSteves1998} and
derive some further solutions. In section 3.1 we review three
types of analytical solutions for the equal mass four body problem
which include the Square, the Equilateral triangle and the
Collinear equilibrium configurations \cite{RoyandSteves1998}. In
section 3.2 we give the solutions for two pairs of equal masses
where the ratio between the two pairs is reduced from 1 to 0 in
order to obtain equilibrium configurations which involve the five
Lagrange points of the Copenhagen problem. We discuss four kinds
of equilibrium configurations both symmetric and non-symmetric
which include the Trapezoidal, the Diamond, the Triangular and the
Collinear equilibrium configurations. The derivation of the
Trapezoidal, the Diamond and the Collinear equilibrium
configurations are a review from \citeasnoun{RoyandSteves1998},
while the derivation of the two Triangular equilibrium
configurations is original work.

It would be useful to broaden this understanding to include the
stability of general solutions to restricted four body problems.
In chapter 4 we look at the stability of a restricted case of the
four body problem called the Caledonian Symmetric Four Body
Problem (CSFBP). This symmetrically restricted four body problem
was first developed by Steves and Roy (1998).

Steves and Roy (2001) later derived an analytical stability
criterion valid for all time for the CSFBP. They show that the
hierarchical stability of the CSFBP solely depends on a parameter
they call the Szebehely constant, $C_0$, which is a function of
the total energy and angular momentum of the system. This
stability criterion has been verified numerically by Sz\'ell,
Steves and \'Erdi (2003a, 2003b). \citeasnoun{AndrasMNRAS} analyze
the connection between the chaotic behavior of the phase space and
the global stability given by the Szebehely constant using the
Relative Lyapunov indicator (RLI) and the smallest alignment
indices (SALI) methods. They found that as the Szebehely constant
is increased, making the system hierarchically stable from a
global point of view, the corresponding phase space becomes
increasingly more regular.

In chapter 4 we review their research on the CSFBP and then
investigate the CSFBP phase space using nearly symmetric,
perturbed, initial conditions and the general four body equations
in order to study if the CSFBP system will remain symmetric under
slightly perturbed initial conditions. We use long time
integrations of a million time-steps to study the behavior of the
slightly perturbed CSFBP phase space. An integrator is
specifically developed for this purpose using the Microsoft Visual
C++ Software. The results of integrations are processed using
Matlab 6.5. During the integrations, we record the following
observations:\begin{enumerate} \item For all orbits which have a
close encounter, we stop the integration  and record the type of
the collision or the close encounter.\item For all orbits which
fail the symmetry breaking criterion, we stop the integration and
record the symmetry as broken .\item For all orbits where there
are no collisions or breaking of the symmetry, the integration
continues to the end of 1 million time-steps.\end{enumerate}

In Chapter 5 we introduce a stationary mass to the centre of mass
of the CSFBP and derive an analytical stability criterion for this
new five body symmetrical system similar to the one of the CSFBP.
This stability criterion enables us to analyze the effect of the
additional central body on the stability of the whole system. We
call this new problem the Caledonian Symmetric Five body Problem
(CS5BP). The  CS5BP has direct applications in dynamical systems
where a very large mass exists at the centre of mass with four
smaller masses moving in dynamical symmetry about it. The four
small masses are affected by the central mass but are small enough
that they do not dynamically affect the central body. Hence the
central body remains stationary and dynamical symmetry is
maintained. This could occur, for example, in exoplanetary systems
of a star with four planets or a planetary system with four
satellites.

For completion we analyze the full range of mass ratios of central
body to the other four bodies: from the large central body system
described above to a five body system of equal masses to a small
central body with four large surrounding bodies. In the case of
five equal masses or a small central body with four surrounding
large masses, the central body is unlikely to remain stationary as
is required by the CS5BP. The model may, however, still be
applicable to real systems in which the outer bodies are well
spaced and stationed far away from the central body so that they
have minimal effect on the central body.

In chapter 6 we continue our analysis of the CS5BP. We investigate
numerically the hierarchical stability of the CS5BP. The main
objective of this exercise is to validate the hierarchical
stability criterion derived in chapter 6 and to study the
relationship between the number of hierarchy changes and the value
of $C_0$.

\citeasnoun{andras2} provide a numerical investigation of the
hierarchical stability of the CSFBP which covered half of the
phase space. We provide a brief review of these results in section
6.1 before deriving the equations of motion for the CS5BP in
section 6.2. The comprehensive numerical exploration was completed
using an integrator specially developed for the CS5BP in Fortran
and C++. We used Matlab 6.5 software to process the results. 3000
orbits were integrated to 1 million time steps of integrations for
each of 63 values of $C_0$. The total number of orbits integrated
were $3000 \times 63$ which took about 70 days of CPU time. During
the integrations the following information was recorded for each
value of $C_0$ to be able to comment on the hierarchical stability
of the system under discussion:
\begin{enumerate}\item The total number of hierarchy changes and \item
The types of hierarchy changes \end{enumerate}   In section 6.5,
we give a complete analysis of the hierarchical stability of the
CS5BP for a whole range of mass ratios. This includes the
completion of the analysis of \citeasnoun{andras2} and
\citeasnoun{anrasHieStab} as the CSFBP is a special case of the
CS5BP.

In chapter 7, we generalize the Caledonian Symmetric 5 Body
Problem (CS5BP) to the Caledonian Symmetric N Body Problem (CSNBP)
to derive an analytical stability criterion for the Symmetric N
Body problem where $N \ge 4$.

Finally in chapter 8 we give a brief summary of the results
obtained in this thesis and in chapter 9 we point out some
possible areas for future exploration.

In this thesis the original research involves: the determination
of the triangular equilibrium solutions of the four body problem
(sections 3.3.3 and 3.3.4); the stability analysis of the near
symmetric coplanar CSFBP (sections 4.4 and 4.5); the derivation of
the analytical stability criterion valid for all time  for the
symmetric configuration of the general five-body problem, the
CS5BP, (Chapter 5); the numerical investigation of the
hierarchical stability of the CS5BP (sections 6.2 to 6.7) and the
derivation of the stability criterion for the CSNBP (Chapter 7).

Now in chapter 2 we present a brief review of the current state of
the literature on analytical studies of the four and five body
problems.

\chapter{The Four and Five Body Problem : A literature review}

In our galaxy it is estimated that roughly two-thirds of all stars
exist in binary systems. Furthermore it is estimated that about
one-fifth of these systems actually exist in triple systems, while
a further one-fifth of these triple systems are believed to exist
in quadruple or larger multiple systems. Therefore, out of the
$10^{11}$ stars in the Galaxy, of the order of $2\times{10^9}$ of
them exist in quadruple stellar systems \cite{BonnieRoy1998}. It
is therefore of interest to study the dynamical behavior of such
systems.

The classical equation of motion for the n-body problem assumes
the form
\begin{equation}
m_{i}\frac{d^{2}\mathbf{r}_{i}}{dt^{2}}=\frac{\partial U}{\partial
r_{i}}=\sum_{j\neq i}\frac{m_{i}m_{j}\left(
\mathbf{r}_{j}-\mathbf{r}_{i}\right)
}{|\mathbf{r}_{i}-\mathbf{r}_{j}|^{3}}\qquad i=1,2,...,n,
\end{equation}
where the units are chosen so that the gravitational constant is
equal to one, $ \mathbf{r}_{i}$ is a vector in three space,
\begin{equation}
U=\sum_{1\preceq i<j\preceq
n}\frac{m_{i}m_{j}}{|\mathbf{r}_{i}-\mathbf {r}_{j}|}
\end{equation}
is the self-potential, $\vec{r}_i$ is the location vector of the
$i$th body and $m_i$ is the mass of the $i$th body.

Because of the greater complexity of higher number of bodies, the
main focus of the literature for four and five body problems has
been on the analytical study of the Equilibrium Configurations. We
therefore largely concentrate our review on this research. We also
review the few existing papers which address the general four body
problem: In particular the papers of \citeasnoun{loks1985} and
\citeasnoun{loks1987}. Publications on the equilibrium solutions
of the four body problems and the stability of the Caledonian
Symmetric Four Body Problem ( Steves and Roy (1998, 2000, 2001),
\citeasnoun{RoyandSteves1998} and \citeasnoun{Andras1}) form the
main source of knowledge relevant to the research of this thesis.

In this chapter we review the analytical research on both four and
five body problems. In section 2.1 a literature review is given
for the analytical solutions and stability for four body problems
and in section 2.2 for the five body problems.

\section{The Four Body Problem : A literature review}
We divide this section further into four subsections. The first
subsection deals with particular solutions, the Central
Configurations, of the Four Body Problem. The remaining three
subsections deal with the stability analysis of various degrees of
restricted four body problem: the restricted four body problem,
the symmetric four body problem and the general four body problem.

\subsection{Central Configurations of the Four Body Problem}

To understand the dynamics presented by a total collision of the
masses or the equilibrium state for a rotating system, we are led
to the concept of a central configuration. For a system to be the
central configuration, the acceleration of the $i$th mass must be
proportional to its position (relative to the centre of mass of
the system), thus $\ddot{r_{i}}=\lambda r_{i}$ for all
$i=1,2,...,n$. A Central Configuration can also be expressed as a
critical point for the function $U^{2}I$, where I is the moment of
inertia. Though one has two different formulations of a central
configuration, a number of basic questions concerning them remain
unanswered, such as ``for a fixed number of particles with equal
mass, do there exist a finite number of Central Configurations?

In this section a review of papers concerning central
configurations and hence equilibrium solutions which is a special
case of central configurations is provided. \emph{An equilibrium
configuration of four-bodies is a geometric configuration of four
bodies in which the gravitational forces are balanced in such a
way that the four bodies rotate their centre of mass and thus the
geometric configuration is maintained for all time.} Equilibrium
solutions of the symmetric four body problem are studied in
chapter 3 and chapter 4 in greater detail. These chapters include
a review of \citeasnoun{BonnieRoy1998} along with original work.

The straight line solutions of the $n$-body problem were first
published by \citeasnoun{moulton1910}. He arranged $n$ masses on a
straight line so that they always remained collinear and then
solved the problem of the values of the masses at $n$ arbitrary
collinear points so that they remained collinear under proper
initial projections. This paper is not particularly concerned
about the specific case of the four body problem but it can be
deduced that there are 12 collinear solutions in the equal mass
four body problem. These solutions are sometimes referred to as
Moulton Solutions.

The two papers on the Classification of Relative Equilibria by
Palmore(1975, 1982) presented several theorems on the
classification of equilibrium points in the planar $n$-body
problem. He then applied his results to the cases of three and
four body equal mass systems. These highly technical papers show
that the theorems lead to the existence of 120 possible
configurations of relative equilibrium for four equal masses, of
which 24 exist in the isosceles configuration. This number takes
into account all of the possible permutations of the masses in the
same configurations including the 12 Moulton collinear solutions.

\citeasnoun{yanZhiming} in their article under the title, 'The
Central Configurations Of the General 4-Body Problem', obtain the
equations of Central Configurations using a method different from
that introduced by \citeasnoun{Palmore}. \citeasnoun{yanZhiming}
also investigate the finiteness of Central Configurations for the
general four-body problem. They show that for the collinear
four-body problem there are twelve central Configurations for each
set of masses. This agrees with the existence of 12
\citeasnoun{moulton1910} solutions. They also prove the following
very important theorems: 1. {\it There is not any point-line type
Central Configuration, where three of the four mass-points lie on
a line while the fourth does not, in the general four-body
problem. } 2.{\it \ For any four-body Central Configurations if
there is a mass point which keeps the same distance from two of
the other three mass points, the fourth one also keeps the same
distance from the two. Such a Central Configuration is symmetric.}

Using algebraic and geometric methods, \citeasnoun{Arenstrof}
investigates the number of equivalence classes of central
configurations in the planar four-body problem with three
arbitrary sized masses and a fourth small mass $m_4$. For example
the following system of algebraic equations is algebraically
reduced to three equations in three unknowns and then the problem
is reduced to the case $m_{4}=0.$
\[
f_{1}=f_{2}=...=f_{N}=0,\qquad
f_{k}=\sum\limits_{j=1}^{N}\frac{m_{j}}{{\bf r}_{ij}^{3}}\left(
q_{k}-q_{j}\right) ,\qquad r_{jk}=|q_{j}-q_{k}|>0.
\] His results
show that each three-body collinear Central Configuration
generates exactly two non-collinear Central Configurations
(besides four collinear ones) of four bodies with small $m_{4}\geq
0$ ; and that the three body equilateral triangle Central
Configuration generates exactly eight, nine or ten planar
four-body Central Configurations with $ m_{4}=0$.

\citeasnoun{Kathryn} studies the Central Configurations of the
classical N-body problem and the asymptotic properties of a system
of repelling particles. An asymmetric configuration obtained in
the eight-particle system is described and a bifurcation in the
four-particle system is investigated. An asymmetric configuration
with eight particles, which is the smallest number of particles
with no axis of symmetry, is found. Bifurcation occurs in the
general system of particles as the parameter $\beta $ is varied,
where $U^{\prime }(r)=U^{\prime }(1)/r^{1+\beta }$ . A particular
example with four particles is discussed and the following four
equilibrium configurations for this system with $\beta =1$ is
given.\begin{enumerate}\item A configuration with four particles
lying along a line through the origin.\item A configuration with
particles at the vertices of a square.\item A configuration with
particles at the vertices of an equilateral triangle and one
particle at the origin.\item A configuration with particles at the
vertices of an isosceles (but not equilateral) triangle and one
particle within the triangle and on its symmetry
axis.\end{enumerate}

\citeasnoun{simo} presents the classification and solutions of the
central configurations of the four body problem using topological
proofs. Sim$\acute{o}$ first considered the restricted problem of
three finite masses plus one infinitesimal mass. He showed
diagrammatically the location of equipotential curves and the nine
equilibrium points in the case of three equal masses and in the
case of masses $m_1=0.2, m_2=0.3$ and $m_3=0.5$. He then moves on
to the problem of four arbitrary masses. He found (a) a division
of the mass space in regions for which there is a different number
of relative equilibria, (b) masses for which degeneration occurs,
and (c) for given masses, the determination of the relative
equilibria and the relation with the corresponding relative
equilibria of the equal mass case.

Several papers (\citeasnoun{Majorana}, \citeasnoun{Kozak},
\citeasnoun{JaganBonnie}) derive the equilibrium solutions and
analyze their stability for different types of four body problems.
\citeasnoun{Majorana} studies the linear stability of the
equilibrium points in the restricted four body problem, where
three bodies, of masses $\mu$, $\mu$ and $1-2\mu$ rotate in an
equilateral triangular configuration (the Lagrange solution),
whilst the fourth body of negligible mass moves in the same plane.
The equations of motion of the particle under the influence of the
other three bodies were derived which led to the determination of
eight equilibrium points in the problem. The author found
numerically that the linear stability of some of these equilibrium
points depended on the masses of the three bodies. Five of the
points were found to be unstable for all values of $\mu$, whilst
two of the other points were found to be stable for small values
of $\mu$.

\citeasnoun{Kozak} studied the motion of a negligible mass in the
gravitational field generated by a collinear configuration of
three bodies (of masses $m_{0}\neq m_{1}=m_{2}=m$). To study the
linear stability of the equilibrium points, the authors write the
equations of motion in their Cauchy form following the definition
of the equilibrium points. Because it is impossible to find a
compact analytic form for the solutions, they find the approximate
solutions by: \begin{enumerate}\item expanding terms as a power
series in $m/m_{0}$ and $m_{0}/m$.\item using Newton's Iterative
method for given values of $m/m_{0}.$\end{enumerate}

They proved that the straight line (collinear) configurations are
linearly unstable for any value $m$ as in the Lagrange case of the
three-body problem. Intervals are given for the stable and
unstable regions of the triangular configurations.

This review on central configurations gives the context for the
research of chapters 3, where the work by
\citeasnoun{BonnieRoy1998} on equilibrium solutions of symmetric
four body problem is reviewed and extended to include derivations
of the triangular equilibrium solutions.
\citeasnoun{RoyandSteves1998} discuss some special analytical
solutions of four body problems. They show that these solutions
reduce to the Lagrange solutions of the Copenhagen problem when
two of the masses are equally reduced. The content of this paper
will be discussed in chapter 3 in detail.

\subsection{Symmetric Four Body Problem}
It is well known that it is not possible to find analytical
solutions for the general four body problem. Therefore restriction
methods utilizing assumptions of neglecting the masses of some of
the bodies or assumptions which require specific conditions of
symmetry have been used to reduce the dimensions of the phase
space to manageable levels while still producing results which are
meaningful to real physical systems. For example a four body model
requiring symmetrical restrictions was used by
\citeasnoun{mikolaetall1984} as a means of understanding
multi-star formation in which symmetries produced in the initial
formation of the star system were maintained under the evolution
of the system.
In this section papers which used specific conditions of symmetry
to simplify the four body problem analysis will be reviewed.These
include a series of papers on the Caledonian Symmetric Four Body
Problem (CSFBP) which are very relevant to the research presented
in this thesis, in chapters 5 to 8.

\citeasnoun{Roysteves2000} first introduced a symmetric problem
which they called the Caledonian Symmetric Double Binary Problem
(CSDBP). It is a special case of the Caledonian four-body problem
introduced earlier by the same authors \cite{RoyandSteves1998}. To
form the CSDBP, they utilized all possible symmetries. The CSDBP
is a three dimensional problem, with two pairs of masses, each
pair binary having unequal masses but the same two masses as the
other pair. They have shown that the simplicity of the model
enables zero-velocity surfaces to be found from the energy
integral and expressed in a three dimensional space in terms of
three distances $r_1, r_2$ and $r_{12}$, where $r_1$ and $r_2$ are
the distances of the two bodies which form the pair from the
centre of mass of the four body system. $r_{12}$ is the separation
between the two bodies. This problem was later renamed as the
Caledonian Symmetric Four Body Problem (CSFBP). Further details of
the CSFBP are given in chapter 5.

Steves and Roy (2001) derived an analytical stability criterion
for the CSFBP valid for all time. They show that the hierarchical
stability of the CSFBP depends solely on a parameter they call the
Szebehely Constant, $C_0$, which is a function of the total energy
and angular momentum of the system.\emph{A four body system is
said to be hierarchically stable if it maintains its initial
hierarchy state for all time.} This analytical stability criterion
was numerically verified by Sz\'ell, B. Steves and Roy (2002) and
Sz\'ell,  Steves and  \'erdi (2004a) for the coplanar CSFBP. They
have performed a wide range of numerical integrations for a
variety of values of the Szebehely constant $C_0$. They also show
that for $C_0$ greater than a critical value, the system is
hierarchically stable for all time and will undergo no change in
its hierarchical arrangement. They also show that the number of
hierarchy changes decreases with the increasing value of $C_0$.

\citeasnoun{AndrasMNRAS} investigate the chaotic and stable
behavior of the CSFBP using the relative Lyapunov indicator (RLI)
and the smallest alignment indices (SALI) which are fast chaos
detection methods. They analyze the phase space of the CSFBP in
detail for different mass ratios and for each mass ratio a range
of values $C_0$. The color coded points on their graphs give the
nature of the orbits (chaotic, regular, ending in collisions)
defined by different initial conditions of the CSFBP. They show
that the regular and chaotic behavior of the phase space is
closely connected with the Szebehely constant. The larger the
Szebehely constant, the more regular the phase space becomes.

\citeasnoun{anrasEsc} give a numerical escape criterion for the
CSFBP. By approximating the symmetrical four body problem on the
verge of break up as a two body problem, they derive four energy
like parameters $E_1, E_2, E_3$ and $E_4$ which are related to
each of the four types of possible escapes. Numerous orbits were
investigated with the help of the numerical escape criterion. They
found that for stellar systems of nearly equal mass the most
likely escape configuration is the double binary escape, i.e. the
system falls apart into two binary-star configurations. For small
values of the mass ratio, $\mu$, the system can be a model for two
stars two planets systems. The integrations show that the most
likely escape in this case is the escape of the two planets.

\citeasnoun{winston} discusses a symmetrical four body system
where the bodies are distributed symmetrically about the centre of
mass. He also restrict the bodies to move upon a line undergoing
elastic collisions. It is shown by numerical simulations that
there is a similarity between this system and the one dimensional
newtonian three-body problem. The non-linear stability of the
Schubart-like orbit is studied using long numerical integration.
It is shown that this orbit is unstable in three dimensions.

\citeasnoun{JIanLiao} study a particular case of the 4-body
problem in the plane with a rhomboidal configuration. The
rhomboidal four body configuration means the four particles with
masses $m_1,m_2,m_3,m_4$ respectively are located in the plane at
the vertices of a rhombus. The particles are given symmetric
initial conditions in position and velocities with respect to the
axes in the plane and they thus always keep a symmetric rhomboidal
configuration under the law of newtonian attraction.
\citeasnoun{JIanLiao} show that the larger the mass ratio
($\alpha=m_3/m_1$), the more extended is the region of stable
motion in the phase diagram.


\citeasnoun{Claudio} studies a symmetric equal mass four body
problem, The Tetrahedral 4-Body Problem with Rotation, which is a
particular case of the general four body problem in space. It is
obtained when two bodies form a binary pair $m_1$ and $m_2$ that
are always symmetric to one another with respect to the vertical
axis, and the other two bodies form a pair $m_3$ and $m_4$ that
are symmetric to one another with respect to the same axis, with
all the masses equal to each other. The total angular momentum is
then forced to be zero by making the angular momentum of the
binary ($m_1-m_2$) equal and opposite to the angular momentum of
the other binary ($m_3-m_4$). He proves that all singularities of
this model are due to collision. He also proves that the
singularities due to simultaneous double collisions are
regularizable. The set of equilibrium points on the total
collision manifold is studied, as well as the possible connections
among them. It is shown that the set of initial conditions on a
given energy surface going to quadruple collision is a union of
twenty submanifolds.

\subsection{Restricted Four Body Problem}
Four body problems found in nature can often be simplified using
the reduction of variables given by having some masses which are
negligible compared to others. Thus they have no gravitational
effect on the finite masses, yet all finite masses have an effect
on them. A restricted four-body system of this type usually
contains three bodies with non-zero masses (primaries) and a
fourth massless particle. Here the time evolution of the massless
particle's orbit can be studied. In the case of the restricted
three-body problem,(problem with two primaries and a third
massless particle) the orbits of the two primaries were well known
but in the case of the restricted four-body problem the motions of
the three primaries cannot be given analytically. Thus some
assumptions on their motions must be made.

The "very restricted four-body problem" was introduced by
\citeasnoun{huang1960} being a circular restricted four-body
system. \emph{In the circular restricted four-body system, it is
assumed that the primaries have circular orbits i.e. orbit on a
circle.} He defined the masses of the bodies in the system be
$m_1$, $m_2$, $m_3$ and $m_4$ such that $m_1 \gg m_2 + m_3$ and
$m_4 = 0$ the massless particle. He assumed that $r_{23} \ll
r_{12}, r_{13}$ for all time,  where $r_{ij}$ denotes the distance
between the $i$th and $j$th bodies. If $m_2$ and $m_3$ revolve on
a circular orbit and their barycenter revolves on a circular orbit
about the $m_1$ body and all bodies are in the same plane then the
system is defined to be a "very restricted four-body system".

Similarly to the restricted three-body problem, zero velocity
curves can be defined in this case (Huang, 1960), i.e. surfaces
can be determined within which motion can take place. Outside
these surfaces, motion is impossible; however in the "very
restricted four body problem", the surfaces change their shape
under the evolution of the system and so osculating zero velocity
curves must be introduced. Applying this method to the
Sun-Earth-Moon-satellite very restricted four-body system
($m_1=m_{Sun}$, $m_2=m_{Earth}$, $m_3 = m_{Moon}$, $m_4 =
m_{satellite}=0$) it can be found that any artificial satellite
orbiting the Moon must have a closed orbit in order to be stable
(Huang, 1960).

 \citeasnoun{matas} studies a special restricted four body
problem. He considers four material points, three of them have
finite non-zero mass and the fourth represents an infinitesimal
body. He also requires a constant configuration for the three
finite masses i.e. the motion of the first three points is given
by the particular Lagrange's solution of the problem. He obtains
an integral which is a generalization of the Jacobi's integral. As
a consequence of this integral, the equation of the surface of
zero relative velocity (a generalization of Hill's surfaces) is
derived.

\citeasnoun{Mohn} consider the motion of a particle in the
gravitational field of three mass points (Sun-Earth-Moon) that
obey the restricted three body equations. In particular, they
concentrate on the two cases where the three primaries move
according to Hills theory for periodic lunar orbits. In these
cases the leading terms in the solar perturbation are included in
a systematic and dynamically consistent manner so that the results
can then be derived in a form where, for very close configurations
of earth, moon, and particle, the solar influence introduces a
perturbation upon a motion which would otherwise obey the
restricted three-body problem for earth, moon, and particle.

\citeasnoun{hadji1980} studies the motion of a small planet moving
in the gravitational field of the Sun-Jupiter-Saturn system. He
first gives the initial conditions of a periodic orbit in the
general three body problem of the Sun-Jupiter-Saturn system, using
a suitable rotating frame of reference. He then derives the
equations of motion for a massless body moving in the
Sun-Jupiter-Saturn system in the rotating frame. The approximate
initial conditions for a periodic orbit of the massless body are
then determined. The stabilities of the periodic orbits are
examined using numerical techniques.

\citeasnoun{michalo} introduces the restricted four body problem
by generalizing the restricted three body problem. He studies a
special case of the restricted four body problem, the circular
restricted four body problem, which can be considered as a
generalization of the circular restricted three body problem.
Using the method of \citeasnoun{szebehe1967} he finds some
equilibrium points including four on the x-axis (collinear points)
and two on the y-axis where all the equilibrium points are
symmetrically located with respect to the origin. He gives several
monoparametric families of periodic orbits.

\citeasnoun{giorgio} gives the regularization of the restricted
four body problem, where the same transformations as in the
restricted three body problem \cite{szebegiacag} were used to
remove the singularities. The author then extends the
transformation of the equations of motion to the problem of a
particle moving under the influence of $n$ equal mass bodies which
move unperturbed on the vertices of an $n$-polygon.
\subsection{General Four Body Problem}

The general four body problem is the most general case without any
restrictions. The Newtonian equations of motion may be written
$$
\ddot{\vec{r}}_i = -G \sum_{j\neq i}^{4} \frac{m_j}{|\vec{r}_i -
\vec{r}_j|^3} (\vec{r}_i - \vec{r}_j) \;, \quad  j=0,1,2,3;
$$
They are second order ordinary differential equations for the
unknown $\vec{r}_i (t)$ functions. It is a 24th degree system
since there are four bodies with three location coordinates and
three velocity coordinates. With the help of the energy integral,
momenta integrals and the elimination of the node it can be
reduced to a 14th degree system.

The equations can not be solved analytically; however some
analytical statement can be made by examining the energy and
momenta manifolds of the system. Using the concept of zero
velocity curves, some hyper-surfaces of zero velocity can be
constructed which determine the regions of possible motion (Loks
and Sergysels, 1985, Sergysels and Loks, 1987). Since these
hypersurfaces are many dimensional, the visualization of the
results are very difficult. In the general three body problem,
\citeasnoun{zare76}, \citeasnoun{zare77} and
\citeasnoun{MarchalSari} showed that the energy $H$ and angular
momentum $c$ integrals can combine to form a stability criterion
$c^2H$ which, for a given critical value $c^2H_{crit}$, produces a
phase space for the three body system which contain topologically
separate subregions. Thus the hierarchically arrangement of the
system which exists in one subregion of the real motion cannot
physically evolve into the hierarchical arrangement of the system
existing in another subregion of the real motion.

In the two papers by \citeasnoun{loks1985} and
\citeasnoun{loks1987}, the $c^2H$ stability criterion in the three
body problem was extended to the general four body problem by
finding the zero velocity hypersurfaces which define the regions
in which motion can take place. In their first paper, the authors
began by describing the kinetic energy and the angular momentum of
four planar bodies, using generalized Jacobi variables. By
transforming the equations they then derived expressions for the
kinetic and potential energies of the system. From these equations
an expression for zero velocity hypersurfaces was found. In their
second paper, the authors applied Sundman's inequality for $n$
bodies to the four body problem in the generalized Jacobi
coordinates in order to derive zero velocity surfaces in three
dimensional space. The zero velocity surfaces then define the
limits of three dimensional regions in which the motion is
constrained. The cross section of these regions were then plotted
for an equal mass four body problem. These papers are highly
relevant to the work done by Roy, Steves and the author.

\citeasnoun{ReginaSimo} consider simultaneous binary collisions in
the general planar four body problem. They prove that these
collision are regularizable in the sense of continuity with
respect to initial conditions using a blow-up of the singularity.
They discuss subproblems of the general case which include the
particular example of the trapezoidal solution of the equal mass
four body problem and double isosceles four body problem.
Numerical evidence is given that the differentiability of the
regularization should be, in general, less than ${8/3}$.

Asteroids can be found around the $\gamma_6$ secular resonance
with Saturn. The longitudes of perihelion of these asteroids
rotate with the same angular velocity value as Saturn's longitude
of perihelion. Williams (1979) listed the names of the supposed
$\gamma_6$ asteroids. Performing more than a 1 Myr numerical
integration on the elliptical restricted four-body model i.e. the
restricted four body problem where the three primaries' orbits are
ellipses, so that the asteroid is allowed to move in an elliptical
orbit. Froeschl\'e and Scholl (1987) found that there were only
two asteroids from Williams's list located deeply in the
$\gamma_6$ resonant region. In order to check the validity of
their results, they repeated the integration including the effects
of all planets (beside Pluto) in the Solar System. They found some
differences which suggests that in some cases neither the
restricted three-body model nor the restricted four-body model can
describe the complete behavior of the asteroid belt. Thus the
investigations of the five or more body systems is very important
to consider.

\section{The Five Body Problem: A literature Review}

In the past Celestial Mechanics has been mainly devoted to study
the three and four body problems. Very little has been published
with regards to the five body problem. Because of the greater
complexity of higher numbers of bodies, the main focus of the
literature has been on the analytical study of central
configurations and relative equilibria of the five body problem.
\emph{A relative equilibrium is a special configuration of masses
of the n-body problem which rotates about its centre of mass if
given the correct initial momenta. In rotating coordinates these
special solutions become fixed points, hence the name relative
equilibria}. After an extensive search for literature on the five
body problem the author is led to believe that our work in chapter
5 and 6 is the first to address the issue of the analytical
stability of five body problem. Therefore all the literature
review is on equilibrium solutions and relative equilibria
configurations. The following is a review of the most relevant,
perhaps all the papers on the five body problem.

\begin{figure}
\centerline{\epsfig{file=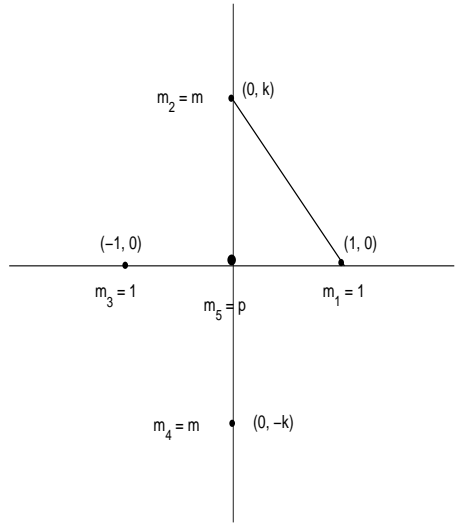,width=6cm,height=7cm}}
\caption{Set up for the 1+rhombus relative equilibria}
\label{Robertsge}
\end{figure}

The easiest and most accessible relative equilibria are those
configurations which contain large amounts of symmetry.
\citeasnoun{GERoberts} discusses relative equilibria for a special
case of the 5-body problem. He considers a configuration which
consists of four bodies at the vertices of a rhombus, with
opposite vertices having the same mass, and a central body, see
figure (\ref{Robertsge}). He shows that in this five body problem
for the masses $(m_1,m_2,m_3,m_4,m_5)=(1,1,1,1,-1/4)$, there exist
a one parameter family of degenerate relative equilibria where the
four equal masses are positioned at the vertices of a rhombus with
the remaining body located at the centre. As the parameter $k$
varies, one pair of opposite vertices move away from each other
while the other pair move closer, maintaining a fixed length
between consecutive vertices. He also shows that the number of
relative equilibria equivalence classes is not finite.

\citeasnoun{MiocBlaga} discusses the same problem introduced above
in the post Newtonian field of Manev\footnote{In the 1920's, Manev
proposed a gravitational model based on the $A/r+B/r^2$ potential,
with $A=\mu$, a gravitational parameter of the field generating
body, $B=3\mu^2/(2c^2)$, with $c=$ speed of light. Manev
considered that this model could be used as a substitute to
general Relativity \cite{ManevRelativity}}. They prove the
existence of monoparametric families of relative equilibria for
the masses $(m_0,1,m,1,m)$, where $m_0$ is the central mass, and
prove that the Manev square five body problem, where $k=1$ with
masses $(m_0,1,m,1,m)$ admits relative equilibria regardless of
the value of the mass of the central body. A continuum of such
equilibria (as in the Newtonian field) does not exist in the Manev
rhomboidal five-body problem.

\citeasnoun{AlanAlbouy2002} deals with the central configurations
of the 1+4 body problem i.e. they study the central configurations
without collisions that are the limit of the central
configurations of the five body problem in space, when the mass of
one of the bodies goes to infinity. He considers four equal masses
on a sphere whose centre is the 'big' mass. They find four
symmetric central configurations and prove that they all have at
least one plane of symmetry. He also conjectures that there are
exactly five classes of central configurations, as one can always
exchange some bodies and apply some isometry or change of scale.

\citeasnoun{Ollongren} studies a restricted five body problem
having three bodies of equal mass $m$ placed on the vertices of
the equilateral triangle; they revolve in the plane of the
triangle around their gravitational centre in circular orbits
under the influence of their mutual gravitational attraction; at
the centre a mass of $\beta m$ is present where $\beta \geq 0$. A
fifth body of negligible mass compared to $m$ moves in the plane
under the gravitational attraction of the other bodies. All bodies
are considered to be point masses. He discuss the existence and
location of the Lagrangian equilibrium points.

Ollongren shows that there are 9 Lagrangian equilibrium points.
There is a critical value of $\beta = 43.181$ for which the
outer lagrangian points on the negative x-axis become stable. For
$\beta$ less than this value, all Lagrangian points in the plane
are unstable and for $\beta$ larger than this value, the 3 outer
Lagrangian points are stable and the fifth body will carry out a
libration motion around them. They also conclude that the central
mass has a stabilizing effect on the motion of the fifth small
body provided the mass of the central body is large enough.

\citeasnoun{Kalvou} considers an $n$-body ring problem with $n-1$
primaries of equal mass ($m$) arranged in equal arcs on an ideal
ring and a central body of a different mass ($\beta m$) located at
the centre of mass of the system. He adds a further negligible
mass to the system and derives the zero velocity curves and zones
of stationary solutions for the negligible mass. He does this for
different values of $n$ which include the five body case and the
Copenhagen case. He shows that the stationary solutions are
arranged on the crossing points of concentric circles with radial
lines forming equal angles between them. Their number depends on
the specific values of the parameter $v = n-1$ and $\beta$

\section{Summary}

Because of the greater complexity of higher number of bodies, the
main focus of the literature has been on the analytical study of
the Equilibrium Configurations of the four and five body problems.
Specially for the five body problem nothing is published yet on
the analytical stability of the five body problem and the authors
work in chapter 6 and 7 may be the first of its kind.

A review of the papers which address the four body problems
relevant to our work is given in section 2.1. In particular the
papers of \citeasnoun{loks1985} and \citeasnoun{loks1987} are
amongst the most relevant. Publications on the equilibrium
solutions and the stability of the Caledonian Symmetric Four Body
Problem ( Steves and Roy (1998, 2000, 2001), Roy and Steves (1998)
and \citeasnoun{Andras1}) form the main source of knowledge
relevant to the research of this thesis. In section 2.2 a
comprehensive literature review is given for the five body
problem. The work of \citeasnoun{GERoberts} is of more interest to
us as he discuses a symmetric five body problem which is similar
to our problem. In future it will be interesting to search for the
relative equilibria of the CS5BP using the method of
\citeasnoun{GERoberts}. It is also possible to find equilibrium
solutions for the five body problem using the same method as Roy
and Steves (1998).

\chapter{Equilibrium Configurations of the Four Body Problem}

In this chapter we discuss Equilibrium Configurations of four-body
problems. \emph{An equilibrium configuration of four-bodies is a
geometric configuration of four bodies in which the gravitational
forces are balanced in such a way that the four bodies rotate
together about their centre of mass and thus the geometric
configuration is maintained for all time.}

In this chapter we present a detailed summary of the method of
derivation of the equilibrium configurations given by
\citeasnoun{RoyandSteves1998}. We give all the analytical
solutions given by them including the equal mass cases and the
cases with two pairs of equal masses. We also give two analytical
solutions of our own called the triangular solutions of the four
body problem. Section 3.1 contains a review of the Lagrangian
solutions of the restricted three body problem, in particular the
solutions to the Copenhagen problem. The Copenhagen solutions are
useful for the analysis of equilibrium solutions in the four body
problem as they are the solutions which occur when two of the
masses are reduced to zero. In section 3.2 we review three types
of analytical solutions of the equal mass four body problems which
include the Square, the Equilateral Triangle and the Collinear
equilibrium configurations \cite{RoyandSteves1998}. In section 3.3
we give solutions for two pairs of equal masses, where the ratio
between the two pairs is reduced from 1 to 0 in order to obtain
the five Lagrange points of the Copenhagen problem. We discuss
four kinds of equilibrium configurations both symmetric and
non-symmetric which include the Trapezoidal, the Diamond, the
Triangular and the Collinear equilibrium configurations. The
Trapezoidal, the Diamond and
 the Collinear equilibrium configurations are a review of
\citeasnoun{RoyandSteves1998}, while the Triangular equilibrium
configurations are original work derived as an extension of
\citeasnoun{RoyandSteves1998}. In section 3.4 we summarize our
results.
\section{The Lagrangian Solutions of the Restricted three body problem}

\begin{figure}
\centerline{\epsfig{file=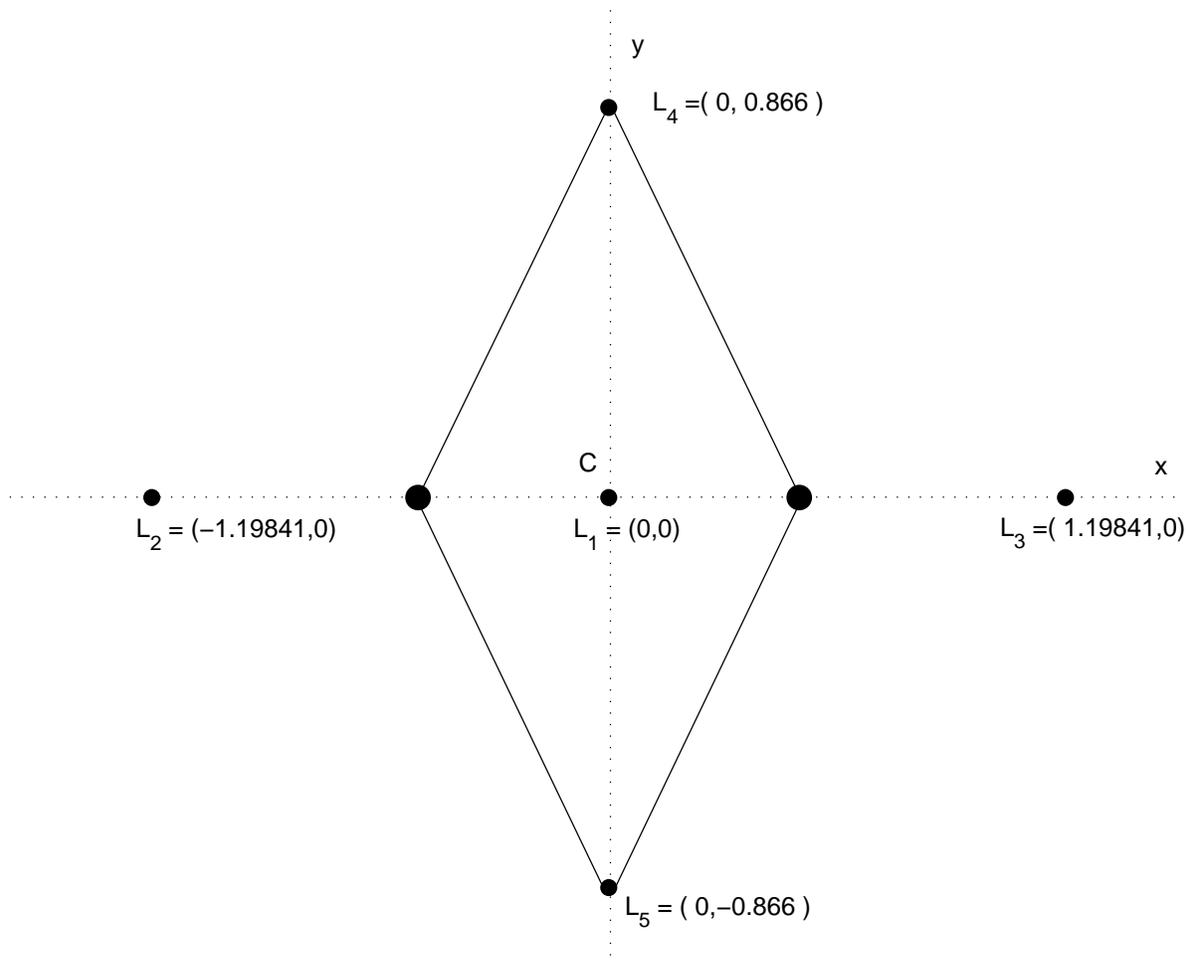,height=13cm,angle =
0}}

 \caption{The Lagrange equilibrium Solutions to the Copenhagen problem}
 \label{LagSolutions}
\end{figure}

It is well known that in the general three-body problem there
exists five special equilibrium configurations, where either the
three masses $m_1$, $m_2$, $m_3$ are collinear or they occupy the
vertices of an equilateral triangle.

The Restricted three body problem has five corresponding special
equilibrium solutions where the particle of infinitesimal mass
resides at one of five points, usually denoted by the letters
$L_1$, $L_2$, $L_3$, $L_4$ and $L_5$. The first three points are
collinear with the two primary masses while the last two points
form equilateral triangles with the two primary masses (Figure
\ref{LagSolutions}).

In the Copenhagen problem (Figure \ref{LagSolutions}), with
$m_1=m_2=m$, if a frame of coordinates rotating with constant
angular velocity $n$ is centred on the centre of mass C of the two
finite mass bodies and $P_1$ and $P_2$ are placed at (-0.5,0) and
(0.5,0) respectively, the lagrange solutions in the rotating frame
become $L_3=-L_2=(1.19841,0)$, $L_1=(0,0)$, $L_4=-L_5=(0,0.866)$
(See \citeasnoun{RoyBook}).

\section{The Equal mass cases- A Review}

In this section we discuss equilibrium configurations for the
four-body problem where the four bodies are of equal mass. These
were discussed by \citeasnoun{RoyandSteves1998}. We consider the
following cases: A) Four equal masses arranged in a square; B)
Four equal masses arranged in an equilateral triangle; C)Four
equal masses arranged in a line. In all the three cases, the
origin is taken to be the centre of mass of the system and is
assumed to be stationary.

\subsection{The Equilibrium Configuration of the four body problem with four equal
masses making a square}

\begin{figure}[tbp]
\begin{center}
\epsfig{file=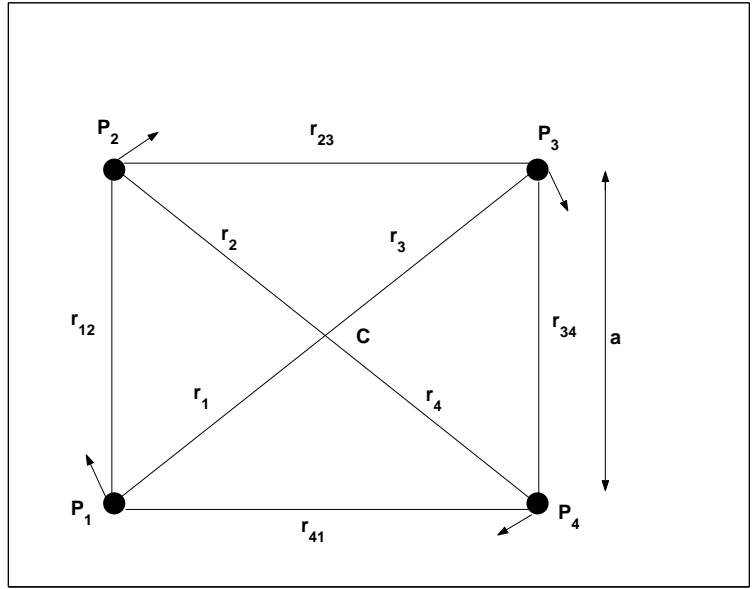, width=10cm}
\end{center}
\caption{Square Equilibrium Configuration of the Four Body Problem
} \label{figsquare}
\end{figure}

First we find the equations of motion for four particles of mass
$m_{i}$ $ (i=1,2,\cdots 4)$ whose radius vectors from an
unaccelerated point C are $ r_{i}$ , while the distances between
the particles are given by $r_{ij}$ (see any standard text for
more detailed discussion, for example \citeasnoun{RoyBook}) where

\begin{equation}
{\bf r}_{ij}{\bf =r}_{j}{\bf -r}_{i}
\end{equation}
Newton laws of motion are then written as
\begin{equation}
m_{i}\stackrel{..}{{\bf r}_{i}}=G\sum\limits_{j=1, j\neq
i}^{4}\frac{m_{i}m_{j}}{ r_{ij}^{3}}{\bf r}_{ij}\qquad  i=1,2,3,4.
\label{2.2}
\end{equation}
We choose our units so that $G$ becomes unity. Also let
$\frac{{\bf r}_{ij} }{r_{ij}^{3}}={ \mathbf{\rho}}_{ij}$, then
(\ref{2.2}) becomes
\begin{equation}
m_{i}\stackrel{..}{{\bf r}_{i}}=G\sum\limits_{j=1,j\neq
i}^{4}m_{i}m_{j}{\bf \rho } _{ij}\qquad i=1,2,3,4.  \label{2.3}
\end{equation}
Equation (\ref{2.3}) is the form of the equations of motion used
for all the derivations of equilibrium configurations found in
chapter 3. Simplifying equation (\ref{2.3}) and putting $m_{i}=M$,
we get the following four equations of motion for the equal mass
four body problem.

\begin{equation}
\left.
\begin{array}{c}
\stackrel{..}{{\bf r}_{1}}=M({\bf \rho }_{12}+{\bf \rho
}_{13}+{\bf \rho }
_{14}), \\
\stackrel{..}{{\bf r}_{2}}=M({\bf \rho }_{21}+{\bf \rho
}_{23}+{\bf \rho }
_{24}), \\
\stackrel{..}{{\bf r}_{3}}=M({\bf \rho }_{31}+{\bf \rho
}_{32}+{\bf \rho }
_{34}), \\
\stackrel{..}{{\bf r}_{4}}=M({\bf \rho }_{41}+{\bf \rho }_{42}+{\bf \rho }
_{43}).
\end{array}
\right\}  \label{2.4}
\end{equation}
Now specific to deriving the square equilibrium configuration, let
the four particles of mass $M$ lie at the vertices of a square of
side of length $a$, as shown in Figure (\ref{figsquare}) It is
clear from the geometry that
\begin{equation}
\left.
\begin{array}{c}
{\bf r}_{4}=-{\bf r}_{2}, \\
{\bf r}_{1}=-{\bf r}_{3}, \\
{\bf r}_{34}=-{\bf r}_{12}, \\
{\bf r}_{23}=-{\bf r}_{14.}
\end{array}
\right\}  \label{2.5}
\end{equation}
Also
\begin{equation}
r_{12}=r_{23}=r_{34}=r_{41}=a.  \label{2.6}
\end{equation}
From Figure (\ref{figsquare}) we can easily find that

\begin{equation}
r_{13}=r_{24}=b=\sqrt{2}a.  \label{2.7}
\end{equation}

Using equations (\ref{2.4}), (\ref{2.5}), (\ref{2.6}) and
(\ref{2.7}), we get the following equations of motion for the
four-body problem under discussion.
\begin{equation}
\left.
\begin{array}{c}
\stackrel{..}{{\bf
r}_{1}}=-\frac{M}{a^{3}}(2+\frac{1}{\sqrt{2}}){\bf r}
_{1}, \\
\stackrel{..}{{\bf
r}_{2}}=-\frac{M}{a^{3}}(2+\frac{1}{\sqrt{2}}){\bf r}
_{2}, \\
\stackrel{..}{{\bf
r}_{3}}=-\frac{M}{a^{3}}(2+\frac{1}{\sqrt{2}}){\bf r}
_{3}, \\
\stackrel{..}{{\bf
r}_{4}}=-\frac{M}{a^{3}}(2+\frac{1}{\sqrt{2}}){\bf r} _{4}.
\end{array}
\right\}
\end{equation}
Since the co-efficient of all the differential equations are the
same and negative a simple harmonic motion solution with constant
angular velocity for all masses is possible. We can write all the
equations of motion in a compact form as follows,
\begin{equation}
\ddot\mathbf{r}_{i}=-n^{2}\mathbf{r}_{i},\qquad i=1,2,3,4.
\label{2.9}
\end{equation}
where
\begin{equation}
n^{2}=\frac{M}{a^{3}}(2+\frac{1}{\sqrt{2}}).
\end{equation}

Equation (\ref{2.9}) is a second order linear differential
equation with the following solution

\begin{equation}
\mathbf{r}_{i}=\mathbf{r}_{i0}\cos
nt+\frac{{\dot\mathbf{r}_{i0}}}{n}\sin nt,
\end{equation}

where $\mathbf{r}_{i0}$ and $\dot\mathbf{r}_{i0}$ are the radius
and velocity vectors respectively at $t=0$ and $n$ is the angular
velocity at which the square configuration rotates about its
centre of mass.

\subsection{Equilibrium Configuration of the four body problem with four equal
masses making an equilateral triangle}

\begin{figure}
\centerline{ \epsfig{file=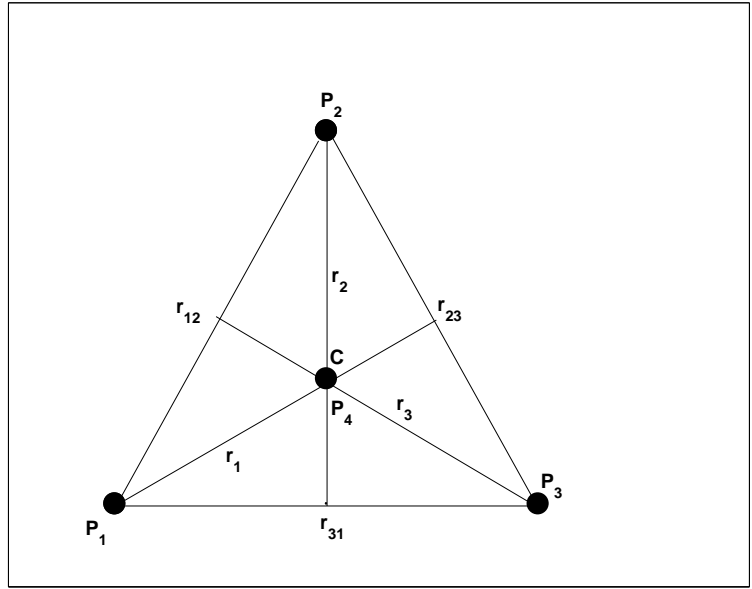,width=10cm}}
\caption{Equilibrium Configuration of Four Body Problem making an
equilateral triangle} \label{equilateraltriangle}
\end{figure}

We now consider the equilateral triangle equilibrium configuration
of the four-body equal mass problem \cite{RoyandSteves1998}. Let
three particles of mass $M$ lie at the vertices of an equilateral
triangle, with the fourth particle of the same mass at the
centroid of the triangle. (See Figure (\ref{equilateraltriangle}))
As this is an equilateral triangle therefore
\begin{equation}
r_{12}=r_{23}=r_{31}=a.  \label{2.10}
\end{equation}
Also, as the fourth mass lies at the centroid of the triangle
which is also the centre of mass of the system,

\begin{equation}
r_{1}=r_{2}=r_{3}=\frac{a}{\sqrt{3}}\textrm{ and }r_{4}=0.
\label{2.13}
\end{equation}

Using equations (\ref{2.4}) , (\ref{2.10}) and (\ref{2.13}), we
get the following second order differential equations as our
equations of motion.

\begin{equation}
\left.
\begin{array}{l}
\stackrel{..}{{\bf r}_{1}}=-\frac{M}{a^{3}}\left( 3+3^{3/2}\right)
{\bf r}
_{1} \\
\stackrel{..}{{\bf r}_{2}}=-\frac{M}{a^{3}}\left( 3+3^{3/2}\right)
{\bf r}
_{2} \\
\stackrel{..}{{\bf r}_{3}}=-\frac{M}{a^{3}}\left( 3+3^{3/2}\right)
{\bf r} _{3}\\
\stackrel{..}{{\bf r}_{4}}=0
\end{array}
\right\}  \label{2.14}
\end{equation}
As we have the same co-efficients for all differential equations
which are also negative, again simple harmonic motion solution is
possible. The following is the compact form of equations
(\ref{2.14}).

\begin{equation}
\stackrel{..}{{\bf r}_{i}}=-n^{2}{\bf r}_{i}\textrm{ \ \ \
}(i=1,2,3) \label{x2.15}
\end{equation}
where
\begin{equation}
n^{2}=\frac{M}{a^{3}}\left( 3+3^{3/2}\right) .
\end{equation}
Equation (\ref{x2.15}) is a second order linear differential
equation and has the following solution

\begin{equation}
\mathbf{r}_{i}=\mathbf{r}_{i0}\cos
nt+\frac{{\dot\mathbf{r}_{i0}}}{n}\sin nt,
\end{equation}
where $\mathbf{r}_{i0}$ and $\dot\mathbf{r}_{i0}$ are the radius
and velocity vectors respectively at $t=0$ and $n$ is the angular
velocity at which the equilateral triangle configuration rotates
about its centre of mass.

\subsection{Equilibrium Configuration of the four body problem with four equal
masses lying along a straight line}

We now consider the collinear equilibrium configuration of the
four-body equal mass problem. In this case, the solution must be
symmetrical about the centre of mass $ C $. Then
\begin{equation}
{\bf r}_{4}=-{\bf r}_{1}\textrm{, }{\bf r}_{3}=-{\bf
r}_{2}\textrm{ and let } {\bf r}_{2}=\alpha {\bf r}_{1}.
\label{x2.18}
\end{equation}
Using equations (\ref{2.4}) and (\ref{x2.18}) we get the following
equations of motion

\begin{equation}
\left.
\begin{array}{c}
\stackrel{..}{{\bf r}_{1}}=-\frac{M}{r_{1}^{3}}\left(
\frac{1}{4}+\frac{ 2(1+\alpha ^{2})}{(1-\alpha ^{2})^{2}}\right)
{\bf r}_{1},\qquad \\
=-\frac{M}{r_{1}^{3}}R_{1}{\bf r}_{1},\qquad \\
\stackrel{..}{{\bf r}_{2}}=-\frac{M}{r_{1}^{3}}\left(
\frac{1}{\alpha } \left( \frac{1}{4\alpha ^{2}}-\frac{4\alpha
}{\left( 1-\alpha ^{2}\right)
^{2}}\right) \right) {\bf r}_{2},\qquad \\
=-\frac{M}{r_{1}^{3}}R_{2}{\bf r}_{2}, \\
\stackrel{..}{{\bf r}_{3}}=-\frac{M}{r_{1}^{3}}\left(
\frac{1}{\alpha } \left( \frac{1}{4\alpha ^{2}}-\frac{4\alpha
}{\left( 1-\alpha ^{2}\right)
^{2}}\right) \right) {\bf r}_{3},\textrm{ \ } \\
=-\frac{M}{r_{1}^{3}}R_{3}{\bf r}_{3}, \\
\stackrel{..}{{\bf r}_{4}}=-\frac{M}{r_{1}^{3}}\left(
\frac{1}{4}+\frac{
2(1+\alpha ^{2})}{(1-\alpha ^{2})^{2}}\right) {\bf r}_{4},\qquad \\
=-\frac{M}{r_{1}^{3}}R_{4}{\bf r}_{4},
\end{array}
\right\}   
\end{equation}
where $R_{1}=R_{4}$ and $R_{2}=R_{3}$.

For a rigid rotating geometry we must have $R_{1}=R_{2}=R_{3}=R_{4}.$ Now
equating $R_{1}$ to $R_{2}$ we get

\begin{equation}
\frac{1}{4}\left( 1-\frac{1}{\alpha ^{3}}\right) +\frac{6+2\alpha ^{2}}{%
\left( 1-\alpha ^{2}\right) ^{2}}=0.
\end{equation}
After further simplification we get

\begin{equation}
\left( \alpha ^{7}+6\alpha ^{5}-\alpha ^{4}+25\alpha ^{3}+2\alpha
^{2}-1\right) /\left( 4\alpha ^{2}\left( 1-\alpha ^{2}\right) ^{2}\right) =0,
\end{equation}
Therefore
\begin{equation}
\left( \alpha ^{7}+6\alpha ^{5}-\alpha ^{4}+25\alpha ^{3}+2\alpha
^{2}-1\right) =0.
\end{equation}
$\alpha =0.3162$ is the only real solution for the above
equation.This value of $\alpha $ gives $R_{i}=2.966$ $
(i=1,2,3,4)$. Hence
\begin{equation}
{\bf r}_{i}={\bf r}_{i0}\cos nt+\frac{\stackrel{.}{{\bf r}_{i0}}}{n}\sin nt
\end{equation}
is the required solution, where the angular velocity $n$ is given
by:

\begin{equation}
n^{2}=\frac{M}{r_{1}^{3}}R_{i}.
\end{equation}

\begin{figure}[tbp]
\begin{center}
\mbox{\epsfig{file=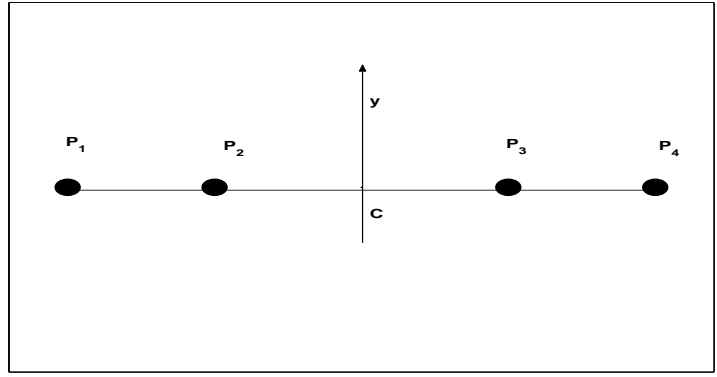,height=5cm, width=10cm}}
\end{center}
\caption{Collinear Equilibrium Configuration of the Four Body
Problem} \label{collinearequalmass}
\end{figure}

\section{Solution for two pairs of equal masses}

In this section we discuss equilibrium configurations of two pairs
of equal masses for the four-body problem. The first two
subsections will present a review of \citeasnoun{RoyandSteves1998}
work i.e. the Trapezoidal equilibrium configuration and Diamond
equilibrium configurations, while the last two subsections
introducing the triangular equilibrium configurations are new
analysis following the procedure of Roy and Steves (2001).

\subsection{The Trapezoidal equilibrium configuration for the four-body problem- A Review}

\begin{figure}
\centerline{ \epsfig{file=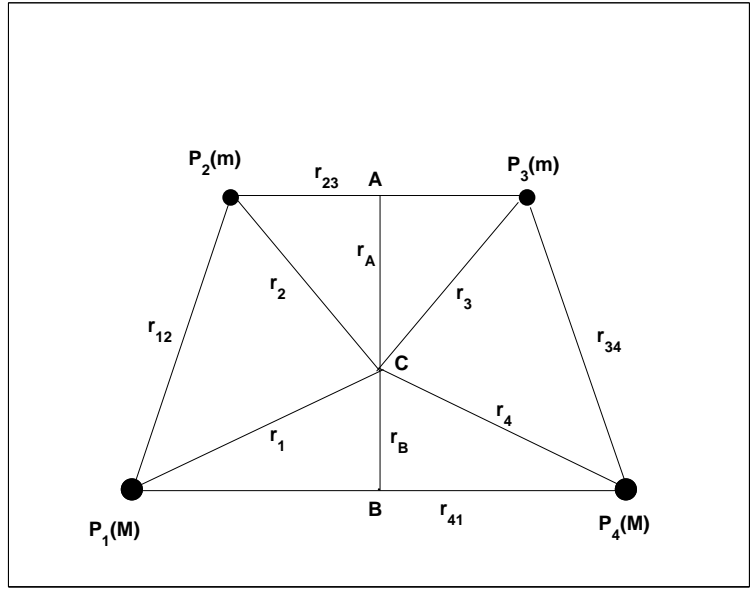, width=10cm }}
\caption{Trapezoidal Equilibrium Configuration of Four Body
Problem} \label{trapezoidal}
\end{figure}

We now consider the Trapezoidal equilibrium configuration of the
four-body problem \cite{RoyandSteves1998}. It has two pairs of
equal masses i.e. $M$ and $m.$ In Figure (\ref{trapezoidal}), the
geometry is taken to be symmetrical about the line $\bar{AB}$
joining the centre of mass $A$ of the pair $P_{2}$ and $P_{3}$,
each of mass $m$, and the centre of mass $B$ of the pair $P_{1}$
and $P_{4}$, each of mass $M.$

Let $\mu =m/M<1.$ Let C be the centre of mass of the system then
$\mu \mathbf{r}_{A}+\mathbf{r}_{B}=0.$ Let
\begin{equation}
{\bf r}_{23}=-\alpha {\bf r}_{14};\left| {\bf r}_{A}-{\bf
r}_{B}\right| =\beta r_{14}\textrm{ and }{\bf r=r}_{A}-{\bf
r}_{B}. \label{y2.14}
\end{equation}
We can easily see from the Figure (\ref{trapezoidal}) that
\begin{equation}
\left.
\begin{array}{l}
r_{12}=r_{34}=\left( \beta ^{2}+\frac{1}{4}\left( 1-\alpha
\right)^{2}
\right) ^{1/2}r_{14}, \\
r_{24}=r_{13}=\left( \beta ^{2}+\frac{1}{4}\left( 1+\alpha\right)
^{2} \right) ^{1/2}r_{14}, \\
r_{A}=\left( \frac{\beta }{1+\mu }\right) r_{14}, \\
r_{B}=\left( \frac{\mu \beta }{1+\mu }\right) r_{14}.

\end{array}
\right\}  \label{2.14}
\end{equation}

Using the relation given in equations (\ref{y2.14}) in conjunction
with the geometry (\ref{2.14}), we get the following vectors which
describe the locations of the four bodies,
\begin{equation}
\left.
\begin{array}{c}
\mathbf{r}_{1}=-\frac{\mu }{1+\mu }{\bf r+}\frac{1}{2}{\bf r}_{41}, \\
\mathbf{r}_{2}=\frac{1}{1+\mu }{\bf r+}\frac{\alpha }{2}{\bf
r}_{41},\textrm{ \ }
\\
\mathbf{r}_{3}=\frac{1}{1+\mu }{\bf r-}\frac{\alpha }{2}{\bf
r}_{41},\textrm{ \ }
\\
\mathbf{r}_{4}=-\frac{\mu }{1+\mu }{\bf r-}\frac{1}{2}{\bf
r}_{41}.
\end{array}
\right \}  \label{2.15}
\end{equation}
The magnitudes of the vectors given in equations (\ref{2.15}) are
the following
\begin{equation}
\left.
\begin{array}{l}
r_{1}=\left( \frac{1}{4}+\left( \frac{\mu }{1+\mu }\right) ^{2}\beta
^{2}\right) ^{1/2}r_{14}, \\
r_{2}=\left( \frac{\alpha ^{2}}{4}+\left( \frac{\beta }{1+\mu }\right)
^{2}\right) ^{1/2}r_{14}, \\
r_{3}=r_{2},\\%
r_{4}=r_{1}\\%
\end{array}
\right\}  \label{2.16}
\end{equation}
The equations of motion of the system are the following
\begin{equation}
\left.
\begin{array}{c}
\stackrel{..}{{\bf r}_{1}}=M\left[ \mu \left( {\bf \rho
}_{12}+{\bf \rho }
_{13}\right) {\bf +\rho }_{14}\right] , \\
\stackrel{..}{{\bf r}_{2}}=M\left[ {\bf \rho }_{21}+\mu {\bf \rho
}_{23}{\bf
+\rho }_{24}\right] ,\textrm{ \ \ } \\
\stackrel{..}{{\bf r}_{3}}=M\left[ {\bf \rho }_{31}+\mu {\bf \rho
}_{32}{\bf
+\rho }_{34}\right] ,\textrm{ \ \ } \\
\stackrel{..}{{\bf r}_{4}}=M\left[ {\bf \rho }_{41}+\mu \left( {\bf \rho }%
_{42}{\bf +\rho }_{43}\right) \right] .
\end{array}
\right\}  \label{2.17}
\end{equation}

We use equations (\ref{2.14}),  (\ref{2.15}) and  (\ref{2.17}) to
obtain differential equations for ${\bf r}$ and ${\bf r}_{14}$.
\begin{equation}
\stackrel{..}{{\bf r}}_{14}=-\frac{M}{r_{14}^{3}}\left( 2+\mu
\left( \frac{ 1-\alpha }{a^{3}} +\frac{\alpha
+1}{b^{3}}\right)\right) {\bf r}_{14}, \label{2.18}
\end{equation}
where
\begin{equation}
a=(\beta ^{2}+\frac{1}{4}\left( 1-\alpha \right) ^{2})^{1/2},  \label{18a}
\end{equation}
and
\begin{equation}
b=(\beta ^{2}+\frac{1}{4}\left( 1+\alpha \right) ^{2})^{1/2}.  \label{18b}
\end{equation}
Now for ${\bf r}$ we proceed as follows,
\begin{equation}
{\bf r}_{4}+{\bf r}_{1}=-\frac{2\mu }{1+\mu }{\bf r.}
\end{equation}
Thus
\begin{equation}
\stackrel{..}{{\bf r}}{\bf =-}\frac{M}{r_{14}^{3}}\left( \left(
1+\mu \right) \left( \frac{1}{a^{3}}+\frac{1}{b^{3}}\right)
\right) {\bf r.} \label{2.19}
\end{equation}
If the coefficients of ${\bf r}$ and ${\bf r}_{41}$ are equal and
negative in equations (\ref{2.18}) and (\ref{2.19}), then the
solutions to these equations give a rigid and rotating geometry.
Therefore equating the coefficients of ${\bf r}$ and ${\bf
r}_{41}$ in equations (\ref{2.18}) and (\ref{2.19}) we get the
following
\begin{equation}
\left( 1+\mu \right) \left( \frac{1}{a^{3}}+\frac{1}{b^{3}}\right) -2-\mu
\left( \frac{1-\alpha }{a^{3}}+\frac{1+\alpha }{b^{3}}\right) =0,
\end{equation}
or
\begin{equation}
\frac{1+\mu \alpha }{a^{3}}+\frac{1-\mu \alpha }{b^{3}} -2=0,
\label{2.20}
\end{equation}
where $a$ and $b$ are functions defined by equations (\ref{18a})
and (\ref{18b}).

To obtain a unique solution for a given value of the mass ratio $\mu $, we
require a second relation in $\alpha $ and $\beta $, which is obtained by
using
\begin{equation}
{\bf r}_{23}=-\alpha {\bf r}_{14},
\end{equation}
or
\begin{equation}
{\bf r}_{3}-{\bf r}_{2}+\alpha ({\bf r}_{1}-{\bf r}_{4})=0.
\end{equation}
Differentiating twice and using the equations of motion, it is found that
\begin{equation}
\left( \frac{1}{b^{3}}\left( 1-\alpha \mu \right) \left( \alpha
+1\right) + \frac{1}{a^{3}}\left( 1+\alpha \mu \right) \left(
\alpha -1\right) -2\alpha \left( 1-\frac{\mu }{\alpha ^{3}}\right)
\right) r_{41}=0,
\end{equation}
as $r_{41}\neq 0$, hence we can write
\begin{equation}
\left( \frac{1}{b^{3}}\left( 1-\alpha \mu \right) \left( \alpha
+1\right) + \frac{1}{a^{3}}\left( 1+\alpha \mu \right) \left(
\alpha -1\right) -2\alpha \left( 1-\frac{\mu }{\alpha ^{3}}\right)
\right) =0.  \label{2.21}
\end{equation}
Solving equations (\ref{2.20}) and (\ref{2.21}) simultaneously for
a given value of $ \mu $ ($0\leq \mu \leq 1)$ we can find $\alpha
$ and $\beta .$

The following two special solutions give the boundary of the family of
solutions.

\begin{enumerate}
\item  $\mu =1$ (Case of four equal masses)

Solving equations (\ref{2.20}) and (\ref{2.21}) simultaneously for
$\mu =1$ gives $ \alpha =1$ and $\beta =1.$ This is the square
solution for the four-body problem.

\item  $\mu =0.$ (Case of the Lagrange solution of the Copenhagen
Problem)

Solving equations (\ref{2.20}) and (\ref{2.21}) simultaneously for
$\mu =0$ (i.e. the two smaller masses become equal to zero) gives
$\alpha =0$ and $\beta = \sqrt{3/4}$ and hence the problem reduces
to the famous Copenhagen Problem and the Lagrange point $L_4$ is
reached.
\end{enumerate}

\begin{figure}
\centerline{ \epsfig{file=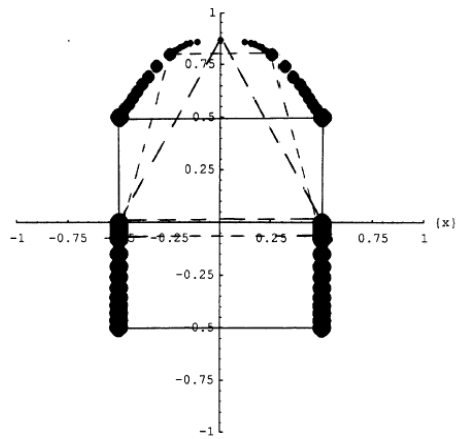 ,height=10cm}}
\caption{The evolution of all the four masses when $\mu $ is
varied from 1 to 0, for the Trapezoidal equilibrium solution. (The
Origin is located at the centre of mass of the system).}
\label{trapgraprep}
\end{figure}

Figure (\ref{trapgraprep}) gives the locations of the four bodies
for a range of $\mu$ from 1 to 0. It shows there is a continuous
family of solutions for  $\mu \in \,(0,1)$. The bigger points in
the figure indicate the locations of the two large masses, which
remain at size $M$ and the location of the other two bodies when
their masses are also $M\,(\mu =1).$ The smaller points indicate
the location of the other two bodies when $\mu $ reduces to zero.
As the origin of the system is at the centre of mass of the
system, the two primaries $M$ appear to move up in the graph to
their positions at $(0.5,0),(-0.5,0)$ as $\mu$ approaches 0. The
arrow shows the change in the equilibrium solution as $\mu$ is
varied from 1 to 0, i.e. from the square solution to the
triangular.

\subsection{The Diamond Equilibrium configuration of the four-body problem- A Review}

\begin{figure}
\centerline{ \epsfig{file=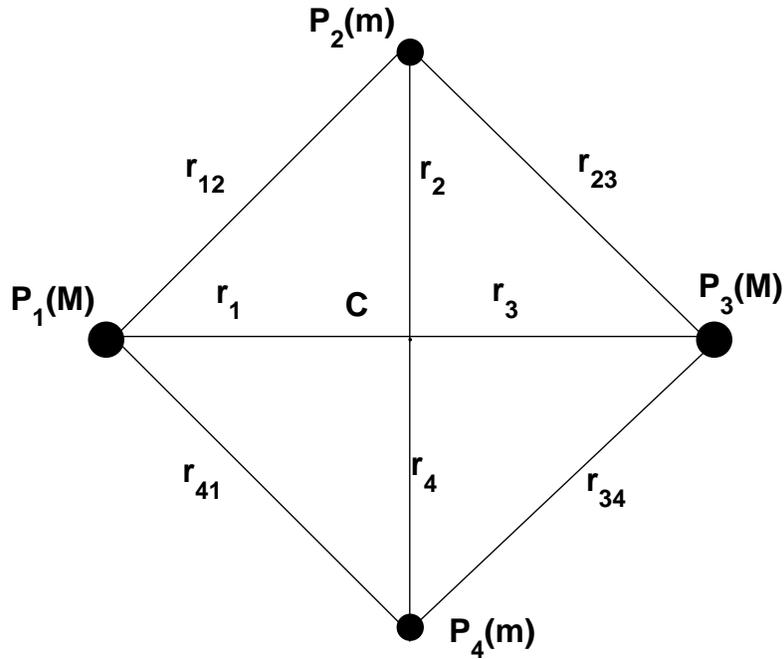 ,height=10cm}} \caption{The
Diamond Equilibrium Configuration of the Four Body Problem}
\label{diamond}
\end{figure}

\begin{figure}
\begin{center}
\mbox{\epsfig{file=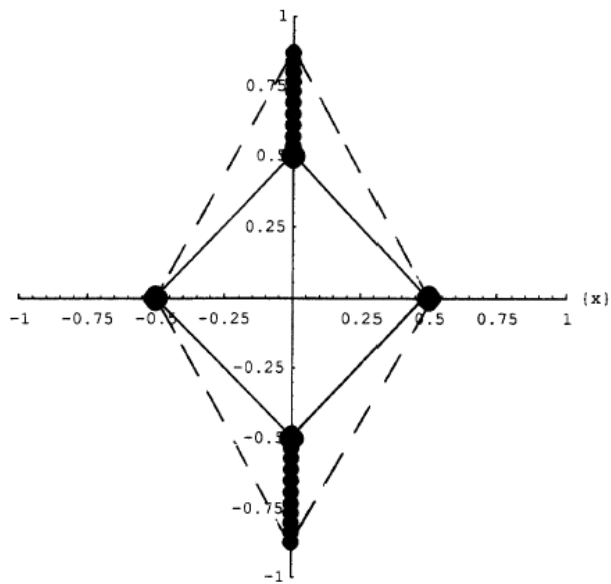,width=10cm}}
\end{center}
\caption{The evolution of all the four masses when $\protect\mu $
is varied from 1 to 0 for the diamond equilibrium solution. (The
origin is located at the centre of mass, which is also always the
halfway point between two primaries.)} \label{diamondgraphrep}
\end{figure}

We now consider the diamond equilibrium configuration of the
four-body problem \cite{RoyandSteves1998}. It has two pairs of
equal masses: a pair of larger masses $M$ and another pair of
smaller masses $m$. It is clear from Figure (\ref{diamond}) that
the geometry is symmetrical about the line $P_{2}P_{4}$ and
$P_{1}P_{3}$. Because of the high degree of symmetry only one
parameter needs to be introduced, namely $\alpha $, where
$r_{2}=\alpha r_{1}.$ Thus

\begin{equation}
{\bf r}_{3}=-{\bf r}_{1}\textrm{ and }{\bf r}_{4}=-{\bf r}_{2}.
\end{equation}
Therefore only the equations of motion for  ${\bf r}_{1}$ and
${\bf r}_{2}$ need to be studied.
\begin{equation}
\stackrel{..}{{\bf r}_{1}}=M\left[ \mu \left( {\bf \rho
}_{12}+{\bf \rho } _{14}\right) +{\bf \rho }_{13}\right],
\end{equation}
\begin{equation}
\stackrel{..}{{\bf r}_{2}}=M\left[ {\bf \rho }_{21}+{\bf \rho
}_{23}+\mu {\bf \rho }_{24}\right].
\end{equation}
After some reduction, by utilizing the symmetry conditions, we obtain
\begin{equation}
\stackrel{..}{{\bf r}_{1}}=-\frac{M}{r_{1}^{3}}\left[ \frac{1}{4}+\frac{2\mu
}{\left( 1+\alpha ^{2}\right) ^{3/2}}\right] {\bf r}_{1},
\end{equation}
\begin{equation}
\stackrel{..}{{\bf r}_{2}}=-\frac{M}{r_{1}^{3}}\left[ \frac{\mu }{4\alpha
^{3}}+\frac{2}{\left( 1+\alpha ^{2}\right) ^{3/2}}\right] {\bf r}_{2}.
\end{equation}
Now proceeding along the same lines as for the Trapezoidal
solution and searching for solutions as the masses $m$ are reduced
to zero, we find at $\mu =1$, the square solution, and at $\mu
=0$, which is the equilateral triangle solution of the Copenhagen
problem. See Figure (\ref{diamondgraphrep}). Between $\mu=1$ and
$\mu=0$ there lies a continuous family of solutions.

\subsection{Triangular Equilibrium Configuration of four-body problem: Case I
}

\begin{figure}\label{}
\centerline{ \epsfig{file=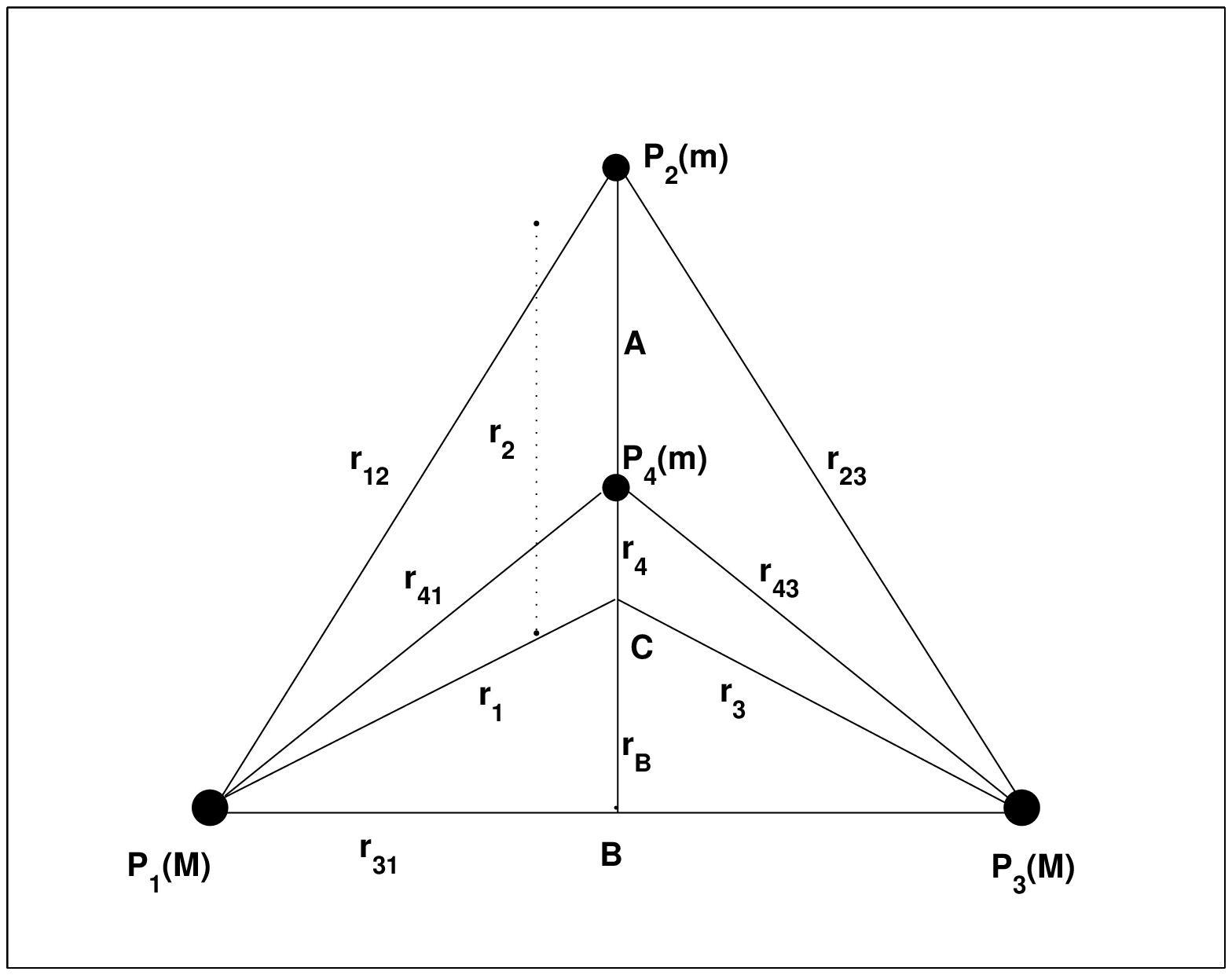, width=10cm}}
\caption{Triangular Equilibrium Configuration of the Four Body
Problem. Case-I}\label{triangle1}
\end{figure}

Consider the Triangular Equilibrium Configuration given in Figure
(\ref{triangle1}), where two large masses $M$ lie at the two
vertices,$P_1,P_3$, of a triangle and two smaller masses $m$ lie
at the vertex $P_2$ and at $P_4$ on the line of symmetry of the
triangle. Let $A$ and $B$ be, respectively, the centre of mass of
the pair $P_{2}$ and $P_{4}$ and the pair $P_{1}$ and $P_{3}.$ Let
$\bar{CA}=r_{A}$ and $\bar{CB}=r_{B}.$ From the centre of mass
relation
\begin{equation}
r_{B}+\mu r_{A}=0,
\end{equation}
where
\begin{equation}
\mu =\frac{m}{M}\leq 1.
\end{equation}

Let
\begin{equation}
r_{12}=\alpha r_{13}\textrm{ and }r_{14}=\beta r_{13}.
\end{equation}
We also have from the symmetry conditions that
\begin{equation}
r_{12}=r_{23};\textrm{ }r_{41}=r_{43};\textrm{ and }r_{1}=r_{3}.
\end{equation}
Now its easy to show that
\begin{equation}
\left.
\begin{array}{l}
{\bf r}_{1}=-\mu {\bf r}_{A}-\frac{1}{2}{\bf r}_{13}, \\
{\bf r}_{2}=C_{2}{\bf r}_{A},\\
{\bf r}_{3}=-\mu {\bf r}_{A}+\frac{1}{2}{\bf r}_{13}, \\
{\bf r}_{4}=C_{4}{\bf r}_{A},
\end{array}
\right\}  \label{22}
\end{equation}
where
\begin{equation}
C_{2}=\left( \frac{-1+4\alpha ^{2}-\mu +2\mu \alpha ^{2}+2\mu \beta
^{2}-\left( 1+\mu \right) \sqrt{\left( -1+4\alpha ^{2}\right) \left(
-1+4\beta ^{2}\right) }}{2\left( \alpha ^{2}-\beta ^{2}\right) }\right) ,
\label{22a}
\end{equation}
and
\begin{equation}
C_{4}=\left( \frac{1-4\beta ^{2}+\mu -2\mu \alpha ^{2}-2\mu \beta
^{2}+\left( 1+\mu \right) \sqrt{\left( -1+4\alpha ^{2}\right) \left(
-1+4\beta ^{2}\right) }}{2\left( \alpha ^{2}-\beta ^{2}\right) }\right) .
\label{22b}
\end{equation}
Now using the symmetry conditions we get the magnitudes of all the vectors
involved, which are listed below
\begin{equation}
\left.
\begin{array}{l}
r_{1}=\left( \mu ^{2}+\frac{\left( c_{4}+\mu \right) ^{2}}{4\beta
^{2}-1}
\right) ^{1/2}r_{A}, \\
r_{2}=C_{2}r_{A}, \\
r_3=r_1\\
r_{4}=C_{4}r_{A,} \\
r_{12}=\left( \frac{\left( C_{4}+\mu \right) ^{2}}{4\beta ^{2}-1}+\left(
C_{2}+\mu \right) ^{2}\right) ^{1/2}r_{A}, \\
r_{14}=\frac{C_{4}+\mu }{\left( \beta ^{2}-1/4\right) ^{1/2}}\beta r_{A},%
\textrm{ \ \ \ \ \ \ \ \ \ \ \ \ } \\
r_{13}=\frac{C_{4}+\mu }{\left( \beta ^{2}-1/4\right)
^{1/2}}r_{A},
 \\
r_{24}=\left( C_{2}-C_{4}\right) r_{A}.
\end{array}
\right\}
\end{equation}

\begin{figure}\label{}
\centerline{ \epsfig{file=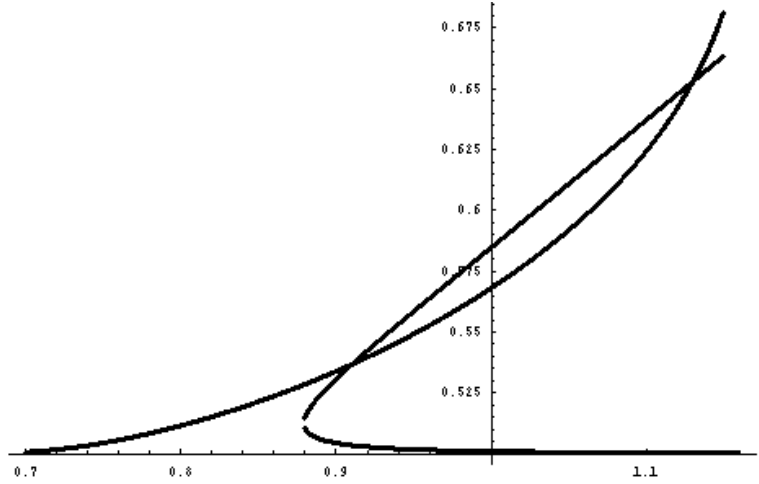, width=10cm}}
\caption{Graphs of $F_1(\alpha,\beta)=0$ and $F_2(\alpha,\beta)=0$
for $\mu=0.9$, showing two intersections.}\label{tworoots}
\end{figure}
Equations (\ref{22}) give
\begin{equation}
{\bf r}_{A}{\bf =-}\frac{1}{2\mu }{\bf (r}_{1}{\bf +r}_{3}{\bf ).}
\label{23}
\end{equation}
Differentiating equation (\ref{23}) twice and using equations
(\ref{2.17}) we get the
following equation of motion for ${\bf r}_{A}$%
\begin{eqnarray}
\stackrel{..}{{\bf r}}_{A} &=&-\frac{M}{r_{13}^{3}}\left[
\frac{C_{2}+\mu }{
\alpha ^{3}}+\frac{C_{4}+\mu }{\beta ^{3}}\right] {\bf r}_{A},  \label{24} \\
&=&C_{A}{\bf r}_{A}.  \nonumber
\end{eqnarray}
Now further using equations (\ref{2.17}) we get the following
equation of motion for ${\bf r}_{13}$
\begin{eqnarray}
\stackrel{..}{{\bf r}}_{13} &=&-\frac{M}{r_{13}^{3}}\left[ 2+\mu \left(
\frac{1}{\alpha ^{3}}+\frac{1}{\beta ^{3}}\right) \right] {\bf r}_{13},
\label{25} \\
&=&C_{B}{\bf r}_{13}. \nonumber
\end{eqnarray}
Other equations of motion derived in a similar manner, which will
be needed for further analysis, are the following
\begin{equation}
\left.
\begin{array}{c}
\stackrel{..}{{\bf r}}_{2}=-\frac{M}{C_{2}r_{13}^{3}}\left[ \mu \frac{\left(
c_{4}+\mu \right) ^{3}}{\left( C_{2}-C_{4}\right) ^{2}\left( \beta
^{2}-1/4\right) 3/2}+2\frac{C_{2}+\mu }{\alpha ^{3}}\right] {\bf r}_{2}, \\
\stackrel{..}{{\bf r}_{4}}=-\frac{M\left( C_{4}+\mu \right) }{r_{13}^{3}C_{4}%
}\left[ \mu \frac{\left( C_{4}+\mu \right) ^{2}}{\left( C_{2}-C_{4}\right)
^{2}\left( \beta ^{2}-1/4\right) 3/2}-\frac{2}{\beta ^{3}}\right] {\bf r}%
_{4}.
\end{array}
\right\}  \label{26}
\end{equation}

For a rigid, rotating system solution we require
\begin{equation}
C_{A}-C_{B}=0,
\end{equation}
which gives
\begin{equation}
F_1(\alpha,\beta)=\frac{C_{2}}{\alpha ^{3}}+\frac{C_{4}}{\beta
^{3}}-2=0. \label{27}
\end{equation}
For a second relation we proceed as follows. We know from the
centre of mass relation that
\begin{equation}
{\bf r}_{A}=\frac{1}{2}({\bf r}_{2}{\bf +r}_{4}),  \label{28}
\end{equation}
and also
\begin{equation}
{\bf r}_{A}=\frac{{\bf r}_{2}-{\bf r}_{4}}{C_{2}-C_{4}}.
\label{29}
\end{equation}
Equating (\ref{28}) and (\ref{29}) and then differentiating the resulting
equation twice with respect to ''time'' we get
\begin{equation}
(C_{4}-C_{2}+2)\stackrel{..}{{\bf
r}}_{2}+(C_{4}-C_{2}-2)\stackrel{..}{{\bf r}}_{4}=0.
\end{equation}
Now using equations (\ref{26}) in conjunction with the above
equation we obtain the second relation needed for the rigid body
motion,
\begin{eqnarray}
F_2(\alpha,\beta)=\frac{2\mu (C_{4}+\mu )^{3}}{\left(
C_{2}-C_{4}\right) ^{2}(\beta
^{2}-1/4)^{3/2}}+\frac{(C_{4}-C_{2}+2)(C_{2}+\mu )}{\alpha
^{3}}+\\  \frac{\mu (C_{4}-C_{2}-2)\left( C_{4}+\mu \right)
}{\beta ^{3}}=0. \nonumber \label{30}
\end{eqnarray}

\begin{figure}[hbtp]
  \centerline{
    \epsfxsize=7.0cm
    \epsffile{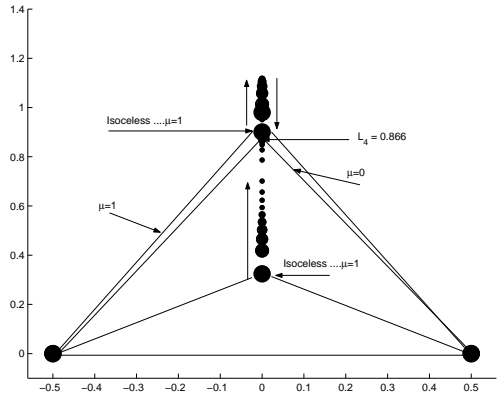}
  }
  \vspace{2pt}
   \centerline{(a)}
  \vspace{2pt}

  \centerline{
    \epsfxsize=7.0cm
    \epsffile{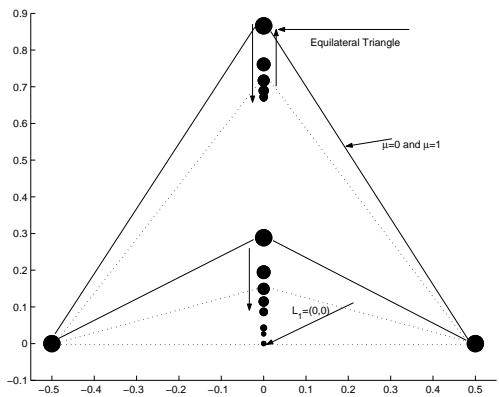}
  }
 \vspace{2pt}
  \centerline{(b)}
  \vspace{2pt}

  \caption{The evolution of all four masses when $\mu$ is varied from
1 to zero in the triangular equilibrium case-1 (a). Solution
1-Isosceles triangle (b) Solution 2-Equilateral triangle (The
Origin is located halfway between the two primaries and thus the
centre of mass moves as $\mu$ is varied.)}
\label{combinedgraphfirsttriangle}
\end{figure}

To find the rigid body solution we need to solve equations
(\ref{27}) and (3.63), simultaneously, for all values of $\alpha $
and $\beta $. These two equations are highly non-linear and
therefore its algebraic solution is not possible. We used
Mathematica to find numerical solutions for it. Figure
(\ref{tworoots}), which is the graph of $F_1(\alpha,\beta)=0$ and
$F_2(\alpha,\beta)=0$, given by equations (3.59) and (3.63)
respectively, for $\mu=0.9$ shows that there are two solutions for
each $\mu$ value. These two $(\alpha,\beta)$ solutions correspond
to two different triangular equilibrium configurations possible
for the given $\mu$. Two special solutions give the boundary of
the family of solutions:
\begin{enumerate}
\item $\mu=1$ (Case of four equal masses). We get the two well
known solutions namely the isosceles triangle solution (see figure
(\ref{combinedgraphfirsttriangle}a)) and the equilateral triangle
solution \cite{simo}, (see figure
(\ref{combinedgraphfirsttriangle}b)). The equilibrium solutions
given by $\alpha$ and $\beta$ are listed in Tables (\ref{x}) and
(3.2). We call the case 1 solutions that start with the isosceles
triangle at $\mu=1$, solution 1, and the case 1 solutions that
start with the equilateral triangle at $\mu=1$, solution 2.

\item $\mu=0$ (Case of Lagrange solutions of the Copenhagen
problem). In  case 1 solution 1-isosceles triangle, as
$\mu\rightarrow 0$, both $P_2$  and $P_4$ go to $L_4$ (see Table
(\ref{x})). In  case 1 solution 2-equilateral triangle, as
$\mu\rightarrow 0$, $P_2$ goes to $L_4$ and $P_4$ goes to $L_1$
(see Table (3.2).
 \end{enumerate}

 Between $\mu=1$ and 0, there is a continuity of solutions, each
$\mu$ having two equilibrium solutions. Please note that there are
some numerical errors in table 3.1 and 3.2 of the order of
$10^{-7}$ as $\mu$ goes to 1. These are given in Tables (\ref{x})
and (3.2) and two families of the equilibrium configurations are
shown in Figure (3.11a) and  (3.11b). Note that Figures (3.11)
have as their origin the point halfway between the two primaries
of masses $M$. Thus unlike in Figure 3.6, the centre of mass is a
point that moves as $\mu$ is reduced from 1 to 0.

\renewcommand{\baselinestretch}{1.5}
\begin{table*}
\begin{center}
  \caption{Equilibrium solutions in the triangular case 1, solution 1, Isosceles triangle.}\label{x}
\bigskip
  \begin{tabular}{|c|c|c|c|c|c|c|c|} \hline
$ \mu $ & $\alpha_1$ & $\beta_1$ & $r_1=r_3$& $r_4$& $r_2$ \\
\hline \hline 1 & 1.03  &   0.596 & 0.58631831 & 0.018152654&
0.594282356
      \\
0.9     & 1.13   & 0.652   & 0.571069502& 0.079340209& 0.674246656
     \\
0.8     & 1.17    & 0.683   &  0.545071889 &0.126825011&
0.719321922
      \\
0.7   &  1.196        &  0.709  & 0.518599486& 0.175497276&
0.759292937
      \\
0.6       &     1.213 &  0.732 &  0.492688262& 0.227166212&
0.797696755
     \\
0.5     &1.22256 &0.754&   0.46877787 & 0.284370156& 0.835638092
      \\
0.4     &  1.224 &  0.776 &  0.448824987& 0.349063876 &
0.872837665
     \\
0.3     &  1.217   &0.799&   0.436231551& 0.423284096 &0.909610336
     \\
0.2     &   1.199 &  0.825 &  0.436287216& 0.510720961&
0.944271801
     \\
0.1     &   1.162 &  0.861  & 0.455927961& 0.621403477&
0.969385705
     \\
0.01     &  1.074  & 0.932 &  0.494662843& 0.777927333&
0.941914337
     \\
0.001    &     1.034  & 0.967 &  0.499478426& 0.826835512&
0.904206852
      \\
0.0001  &    1.015  & 0.985 &  0.499948997 &0.84857412 &
0.883216869
    \\
0.00001  &   1.007 &  0.993 &  0.499994953 & 0.857924321 &
0.874090305
     \\ \hline \hline

 \end{tabular}
 \end{center}
\end{table*}

\begin{table*}
\begin{center}
  \caption{Equilibrium solutions in the triangular case 1, solution 2-Equilateral triangle}\label{x2}
\bigskip
  \begin{tabular}{|c|c|c|c|c|c|c|c|} \hline
$ \mu $ & $\alpha_2$ & $\beta_2$ & $r_1$& $r_4$& $r_2$ \\ \hline
\hline
1   &1&   0.57735& 0.577350202 &$4.03\times 10^{-07}$&    0.577350404\\
0.9 & 0.910515913& 0.536533511 &0.496437204& 0.031715456&
0.53463326\\
0.8 & 0.874052 &   0.521807&    0.418007111& 0.043212884 &
0.524428572 \\
0.7 & 0.851383  &  0.513011 &   0.328494841 & 0.050703648 &
0.523586719\\
 0.6 &0.837823  &  0.507424 &   0.222436467& 0.055784156&
0.53000428\\
0.5 &0.832221 &   0.503911  &  0.122366589 &0.058662486
&0.543953079\\
 0.4& 0.834933&    0.501822&    0.236552793& 0.05890323&
0.567038183\\
0.3& 0.847573 &   0.500701&    0.589645065& 0.055537618&
0.602359626\\
 0.2 &0.84101 &0.5002 & 1.128549593 &0.043388245 &
0.618706196\\
 0.1& 0.7873491&   0.500194&    0.414519964& 0.01434931
&0.579930294\\ \hline

 \end{tabular}
\end{center}
\end{table*}

\renewcommand{\baselinestretch}{2}

\subsection{Triangular Equilibrium configuration of four-body problem: Case
II}

\begin{figure}
\centerline{ \epsfig{file=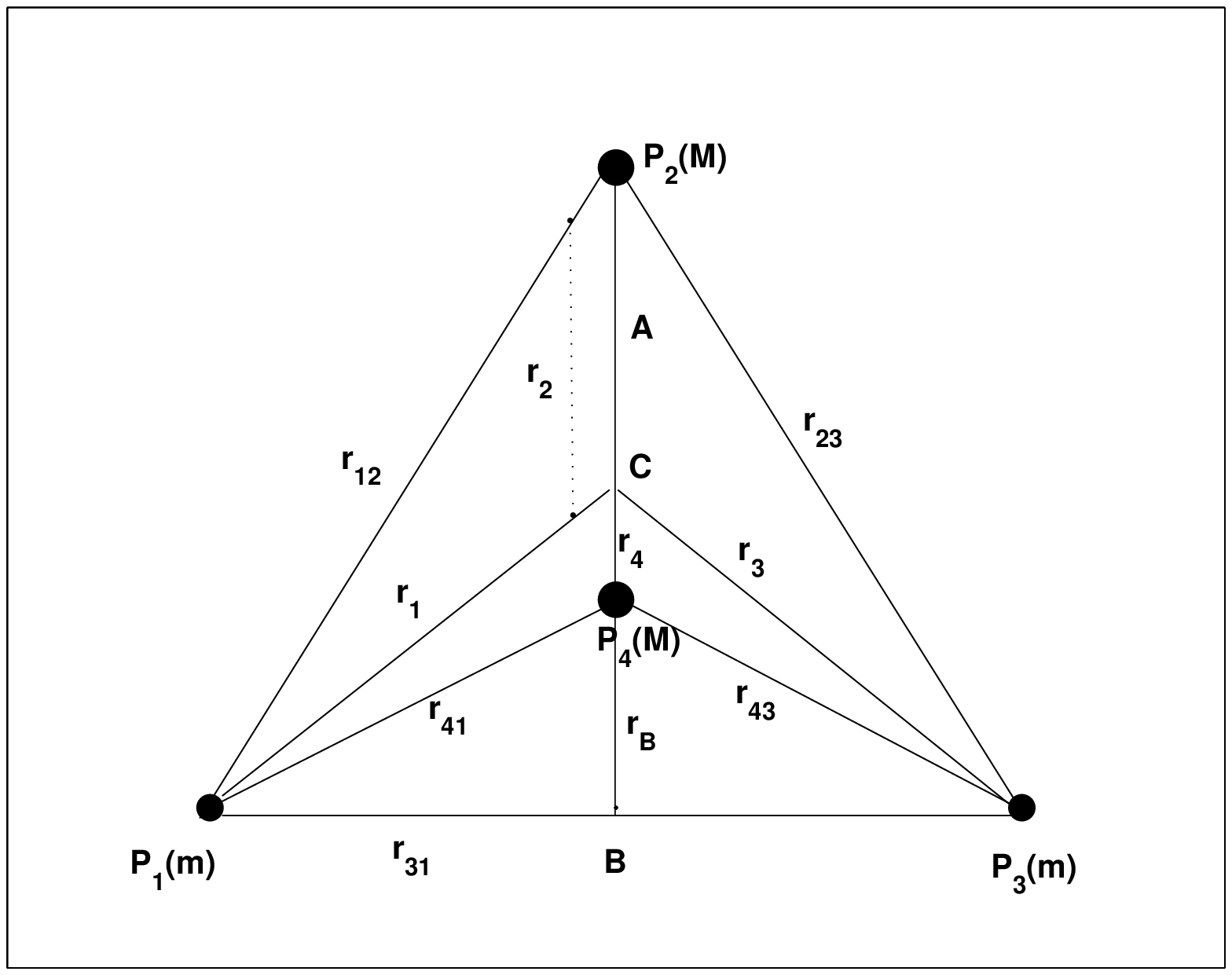 ,height=7cm}}
\caption{Triangular Equilibrium Configuration of Four Body
Problem. Case-II} \label{triangle2}
\end{figure}

We now consider the Triangular Equilibrium Configuration given in
Figure (\ref{triangle2}). The geometry is taken to be symmetrical
about the line $P_{2}B$. In this case it is the masses $P_{2}$ and
$P_{4}$ that retain the original mass $M$ with the masses $\
P_{1}$ and $P_{3}$ decreasing in mass though remaining equal in
mass $m$ so that $\mu =\frac{m}M{}\leq 1.$

Points $A$ and $B$ are the centers of mass of the pair $P_{2}$ and
$P_{4}$
and the pair $P_{1}$ and $P_{2}$. Then letting ${\bf r}_{A}=CA$ and ${\bf r}%
_{B}=CB,$ we have
\begin{equation}
{\bf r}_{A}+\mu {\bf r}_{B}=0.
\end{equation}
Let
\begin{equation}
{\bf r}=P_{2}P_{4}={\bf r}_{24},{\bf r}_{B}=CB=\alpha {\bf
r}\textrm{ and } P_{1}B=\beta {\bf r,}
\end{equation}
then from the centre of mass relations and the geometry of the
figure, we get the following list of vectors representing all the
four bodies
\begin{equation}
\left.
\begin{array}{l}
{\bf r}_{1}=\alpha {\bf r}+\frac{1}{2}{\bf r}_{31},\textrm{ \ \ \ } \\
{\bf r}_{2}=-\left( \frac{1}{2}+\mu \alpha \right) {\bf r}, \\
{\bf r}_{3}=\alpha {\bf r}-\frac{1}{2}{\bf r}_{31},\textrm{ \ \ \ } \\
{\bf r}_{4}=\left( \frac{1}{2}-\mu \alpha \right) {\bf r}.\textrm{
\ \ }
\end{array}
\right\}
\end{equation}

Using Figure (\ref{triangle2}) and the symmetry conditions the
magnitudes of the vectors involved are
\begin{equation}
\left.
\begin{array}{l}
r_{1}=r_{3}=\left( \sqrt{\alpha ^{2}+\beta ^{2}}\right) r, \\
r_{2}=\left( \frac{1}{2}+\mu \alpha \right) r,
\\
r_{4}=\left( \frac{1}{2}-\mu \alpha \right) r,
\end{array}
\right\}
\end{equation}
Also
\begin{equation}
\left.
\begin{array}{l}
r_{12}=\left( \sqrt{\left( \frac{1}{2}+\alpha \left( 1+\mu \right) \right)
^{2}+\beta ^{2}}\right) r=r_{23}, \\
r_{14}=\left( \sqrt{\left( -\frac{1}{2}+\alpha \left( 1+\mu
\right) \right) ^{2}+\beta ^{2}}\right) r.\textrm{ \ \ \ \ \ \ }
\end{array}
\right\}
\end{equation}
Using the same techniques as before we get the following differential
equations as equations of motion,
\begin{eqnarray}
\stackrel{..}{{\bf r}} &=&{\bf -}\frac{M}{2\alpha r^{3}}\left[ \frac{%
1+2\alpha \left( 1+\mu \right) }{\left( \left( \frac{1}{2}+\alpha \left(
1+\mu \right) \right) ^{2}+\beta ^{2}\right) ^{3/2}}+\frac{1-2\alpha \left(
1+\mu \right) }{\left( \left( -\frac{1}{2}+\alpha \left( 1+\mu \right)
\right) ^{2}+\beta ^{2}\right) ^{3/2}}\right] {\bf r,}  \nonumber \\
&=&X{\bf r,}  \label{31}
\end{eqnarray}
\begin{eqnarray}
\stackrel{..}{{\bf r}}_{31} &=&-\frac{M}{r^{3}}\left[ \frac{1}{\left( \left(
\frac{1}{2}+\alpha \left( 1+\mu \right) \right) ^{2}+\beta ^{2}\right) ^{3/2}%
}+\frac{1}{\left( \left( -\frac{1}{2}+\alpha \left( 1+\mu \right) \right)
^{2}+\beta ^{2}\right) ^{3/2}}+\frac{\mu }{4\beta ^{3}}\right] {\bf r}_{31},
\nonumber \\
&=&Y{\bf r}_{31}.  \label{32}
\end{eqnarray}
For a rigid body motion we must have $X-Y=0,$ which gives us
\begin{equation}
G_1(\alpha,\beta)=\left[ \frac{\mu \alpha +1/2}{\left( \left(
\frac{1}{2}+\alpha \left( 1+\mu \right) \right) ^{2}+\beta
^{2}\right) ^{3/2}}+\frac{\mu \alpha -1/2}{\left( \left(
-\frac{1}{2}+\alpha \left( 1+\mu \right) \right) ^{2}+\beta
^{2}\right) ^{3/2}}-\frac{\mu \alpha }{4\beta ^{3}}\right] =0.
\label{Rel1}
\end{equation}

To completely find the rigid body solution, we need a second
relation for which we make use of the following
\begin{equation}
{\bf r}_{A}=-\mu \alpha {\bf r}\textrm{ and }{\bf
r}_{A}=\frac{{\bf r}_{2}+{\bf r}_{4}}{2},
\end{equation}
this gives us
\begin{equation}\label{2.73}
\left( \mu \alpha +1/2\right) {\bf r}_{4}-\left( \mu \alpha
-1/2\right) {\bf r}_{2}=0.
\end{equation}
Differentiating equation (\ref{2.73}) twice with respect to
''time'' and then after some simplifications we get the following
\begin{eqnarray}
G_2(\alpha,\beta)=\left( \frac{1}{2}-\mu \alpha \right) \left(
\frac{\alpha \left( 1+\mu \right) +1/2}{A}\right) +\\
\nonumber\left( \frac{1}{2}+\mu \alpha \right) \left( \frac{
\alpha \left( 1+\mu \right) -1/2}{B}\right) -\alpha =0,
\label{Rel3}
\end{eqnarray}
where
\begin{equation}
A=\left( \left( \frac{1}{2}+\alpha \left( 1+\mu \right) \right) ^{2}+\beta
^{2}\right) ^{3/2},
\end{equation}
and
\begin{equation}
B=\left( \left( -\frac{1}{2}+\alpha \left( 1+\mu \right) \right)
^{2}+ \beta^{2}\right) ^{3/2}.
\end{equation}

\begin{figure}[hbtp]
  \centerline{
    \epsfxsize=7.0cm
    \epsffile{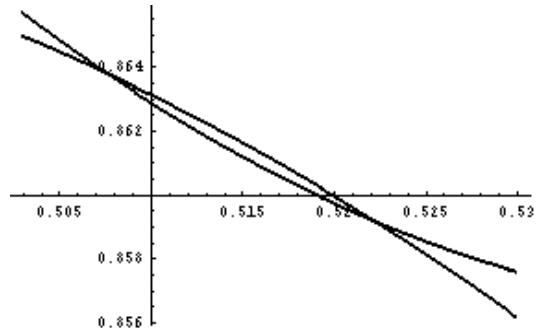}
  }
  \vspace{2pt}
   \centerline{(a)}
  \vspace{2pt}

  \centerline{
    \epsfxsize=7.0cm
    \epsffile{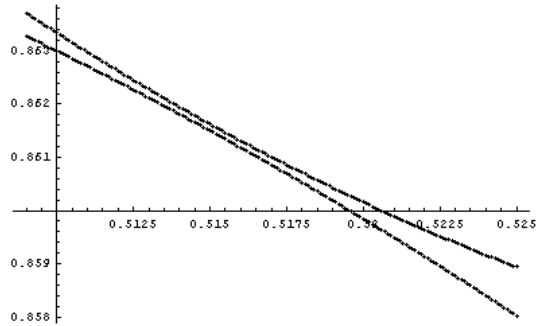}
  }
 \vspace{2pt}
  \centerline{(b)}
  \vspace{2pt}

  \caption{Graphs of the functions  $G_1(\alpha,\beta)=0$ and
$G_2(\alpha,\beta)=0$ for a) $\mu=0.998$ b) $\mu=0.9971$}
\label{tworootscase2}
\end{figure}

\begin{figure}[tbp]
\begin{center}
\epsfig{file=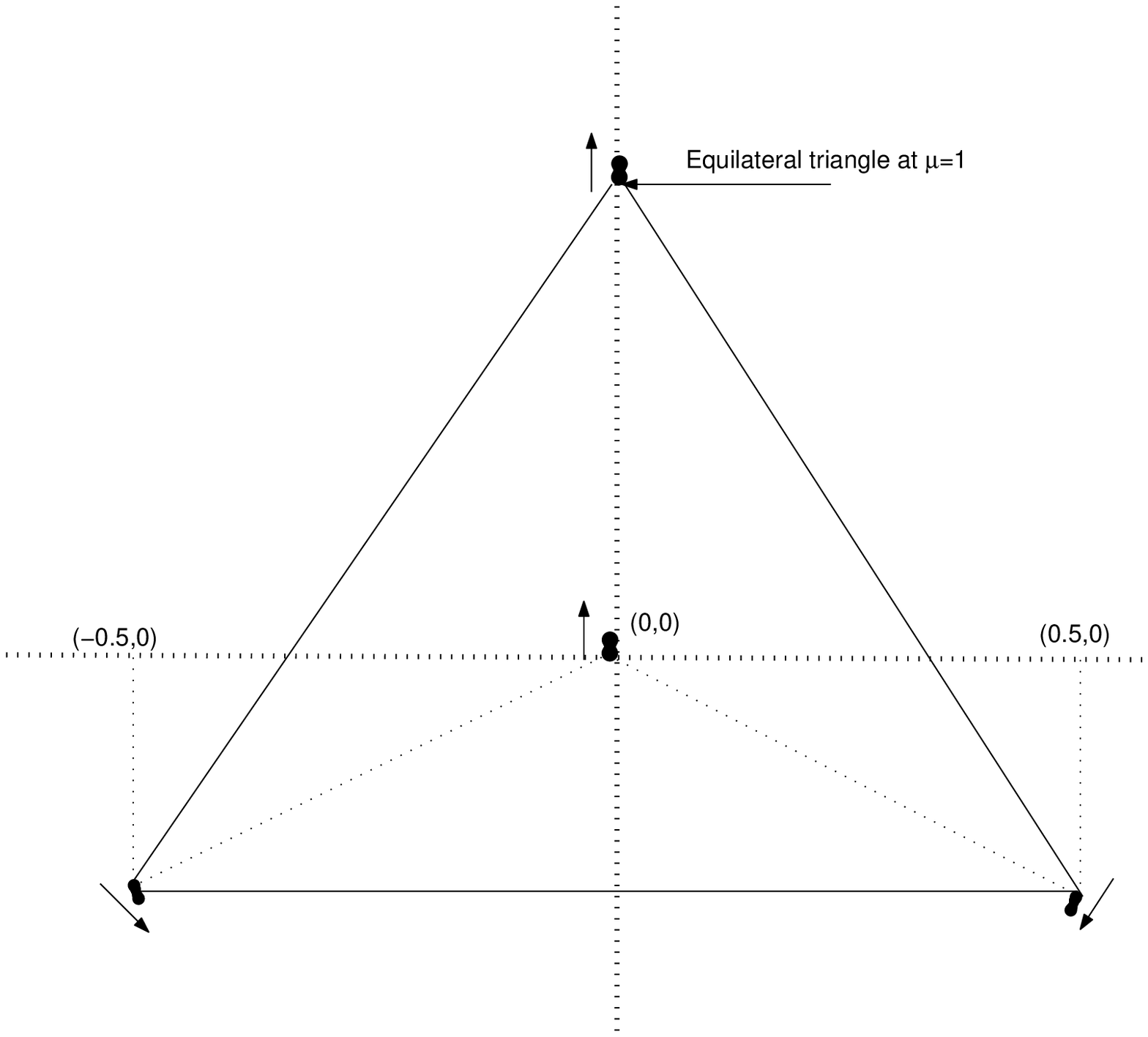,width=7cm,angle=0}
\end{center}
\caption{Evolution of all four masses when $\mu $ varies from 1 to
zero in the second triangular case (The Origin is located at the
Centre of mass)} \label{combinedgraph2ndtriangle}
\end{figure}

To find the equilibrium solutions we need to solve equations
(\ref{Rel1}) and (3.74), simultaneously, for values of $\alpha$
and $\beta$, given a range of $\mu$ from 1 to zero. These two
equations are highly non-linear and cannot be solved analytically.
Therefore we find numerical solutions using Mathematica. There are
two solutions for each $\mu$ value, see figure
(\ref{tworootscase2}a) for the example when $\mu=0.998$. These
solutions are very close to each other as they have a difference
of order $10^{-2}$ at most. Figure (\ref{tworootscase2}) is the
graphs of $G_1(\alpha,\beta)=0$ and $G_2(\alpha,\beta)=0$  for
$\mu=0.998$ and $\mu=0.9971$. Figure (\ref{tworootscase2}b) shows
that there is no solution for $\mu < 0.9972$. The solutions
between $\mu=1$ and $\mu=0.9972$ are shown in Figure
(\ref{combinedgraph2ndtriangle}). At $\mu=1$ we get the
equilateral triangle solution, see Figure
(\ref{combinedgraph2ndtriangle}), as we would expect. In fact
case-ii is a continuation of the family of solutions found in
case-i, where $\mu$ in case-i goes to 1/0.9972.

\renewcommand{\baselinestretch}{1.5}

\begin{table*}
\begin{center}
  \caption{Equilibrium solutions in the triangular case 2}\label{x2}
\bigskip
  \begin{tabular}{|c|c|c|c|c|c|c|c|} \hline
$ \mu $ & $\alpha$ & $\beta$ & $r_1$& $r_2$& $r_4$ \\ \hline
\hline
1 & 0.5 &0.866 &  0.999978  &  1  & 0\\

 & 0.528 &  0.857  & 1.006& 1.028&   0.028
\\ \hline
 0.999&0.503&   0.865&   1.0006&  1.002497&    0.0024
\\
  &0.526 &  0.857&   1.0055& 1.0254&    0.025
\\\hline
   0.998& 0.50757& 0.863  & 1.0012& 1.0066&  0.0066
\\
    & 0.522 &  0.859 &  1.0052&  1.021&    0.021
\\\hline
     0.9975&0.519  & 0.86 &   1.0045& 1.0177&   0.0177
\\
      &0.511 &  0.8627&  1.0027& 1.0097&   0.0098
\\\hline
       0.9974&0.5122 & 0.8624 & 1.0031& 1.0109&  0.01087
\\
        &0.518   &0.8605 & 1.0044& 1.0167&   0.01665
\\\hline
0.9973&0.5158 & 0.8612&  1.0039& 1.0144&  0.0144
\\
 &0.5144 & 0.8617 & 1.0036& 1.0130&  0.0130
\\\hline
  0.99725&0.51272& 0.8622&  1.003& 1.0113&  0.0113
\\
   &0.5174 & 0.8607 & 1.00424& 1.0160&  0.01598
\\\hline
    0.9972&0.5004 & 0.8662&  1.00035& 0.999&  0.001
\\
     &0.5035 & 0.865  & 1.0009& 1.0021& 0.0021
\\ \hline
 \end{tabular}
\end{center}
\end{table*}

\renewcommand{\baselinestretch}{2}

\subsection{Collinear equilibrium configurations for two pairs of equal
masses- A Review}

In this section we discuss all possible arrangements of the two
pairs of equal masses along a straight line
\cite{RoyandSteves1998}. The first two of the four arrangements
are symmetric while the last two arrangements are non-symmetric.

\subsubsection{Case-I}

\begin{figure}[tbp]
\begin{center}
\epsfig{file=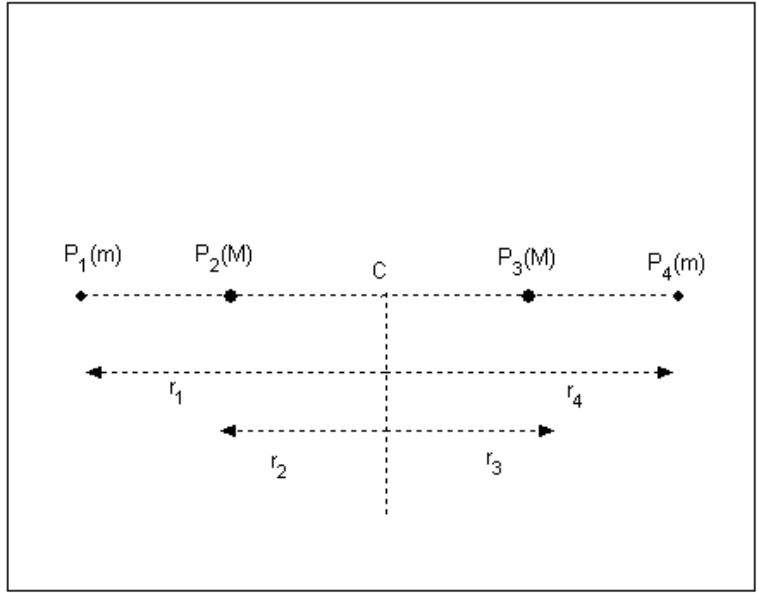,width=10cm}
\end{center}
\caption{Collinear Equilibrium Configuration Case-I}
\label{CollCaseI}
\end{figure}

In this symmetrical arrangement of two pairs of different masses,
the larger masses ($M)$ lie in the middle of the line and the
smaller masses ($m)$ lie at the corners apiece, as shown in the
Figure (\ref{CollCaseI}).

This arrangement is symmetric about the center of mass $C$.

Let $r_{2}=\alpha r_{1}$, $\mu =m/M\leq 1$ and ${\bf \rho }_{ij}={\bf r}%
_{ij}/r_{ij}^{3}.$ By symmetry ${\bf r}_{4}=-{\bf r}_{1}$ and ${\bf r}_{3}=-%
{\bf r}_{2}$.

Now using the general equations of motion in conjunction with the above
considerations we get,
\begin{equation}
\left.
\begin{array}{l}
\stackrel{..}{{\bf r}}_{1}=M({\bf \rho }_{12}+{\bf \rho }_{13}+\mu {\bf \rho
}_{14}), \\
\stackrel{..}{{\bf r}}_{2}=M({\bf \rho }_{21}+{\bf \rho }_{23}+\mu
{\bf \rho }_{24}).
\end{array}
\right\}
\end{equation}
Using all the symmetry conditions we obtain the following final form of the
equations of motion
\begin{equation}
\stackrel{..}{{\bf r}}_{1}=-\frac{M}{r_{1}^{3}}R_{1}{\bf r}_{1},
\end{equation}
where
\begin{equation}
R_{1}=\frac{1}{\left( 1-\alpha \right) ^{2}}+\frac{1}{\left(
1+\alpha \right) ^{2}}+\frac{\mu }{4}.
\end{equation} Also
\begin{equation}
\stackrel{..}{{\bf r}}_{2}=-\frac{M}{r_{1}^{3}}R_{2}{\bf r}_{2},
\end{equation}
where
\begin{equation}
R_{2}=\frac{1}{4\alpha ^{3}}+\frac{\mu }{\alpha }\left(
\frac{1}{\left( 1+\alpha \right) ^{2}}-\frac{1}{\left( 1-\alpha
\right) ^{2}}\right) .
\end{equation}
For a rigid rotating geometry we must have
\begin{equation}
R_{1}-R_{2}=0.
\end{equation}
Therefore
\begin{equation}
\frac{1}{\left( 1-\alpha \right) ^{2}}+\frac{1}{\left( 1+\alpha \right) ^{2}}%
-\frac{1}{4\alpha ^{3}}-\frac{\mu }{\alpha }\left( \frac{1}{\left( 1+\alpha
\right) ^{2}}-\frac{1}{\left( 1-\alpha \right) ^{2}}\right) +\frac{\mu }{4}%
=0.
\end{equation}

\begin{figure}[tbp]
\centerline{
\epsfig{file=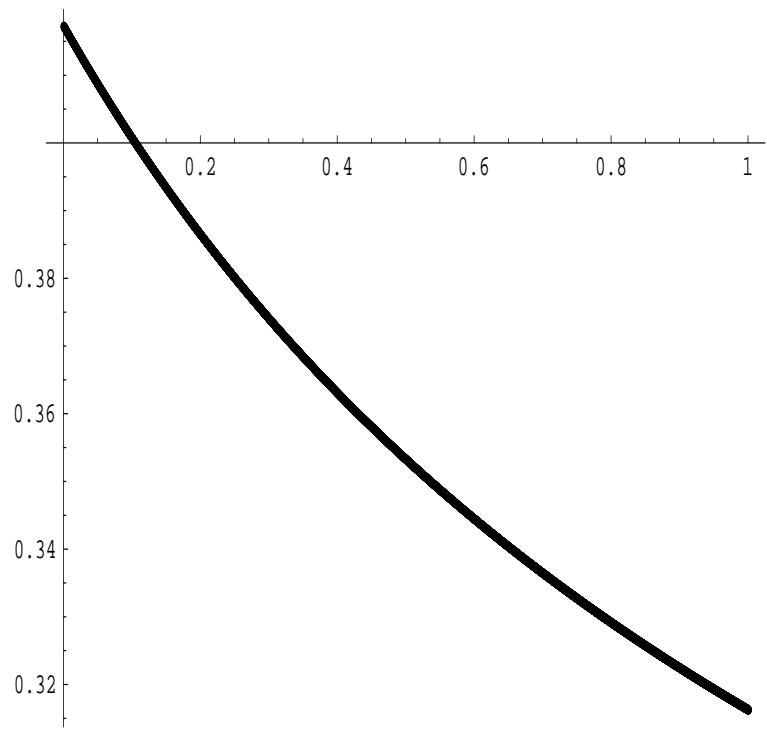,width=10cm,height=10cm}}
\caption{Variation of the parameter $\alpha$ for all values of
$\mu$ in Case-I of the collinear equilibrium configurations of two
pairs of equal masses} \label{equilibSoluCase-I}
\end{figure}

After further simplification we get
\begin{equation}
H_1(\alpha)=\frac{2\left( 1+\alpha ^{2}\right) +4\mu }{\left(
1-\alpha ^{2}\right)^2 }+ \frac{1}{4}\left( \mu -\frac{1}{\alpha
^{3}}\right) =0.
\end{equation}
Solving the above equation for $\mu =1$ (the equal mass case) and
$\mu =0$ (i.e. $ m=0$) we get $\alpha =0.44$ and $\alpha =0.316$
respectively. All other solutions for different mass ratios can be
found easily. See figure (\ref{equilibSoluCase-I}), which is graph
of $H_1(\alpha)$ and shows there is a continuous family of
solutions for  $\mu \in \,(0,1)$.

\subsubsection{Case-II}

In this second symmetrical arrangement of the two pairs of
different masses, the two smaller masses are near the centre of
mass while the two larger masses are farther away.

Again proceeding along the same lines as before, let ${\bf
r}_{2}=\alpha {\bf r} _{1}.$ By symmetry ${\bf r}_{4}=-{\bf
r}_{1}$ and ${\bf r}_{3}=-{\bf r}_{2}.$ The mass ratio is $\mu
=m/M\leq 1$ and ${\bf \rho }_{ij}={\bf r} _{ij}/r_{ij}^{3}.$ The
equations of motion are
\begin{equation}
\left.
\begin{array}{l}
\stackrel{..}{{\bf r}}_{1}=M(\mu {\bf \rho }_{12}+\mu {\bf \rho
}_{13}+{\bf
\rho }_{14}), \\
\stackrel{..}{{\bf r}}_{2}=M({\bf \rho }_{21}+\mu {\bf \rho
}_{23}+{\bf \rho }_{24}).
\end{array}
\right\}
\end{equation}

\begin{figure}
\centerline{ \epsfig{file=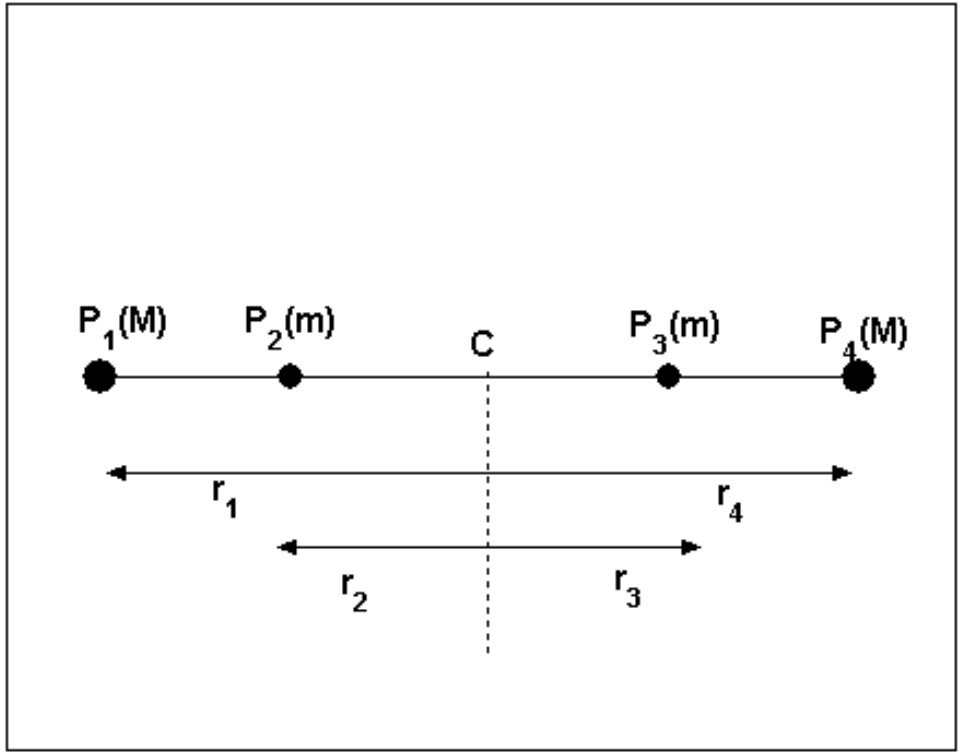 ,height=10cm}}
\caption{Collinear Equilibrium Configuration Case-II}
\label{collcaseii}
\end{figure}

Using the symmetry conditions, we get the following simpler form
of the equations of motion

\begin{equation}
\stackrel{..}{{\bf r}}_{1}=-\frac{M}{r_{1}^{3}}R_{1}{\bf r}_{1},
\end{equation}
where
\begin{equation}
R_{1}=\mu \left( \frac{1}{\left( 1-\alpha \right) ^{2}}+\frac{1}{\left(
1+\alpha \right) ^{2}}\right) +\frac{1}{4}.
\end{equation}
and
\begin{equation}
\stackrel{..}{{\bf r}}_{2}=-\frac{M}{r_{1}^{3}}R_{2}{\bf r}_{2},
\end{equation}
where
\begin{equation}
R_{2}=\frac{1}{\alpha }\left( \frac{1}{\left( 1+\alpha \right) ^{2}}-\frac{1%
}{\left( 1-\alpha \right) ^{2}}+\frac{\mu }{4\alpha ^{2}}\right) .
\end{equation}

For a rigid body motion we must have
\begin{equation}
R_{1}-R_{2}=0.
\end{equation}
Hence
\begin{equation}
\mu \left( \frac{1}{\left( 1-\alpha \right) ^{2}}+\frac{1}{\left(
1+\alpha \right) ^{2}}\right) +\frac{1}{4}-\frac{1}{\alpha }\left(
\frac{1}{\left( 1+\alpha \right) ^{2}}-\frac{1}{\left( 1-\alpha
\right) ^{2}}+\frac{\mu }{ 4\alpha ^{2}}\right) =0.
\end{equation}
After further simplification we get
\begin{equation}
H_2(\alpha)=\frac{1}{4}\left( 1-\frac{\mu }{\alpha ^{3}}\right)
+\frac{2}{\left( 1-\alpha ^{2}\right) ^{2}}(2+\mu(1+\alpha ^2))=0.
\label{33}
\end{equation}

\begin{figure}[tbp]
\centerline{
\epsfig{file=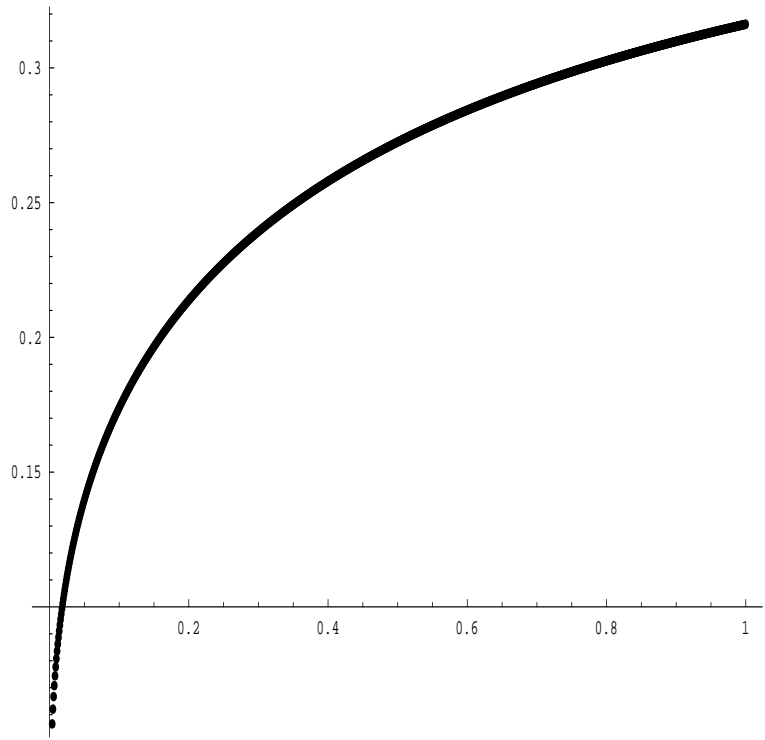,width=10cm,height=10cm}}
\caption{Variation of the parameter $\alpha$ for all values of
$\mu$ in Case-II of the collinear equilibrium configurations of
two pairs of equal masses} \label{equilibSoluCase-II}
\end{figure}
Solving equation (\ref{33}) we get $\alpha = 0.275$ for $\mu =1$,
which is the equal mass case also obtained for case 1, $\mu=1$.
There is no solution for $\mu =0$ but as $\mu $ approaches zero
$\alpha $ converges to zero which is the $L_1$ Lagrange solution.
See figure (\ref{equilibSoluCase-II}), which is the graph of
$H_2(\alpha)$ and shows there is a continuous family of solutions
for $\mu \in \,(0,1)$.

\subsubsection{Case-III}

\begin{figure}
\centerline{ \epsfig{file=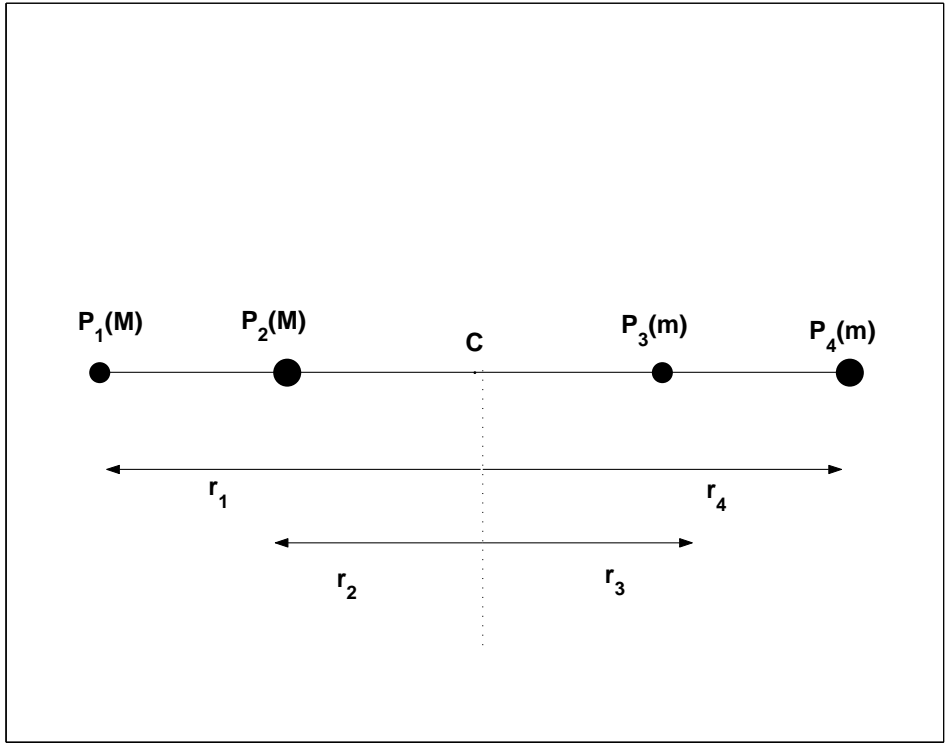 ,height=10cm}}
\caption{Collinear Equilibrium Configuration Case-III}
\label{CollCaseIII}
\end{figure}

From Figure (\ref{CollCaseIII}) we can see that this is a
non-symmetric arrangement of the four bodies. All arrangements
previous to this, have had some kind of symmetry which has
considerably simplified the problem.

Let
\begin{equation}
{\bf r}_{2}=\alpha {\bf r}_{1};{\bf r}_{3}=-\beta {\bf r}_{1};{\bf
r} _{4}=-\gamma {\bf r}_{1}.  \label{40}
\end{equation}
By the centre of mass relation
\begin{equation}
\sum_{i=1}^{4}m_{i}{\bf r}_{i}=0.
\end{equation}
Thus
\begin{equation}
m{\bf r}_{1}+M{\bf r}_{2}+m{\bf r}_{3}+M{\bf r}_{4}=0.
\end{equation}
After substituting $\mu =m/M$ we get,
\begin{equation}
\mu \left( {\bf r}_{1}+{\bf r}_{3}\right) +{\bf r}_{2}+{\bf r}_{4}=0,
\label{41}
\end{equation}
from equations (\ref{40}) and (\ref{41}) we get
\begin{equation}
\left( \mu \left( 1-\beta \right) +\alpha -\gamma \right) {\bf r}_{1}=0.
\end{equation}
As ${\bf r}_{1}\neq 0$ therefore
\begin{equation}
\mu \left( 1-\beta \right) +\alpha -\gamma =0,
\end{equation}
gives
\begin{equation}
\gamma =\mu \left( 1-\beta \right) +\alpha .  \label{42}
\end{equation}
The equations of motion are
\begin{equation}
\left.
\begin{array}{l}
\stackrel{..}{{\bf r}_{1}}=M\left[ {\bf \rho }_{12}+\mu {\bf \rho
}_{13}{\bf +\rho }_{14}\right] ,\\
\ \stackrel{..}{{\bf r}_{2}}=M\left[ \mu {\bf \rho }_{21}+\mu {\bf
\rho }_{23}
{\bf +\rho }_{24}\right] ,\textrm{ \ \ } \\
\stackrel{..}{{\bf r}_{3}}=M\left[ \mu {\bf \rho }_{31}+{\bf \rho
}_{32}{\bf
+\rho }_{34}\right] ,\textrm{ \ \ } \\
\stackrel{..}{{\bf r}_{4}}=M\left[ \mu {\bf \rho }_{41}+{\bf \rho
}_{42}{\bf +\mu \rho }_{43}\right] ,
\end{array}
\right\}
\end{equation}
where ${\bf \rho }_{ij}=\frac{{\bf r}_{ij}}{r_{^{3}ij}}.$

Now using equation (\ref{40}) and the centre of mass relation, it
can easily be shown that
\begin{equation}
\stackrel{..}{{\bf r}}_{1}=-\frac{M}{r_{1}^{3}}R_{1}{\bf r}_{1},
\end{equation}
where
\begin{equation}
R_{1}=\frac{1}{\left( 1-\alpha \right) ^{2}}+\frac{\mu }{\left( 1+\beta
\right) ^{2}}+\frac{1}{\left( 1+\gamma \right) ^{2}}.
\end{equation}
Also

\begin{equation}
\stackrel{..}{{\bf r}}_{2}=-\frac{M}{r_{1}^{3}}R_{2}{\bf r}_{2},
\end{equation}
where
\begin{equation}
R_{2}=\frac{\mu }{\alpha }\left( \frac{1}{\left( \alpha +\beta \right) ^{2}}+%
\frac{1}{\left( 1-\alpha \right) ^{2}}\right) +\frac{1}{\alpha
\left( \alpha +\gamma \right) ^{2}},
\end{equation}

and
\begin{equation}
\stackrel{..}{{\bf r}}_{3}=-\frac{M}{r_{1}^{3}}R_{3}{\bf r}_{3},
\end{equation}
where
\begin{equation}
R_{3}=\frac{1}{\beta }\left( \frac{\mu }{\left( 1+\beta \right) ^{2}}+\frac{1%
}{\left( \alpha +\beta \right) ^{2}}-\frac{1}{\left( \gamma -\beta
\right) ^{2}}\right),
\end{equation}

and
\begin{equation}
\stackrel{..}{{\bf r}}_{4}=-\frac{M}{r_{1}^{3}}R_{4}{\bf r}_{4},
\end{equation}
where
\begin{equation}
R_{4}=\frac{1}{\gamma }\left( \mu \left( \frac{1}{\left( 1+\gamma \right)
^{2}}+\frac{1}{\left( \gamma -\beta \right) ^{2}}\right) +\frac{1}{\left(
\gamma +\alpha \right) ^{2}}\right) .
\end{equation}

For a rigid rotating solution we must have
$R_{1}=R_{2}=R_{3}=R_{4}>0$ i.e. $ R_{i}=R>0$ and constant. We
therefore require $\rho _{i}=R_{i}-R_{1}=0,$ $ i=2,3,4$ for the
set of values of $\alpha ,\beta $ and $\gamma .$ Now from equation
(\ref {42}) each relation $\rho _{i}=0$ is a curve in the $\alpha
,\beta $ plane. It is easy to show that if any two of the $\rho
_{i}^{\prime }s$ are zero the third will also be zero (see
\citeasnoun{RoyandSteves1998}). Thus to find a solution we need to
solve $\rho _{2}=0$ and $\rho _{3}=0$ in the $\alpha ,\beta $
plane to obtain the values of $\alpha $ and $\beta $ and hence
$\gamma$. When the masses of $P_1$ and $P_3$ are reduced to zero
they migrate to the $L_3$ and $L_1$ Lagrange point respectively.
The third $P_4$ of mass $M$ migrates to the point $1/2$ with $P_1$
remaining at $-1/2$.

\subsubsection{Case-IV}

\begin{figure}[tbp]
\begin{center}
\epsfig{file=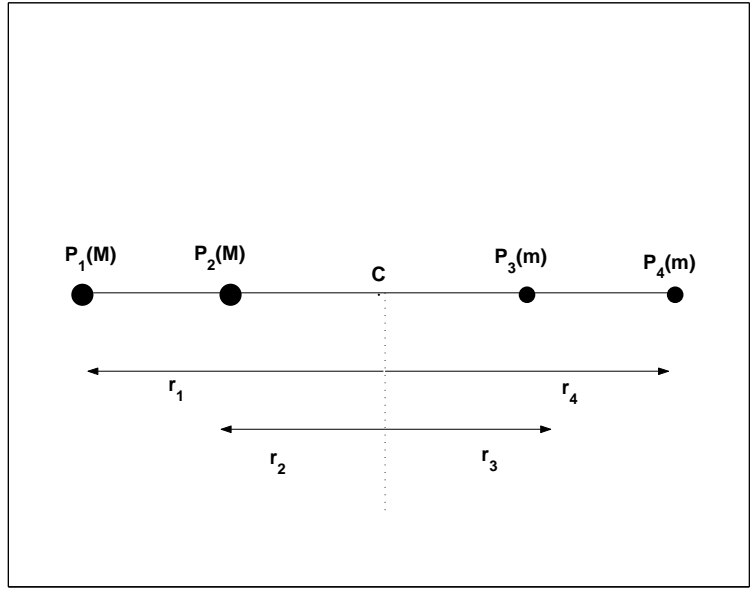,width=10cm}
\end{center}
\caption{Collinear Equilibrium Configuration Case-IV}
\label{CollCaseIV}
\end{figure}

This is the second and last non-symmetric case of four collinear
configurations under discussion. This will be treated the same way as
case-III.

Let
\begin{equation}
{\bf r}_{2}=\alpha {\bf r}_{1};{\bf r}_{3}=-\beta {\bf r}_{1};{\bf
r} _{4}=-\gamma {\bf r}_{1}.  \label{43}
\end{equation}
From the centre of mass relation
\begin{equation}
{\bf r}_{1}+{\bf r}_{2}+\mu \left( {\bf r}_{3}+{\bf r}_{4}\right) =0,
\label{44}
\end{equation}
where $\mu =m/M.$ Now from equations (\ref{43}) and (\ref{44}) we
get
\begin{equation}
\alpha =\mu \left( \beta +\gamma \right) -1.  \label{45}
\end{equation}

The equations of motion are
\begin{equation}
\left.
\begin{array}{l}
\stackrel{..}{{\bf r}_{1}}=M\left[ {\bf \rho }_{12}+\mu \left(
{\bf \rho }
_{13}{\bf +\rho }_{14}\right) \right] , \\
\stackrel{..}{{\bf r}_{2}}=M\left[ {\bf \rho }_{21}+\mu \left(
{\bf \rho }
_{23}{\bf +\rho }_{24}\right) \right] ,\textrm{ \ \ } \\
\stackrel{..}{{\bf r}_{3}}=M\left[ {\bf \rho }_{31}+{\bf \rho
}_{32}{\bf
+\mu \rho }_{34}\right] ,\textrm{ \ \ } \\
\stackrel{..}{{\bf r}_{4}}=M\left[ {\bf \rho }_{41}+{\bf \rho
}_{42}{\bf +\mu \rho }_{43}\right] .
\end{array}
\right\}
\end{equation}
It can easily be shown, as in the previous case, that

\begin{equation}
\stackrel{..}{{\bf r}}_{1}=-\frac{M}{r_{1}^{3}}R_{1}{\bf r}_{1},
\end{equation}
where
\begin{equation}
R_{1}=\frac{1}{\left( 1-\alpha \right) ^{2}}+\mu \left(
\frac{1}{\left( 1+\beta \right) ^{2}}+\frac{1}{\left( 1+\gamma
\right) ^{2}}\right) .\textrm{ }
\end{equation}
Also

\begin{equation}
\stackrel{..}{{\bf r}}_{2}=-\frac{M}{r_{1}^{3}}R_{2}{\bf r}_{2},
\end{equation}
where
\begin{equation}
R_{2}=\frac{1}{\alpha }\left( \frac{\mu }{\left( \alpha +\beta
\right) ^{2}}- \frac{1}{\left( 1-\alpha \right) ^{2}}+\frac{\mu
}{\left( \alpha +\gamma \right) ^{2}}\right),
\end{equation}

and
\begin{equation}
\stackrel{..}{{\bf r}}_{3}=-\frac{M}{r_{1}^{3}}R_{3}{\bf r}_{3},
\end{equation}
where
\begin{equation}
R_{3}=\frac{1}{\beta }\left( \frac{1}{\left( 1+\beta \right)
^{2}}+\frac{1}{ \left( \alpha +\beta \right) ^{2}}-\frac{\mu
}{\left( \gamma -\beta \right) ^{2}}\right),
\end{equation}
and

\begin{equation}
\stackrel{..}{{\bf r}}_{4}=-\frac{M}{r_{1}^{3}}R_{4}{\bf r}_{4},
\end{equation}
where
\begin{equation}
R_{4}=\frac{1}{\gamma }\left( \frac{1}{\left( 1+\gamma \right)
^{2}}+\frac{ \mu }{\left( \gamma -\beta \right)
^{2}}+\frac{1}{\left( \gamma +\alpha \right) ^{2}}\right).
\end{equation}

For a rigid rotating solution we must have
$R_{1}=R_{2}=R_{3}=R_{4}>0$ i.e. $ R_{i}=R>0$ and constant. We
therefore require $\rho _{i}=R_{i}-R_{1}=0,$ $ i=2,3,4$ for the
set of values of $\alpha ,\beta $ and $\gamma .$ From equation
(\ref{45} ) each relation $\rho _{i}=0$ is a curve in  the $\beta
,\gamma $ plane. We will solve the following two equations for
$\beta $ and $\gamma $ to obtain the solution
$(\alpha,\beta)$which gives the rigid rotating geometry.
\begin{equation}
R_{3}-R_{1}=0\textrm{ and }R_{4}-R_{1}=0.
\end{equation}

The two smaller bodies migrate to the $L_2$ Lagrange point when
$\mu\rightarrow 0$ and at $\mu=1$ gives the collinear equal mass
solution.

\section{Summary and Conclusions}

In this chapter we discussed special analytical solutions for the
coplanar four body problem of equal and non-equal masses
\cite{RoyandSteves1998}. The non-equal mass cases have two pairs
of equal masses where the ratio between the two pairs is reduced
from 1 to 0. Three special arrangements are discussed for the four
equal masses which include:
\begin{enumerate}
\item Four equal masses making a square. \item Four equal masses
arranged at the vertices of an equilateral triangle with the
fourth mass at the centroid of the triangle which is also the
centre of mass of the system. \item Four equal masses lying along
a straight line symmetric about the centre of mass.
\end{enumerate}

In section 3.2 we allow two of the masses to reduce symmetrically
to $m$ while the other pair of bodies remain at the original mass
$M$. We define $\mu = m/M \leq 1$. In the cases studied by Steves
and Roy (1998) they found a continuous family of equilibrium
solutions occurred as $\mu$ was reduced from 1 to 0. In all, for
$\mu=1$, the equal mass case solutions were obtained which is a
good check that the equations giving the families of solutions are
correct as $\mu=1$ is a special case. Then in the limit, as $\mu
\rightarrow 0$, they always obtain one of the Lagrange five
equilibrium points $L_1, L_2, L_3, L_4$ and $L_5$ of the
Copenhagen problem \cite{RoyandSteves1998} as locations for the
two masses being reduced to zero. We complete the analysis of
\citeasnoun{RoyandSteves1998} to include two more examples of
equilibrium solutions of the four body problem i.e.
\begin{enumerate}
\item The Triangular equilibrium configuration of the four body
problem with the two bigger masses making the base of the
triangle. \item The Triangular equilibrium configuration of the
four body problem with the two smaller masses making the base of
the triangle.
\end{enumerate}
In the Triangular Case-I for $\mu=1$ we get the two well known
equal mass solutions i.e. the equilateral triangle solution and
the isosceles triangle solution and as $\mu$ is reduced to zero
the isosceles triangle evolves so that $P_2$ and $P_4$ approaches
$L_4$. The equilateral triangle solution evolves so that $P_2$
approaches $L_4$ and $P_4$ evolves to $L_1$. In the Triangular
Case-II for $\mu=1$, the equilateral triangle solution is
obtained. There are no solutions for $\mu<0.9972$. This is the
only case where no continuous  family of solutions exist from
$\mu=1$ to 0.

We conjecture that the equilibrium solutions discussed in this
chapter are all the equilibrium solutions which exist for the
symmetric case where two of the masses reduce symmetrically to $m$
while the other pair of bodies remain at the original masses $M$.

\input{ChapCSFBP}

\include{ChapCS5BP}\include{ChapNumInvist}

\include{ChapCSNBP}\include{SummAndConc}\include{futurework}\input{Appendix1}

\pagebreak \textbf{Accompanying material}

B.A. Steves, M.S. Afridi, A.E. Roy, "The Caledonian Symmetric Five
Body Problem", \emph{Monthly Notices of Royal Astronomical
Society}, submitted for publication
\end{document}

%% file: ChapCSFBP.tex
\chapter{The Stability Analysis of the Nearly Symmetric Caledonian Symmetric Four Body Problem (CSFBP) }

In this chapter we study the stability of a more general case of
the four body problem called the Caledonian Symmetric Four Body
Problem (CSFBP) \cite{BonnieRoy1998}.

To tackle the complicated nature of the general four body problem,
different restriction methods are used, the neglecting of the mass
of some bodies being the most common (see chapter 1 for some
examples). Another method of restriction, not as common but very
effective, is the introduction of some symmetry conditions
\cite{BonnieRoy1998}. This type of restriction method reduces the
dimensions of the phase space very effectively, whilst still
producing a model which can be very close to real systems.

The Caledonian Symmetric Four Body Problem (CSFBP) is a four body
system with a symmetrically reduced phase space. This
symmetrically restricted four body problem was developed by Steves
and Roy (1998) and later they derived an analytical stability
criterion for it (Steves and Roy , 2001). After 2001 different
aspects of this problem were studied by \citeasnoun{Andras1},
\citeasnoun{andras2} and \citeasnoun{AndrasMNRAS}.

In this chapter we investigate the stability of the symmetric
nature of the CSFBP by  using nearly symmetric, slightly
perturbed, initial conditions  and the general four body equations
to be able to see if the CSFBP system remains nearly symmetric.
Covering a comprehensive range of initial conditions, we integrate
each orbit for a million time-steps, to determine the evolution of
the system, recording when and if its symmetry is broken. An
integrator is specifically developed for this purpose using the
Microsoft Visual C++ Software. The results of integrations are
processed using Matlab 6.5. During the integration we record the
following observation. 1. We stop the integration when there is a
close encounter and record the type of the collision or close
encounter which we color code on the graphs shown. 2. We stop the
integration when it fails the symmetry breaking criterion and
color code it to be shown on the graph. 3. If there are no
collision or symmetry breaking then the integration continues
until the 1 million time-steps which we also note and color code
it for the graphs.

In section 4.1 we introduce the CSFBP which is a review of
\citeasnoun{BonnieRoy1998}. In section 4.2 we derive the equations
of motion for the general four body problem to be used for
integrating  the CSFBP. In section 4.3 we discuss the initial
conditions of the CSFBP. Section 4.4 discusses the general four
body integrator, specifically developed for this analysis by the
author to integrate the general four body problem. A comprehensive
set of orbits covering the initial phase space of the CSFBP are
integrated in order to determine their state of symmetry after a
long time. In section 4.5, the procedure of analysis is explained.
In section 4.6 we discuss the results of the different
integrations performed for the equal mass case of the CSFBP. These
results are compared with that of \citeasnoun{AndrasMNRAS} in
section 4.7. In section 4.8, the results of the different
integrations performed for the $\mu=0.1$ case of the symmetric and
nearly symmetric CSFBP's, are analysed. These results are compared
with those of \citeasnoun{AndrasMNRAS} in section 4.9. Finally in
section 4.10 conclusions to the chapter are given.

\section{The Caledonian Symmetric Four Body Problem (CSFBP)- A Review}
Steves and Roy (1998, 2000, 2001) have recently developed a
symmetrically restricted four body problem called the Caledonian
Symmetric Four Body Problem (CSFBP), for which they derive an
analytical stability criterion valid for all time. We will give
its brief introduction in this section.

\begin{figure}
\centerline{\epsfig{file=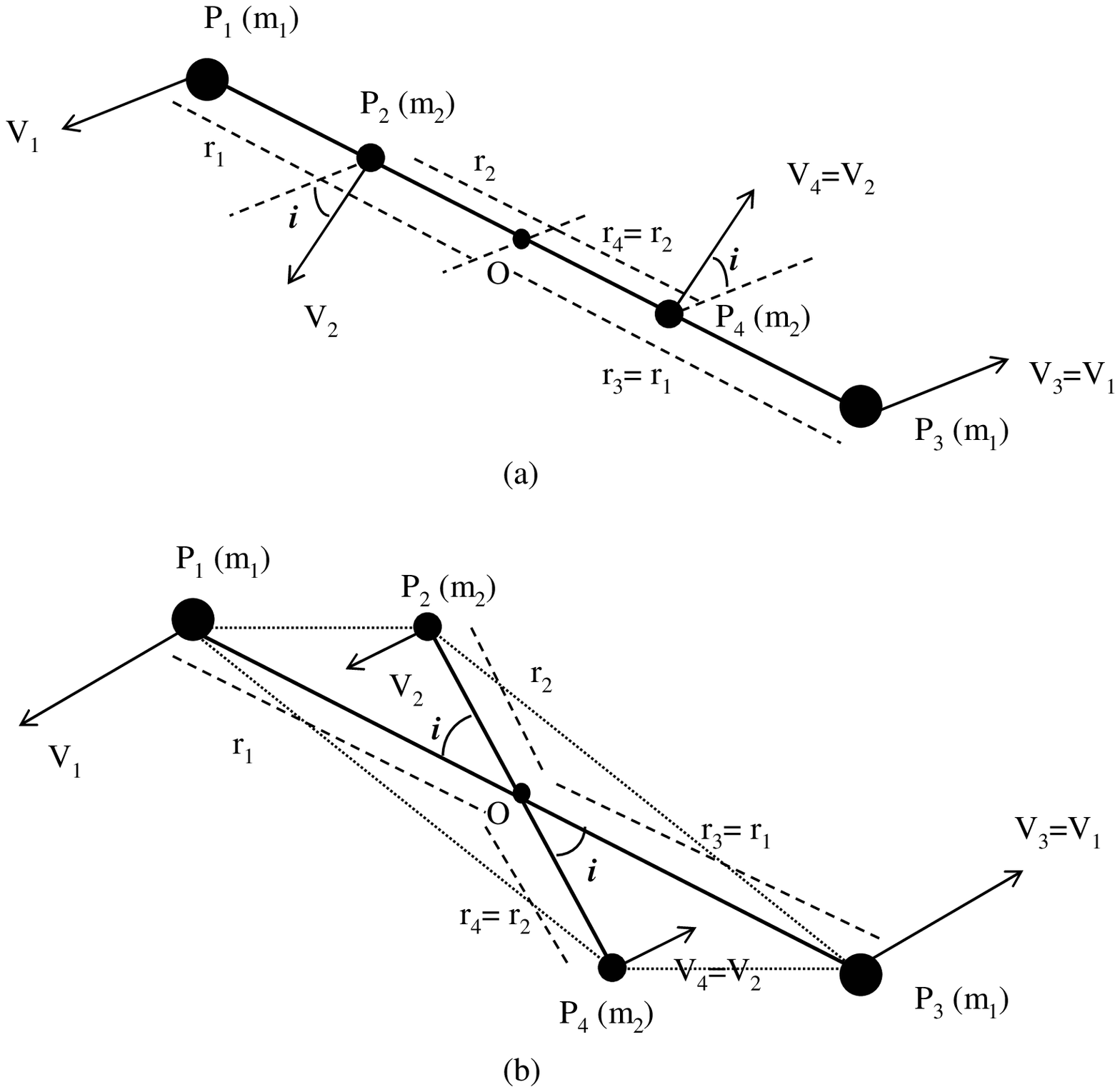,height=18cm,angle=-90}}

 \caption{a. The initial configuration of the CSFBP $\qquad$ b. CSFBP for $t > 0$ }
 \label{modelCSFBPchap4}
\end{figure}

Let us consider four bodies $P_{1}$,$P_{2}$,$P_{3}$,$P_{4}$ of
masses $m_{1}$,$m_{2}$,$m_{3}$,$m_{4}$ respectively existing in
three dimensional Euclidean space. The radius and velocity vectors
of the bodies with respect to the centre of mass of the four body
system are given by $ \mathbf{r}_{i}$ and $\dot{\mathbf{r}}_{i}$
respectively, $i=1,2,3,4$. Let the centre of mass of the system be
denoted by $O$. The CSFBP has the following conditions:
\begin{enumerate}
\item All four bodies are finite point masses with:
\begin{equation}\label{cheq1}
m_{1}=m_{3}=M, \qquad m_{2}=m_{4}=m
\end{equation}
\item  $P_{1}$ and $P_{3}$ are moving symmetrically to each other
with respect to the centre of mass of the system. Likewise $P_{2}$
and $P_{4}$ are moving symmetrically to each other. Thus
\begin{eqnarray}\label{cheq2}
 \mathbf{r}_{1}=-\mathbf{r}_{3},\qquad
\mathbf{r}_{2}=-\mathbf{r}_{4} \nonumber
\\
\mathbf{V}_{1}=\dot{\mathbf{r}}_{1}=-\dot{\mathbf{r}}_{3},\qquad
\mathbf{V}_{2}=\dot{\mathbf{r}}_{2}=-\dot{\mathbf{r}}_{4},
\end{eqnarray}

This dynamical symmetry is maintained for all time $t$. \item At
time $t = 0$ the bodies are collinear with their velocity vectors
perpendicular to their line of position. This ensures past-future
symmetry and is described by:
\begin{eqnarray}
\mathbf{r}_{1}\times \mathbf{r}_{2}=0, \qquad \mathbf{r}_{1}\cdot
\dot{\mathbf{r}}_{1}=0, \qquad \mathbf{r}_{2}\cdot
\dot{\mathbf{r}}_{2}=0
\end{eqnarray}

\end{enumerate}
Figure (\ref{modelCSFBPchap4}a) gives the initial configuration of
the CSFBP.

\section{The Equations of motion}
The classical equations of motion for the general n-body problem
are given by
\begin{equation}\label{4.1}
m_{i}\stackrel{..}{\mathbf{r}_{i}}=\sum_{i\neq
j}\frac{m_{i}m_{j}}{r_{ij}^{3} }\mathbf{r}_{ij},\qquad i=1,2,3...
\end{equation}
where $\mathbf{r_i}=(x_i,y_i)$ and
$\mathbf{r_{ij}}=\mathbf{r_j}-\mathbf{r_i}$. For a general
four-body problem we will get the following equations of motion
from (\ref{4.1}) above

\begin{equation}\label{4.5}
\stackrel{..}{\mathbf{r}}_{1}=\frac{
m_{2}\mathbf{r}_{12}}{r_{12}^{3}}+\frac{m_{3}\mathbf{r}_{13}}{r_{13}^{3}}+
\frac{m_{4}\mathbf{r}_{14}}{r_{14}^{3}}
\end{equation}
\begin{equation}\label{4.6}
\stackrel{..}{\mathbf{r}}_{2}=\frac{
m_{1}\mathbf{r}_{21}}{r_{21}^{3}}+\frac{m_{3}\mathbf{r}_{23}}{r_{23}^{3}}+
\frac{m_{4}\mathbf{r}_{24}}{r_{24}^{3}}
\end{equation}
\begin{equation}\label{4.7}
\stackrel{..}{\mathbf{r}}_{3}=\frac{
m_{1}\mathbf{r}_{31}}{r_{31}^{3}}+\frac{m_{2}\mathbf{r}_{32}}{r_{32}^{3}}+
\frac{m_{4}\mathbf{r}_{34}}{r_{34}^{3}}
\end{equation}
\begin{equation} \label{4.8}
\stackrel{..}{\mathbf{r}}_{4}=\frac{
m_{1}\mathbf{r}_{41}}{r_{41}^{3}}+\frac{m_{2}\mathbf{r}_{42}}{r_{42}^{3}}+
\frac{m_{3}\mathbf{r}_{43}}{r_{43}^{3}}.
\end{equation}

Let $\mu=\frac{m}{M}$ i.e the smaller mass divided by the larger
mass. We take $M=1$ and therefore $\mu = m$. Equations (\ref{4.5})
to (\ref{4.8}) become

\begin{equation}\label{4.9}
\stackrel{..}{\mathbf{r}}_{1}=\mu\left(\frac{
\mathbf{r}_{12}}{r_{12}^{3}}+
\frac{\mathbf{r}_{14}}{r_{14}^{3}}\right)+\frac{\mathbf{r}_{13}}{r_{13}^{3}}
\end{equation}
\begin{equation}\label{4.10}
\stackrel{..}{\mathbf{r}}_{2}=\frac{
\mathbf{r}_{21}}{r_{21}^{3}}+\frac{\mathbf{r}_{23}}{r_{23}^{3}}+
\frac{\mu\mathbf{r}_{24}}{r_{24}^{3}}\qquad
\end{equation}
\begin{equation}\label{4.11}
\stackrel{..}{\mathbf{r}}_{3}=\frac{
\mathbf{r}_{31}}{r_{31}^{3}}+\mu\left(\frac{\mathbf{r}_{32}}{r_{32}^{3}}+
\frac{\mathbf{r}_{34}}{r_{34}^{3}}\right)
\end{equation}
\begin{equation} \label{4.12}
\stackrel{..}{\mathbf{r}}_{4}=\frac{
\mathbf{r}_{41}}{r_{41}^{3}}+\frac{\mu\mathbf{r}_{42}}{r_{42}^{3}}+
\frac{\mathbf{r}_{43}}{r_{43}^{3}}.\qquad
\end{equation}
The above equations can be further simplified by using all
possible symmetries of the CSFBP \cite{Andras1} but we will not
use the symmetric equations of motion as we want to use the most
general form of the equations of motion to be able to see what
happens when we marginally break the symmetry.

The potential function may be written as
\begin{equation}
U=G\sum\limits_{i=1}^{4}\sum\limits_{j=1}^{4}\frac{m_{i}m_{j}}{r_{ij}}
\textrm{ \ \ \ \ }i\neq j,j<i,
\end{equation}
where
\[
{\bf r}_{ij}={\bf r}_{i}-{\bf r}_{j}.
\]

Then the energy $E$ of the system may be written as
\begin{equation}
E=T-U
\end{equation}
where $T$ is the kinetic energy given by
\begin{equation}
T=\frac{1}{2}\sum\limits_{i=1}^{4}m_{i}|\stackrel{.}{{\bf
r}_{i}}|^{2},
\end{equation}

\section{The Initial Conditions}
The numerical integration of (\ref{4.9}) to (\ref{4.12}) require
initial values for $\mathbf{r_1}$, $\mathbf{r_2}$, $\mathbf{r_3}$
and $\mathbf{r_4}$. In order to satisfy the initial conditions of
the CSFBP we immediately have
\begin{equation}
\mathbf{r_1} = (x_1,0)=-\mathbf{r_3} \qquad \textrm{ and } \qquad
\mathbf{r_2} = (x_2,0)=-\mathbf{r_4}
\end{equation}

We provide the values of $x_1$ and $x_2$ by hand by varying $x_1$
and $x_2$ in the intervals (0, 1.5] and (0, 1.5] with a step-size
of 0.008. The initial velocities $\mathbf{V_1}$ of $P_1$ and
$\mathbf{V_2}$  of $P_2$ are calculated using the following
relations which are derived using the energy and angular momentum
equations.

\begin{equation}\label{vel1}
{V_{1y}} =\frac{B-\sqrt{B^2-4AC}}{2A},
\end{equation}
where
\begin{equation}
A = 1+\frac{1}{\mu}\left(\frac{x_1}{x_2}\right)^2
\end{equation}
\begin{equation}
B= \frac{cx_1}{\left(\mu x_2\right)^2}
\end{equation}
\begin{equation}
C = \frac{c^2}{4x_2^2}-U-E_0
\end{equation}

\begin{equation}\label{vel2}
V_{2y} =\frac{c}{2x_2}-\mu V_{1y}\frac{x_1}{x_2}
\end{equation}

Where $V_{1y}$ and $V_{2y}$ are the $y$ components of
$\mathbf{V_1}$ and $\mathbf{V_2}$ respectively, $c =
\sqrt{\frac{C_0}{-E_0}}$ is the angular momentum of the system,
$E_0$ is negative of the energy $E$ and $U$ is the potential given
in section 4.2. The $x$ components of $\mathbf{V_1}$ and
$\mathbf{V_2}$ are set to zero for $t=0$. $C_0$ is the Szebehely
constant (Steves and Roy, 2001). More detailed analysis of how
$C_0$ can be used to determine the stability of the four body
problem can be found in chapter 5. But for now, we simply set
$C_0$ to be key values as used by \citeasnoun{AndrasMNRAS}.

Our aim in this chapter is to examine the motion and stability of
a variety of the CSFBP systems each with a different Szebehely
constant $C_0$. We also want to compare systems with the same
Szebehely constant. Thus we select the initial conditions so that
they result in the same $C_0$. For a given $C_0$, the variables
$V_{1y}$ and $V_{2y}$ can be calculated from equations
(\ref{vel1}) and (\ref{vel2}).

We investigate the following sets of mass ratios of the CSFBP with
several values of $C_0$ for each set of mass ratios.
\begin{enumerate}
\item $\mu = 1$
\begin{enumerate}
\item $C_0= 40, 46, 60$
\end{enumerate}
\item $\mu = 0.1$
\begin{enumerate}
\item $C_0= 0.6, 0.5, 0.9$
\end{enumerate}
\end{enumerate}

\section{The integrator}
The equations of motion for the CSFBP, equations (\ref{4.9}) to
(\ref{4.12}), are highly non-linear second order coupled
differential equations. It is not possible to find their
analytical solution. Therefore we have to use some numerical
technique. To do so we need to develop an integrator which is a
software in which the input data are the initial parameters of the
differential equations and the output data are the solution of
differential equations belonging to the input data. To develop the
integrator we first need to find an accurate and fast numerical
method and then develop an environment where the input and output
data is efficiently handled.

The numerical method, we chose, for this integration project is a
15th order method with a adaptive step size control called the
Radau method of Everhart  \cite{everhart}. This method makes use
of Gauss-Radau spacing. It varies the time-step used according to
the rate at which the variables are changing and thus it makes an
efficient integrator that deals well with close encounters. We
used Microsoft Visual C++ 6.0 to develop the integrator for the
general four body problem and to construct the environment of the
integrator. This integrator can easily be generalized for $n$ body
problems with $n$ being an even integer.

\section{Procedure of Analysis}
\citeasnoun{Andras1} discussed the chaotic and regular structure
of the phase space of the Caledonian Symmetric Four Body Problem.
They gave some very interesting results about the relationship
between the global hierarchical stability of the CSFBP and the
chaotic and regular nature of its phase space. We further their
analysis by perturbing the symmetric initial conditions and using
the general four body equations of motion. We integrated the
system with nearly symmetric initial conditions as well as
symmetric initial conditions, while using the general four body
equations. We studied two indicative CSFBP systems stipulated by
$\mu=1$ (a quadruple stellar system) and $\mu=0.1$ (a binary star
system with two very massive planets or brown dwarf stars). For
each given mass ratio, $\mu$, we kept the energy of the system
constant, effectively providing a relative scale of size for the
system and then chose a range of increasing Szebehely constants
\cite{BonnieRoy1998}.

For any given $\mu$, $C_0$ and $E_0$ each CSFBP orbit is uniquely
determined by its initial values $r_1$ and $r_2$. Therefore the
nature of each orbit can be depicted by integrating it for some
reasonable time. For each orbit, we follow its evolution searching
for collision or the breaking of the symmetry of the problem.
Here,  'collision' is defined to be a close encounter between two
of the bodies such that the conservation of energy fails i.e.
there is a difference between the energy at $t=0$  and the energy
at the current time of the order greater than $10^{-15}$. The
symmetry is defined to be broken when $(x_1+x_3)^2+(y_1+y_3)^2
> 10^{-4}$ or $(x_2+x_4)^2+(y_2+y_4)^2 > 10^{-4}$. We constructed
$r_1r_2$ graphs giving different colors to each type of orbit. See
table 5.1 for a listing of the different color coded final events
possible. We have colored the orbit black if symmetry is broken,
red if there is a collision between $P_1$ and $P_2$ (12
collision), yellow if there is a collision between $P_1$ and $P_3$
(13 collision), magenta if there is a collision between $P_1$ and
$P_4$ (14 collision), blue if there is a collision between $P_2$
and $P_3$ (23 collision), green if there is a collision between
$P_2$ and $P_4$ (24 collision), cyan if there is a collision
between $P_3$ and $P_4$ (34 collision) and grey if nothing happens
during the whole process of integration. We will refer to such
orbits as \emph{stable}. The collisions here are not physical
collisions. It is the situation when any two of the four bodies
come very close to each other and the energy conservation fails.
In such situation we stop the integration and label the event as a
collision orbit.  Please note that we cannot conclude that such
orbits are stable as we do not know what will happen to them if we
integrate it for longer. Also it is not possible to conclude that
orbits which do not remain symmetric are unstable as it is
possible to have stable orbits which are not symmetric.

For $\mu=1$ for each $(C_0, E_0)$  we produce nine $r_1r_2$ graphs
which include:
\begin{enumerate}
\item Three $r_1r_2$ graphs with symmetric initial conditions i.e.
\begin{enumerate}
\item $10^4$ integration time \item $10^5$ integration time \item
$10^6$ integration time
\end{enumerate}
\item Three $r_1r_2$ graphs with a perturbation of $10^{-6}$ in
the $x$ component of $P_1$ to the symmetric initial condition
(a,b,c same as above). \item Three $r_1r_2$ graphs with a
perturbation of $10^{-5}$ in the $x$ component of $P_1$ to the
symmetric initial condition (a,b,c same as above).
\end{enumerate}
For $\mu=0.1$, for each $(C_0, E_0)$ we produce six $r_1r_2$
graphs:

\begin{enumerate}
\item Three $r_1r_2$ graphs with symmetric initial conditions
(a,b,c same as above). \item Three $r_1r_2$ graphs with a
perturbation of $10^{-5}$ in the $x$ component of $P_1$ to the
symmetric initial condition (a,b,c same as above).
\end{enumerate}

\begin{table*}[]
  \begin{center}
  \caption{Criterion and color codes for the different categories of orbits}\label{colorcode}
\bigskip
  \begin{tabular}{c||c||c} \hline
Criterion & Nature of Orbit & Color code  \\\hline \hline
Sundman inequality not true & Forbidden region to real motion & white \\
$(x_1+x_3)^2+(y_1+y_3)^2 > 0$ or&Non-symmetric& Black\\
  $(x_2+x_4)^2+(y_2+y_4)^2 > 0$&&   \\
  $r_{12}\approx 0$& 12 type of collision & red \\
  $r_{13}\approx 0$& 13 type of collision & yellow \\
  $r_{14}\approx 0$& 14 type of collision & magenta \\
  $r_{23}\approx 0$& 23 type of collision & blue \\
  $r_{24}\approx 0$& 24 type of collision & green \\
  $r_{34}\approx 0$& 34 type of collision & cyan \\
  &stable&grey\\
  \end{tabular}
\end{center}
\end{table*}

\pagebreak

\section{Results: The nature of orbits in the $r_1r_2$ space for the equal
mass case of the CSFBP, $\mu=1$}

The value of $E_0$ was fixed to be -7 and $C_0$ was chosen to be
40, 46 and 60. The initial values of $r_1$ and $r_2$ were varied
from 0 to 1.5 with a step size of 0.008.

Recall that we use the general four body integrator and it's not
necessary to have symmetric collision. For example a 12 collision
does not necessarily mean a 34 collision and vice versa. To have
guaranteed symmetric collisions one must use the equations of
motion of the CSFBP \cite{Andras1}. For color coding of the
different types of orbits see table \ref{colorcode}.

In this equal mass case, an interchange of the $r_1$ and $r_2$
produces the same physical orbit. Thus the categorization of the
orbits is symmetric with respect to the $r_1r_2$ line with the
exception that the 24 and 13 collisions are reversed. The initial
ordering of the equal mass case for $r_1\gg r_2$ is 1243 which
becomes 2134 for $r_2\gg r_1$. Therefore the resulting
categorization will be the same except that some of the collisions
will be reordered. Thus comparing the collision types from $r_1\gg
r_2$ space to $r_2\gg r_1$ space, 24 collisions becomes 13
collision and vice versa. 12, 14, 34 and 23 collisions remain the
same.

We can separate the graph into three regions: The double binary
region around $r_1 \approx r_2$ (DB) and two single binary regions
around $r_2 \gg r_1$ (SB1) and $r_1\gg r_2$ (SB2).

\subsubsection{C$_0$ = 40}

Figure (\ref{fig4.2}) shows the results of integrations for
$C_0=40$ with symmetric initial conditions. Figure (\ref{fig4.3})
and (\ref{fig4.4}) shows the results with perturbed initial
conditions with a perturbation of $10^{-6}$ and $10^{-5}$
respectively in the $x$ component of $P_1$.

What is immediately clear from the comparisons of figure
(\ref{fig4.2}a), (\ref{fig4.2}b) and (\ref{fig4.2}c), is that the
longer we integrate this CSFBP system the more unstable orbits we
unearth. The most chaotic region for $C_0=40$ are the single
binary regions, as most of the orbits end up in collisions after a
very small integration time, see figure (\ref{fig4.2}a). Most of
the collisions in the SB1 region are 24 type collisions. There are
also some 12 and 14 type collisions. The orbits in this region
always start in the 24 type hierarchy state which is the reason
for most of the orbits being 24 type collisions. In the SB2 region
we have the same number of 13 type collisions as we had of the 24
type collisions in SB1. The double binary region is surrounded by
12 and 34 type collisions. See Figure (\ref{fig4.2}a).

Figure (\ref{fig4.2}b) shows the results for the same problem but
for a longer integration time i.e. $10^{5}$. The few orbits in the
SB1 and SB2 region which survived the shorter version of
integration have ended up in collisions except for a small grey
island at the bottom i.e. close to the origin. This is a clear
indication of the chaotic nature of these regions as opposed to
the double binary region which is mostly grey. Again most of the
collisions in the SB1 and SB2 regions are of the type 13 and 24.
There are now also a large number of 14 type collisions both in
the SB1 and SB2 region. There is a group of orbits near the island
of the grey region in the SB1 and SB2 regions which fail our
symmetry criterion. The DB region now has many collisions near the
origin but still its mostly grey. The types of collisions we have
in the DB region are either of the type 12 or 34. See figure
(\ref{fig4.2}b).

When integrated for 1 million time steps of integration time, the
grey regions of figure (\ref{fig4.2}b) in the DB region become
black, except for a small island of grey region in the middle
(figure (\ref{fig4.2}c)). The grey areas at the beginning of the
SB1 and SB2 regions of figure (\ref{fig4.2}b) survive this very
long integration time and remain stable.

Figures (\ref{fig4.3}) and  (\ref{fig4.4}) show the analysis of
the same orbits discussed above but with slightly perturbed
initial conditions. It is immediately clear from the comparison of
figures (\ref{fig4.2}), (\ref{fig4.3}) and (\ref{fig4.4}) that
there is no significant difference between the results obtained
using symmetric initial conditions and with perturbed initial
conditions. 

\subsubsection{C$_0$ = 46}

Figure (\ref{fig4.5}) shows the results of integrations for
$C_0=46$ with symmetric initial conditions. Figure (\ref{fig4.6})
and (\ref{fig4.7}) shows the results for the same but with
perturbed initial conditions with a perturbation of $10^{-6}$ and
$10^{-5}$ respectively in the $x$ component of $P_1$.

Like the previous case the shorter version of integration shows a
stable double binary region surrounded by collisions of 12 and 34
types (figure \ref{fig4.5}a). The single binary regions (SB1 and
SB2) has fewer collisions than we had in the previous case but a
large number of orbits fail the symmetry breaking criterion. All
the collision in the SB1 region are the 24 type. The SB2 region
has the same characteristics because of the symmetry, but with the
collisions being 13 type collisions. See figure (\ref{fig4.5}a).

In figure (\ref{fig4.5}b) we show the same results but for a
longer integration time i.e. $10^5$. In the DB region, like the
previous case there are now a lot of collisions near the origin.
All these collisions are either 12 or 34 type collision with 12
being the most frequent. Most of the collisions in the SB1 region
are 24 type. There are a few 12 and 14 type collisions. Most of
the orbits fail the symmetry breaking criterion except for some
near the origin of the graph. Similarly in the SB2 region most of
the collisions are 13 type with the same number of 12 and 14 type
collisions and non-symmetric orbits as in the SB1 region. See
figure (\ref{fig4.5}b).

When integrated for 1 million time steps of integration almost all
of the grey area in the DB region become black except for a small
island of grey region between the collision and the non-symmetric
orbits. Most of the collisions are 12 and 34 types but there are
some 13 type collisions as well. The grey areas at the beginning
of SB1 and SB2 regions survive this long integration and remain
stable. Most of the collisions in the SB1 area are 24 type
collisions. There are some 14, 34 and 12 type collisions. See
figure (\ref{fig4.5}c).

Figure (\ref{fig4.6}) and  (\ref{fig4.7}) shows the analysis of
the same orbits discussed above but with slightly perturbed
initial conditions. It is immediately clear from the comparison of
figures (\ref{fig4.5}), (\ref{fig4.6}) and (\ref{fig4.7}) that
there is no significant difference between the results obtained
using symmetric initial conditions and with perturbed initial
conditions. The small numbers of single binary collisions in the
double binary area and of double binary collisions in the single
binary area  are indications of the fact that we are heading
towards hierarchical stability with increasing values of the
Szebehely constant.

The main characteristics of the orbits for both $C_0=40$ and
$C_0=46$ remain the same. In the single binary regions we have
comparatively more orbits failing the symmetry criterion in the
case of $C_0=46$ than we had for $C_0=40$. For longest integration
time we have slightly bigger island of stable orbits in the double
binary region than the previous case of smaller $C_0$.

\subsubsection{C$_0$ = 60}

Figure (\ref{fig4.8}) shows the results of integrations for
$C_0=60$ with symmetric initial conditions. Figure (\ref{fig4.9})
and (\ref{fig4.10}) shows the results for the same but with
perturbed initial conditions with a perturbation of $10^{-6}$ and
$10^{-5}$ respectively in the $x$ component of $P_1$.

The single binary and double binary regions are completely
disconnected as $C_0=60$ is larger than the critical value (Steves
and Roy, 2001). In the shorter version of integration (figure
\ref{fig4.8}a) the phase space appears to be stable as most of the
regions are grey. Both DB, SB1 and SB2 regions are surrounded by
collisions. In the DB region collisions are either of type 12 or
34. The collisions in the SB1 region are the 24 type collisions as
no other type of collisions are possible because the phase space
is disconnected. Similarly the only type of collisions in the SB2
area are 13 type.

As in all other cases when integrated for a little longer ($10^5$
integration time) the grey regions in the SB1 and SB2 regions turn
black except for a small island near the origin (figure
\ref{fig4.8}b). The SB1 region is surrounded by 24 type collisions
and the SB2 region is surrounded by the 13 type collisions. In the
DB region there are large numbers of collisions around the
$r_1=r_2$ line near the origin. All these collisions are either 12
or 34 type collisions. When integrated further i.e. for 1 million
time steps of integration there is no change in the graph which
shows that the CSFBP system is more stable with larger $C_0$
values.

Figures (\ref{fig4.9}) and  (\ref{fig4.10}) show the analysis of
the same orbits discussed above but with slightly perturbed
initial conditions. It is immediately clear from the comparison of
figures (\ref{fig4.8}), (\ref{fig4.9}) and (\ref{fig4.10}) that
there is no significant difference between the results obtained
using symmetric initial conditions and with perturbed initial
conditions, which shows that perturbing the  symmetric initial
condition does not effect the final evolution by much. Therefore
this equal mass CSFBP system is stable to small perturbations in
the $x-$coordinate of the initial conditions. The absence of
single binary collisions in the double binary area and of double
binary collisions in the single binary area shows that this system
is hierarchically stable.

The main characteristics of the orbits for all the values of the
Szebehely constant i.e. $C_0=40, 46$ and 60 remain the same. In
the current case, $C_0=60$, we have a much larger island of stable
and symmetric orbits in the double binary area.

The main conclusion from the figures is that as the value of $C_0$
increases, the phase space of the CSFBP becomes more stable and
the small perturbation of initial conditions does not effect the
final evolution of the equal mass CSFBP by much. As we have stated
earlier, non-symmetric orbits are not necessarily unstable.
Therefore it will be interesting to see, in the future, the
behavior of these orbits without the symmetry restrictions using
our general four body integrator.

\section{Comparison with Sz\'ell et.al (2004) analysis in $\mu=1$ case}

We used the general four body equations to study the long-term
evolution of the CSFBP system for different $C_0$ values with
symmetric and nearly symmetric initial conditions.
\citeasnoun{AndrasMNRAS} investigated the CSFBP with the symmetric
initial conditions and symmetric equations of motion, searching
for the regular and chaotic regions using the fast chaos detection
methods. Therefore it will be interesting to see if there are any
similarities between the two investigations. Before giving the
comparisons we first review their analysis for the $\mu=1$ case.
Please note that \citeasnoun{AndrasMNRAS} label the bodies in
order 1 2 3 4 which means that $m_1=m_4$ and $m_2=m_3$ while we
label the bodies in the order 1 2 4 3 which means that $m_1=m_3$
and $m_2=m_4$. Because of the difference in labelling we have the
following difference in labelling for the hierarchy states:  13
hierarchy state in \citeasnoun{AndrasMNRAS} case is 14 in our case
and their 23 hierarchy state is 24 in our case. From now on we
shall call the notation arising from 1 2 3 4 ordering of
\citeasnoun{AndrasMNRAS}, the CSFBP notation. All the CSFBP
original research (\cite{RoyandSteves1998},\cite{andras2}. etc)
uses this CSFBP notation. Whenever we compare our work with this
original CSFBP research, we shall indicate clearly which the CSFBP
notation is being use. If no indication is given, it should be
assumed that it is in our new notation.

\subsection{The stable and Chaotic behavior of the CSFBP- A
Review}\label{szell}

\citeasnoun{AndrasMNRAS} used fast chaos detection methods namely
the Relative Lyapunov Indicator (RLI) and the Smaller Alignment
Indices (SALI) to investigate the connection between the chaotic
behavior of the phase space and the global stability given by the
Szebehely Constant.

In order to determine the relationship between the global
hierarchical stability of the CSFBP and the chaotic and regular
nature of its phase space \citeasnoun{AndrasMNRAS} studied two
indicative CSFBP systems stipulated by $\mu = 1$ (a quadruple
stellar system) and $\mu = 0.1$ (a binary star system with two
very massive planets or brown dwarf stars) by choosing a fine grid
containing of the order of $150,000$ pairs of initial values of
$r_1$ and $r_2$ covering all regions of real motion in the
$r_1$$r_2$ plane. The RLI and SALI numbers for each CSFBP orbit,
using $10,000$ iterations for each case were numerically
calculated. They chose a range of increasing Szebehely constants
$C_0$ ie. $C_0 < C_{crit1}$ (unstable); $C_{crit1} < C_0 <
C_{crit2}$ ($23$ hierarchy stable-CSFBP notation) and
$C_0>C_{crit2}$ (stable).

The chaotic and regular nature of the orbits is identified in
Figures (\ref{figmu1}) and (\ref{figmu01}) for both methods of
fast chaos detection.

We now give a brief review of their results for $\mu=1$. A review
of their results $\mu=0.1$ case will be given in section
(\ref{xyz}).
\subsection{Review of Sz\'ell et.al (2004) results for the equal mass case, $\mu=1$}

They chose $C_0=10,40$ and 60 in this case. Note all hierarchies
described in this section are in the CSFBP notation used by
Sz\'ell et.al (2004).

\begin{figure*}
   \centering
   \resizebox{150mm}{!} {
   \includegraphics{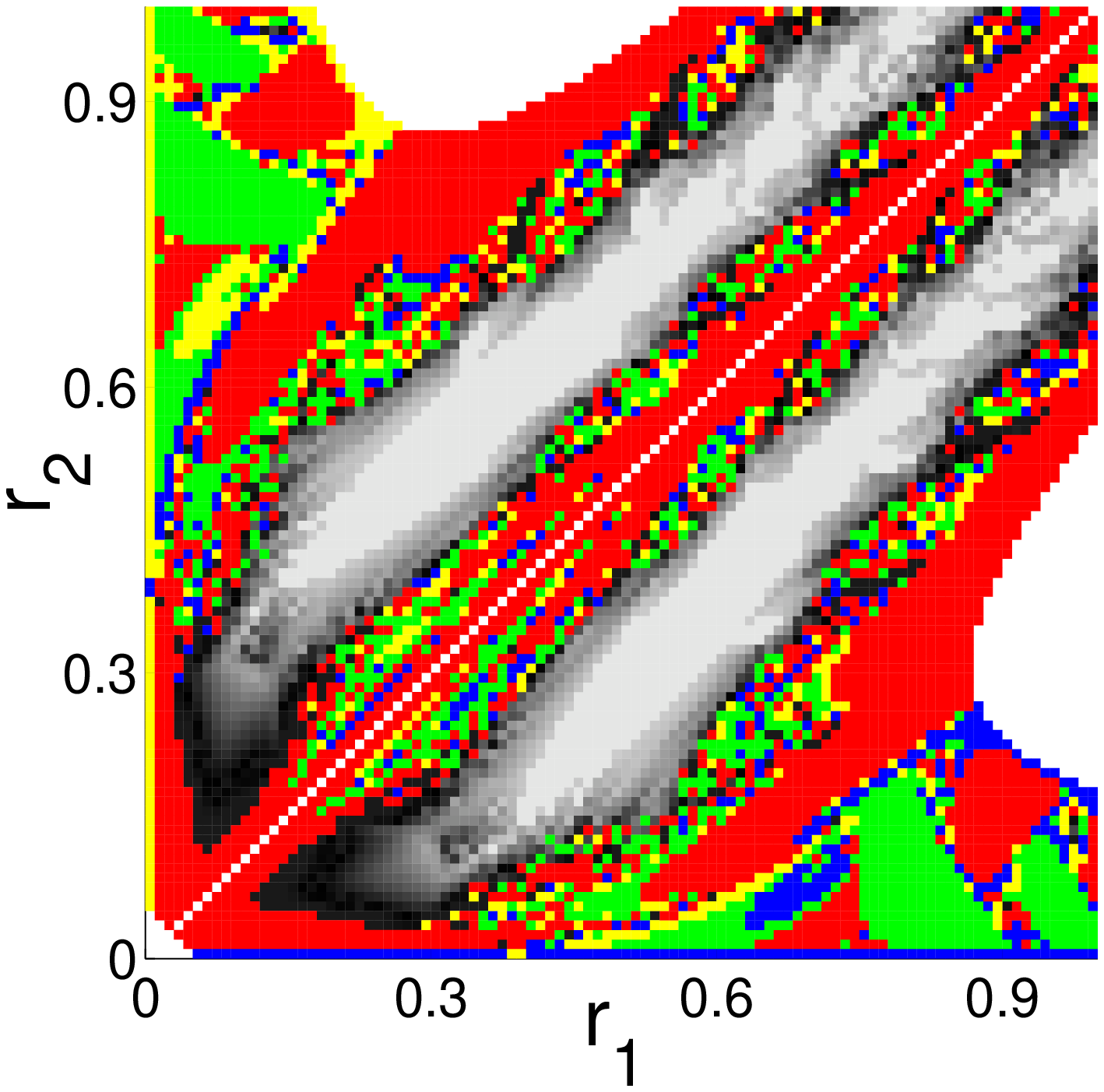}
   \includegraphics{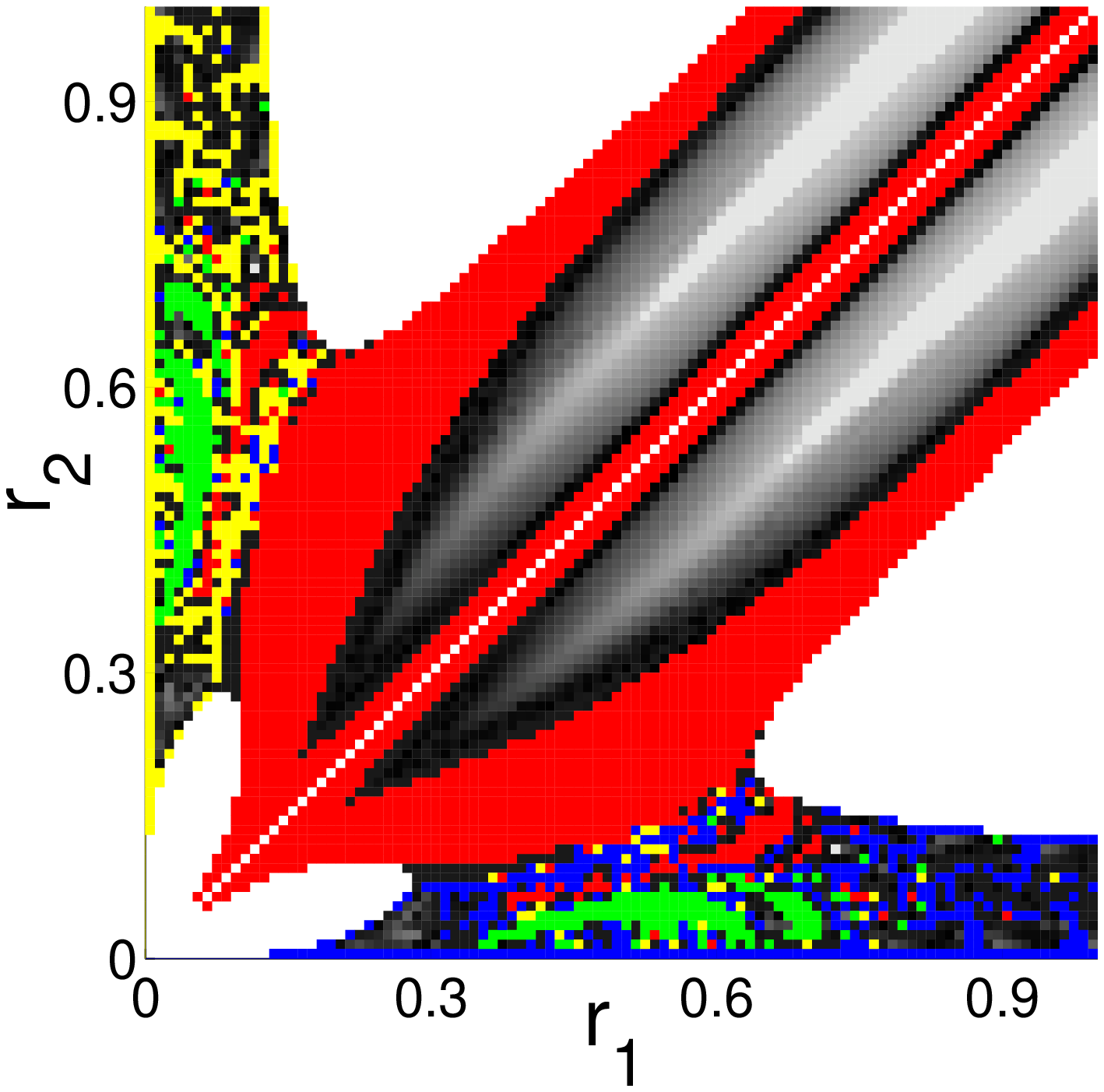}
   \includegraphics{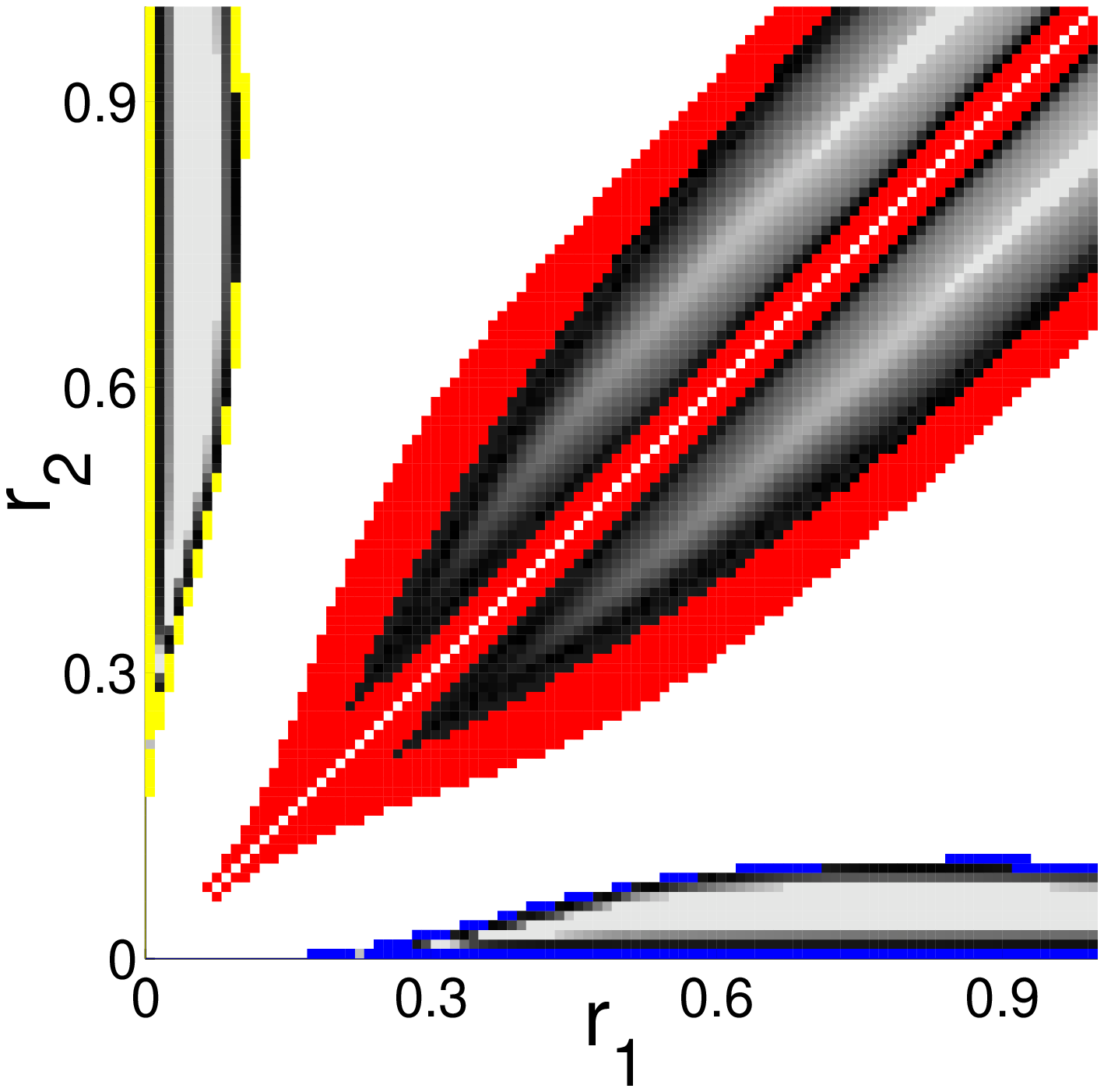}}
              \\
              \resizebox{150mm}{!} {
   \includegraphics{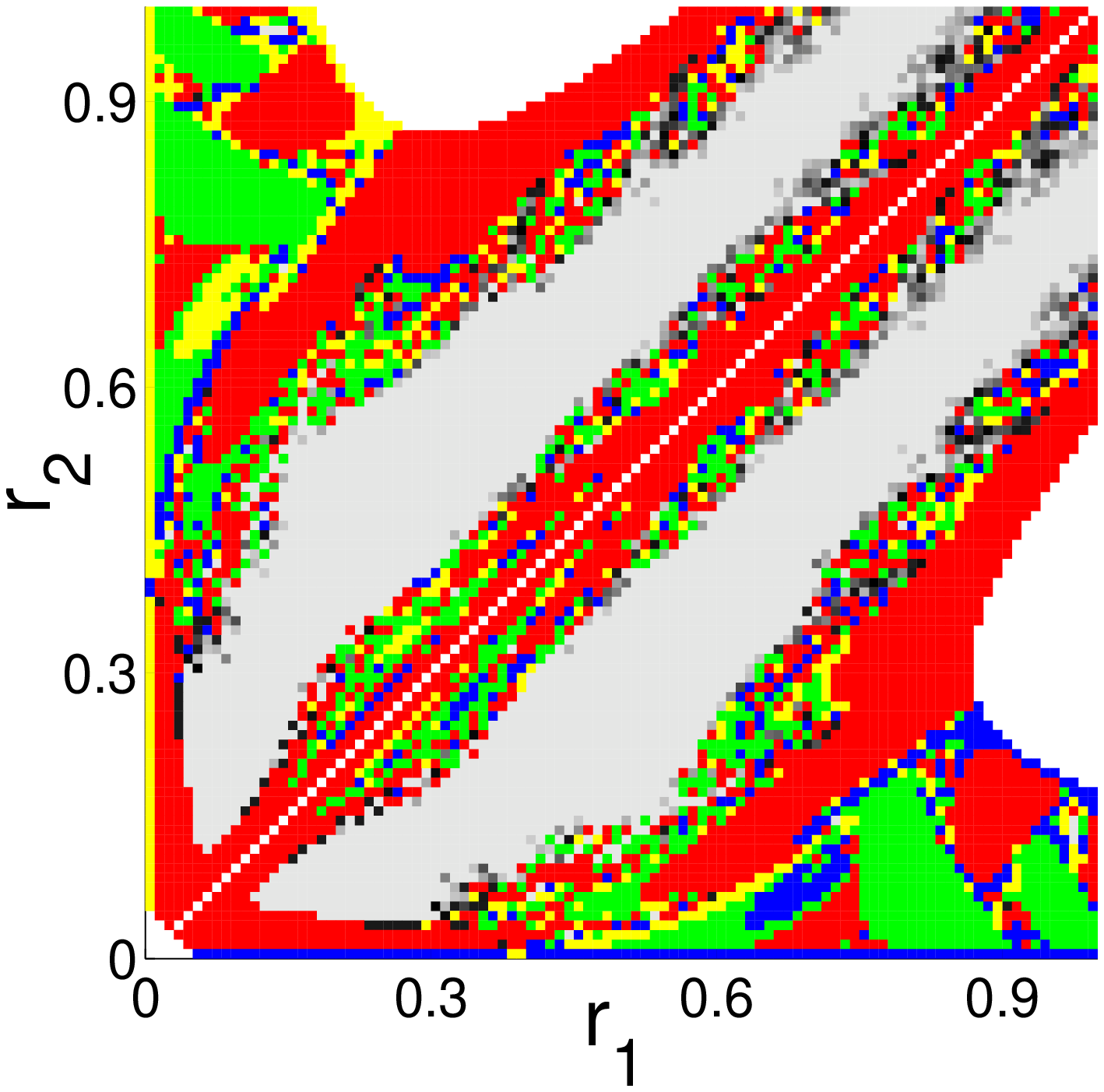}
   \includegraphics{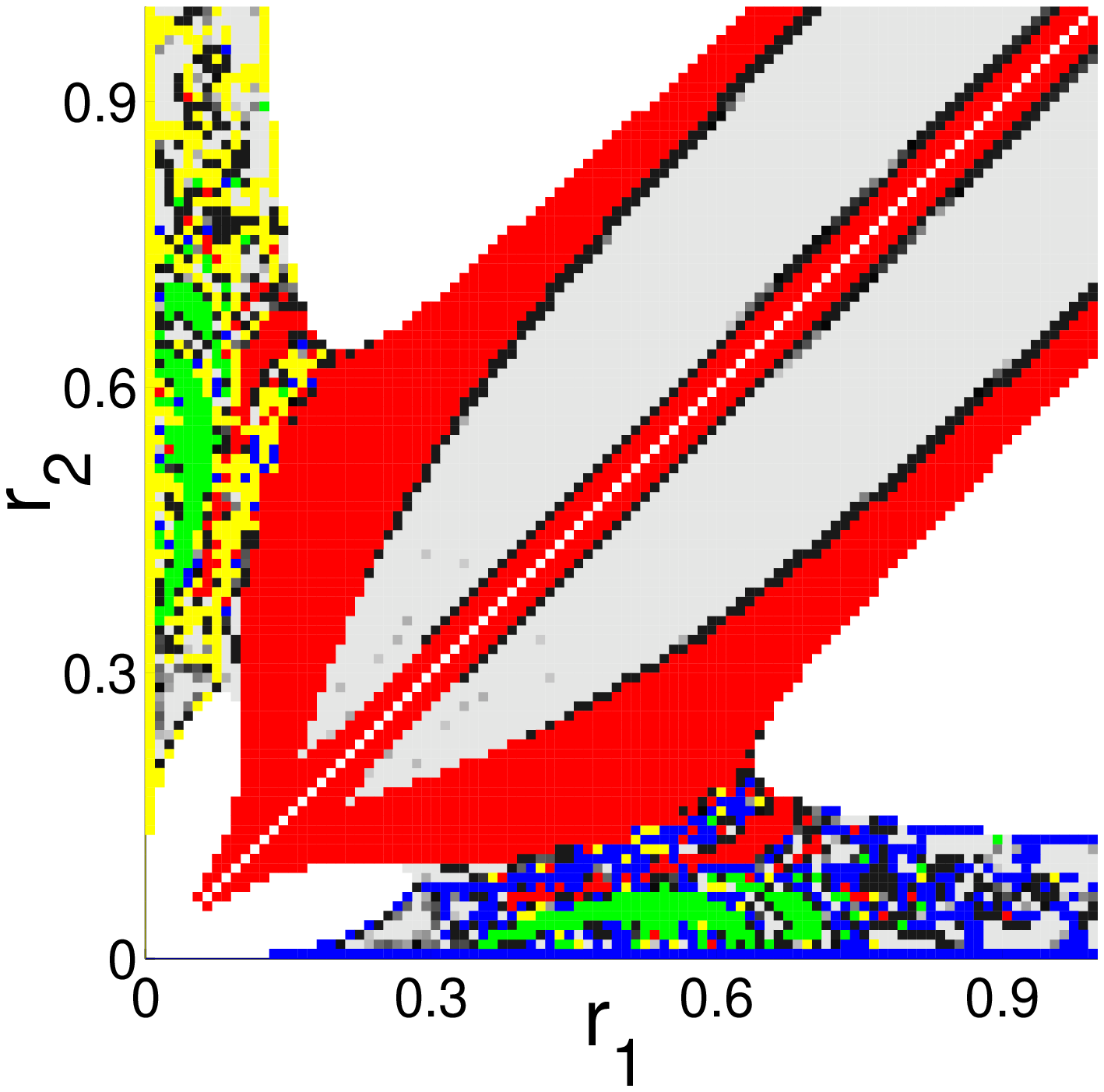}
   \includegraphics{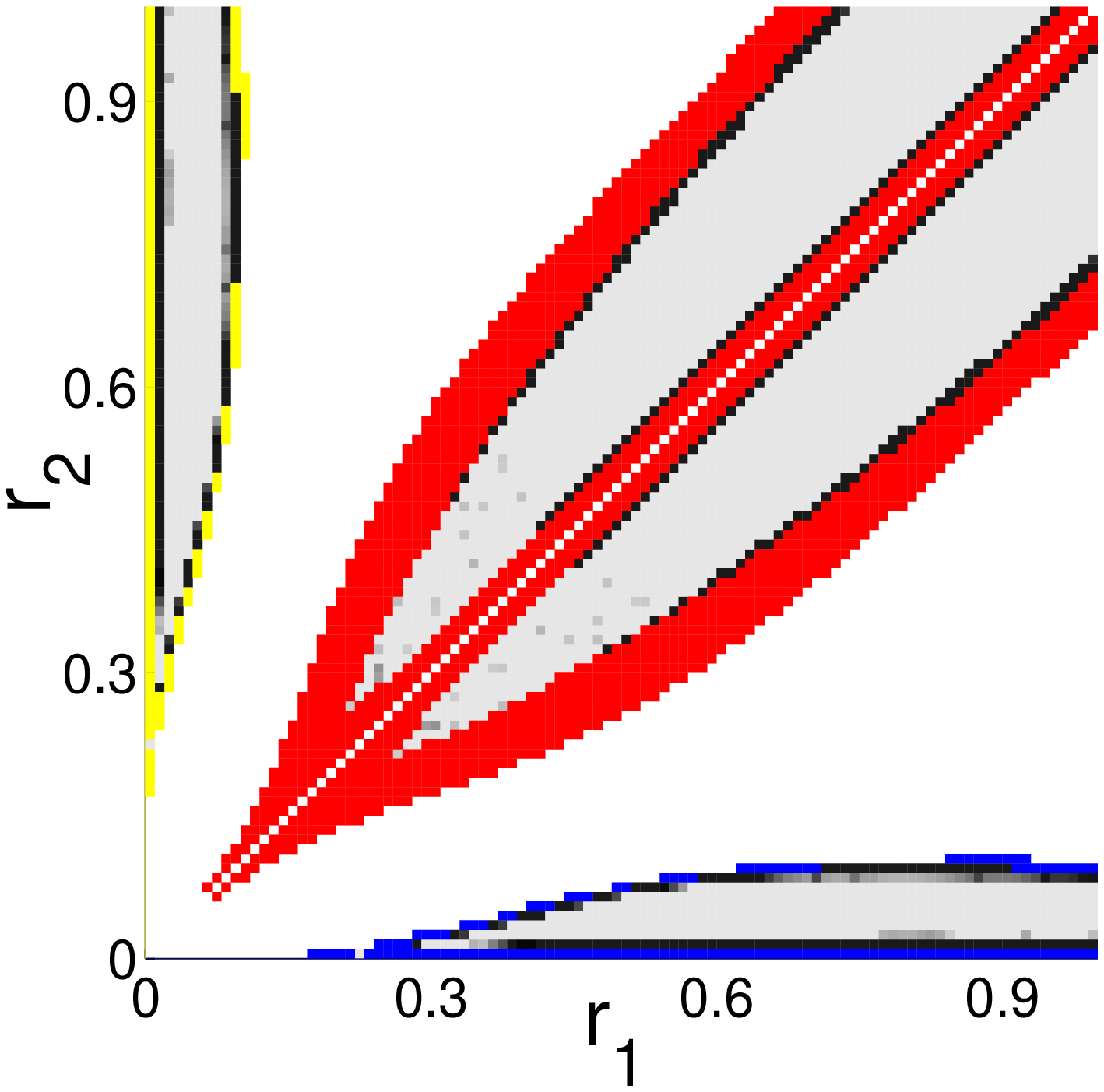}}
     \caption{$\mu = 1$, $E_0 = 7$. Graphs of the $r_1$, $r_2$ initial conditions
     for $C_0 = 10$, $40$, $60$ left to right. The top line of three graphs show the RLI
categorization of the different ($r_1$,$r_2$) orbits, while the
bottom line of three graphs show the SALI categorization for the
same $C_0$ values. The colors indicate collisions: red - "$12$"
type, green - "$13$" type, yellow - "$14$" type, blue - "$23$"
type. The light grey regions are regular, the dark grey regions
are undetermined, the black regions are chaotic, and the white
regions are forbidden for real motion.
              }
         \label{figmu1}
   \end{figure*}

\begin{description}
  \item[$C_0 = 10$] (Figure \ref{figmu1}, left). In this case $C_0$ is well below the critical
  value of $C_{crit1} = C_{crit2}=46.841$.

  In this case the regular part of the double binary region (DB)
  is surrounded by a thick chaotic cloud. There are a large number
  of different type of collisions denoted by different colors. The
  single binary (SB) regions are well separated from the DB region
  by 12 collisions (collisions between $P_1$ and $P_2$) (red) and
  contain only collision orbits indicating chaotic behavior. It can
  be concluded that the
  picture is very chaotic except for the central region of the DB
  area. The collision orbits dominate the phase space.

  \item[$C_0 = 40$] (Figure \ref{figmu1}, middle). In this case $C_0$ is just below the
  critical value of $C_{crit1} = C_{crit2}$.

In this case the SB regions are comparatively more chaotic than
the DB region. Most of the collisions are near the DB area.  For
large $r_1$ values, the orbits are more regular, being light grey
in both the  RLI and SALI graphs. In the SB1 region (where $r_1
\gg r_2$), real physical orbits are likely to be found for greater
$r_1$ values and similarly in the SB2 (where $r_1 \ll r_2$)
region, for greater $r_2$ values.

\item[$C_0 = 60$] (Figure \ref{figmu1}, right). In this case $C_0$
is  much greater than the critical value $C_{crit1} = C_{crit2}$.

  The SB and DB regions are now disconnected. The phase space is
  regular except close to the boundaries of real motion
  where chaotic motion and collisions exist. The DB, SB1, and SB2
  regions are surrounded by 12, 23 (collisions between $P_2$ and $P_3$)
  and 14 (collisions between $P_1$ and $P_4$) type of collisions
  respectively. Only one type of collision is seen in each disconnected
  region, corresponding to the one type of hierarchy state that is possible
  in that region.

  \end{description}

\subsection{A comparison of the Sz\'ell et.al (2004)results with our own CSFBP results}

We now compare the results of Sz\'ell et.al (2004) based on
symmetric initial conditions and symmetric equations of motion
with our results given in section 5.6 based on symmetric initial
conditions and the general four body equations of motion. In both
cases the main features of the CSFBP remain the same.
\begin{enumerate} \item The stability of the CSFBP system
increases as the value of the Szebehely constant increases. \item
The regions of real motion are always surrounded by collision
orbits. \item At the junction of the single binary and double
binary regions almost all of the orbits are collision orbits
\end{enumerate}

\begin{figure}[hbtp]
  \centerline{
    \epsfxsize=7.0cm
    \epsffile{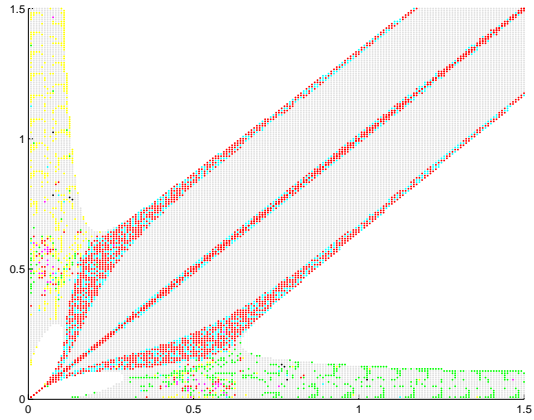}
  }
  \vspace{2pt}
   \centerline{(a)}
  \vspace{2pt}

  \centerline{
    \epsfxsize=7.0cm
    \epsffile{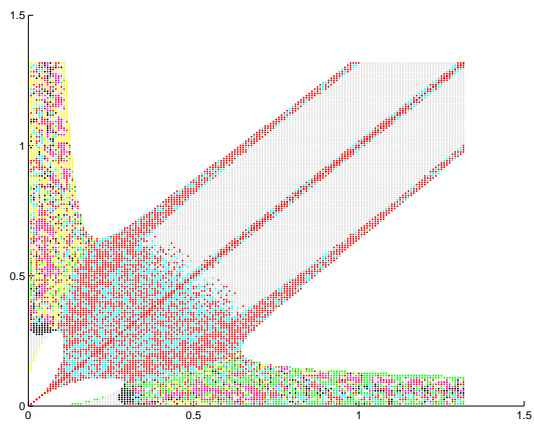}
  }
 \vspace{2pt}
  \centerline{(b)}
  \vspace{2pt}

  \centerline{
    \epsfxsize=7.0cm
    \epsffile{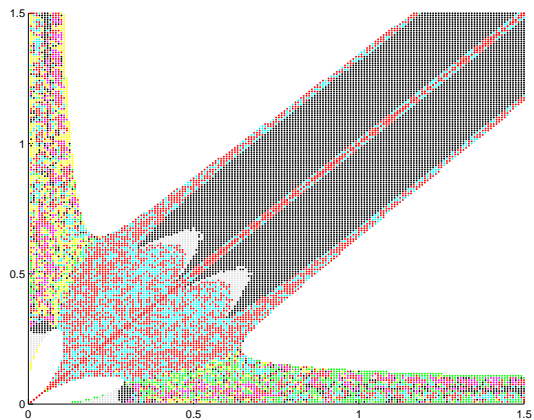}
  }
  \vspace{2pt}
  \centerline{(c)}
  \vspace{2pt}
  \caption{$\mu = 1, C_0=40, E_0= -7$ with no perturbation. Integration time a. $10^4$ time steps
   b. $10^5$ c. $10^6$. The colors indicate categories of orbits: red -12 type, yellow -13 type, magenta -14 type,
   blue -23 type, green -24 type, cyan -34 type, black -symmetry breaking and grey stable.}
  \label{fig4.2}
\end{figure}

From Sz\'ell et.al (2004) at $C_0=40$, (Figure \ref{figmu1},
middle), there is a dark cloud of black regions along the
$r_1=r_2$ line near the origin , which indicate strong chaos. From
our work, in figure (\ref{fig4.2}b and \ref{fig4.2}c) these orbits
end up in collisions. Also in both  cases the SB1 and SB2 regions
contain most of the collision orbits.

Similarly from Sz\'ell et.al (2004) at $C_0=60$, (Figure
\ref{figmu1}, right), there is a dark cloud of black regions
indicating chaos along the $r_1=r_2$ line near the origin. In our
work these regions (\ref{fig4.8}) after a long term integration
ends up in collisions which shows instability in that region. In
both Figures (\ref{figmu1}, right) and (\ref{fig4.8}a) the SB1 and
SB2 regions are shown to be regular and symmetry is maintained at
least at first; however when integrated for longer times most of
the SB1 and SB2 regions becomes black (non-symmetric orbits),
figures (\ref{fig4.8}b) and (\ref{fig4.8}c). This does not mean
that they are no longer regular orbits. It will be interesting to
further investigate these regions in the future. Overall the
double binary region in both the analysis are comparatively more
stable than the single binary regions. We now look at the
evolution of symmetric and nearly symmetric CSFBP for $\mu=0.1$.

\renewcommand{\baselinestretch}{1.5}

\begin{figure}[hbtp]
  \centerline{
    \epsfxsize=7.0cm
    \epsffile{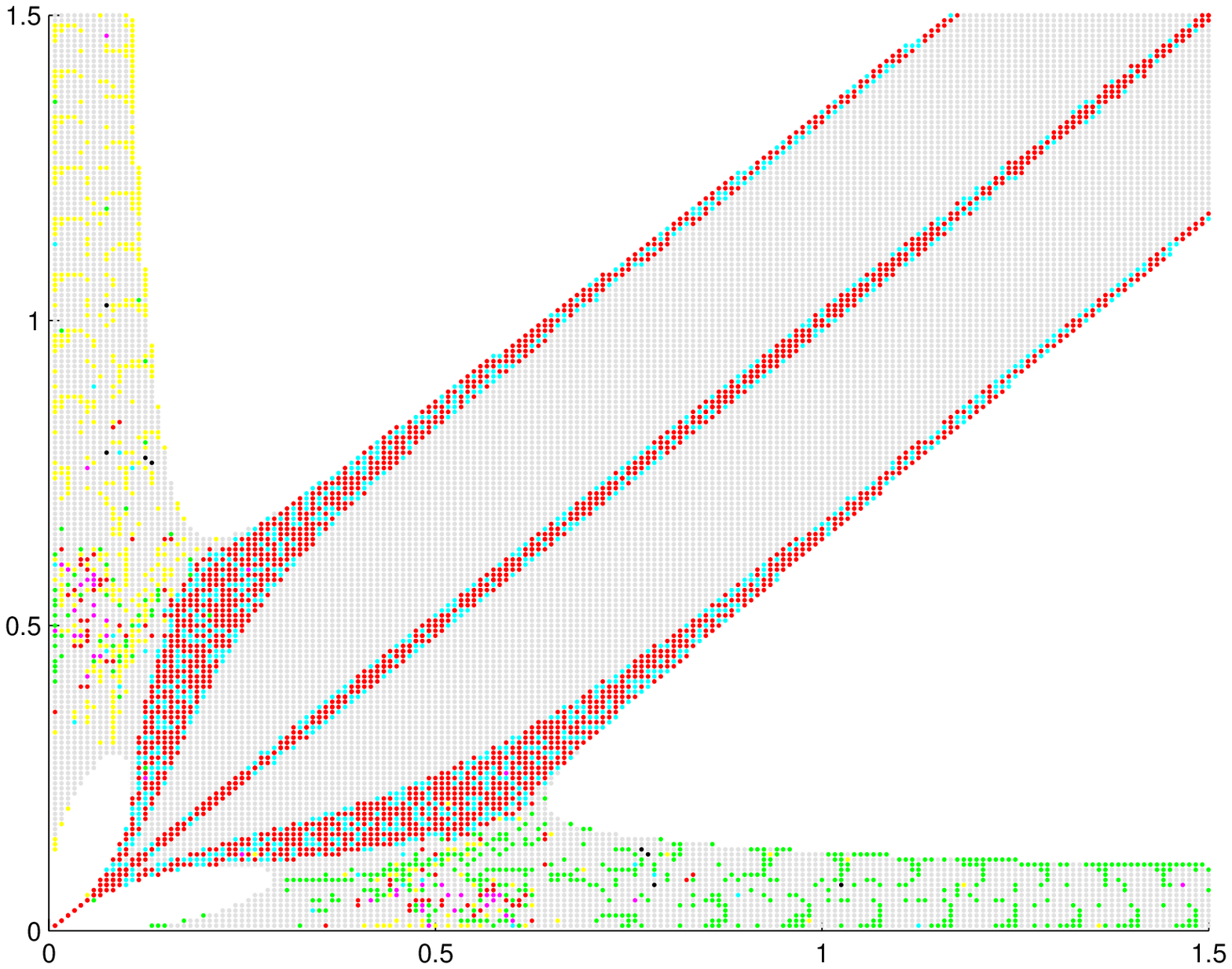}
  }
  \vspace{2pt}
   \centerline{(a)}
  \vspace{2pt}

  \centerline{
    \epsfxsize=7.0cm
    \epsffile{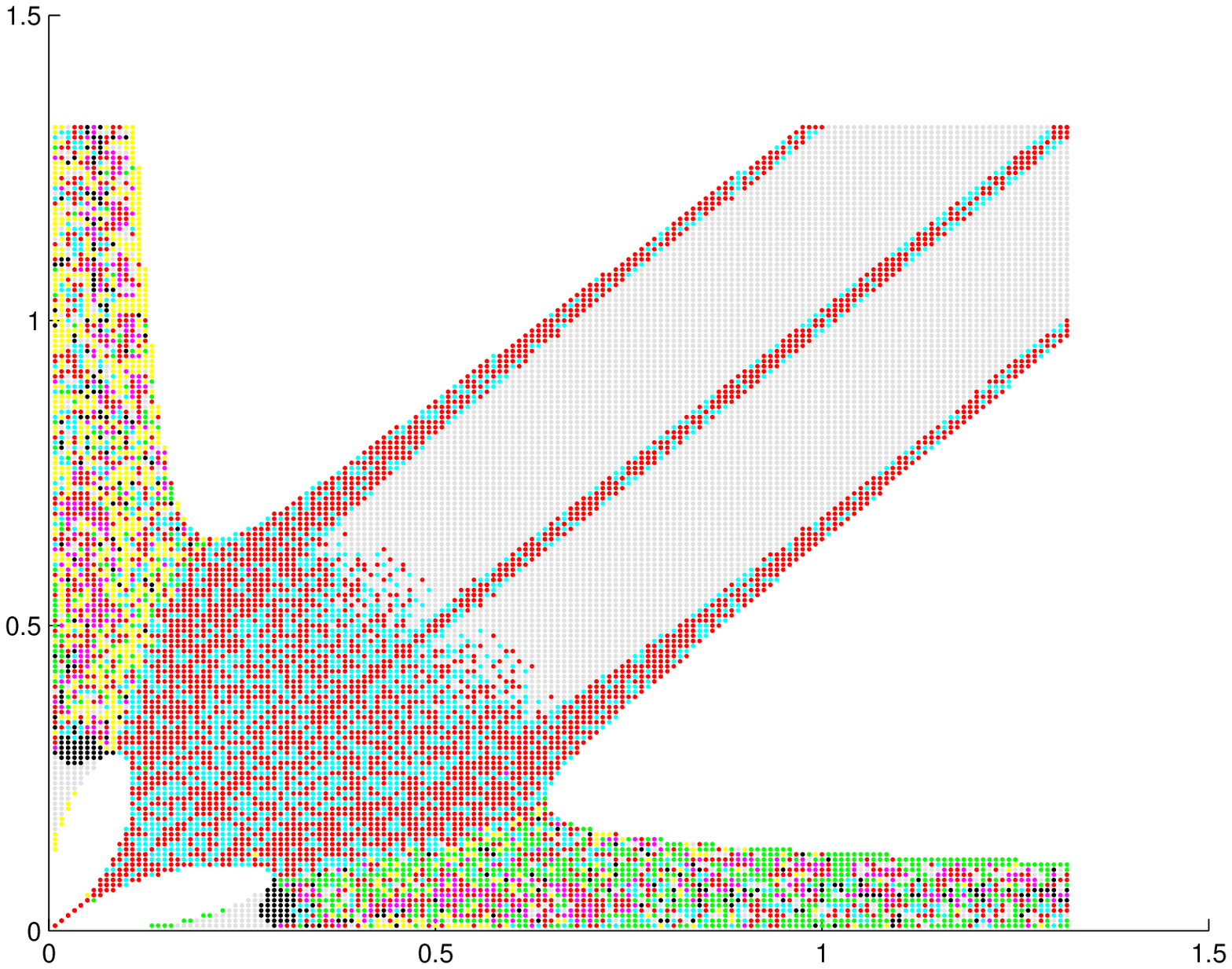}
  }
 \vspace{2pt}
  \centerline{(b)}
  \vspace{2pt}

  \centerline{
    \epsfxsize=7.0cm
    \epsffile{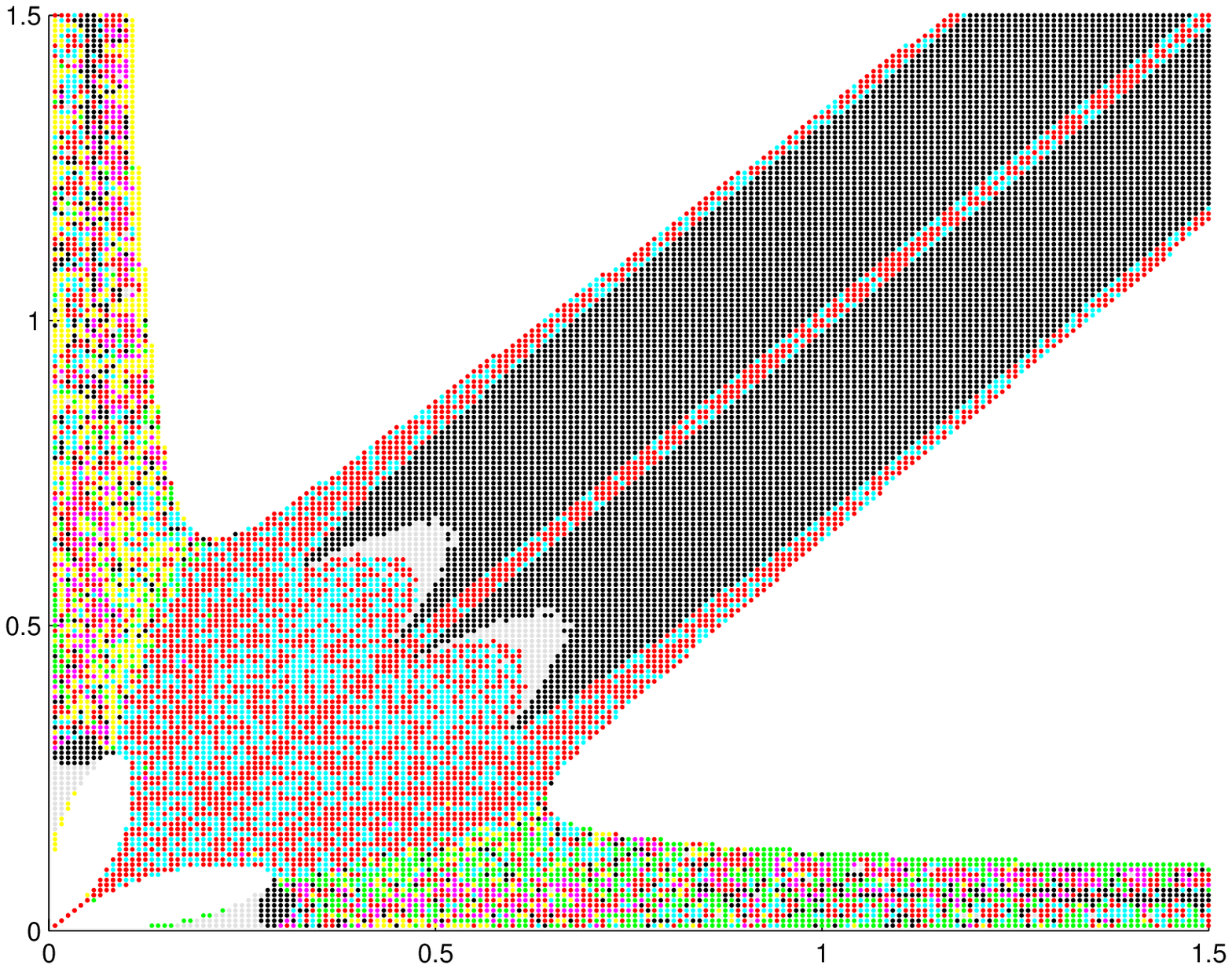}
  }
  \vspace{2pt}
  \centerline{(c)}
  \vspace{2pt}

  \caption{$\mu = 1, C_0=40, E_0= -7$ with perturbation of $10^{-6}$. Integration time a. $10^4$ time steps
   b. $10^5$ c. $10^6$. The colors indicate categories of orbits: red -12 type, yellow -13 type, magenta -14 type,
   blue -23 type, green -24 type, cyan -34 type, black -symmetry breaking and grey stable.}
  \label{fig4.3}
\end{figure}

\begin{figure}[hbtp]
  \centerline{
    \epsfxsize=7.0cm
    \epsffile{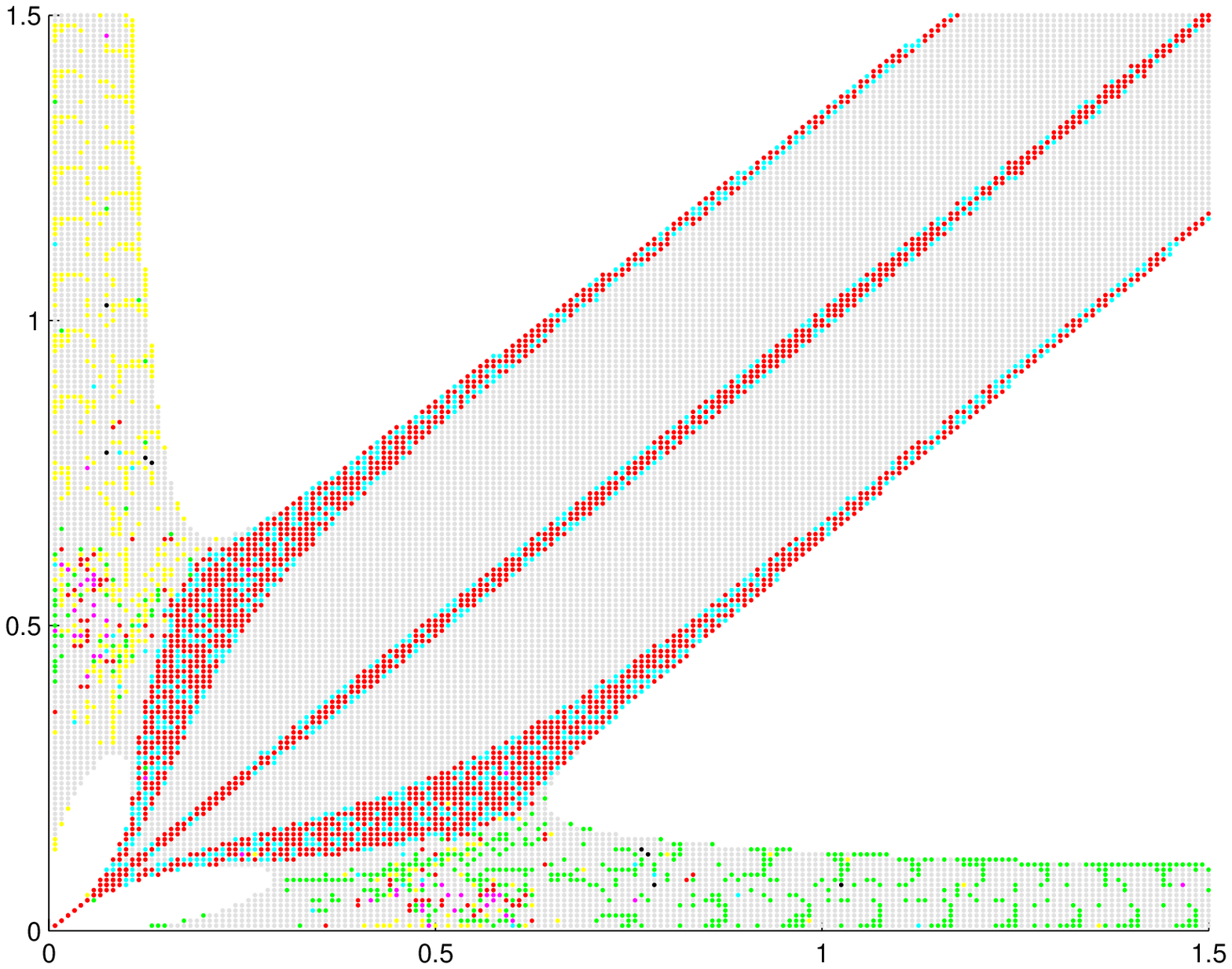}
  }
  \vspace{2pt}
   \centerline{(a)}
  \vspace{2pt}

  \centerline{
    \epsfxsize=7.0cm
    \epsffile{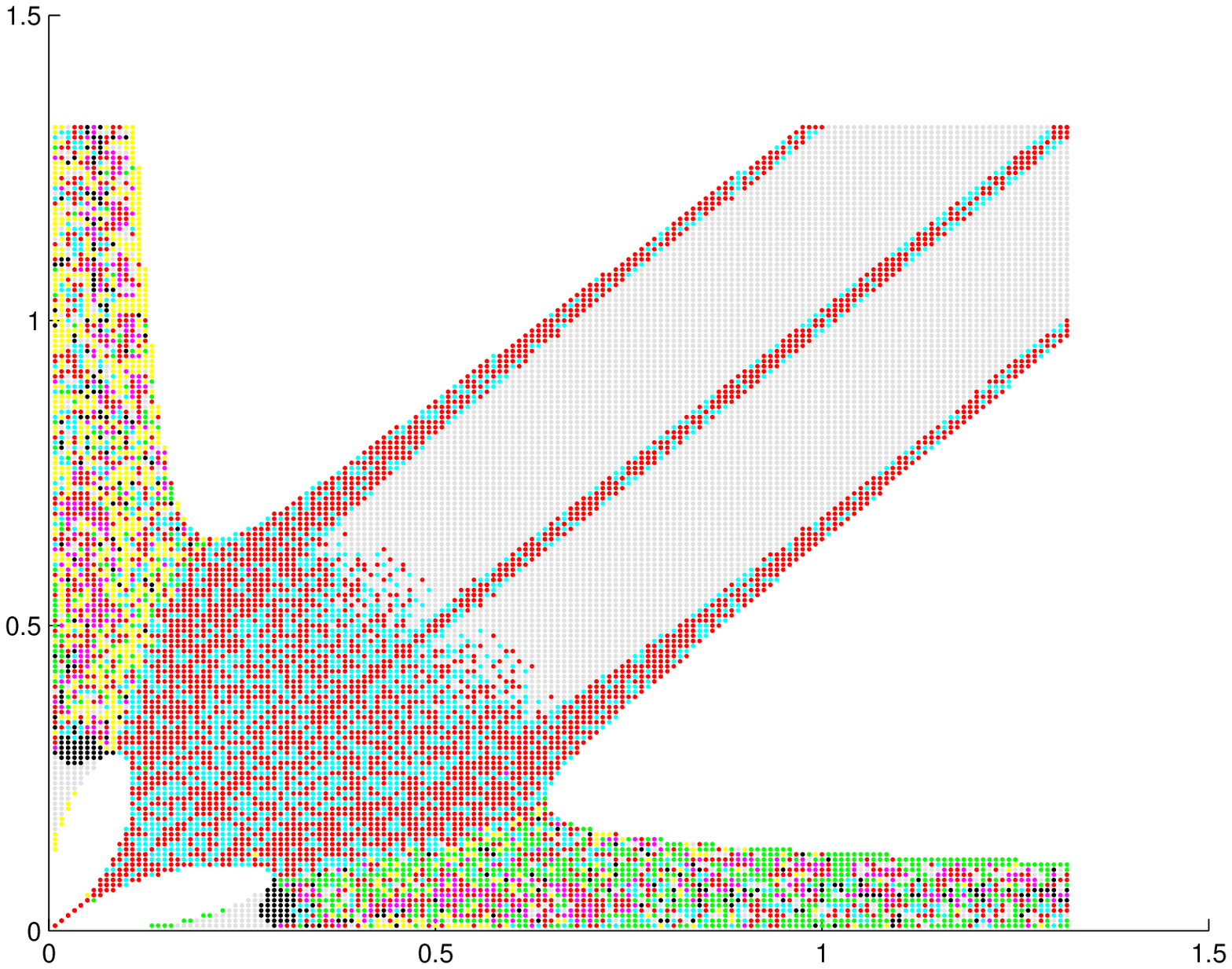}
  }
 \vspace{2pt}
  \centerline{(b)}
  \vspace{2pt}

  \centerline{
    \epsfxsize=7.0cm
    \epsffile{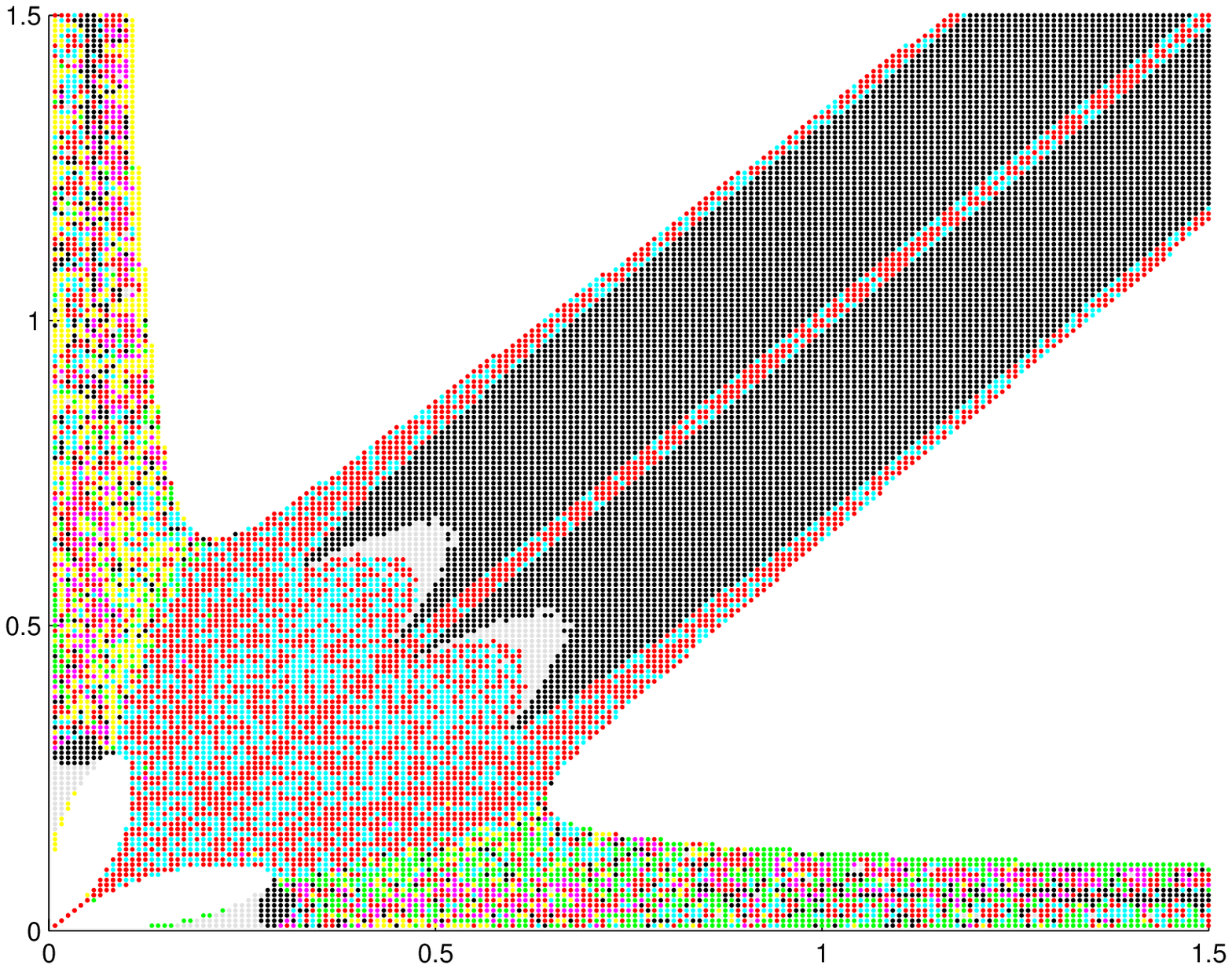}
  }
  \vspace{2pt}
  \centerline{(c)}
  \vspace{2pt}

  \caption{$\mu = 1, C_0=40, E_0= -7$ with perturbation of $10^{-5}$. Integration time a. $10^4$ time steps
   b. $10^5$ c. $10^6$. The colors indicate categories of orbits: red -12 type, yellow -13 type, magenta -14 type,
   blue -23 type, green -24 type, cyan -34 type, black -symmetry breaking and grey stable.}
  \label{fig4.4}
\end{figure}

\begin{figure}[hbtp]
  \centerline{
    \epsfxsize=7.0cm
    \epsffile{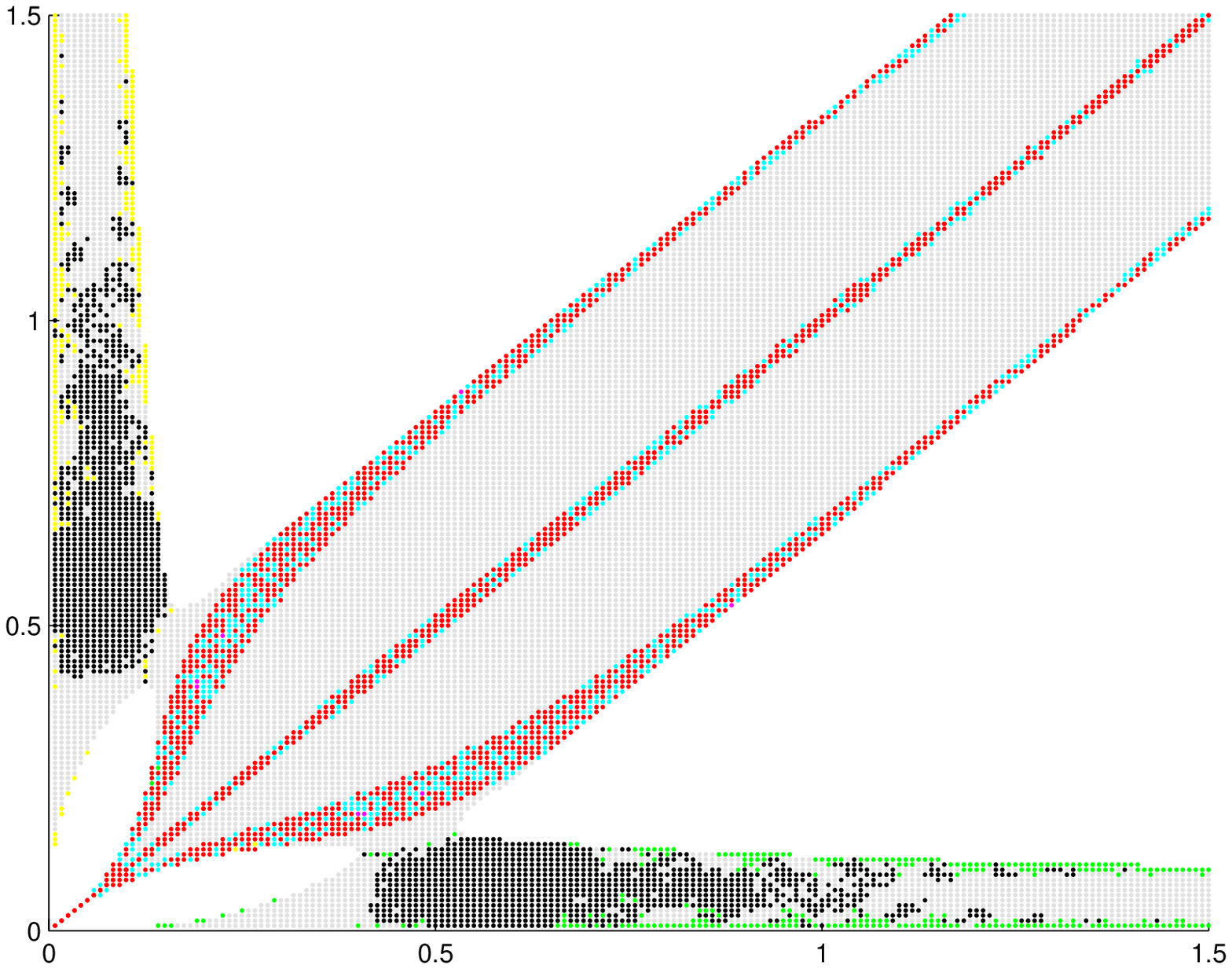}
  }
  \vspace{2pt}
   \centerline{(a)}
  \vspace{2pt}

  \centerline{
    \epsfxsize=7.0cm
    \epsffile{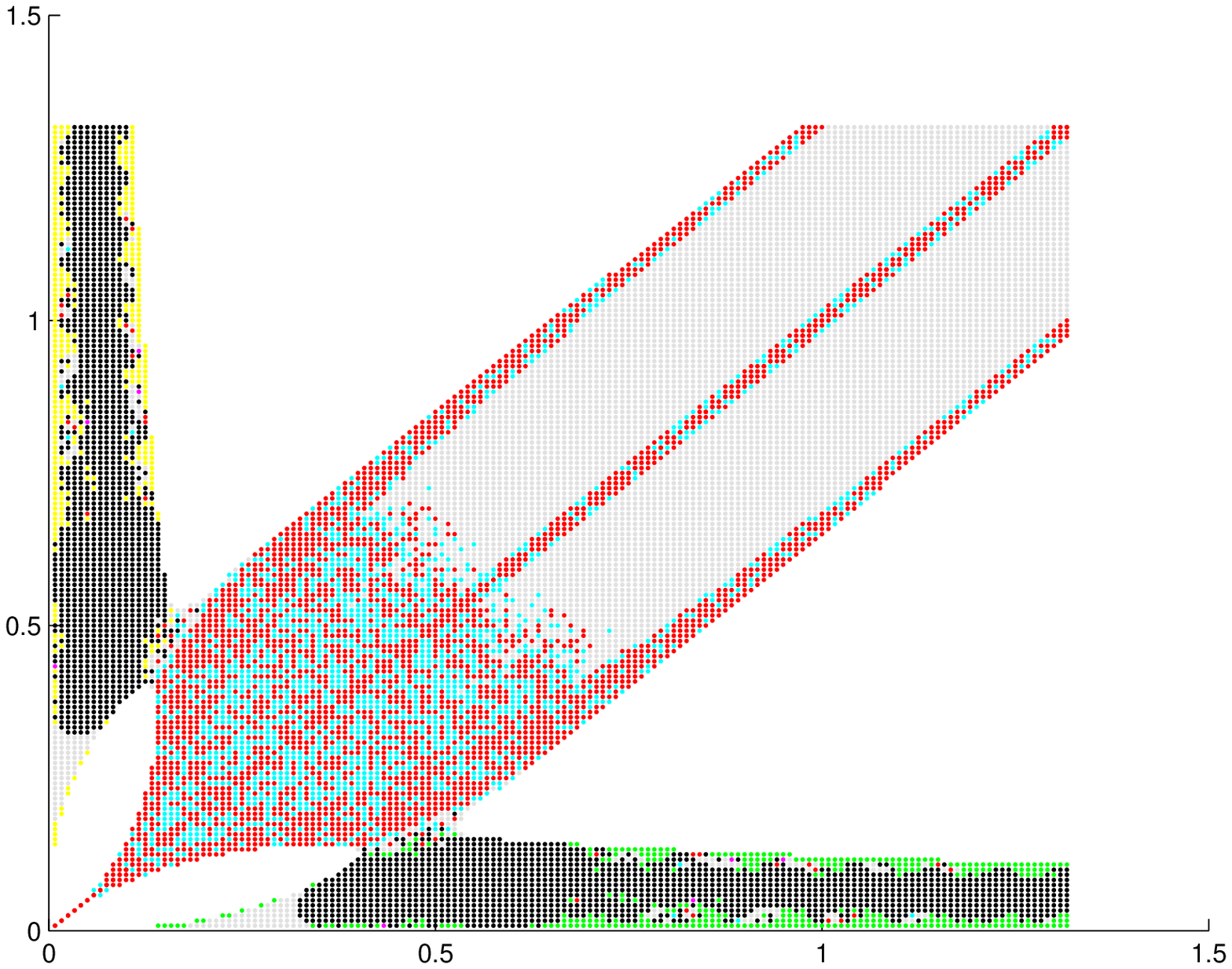}
  }
 \vspace{2pt}
  \centerline{(b)}
  \vspace{2pt}

  \centerline{
    \epsfxsize=7.0cm
    \epsffile{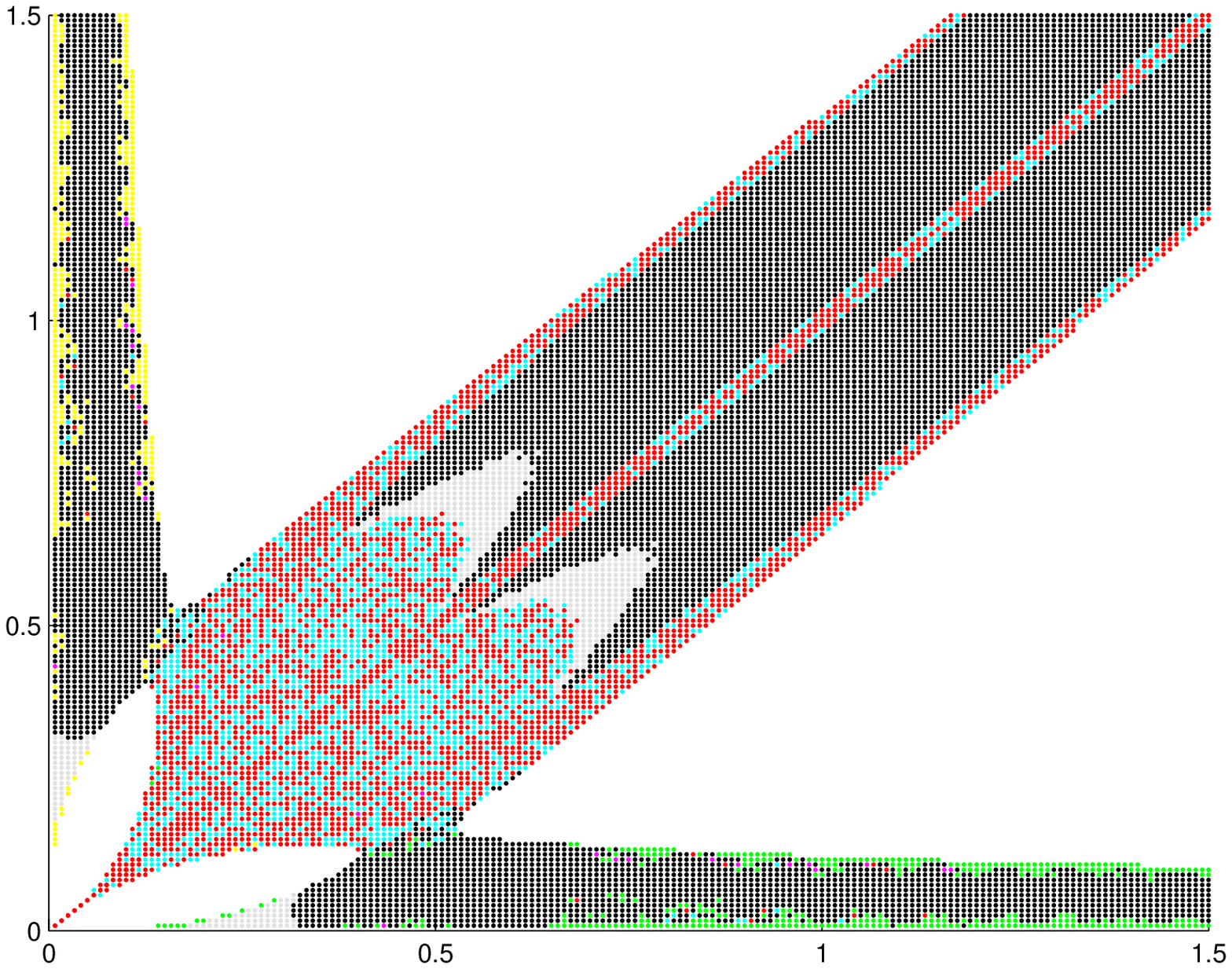}
  }
  \vspace{2pt}
  \centerline{(c)}
  \vspace{2pt}

  \caption{$\mu = 1, C_0=46, E_0= -7$ with no perturbation. Integration time a. $10^4$ time steps
   b. $10^5$ c. $10^6$. The colors indicate categories of orbits: red -12 type, yellow -13 type, magenta -14 type,
   blue -23 type, green -24 type, cyan -34 type, black -symmetry breaking and grey stable.}
  \label{fig4.5}
\end{figure}

\begin{figure}[hbtp]
  \centerline{
    \epsfxsize=7.0cm
    \epsffile{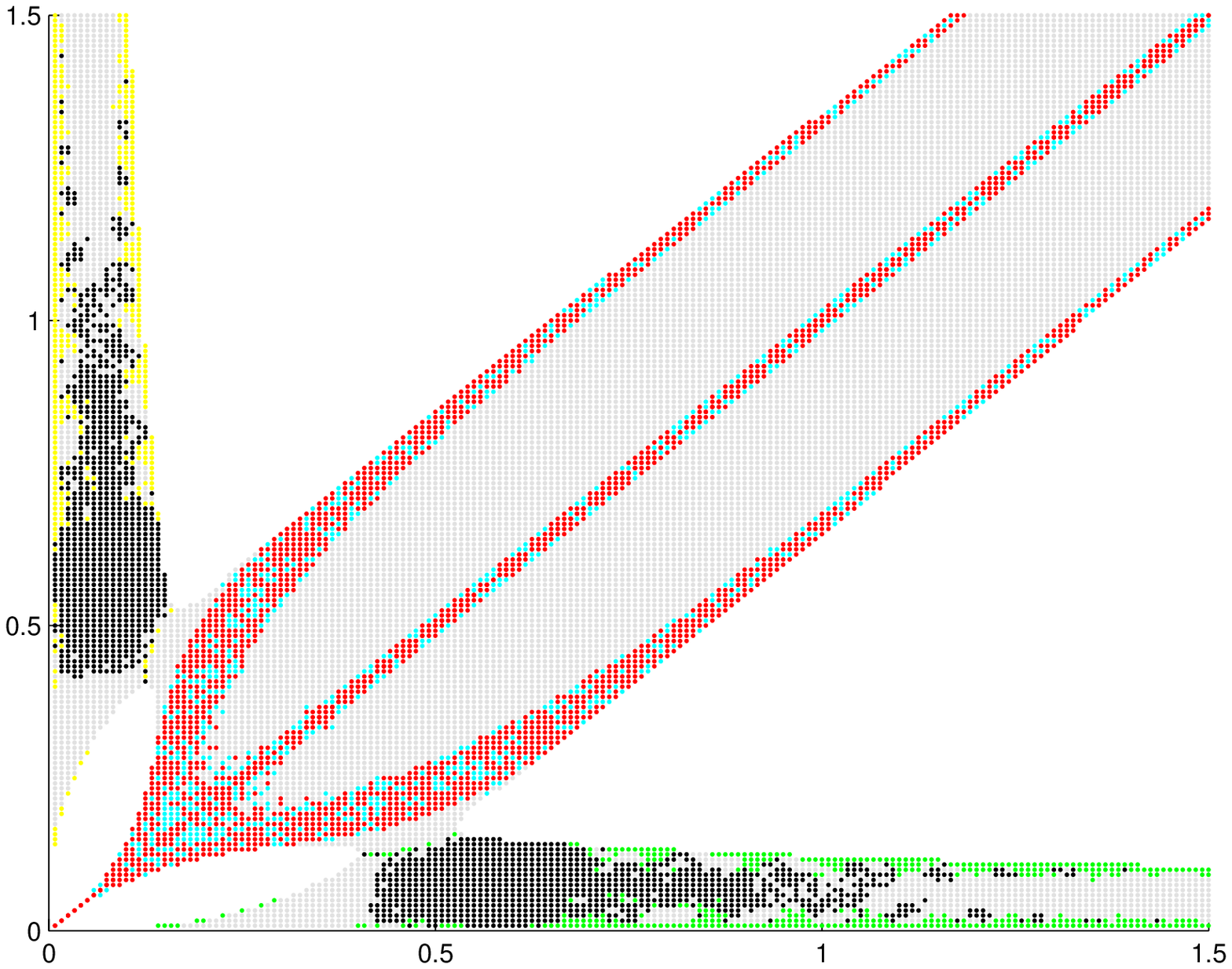}
  }
  \vspace{2pt}
   \centerline{(a)}
  \vspace{2pt}

  \centerline{
    \epsfxsize=7.0cm
    \epsffile{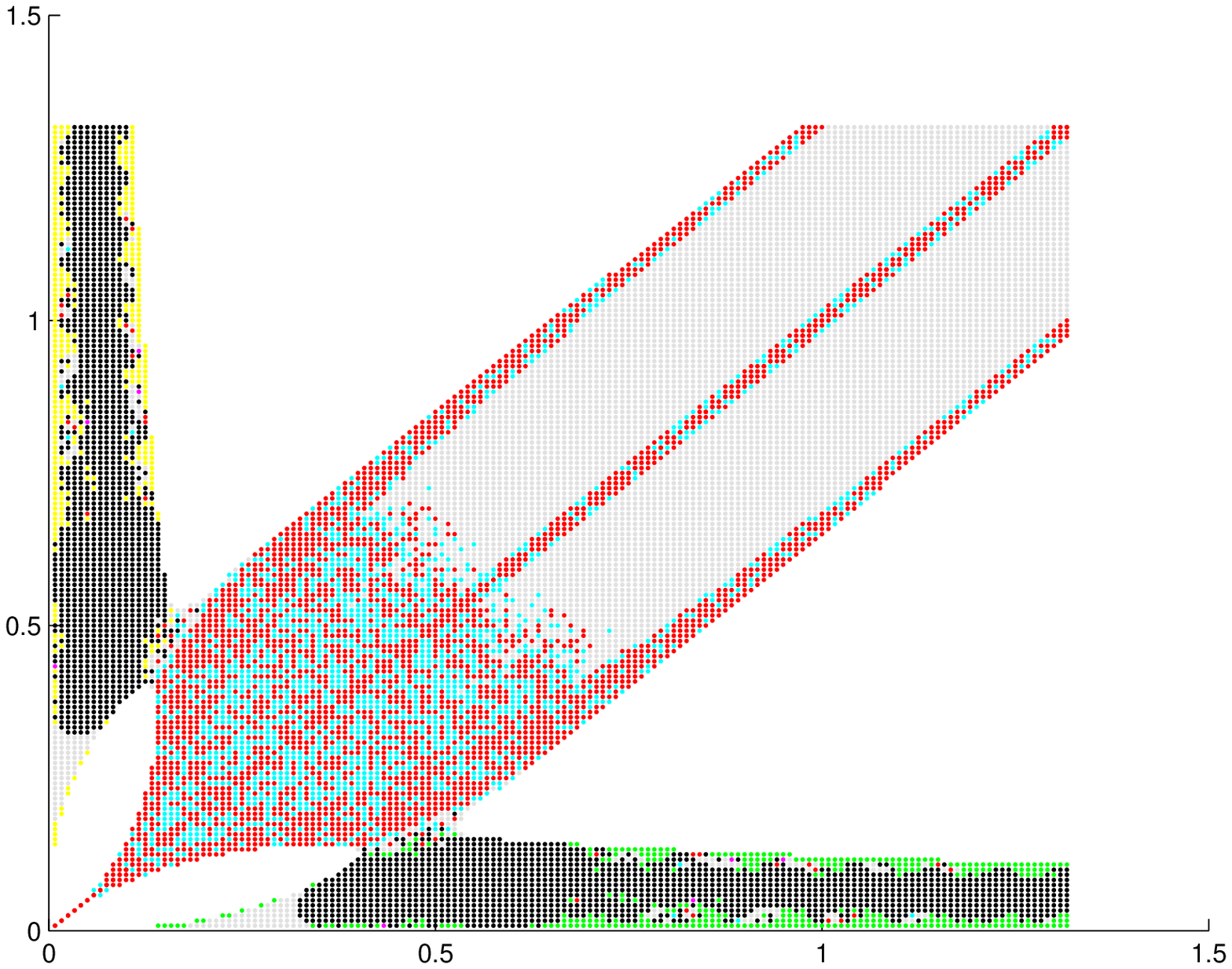}
  }
 \vspace{2pt}
  \centerline{(b)}
  \vspace{2pt}

  \centerline{
    \epsfxsize=7.0cm
    \epsffile{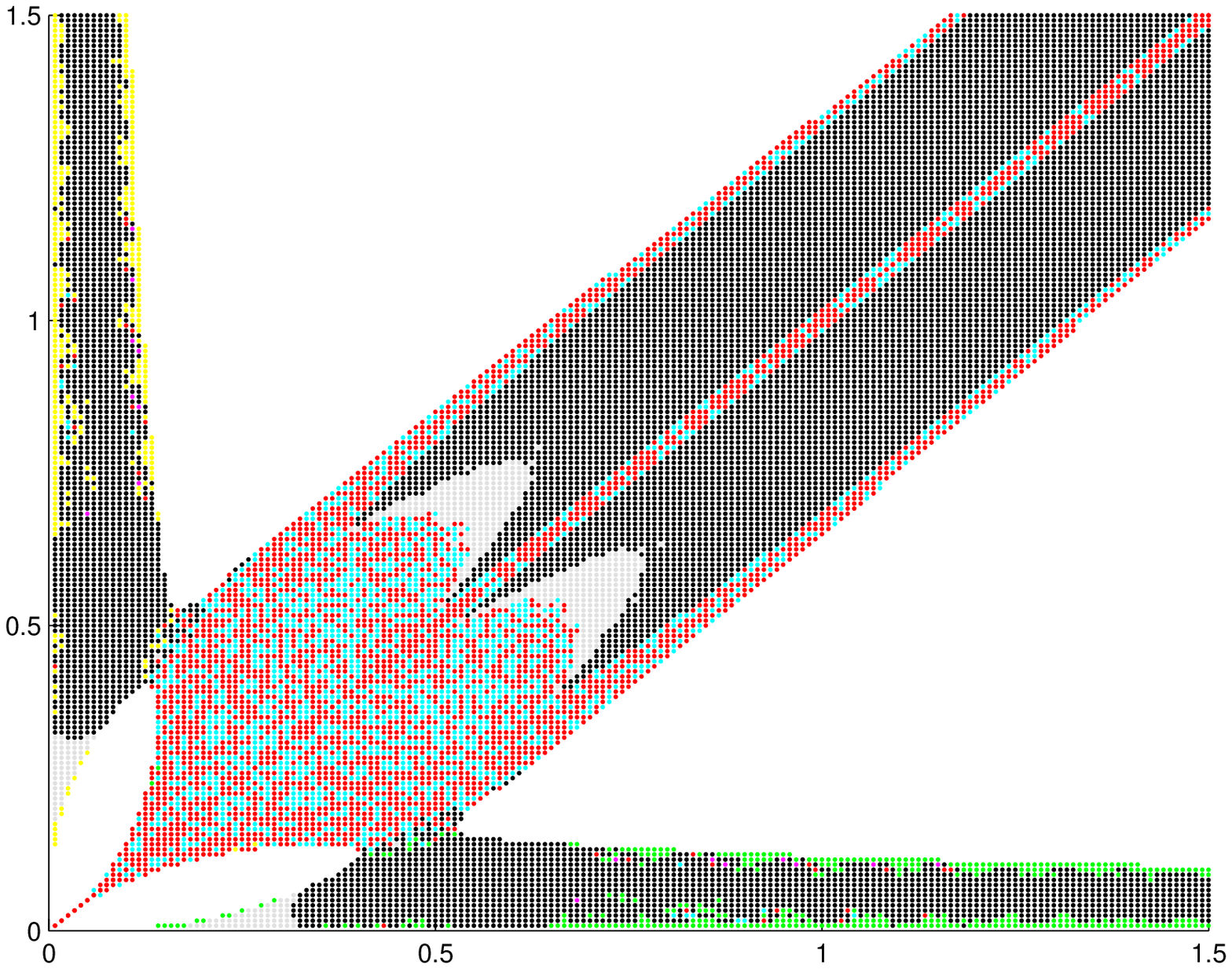}
  }
  \vspace{2pt}
  \centerline{(c)}
  \vspace{2pt}

  \caption{$\mu = 1, C_0=46, E_0= -7$ with perturbation of $10^{-6}$. Integration time a. $10^4$ time steps
   b. $10^5$ c. $10^6$. The colors indicate categories of orbits: red -12 type, yellow -13 type, magenta -14 type,
   blue -23 type, green -24 type, cyan -34 type, black -symmetry breaking and grey stable.}
  \label{fig4.6}
\end{figure}

\begin{figure}[hbtp]
  \centerline{
    \epsfxsize=7.0cm
    \epsffile{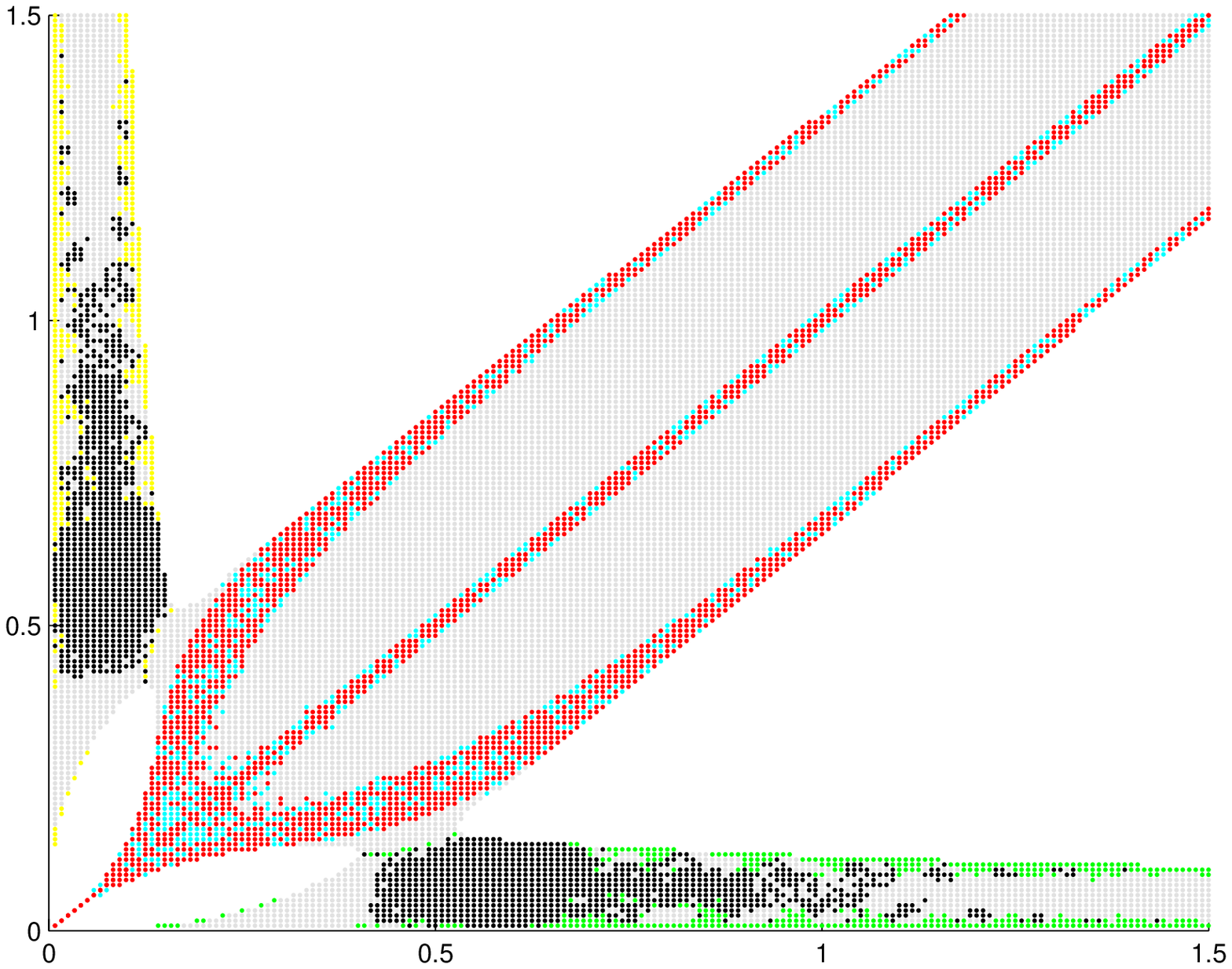}
  }
  \vspace{2pt}
   \centerline{(a)}
  \vspace{2pt}

  \centerline{
    \epsfxsize=7.0cm
    \epsffile{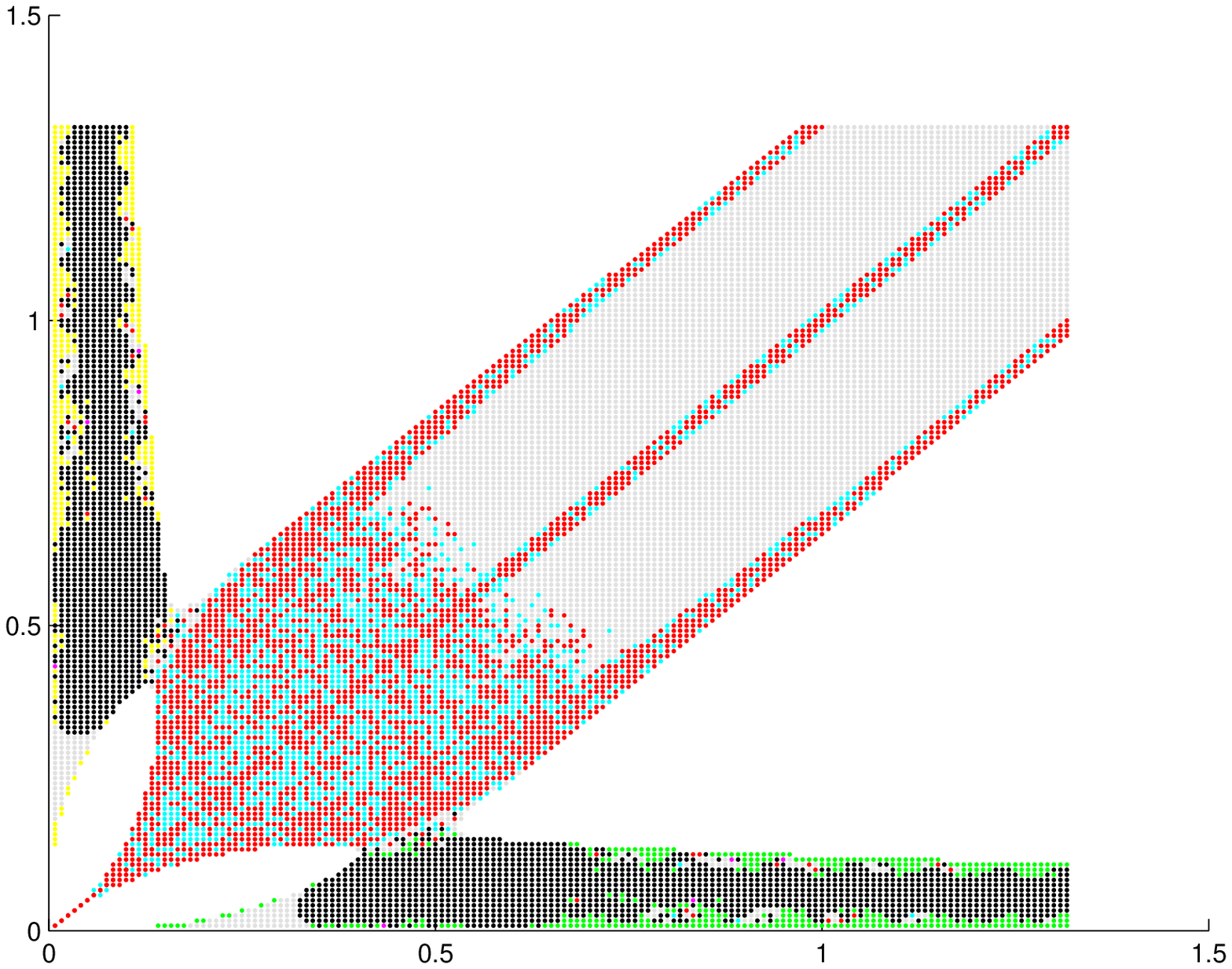}
  }
 \vspace{2pt}
  \centerline{(b)}
  \vspace{2pt}

  \centerline{
    \epsfxsize=7.0cm
    \epsffile{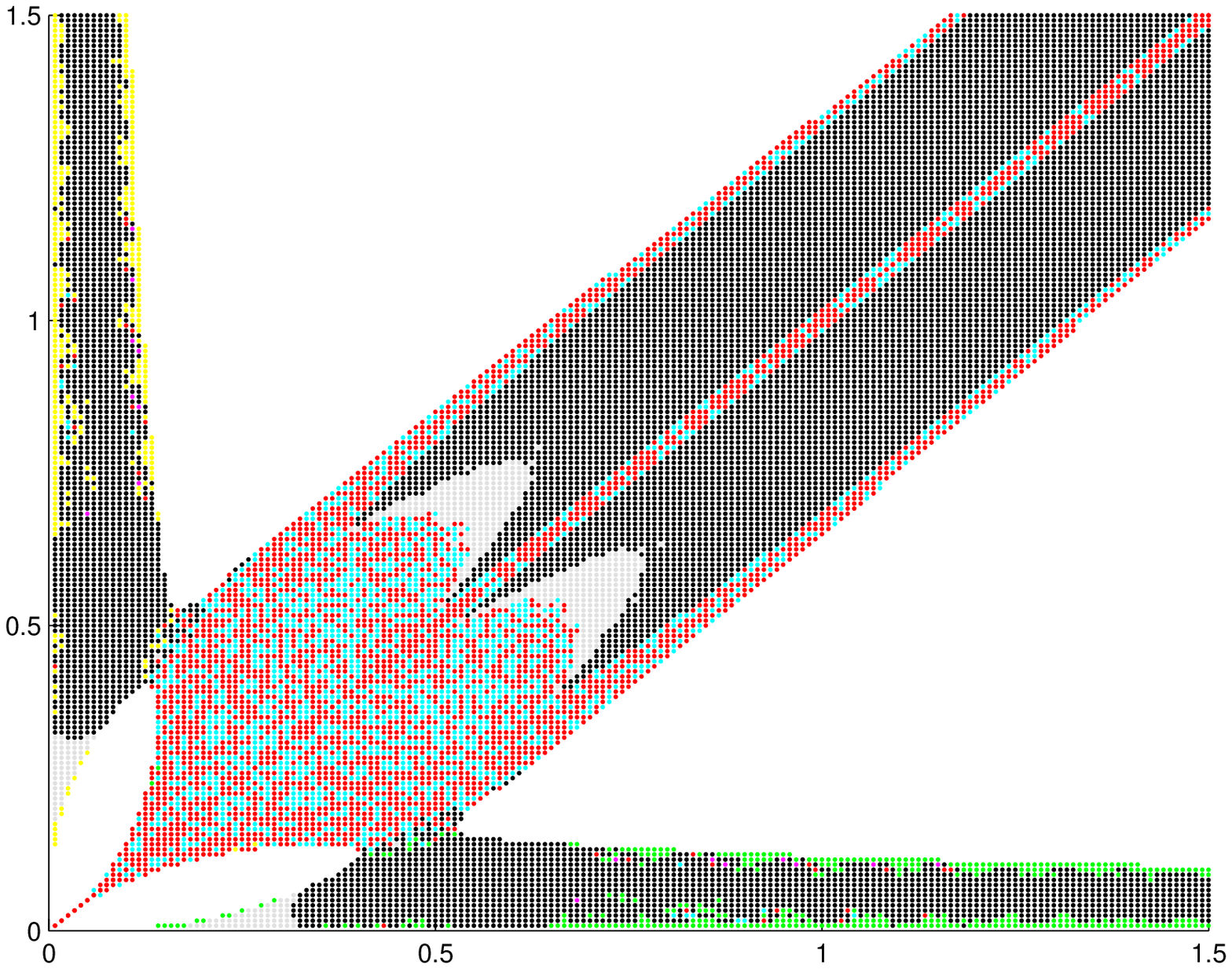}
  }
  \vspace{2pt}
  \centerline{(c)}
  \vspace{2pt}

  \caption{$\mu = 1, C_0=46, E_0= -7$ with perturbation of $10^{-5}$. Integration time a. $10^4$ time steps
   b. $10^5$ c. $10^6$. The colors indicate categories of orbits: red -12 type, yellow -13 type, magenta -14 type,
   blue -23 type, green -24 type, cyan -34 type, black -symmetry breaking and grey stable.}
  \label{fig4.7}
\end{figure}


\begin{figure}[hbtp]
  \centerline{
    \epsfxsize=7.0cm
    \epsffile{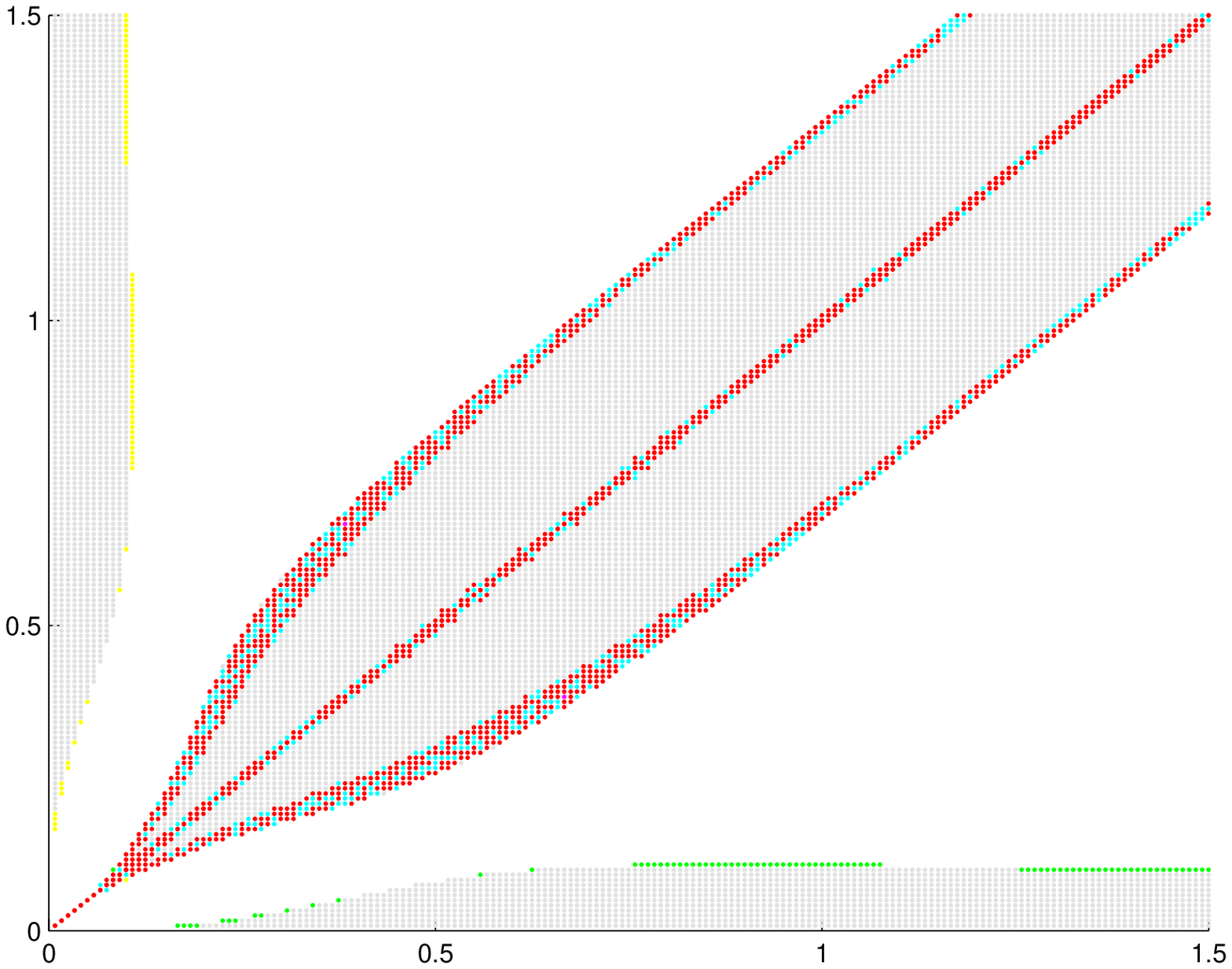}
  }
  \vspace{2pt}
   \centerline{(a)}
  \vspace{2pt}

  \centerline{
    \epsfxsize=7.0cm
    \epsffile{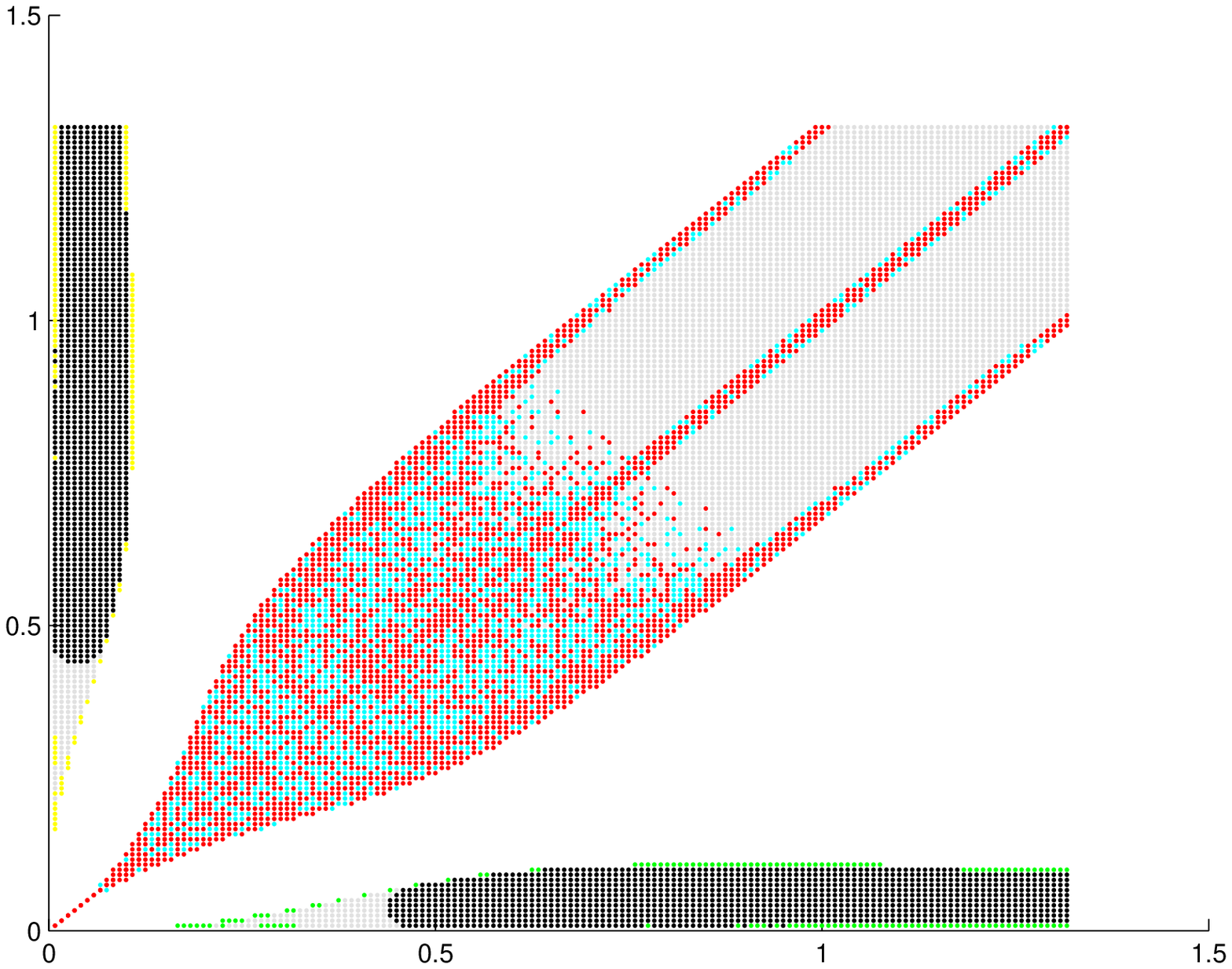}
  }
 \vspace{2pt}
  \centerline{(b)}
  \vspace{2pt}

  \centerline{
    \epsfxsize=7.0cm
    \epsffile{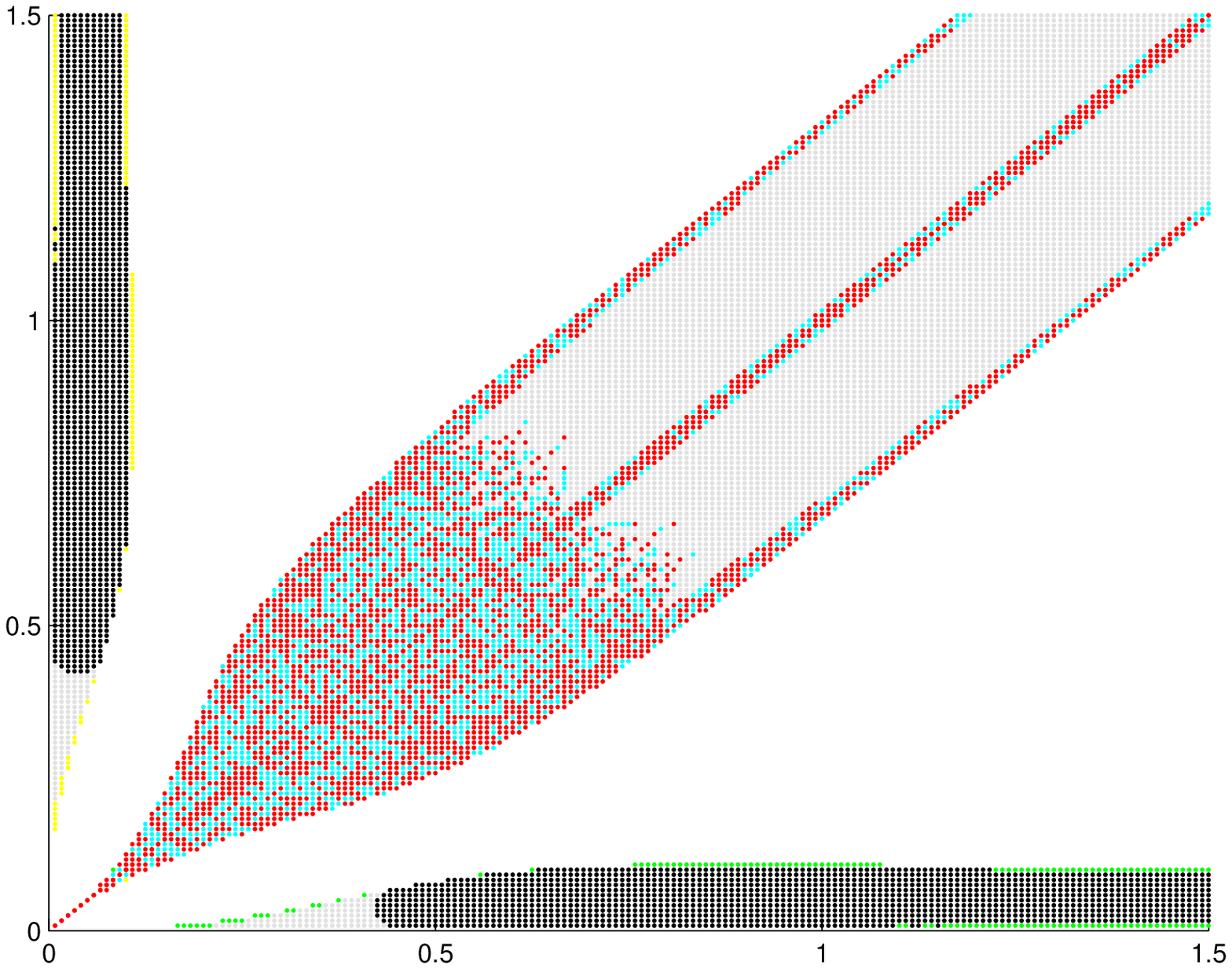}
  }
  \vspace{2pt}
  \centerline{(c)}
  \vspace{2pt}

  \caption{$\mu = 1, C_0=60, E_0= -7$ with no perturbation. Integration time a. $10^4$ time steps
   b. $10^5$ c. $10^6$. The colors indicate categories of orbits: red -12 type, yellow -13 type, magenta -14 type,
   blue -23 type, green -24 type, cyan -34 type, black -symmetry breaking and grey stable.}
  \label{fig4.8}
\end{figure}

\begin{figure}[hbtp]
  \centerline{
    \epsfxsize=7.0cm
    \epsffile{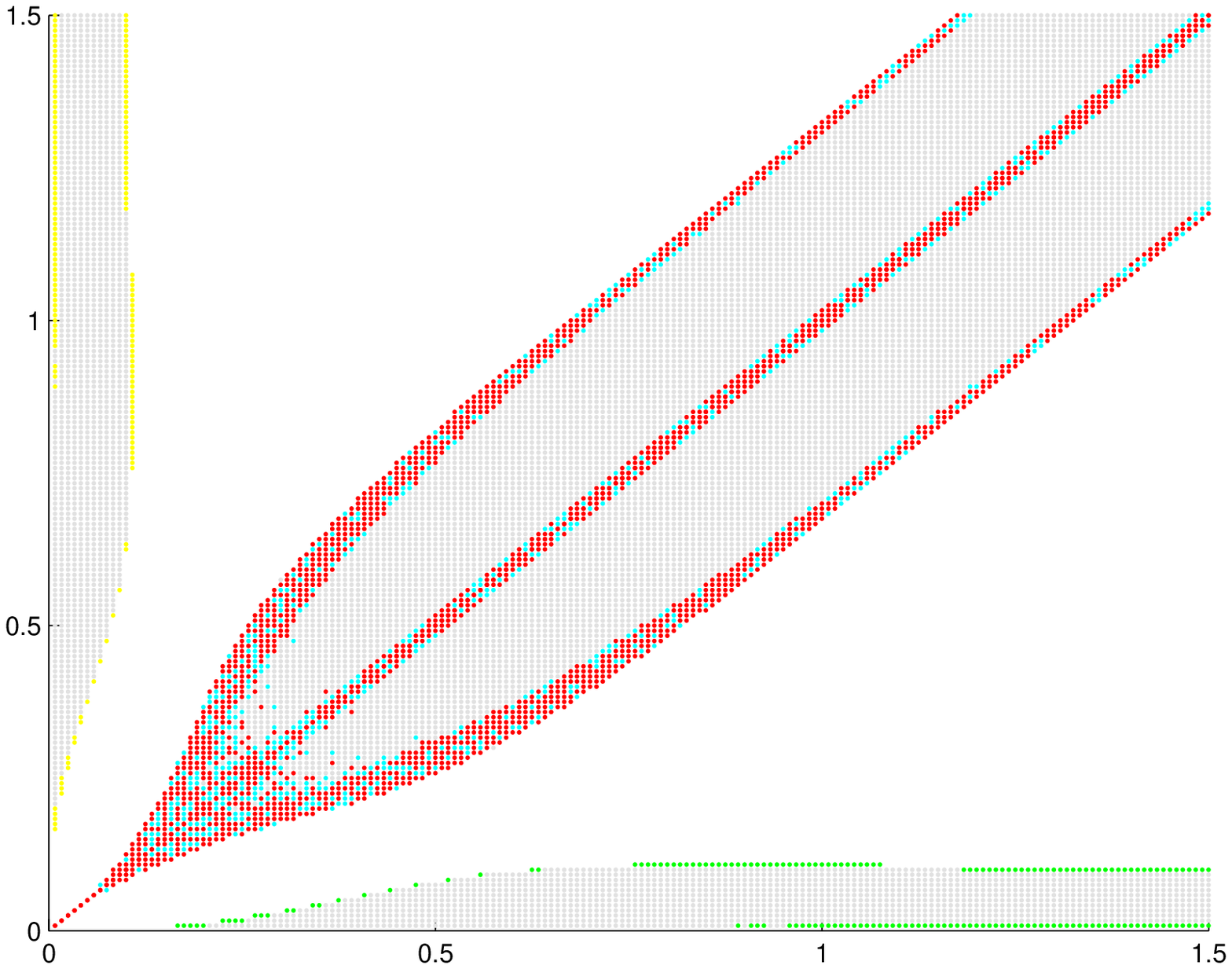}
  }
  \vspace{2pt}
   \centerline{(a)}
  \vspace{2pt}

  \centerline{
    \epsfxsize=7.0cm
    \epsffile{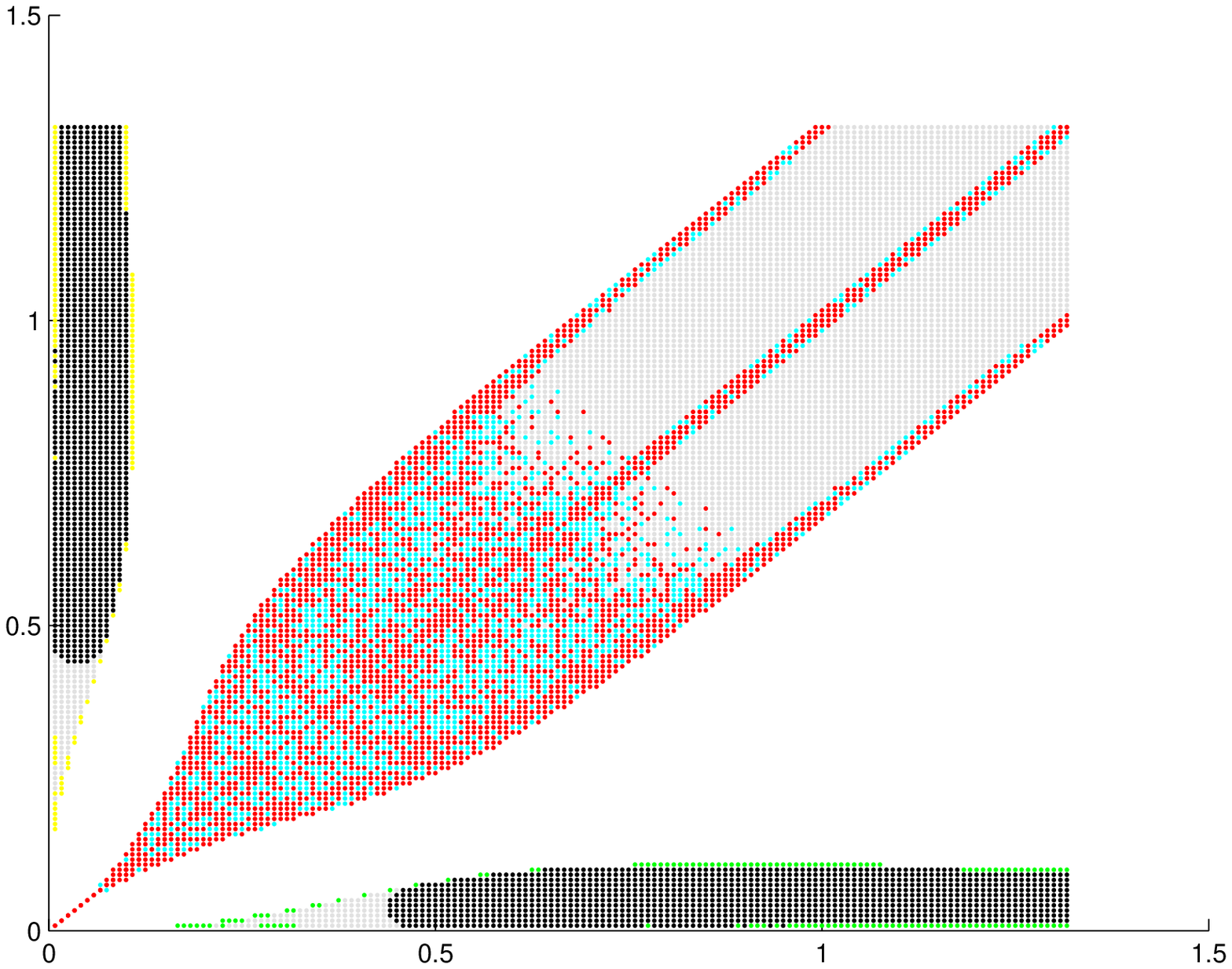}
  }
 \vspace{2pt}
  \centerline{(b)}
  \vspace{2pt}

  \centerline{
    \epsfxsize=7.0cm
    \epsffile{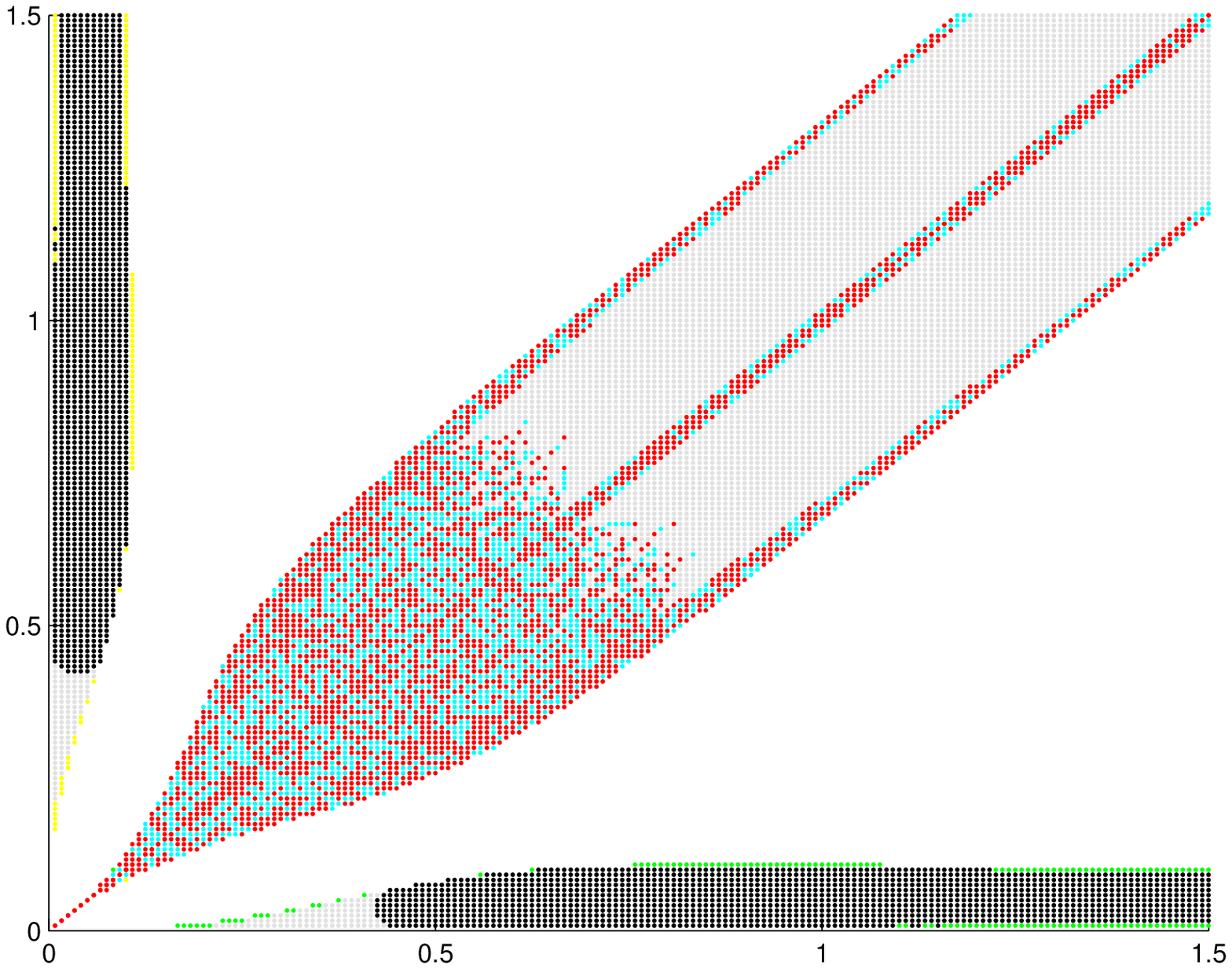}
  }
  \vspace{2pt}
  \centerline{(c)}
  \vspace{2pt}

  \caption{$\mu = 1, C_0=60, E_0= -7$ with perturbation of $10^{-6}$. Integration time a. $10^4$ time steps
   b. $10^5$ c. $10^6$. The colors indicate categories of orbits: red -12 type, yellow -13 type, magenta -14 type,
   blue -23 type, green -24 type, cyan -34 type, black -symmetry breaking and grey stable.}
  \label{fig4.9}
\end{figure}

\begin{figure}[hbtp]
  \centerline{
    \epsfxsize=7.0cm
    \epsffile{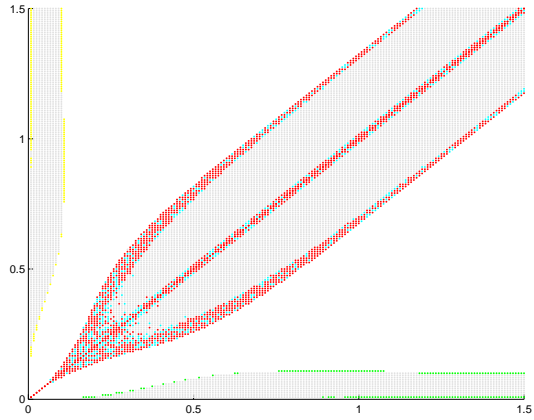}
  }
  \vspace{2pt}
   \centerline{(a)}
  \vspace{2pt}

  \centerline{
    \epsfxsize=7.0cm
    \epsffile{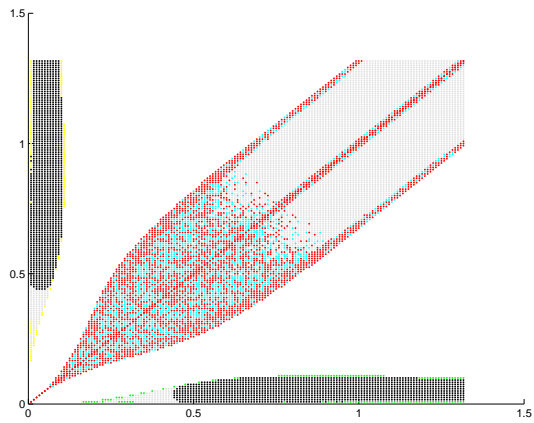}
  }
 \vspace{2pt}
  \centerline{(b)}
  \vspace{2pt}

  \centerline{
    \epsfxsize=7.0cm
    \epsffile{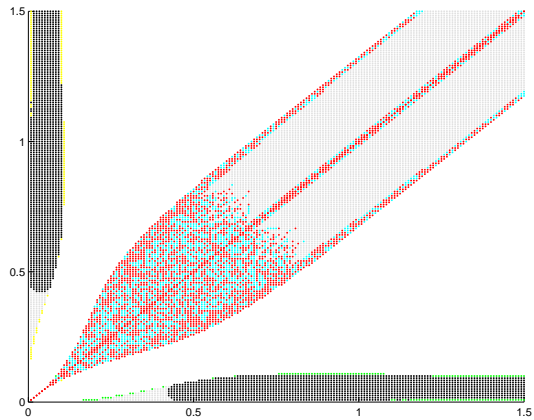}
  }
  \vspace{2pt}
  \centerline{(c)}
  \vspace{2pt}

  \caption{$\mu = 1, C_0=60, E_0= -7$ with perturbation of $10^{-5}$. Integration time a. $10^4$ time steps
   b. $10^5$ c. $10^6$. The colors indicate categories of orbits: red -12 type, yellow -13 type, magenta -14 type,
   blue -23 type, green -24 type, cyan -34 type, black -symmetry breaking and grey stable.}
  \label{fig4.10}
\end{figure}

\pagebreak
\renewcommand{\baselinestretch}{2}
\section{Results: The nature of the orbits in the initial $r_1r_2$ space
for $\mu=0.1$} We now return to our investigation of the stability
of the symmetric nature of the CSFBP when nearly symmetric,
slightly perturbed initial conditions and the general four body
equations are used. Here we study the case of $\mu=0.1$. We aim to
understand which regions of the phase space maintain dynamical
symmetry for long-time. The value of $E_0$ was fixed to be -1.2
and $C_0$ was chosen to be 0.6, 0.7 and 0.9. The initial values of
$r_1$ and $r_2$ were varied from 0 to 1.5 with a step size of
0.008.

Recall that we use the general four body integrator and its not
necessary to have symmetric collision. For example a 12 collision
does not necessarily mean a 34 collision and vice versa. To have
guaranteed symmetric collision one must use the symmetric
equations of motion of the CSFBP \cite{Andras1}. For color coding
of the different types of orbits possible see table
\ref{colorcode}.

Unlike the previous case an interchange of $r_1$ and $r_2$ does
not produce the same physical orbit. Thus the categorization of
the orbits is not symmetric with respect to the $r_1=r_2$ line in
this case. The graphs can be seen in Figure (\ref{fig4.11}) to
(\ref{fig4.16}). Two sets of graphs are given for each $C_0$
value, one with symmetric initial condition and one with perturbed
initial conditions.

The graphs in this case can be separated into four regions: two
double binary regions around $r_1 \approx r_2$, $r_2<r_1$ (DB1)
and $r_2>r_1$ (DB2) and two single binary regions, $r_2\ll r_1$
(SB1) and $r_1 \ll r_2$ (SB2). The SB2 region is very small as
compared to all other regions. Also the SB2 region is very chaotic
as most of the orbits end up in collisions.

\subsubsection{C$_0$ = 0.6}

Figure (\ref{fig4.11}) shows the results of integration for
$C_0=0.6$ with symmetric initial conditions. Figure
(\ref{fig4.12}) shows the same results with perturbed initial
conditions with a perturbation of $10^{-5}$ in the $x$-component
of $P_1$.

What is immediately clear from the comparison of Figures
(\ref{fig4.11}a), (\ref{fig4.11}b) and (\ref{fig4.11}c) is that
the longer we integrate this CSFBP system the more unstable orbits
we unearth. The most chaotic region for $C_0=0.6$ is the SB2
region where every orbit ends up in collision or fails the
symmetry breaking criterion. The region between $r_1=0$ and
$r_1=0.2$ is also very chaotic as most of the orbits are collision
orbits. The collision types in this area are 12 and 14 type
collisions. There are some collisions along the $r_1$-axis which
are mostly of type 24. There are also some 12, 13, 14 and 34 type
collisions. The SB1 and DB1 regions are separated by a thick cloud
of collisions. The majority of these collisions are 12 and 24 type
collisions. There are also some 14 and 34 type collisions. All of
the collisions in SB2 region are of type 13. The double binary
region is surrounded by 12 and 34 type collisions. The collisions
around the $r_1=r_2$ line are also 12 and 34.

When we integrate longer, Figure (4.13b) the number of collisions
increases in all regions except the SB2 region which didn't have
any stable orbits. The double binary region is now surrounded by a
thicker cloud of collisions which are all of 12 or 34 type
collisions, See Figure (\ref{fig4.11}b) and (\ref{fig4.11}c). When
integrated for 1 million time steps of integration we get some
orbits near the meeting point of the SB2 and DB2 region which fail
the symmetry breaking criterion, see Figure (\ref{fig4.11}c). The
number of these orbits is very small as compared to the equal mass
case. The most stable region in this case is the SB1 region. In
the DB regions the number of non-collision orbits are much higher
than the collision orbits. We can therefore conclude that this
case of CSFBP is more stable than the equal mass case.

Figure (\ref{fig4.12}) shows the analysis of the same orbits
discussed above but with perturbed initial conditions. Most of the
orbits give the same results as before i.e. with symmetric initial
conditions except the type of collision change along the $r_1=r_2$
line in Figure (\ref{fig4.12}b). In the perturbed case we get some
24 type collisions along $r_1=r_2$ line which were only 12 and 34
type for the symmetric case.

\subsubsection{C$_0$ = 0.7}

Figure (\ref{fig4.13}) shows the results of integration for
$C_0=0.7$ with symmetric initial conditions. Figure
(\ref{fig4.14}) shows the same results with perturbed initial
conditions with a perturbation of $10^{-5}$ in the $x$-component
of $P_1$.

The behavior of this case is similar to the previous case. Overall
we have fewer number of collisions than we had in the previous
case but the difference is not very big. The only noticeable
difference is in the area which separates the SB1 and DB1 regions.
The number of collisions drops by third from the previous case.
See Figures (\ref{fig4.13}) and (\ref{fig4.14})

\subsubsection{C$_0$ = 0.9}

Figure (\ref{fig4.15}) shows the results of integration for
$C_0=0.9$ with symmetric initial conditions. Figure
(\ref{fig4.16}) shows the same results with perturbed initial
conditions with a perturbation of $10^{-5}$ in the $x$-component
of $P_1$.

The single binary and double binary regions are completely
disconnected as $C_0=0.9$ is greater than the critical value
(Steves and Roy, 2001). There are no collision orbits in the SB1
region for all versions of integrations. In the longest version of
the integration there appears an island of orbits between $r_1=1$
and $r_1=1.5$ which fail the symmetry breaking criterion. In
summary, despite a number of symmetry breakings, the SB1 region is
the most stable in this case too. See Figure (\ref{fig4.15}). The
double binary region is surrounded by collisions of 12 and 34
type. In the shorter version of integration we have a very few
number of collisions but when integrated for longer the number of
collisions increases on the boundaries of the double binary region
and also along the $r_1=r_2$ line in the DB2 region. Surprisingly
there is no symmetry breaking which may be because of the
hierarchical stability we have in this case. The SB2 region as in
all other cases is the most unstable and none of the orbits are
stable. The collisions in the SB2 region are all 13 type
collisions, as no other types of collisions are possible. See
Figure (\ref{fig4.15}).

Figure (\ref{fig4.16}) shows the analysis of the same orbits but
with perturbed initial conditions. There is no significant
difference between the results obtained using symmetric initial
conditions and with perturbed initial conditions. The only
difference is the increase in the number of non-symmetric orbits.
For example Figure (\ref{fig4.15}b) has no such orbits and the SB1
region is completely grey but we have a big island of such orbits
in Figure (\ref{fig4.16}b) which is almost of the same size as we
have in Figure (\ref{fig4.15}c). Also the island of these orbits
in the SB1 region in Figure (\ref{fig4.16}c) is thicker than in
Figure (\ref{fig4.15}c).

The absence of single binary collisions in the double binary area
and of double binary collisions in single binary areas show that
this system is hierarchically stable. The main characteristics of
the orbits for all values of the Szebehely constant i.e. $C_0=0.6,
0.7$ and 0.9 remains the same. In the current case, $C_0=0.9$, we
have much less number of collision than we had in any other case
of the CSFBP discussed in this chapter.

\section{Comparison with Sz\'ell et.al (2004) analysis in the
$\mu=0.1$ case}

Before comparing the results we give a brief review of the results
of \citeasnoun{AndrasMNRAS} in the $\mu=0.1$ case.

\subsection{Review of Sz\'ell et.al (2004) results for the non-equal mass case, $\mu=0.1$}\label{xyz}

In this case $C_0$ was chosen to be $0.4$, $0.8$ and $0.9$. For
$\mu = 0.1$ $C_{crit1} = 0.7792$ and $C_{crit2} = 0.8886$. Note
again all hierarchies described in this section are the CSFBP
notation used by Sz\'ell et.al (2004).

The graphs can be separated into four regions: two double binary
regions around $r_2 \approx r_1$, $r_2 < r_1$ (DB1) and $r_2 >
r_1$ (DB2), and two single binary regions, $r_2 \ll r_1$ (SB1) and
$r_1 \ll r_2$ (SB2). In the following each region is analyzed for
the given values of $C_0$.

\begin{figure*}
   \centering
   \resizebox{150mm}{!} {
   \includegraphics{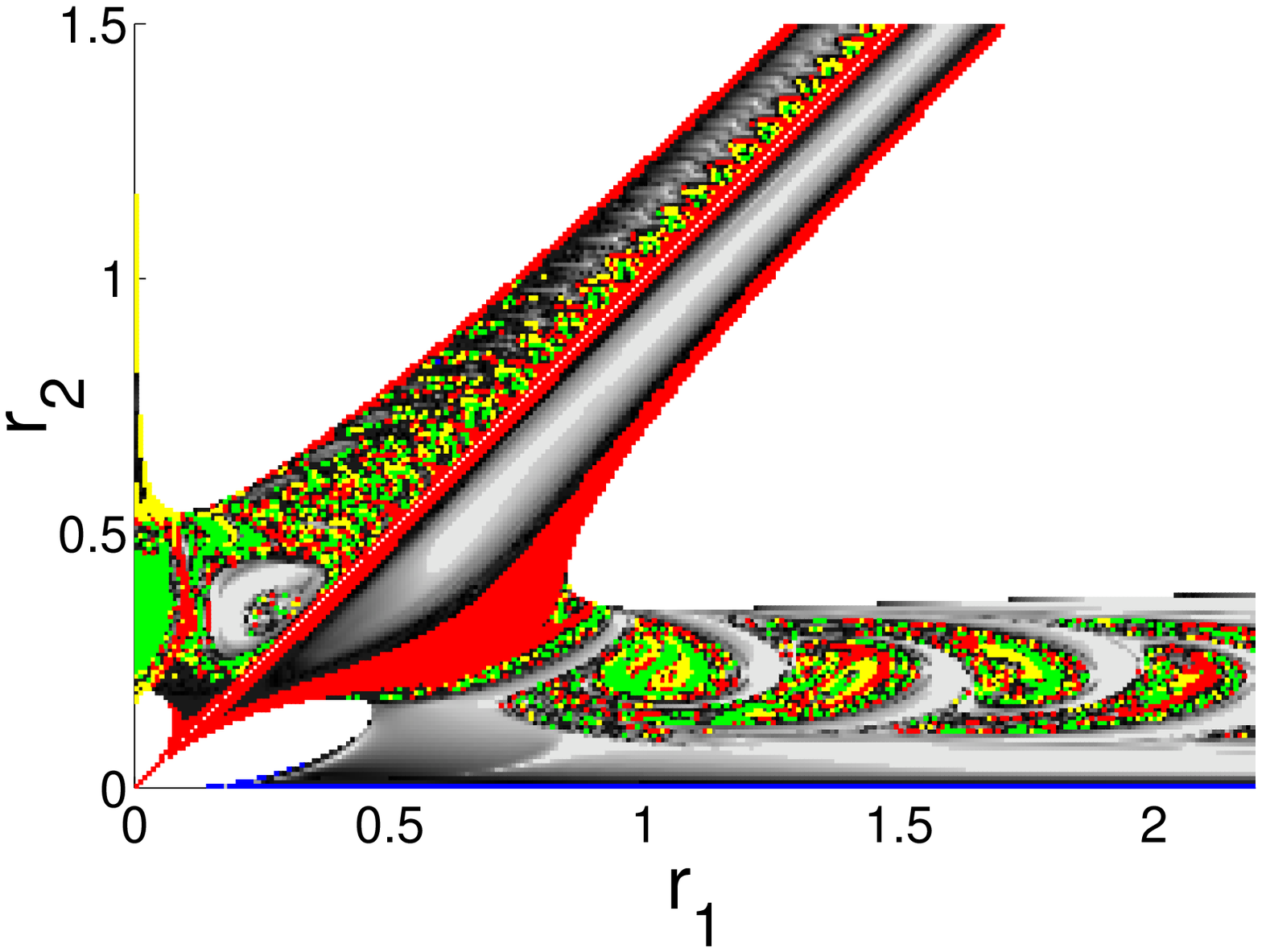}
   \includegraphics{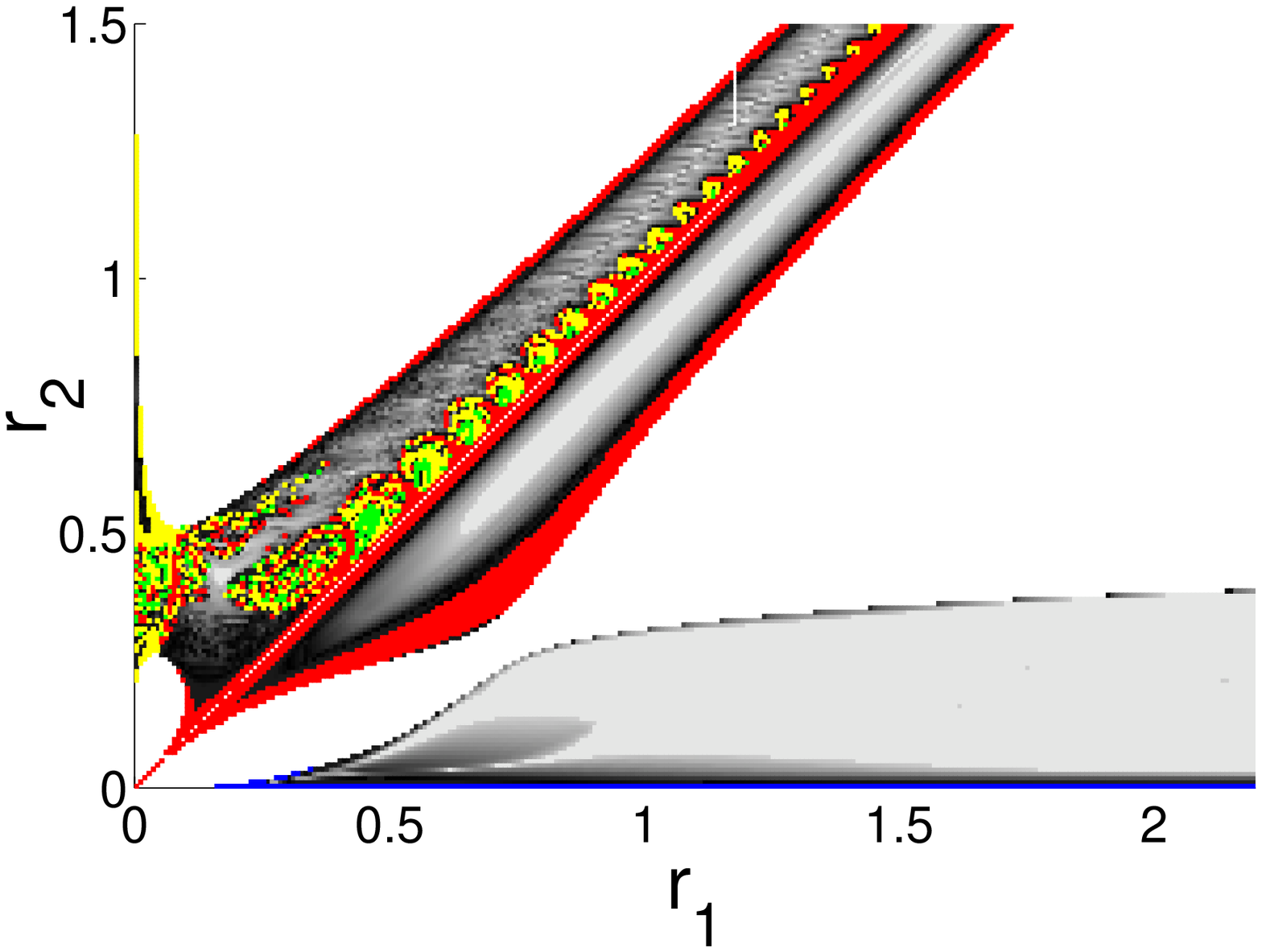}
   \includegraphics{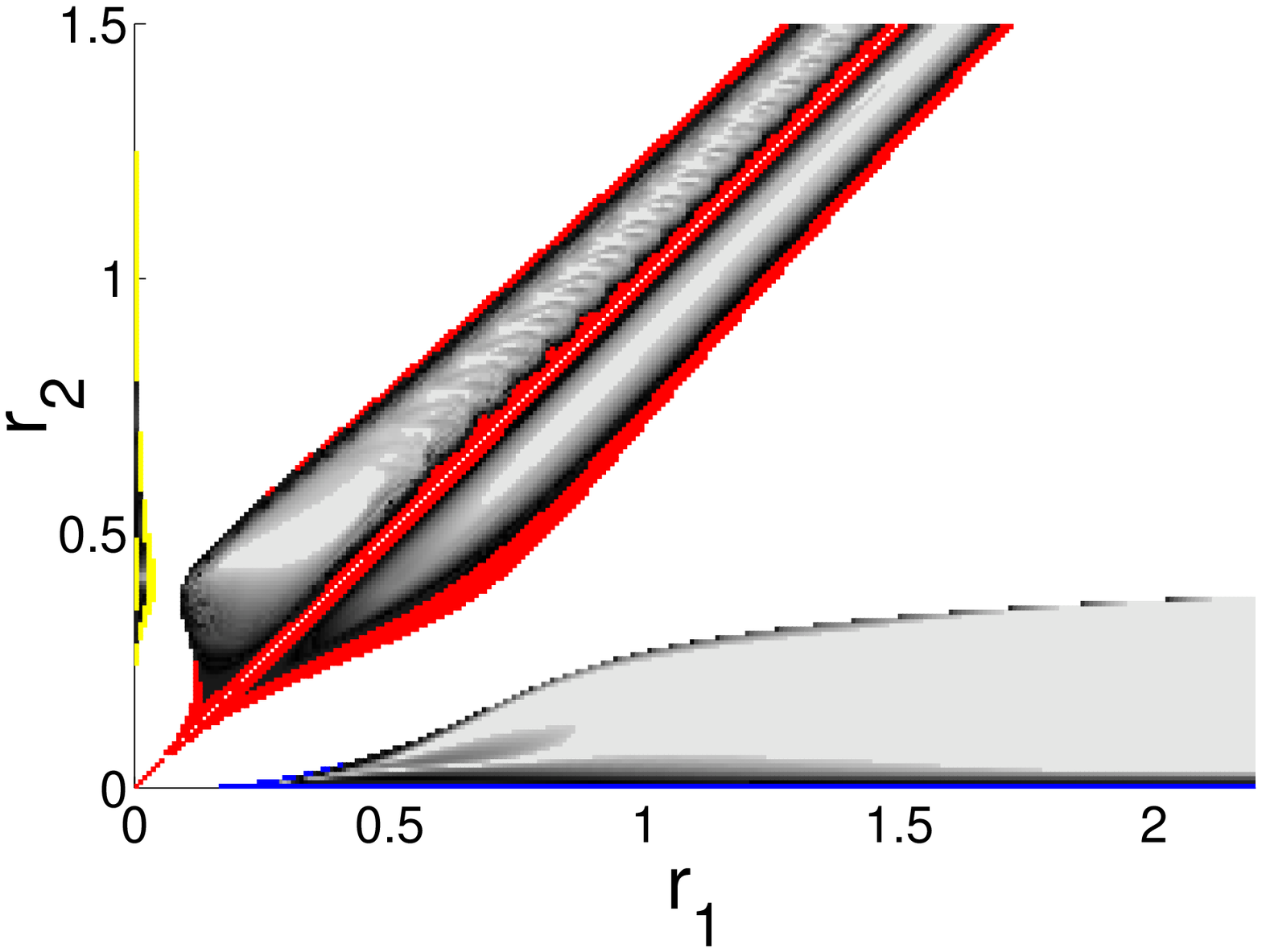}}
              \\
              \resizebox{150mm}{!} {
   \includegraphics{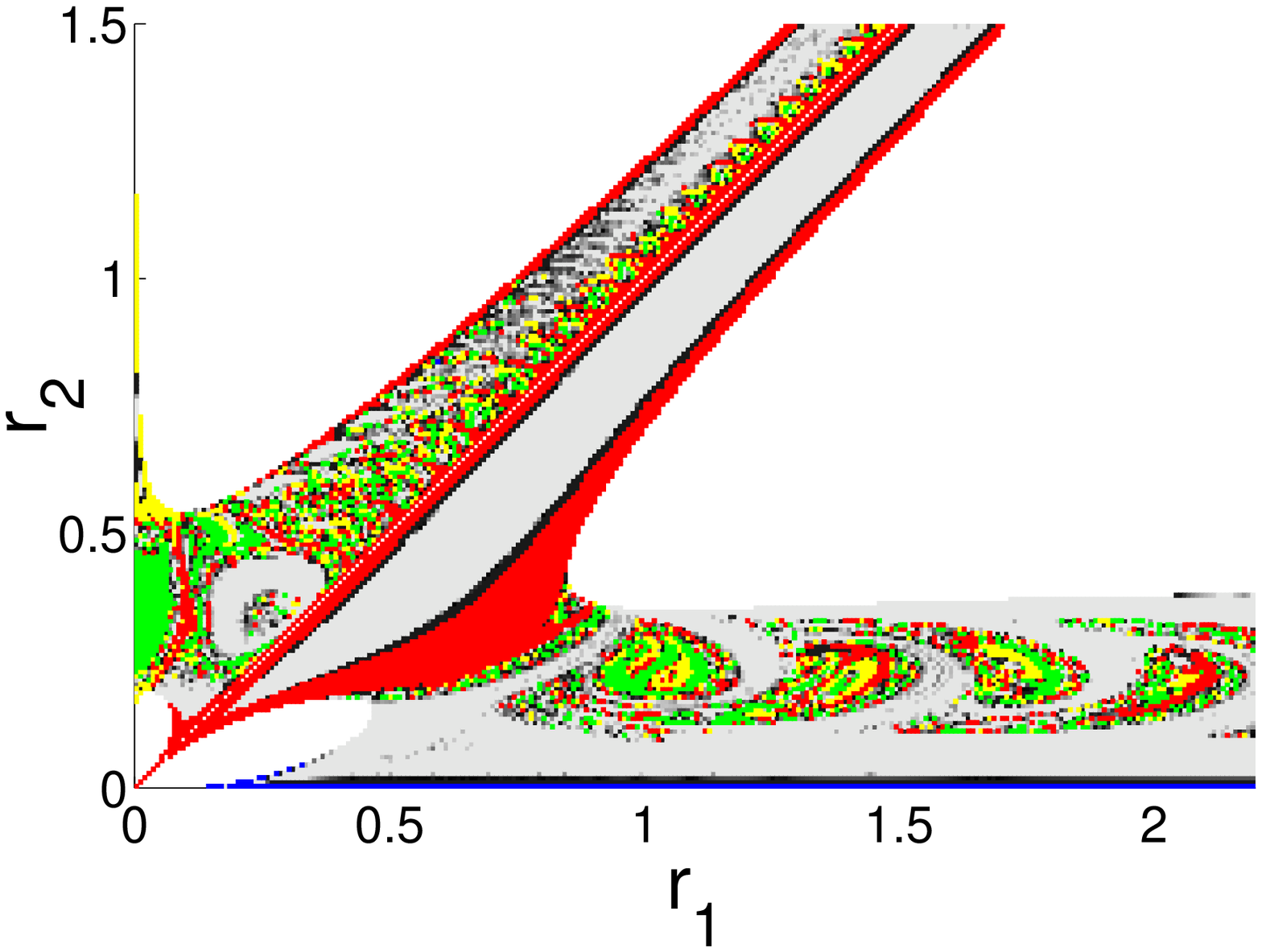}
   \includegraphics{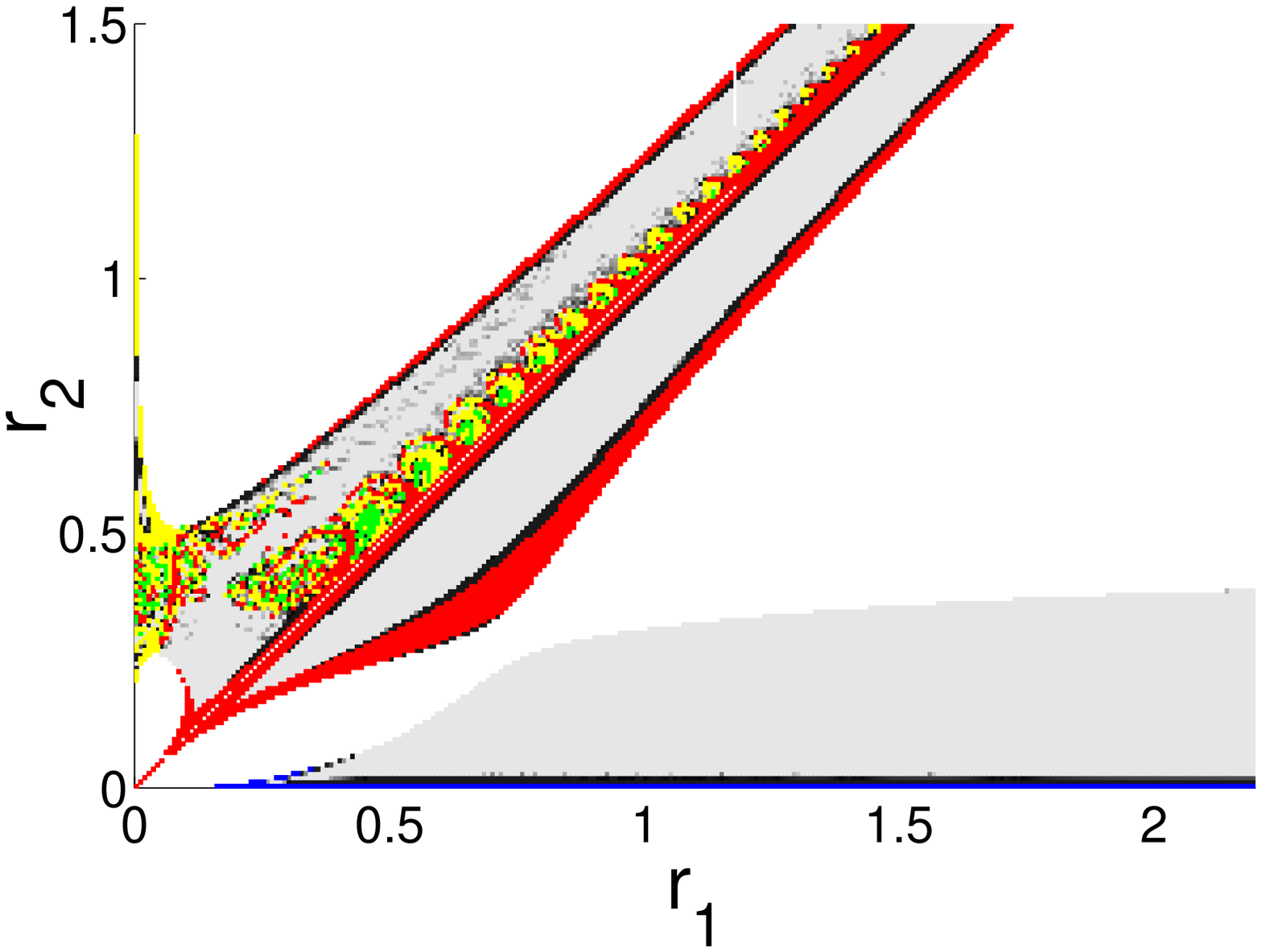}
   \includegraphics{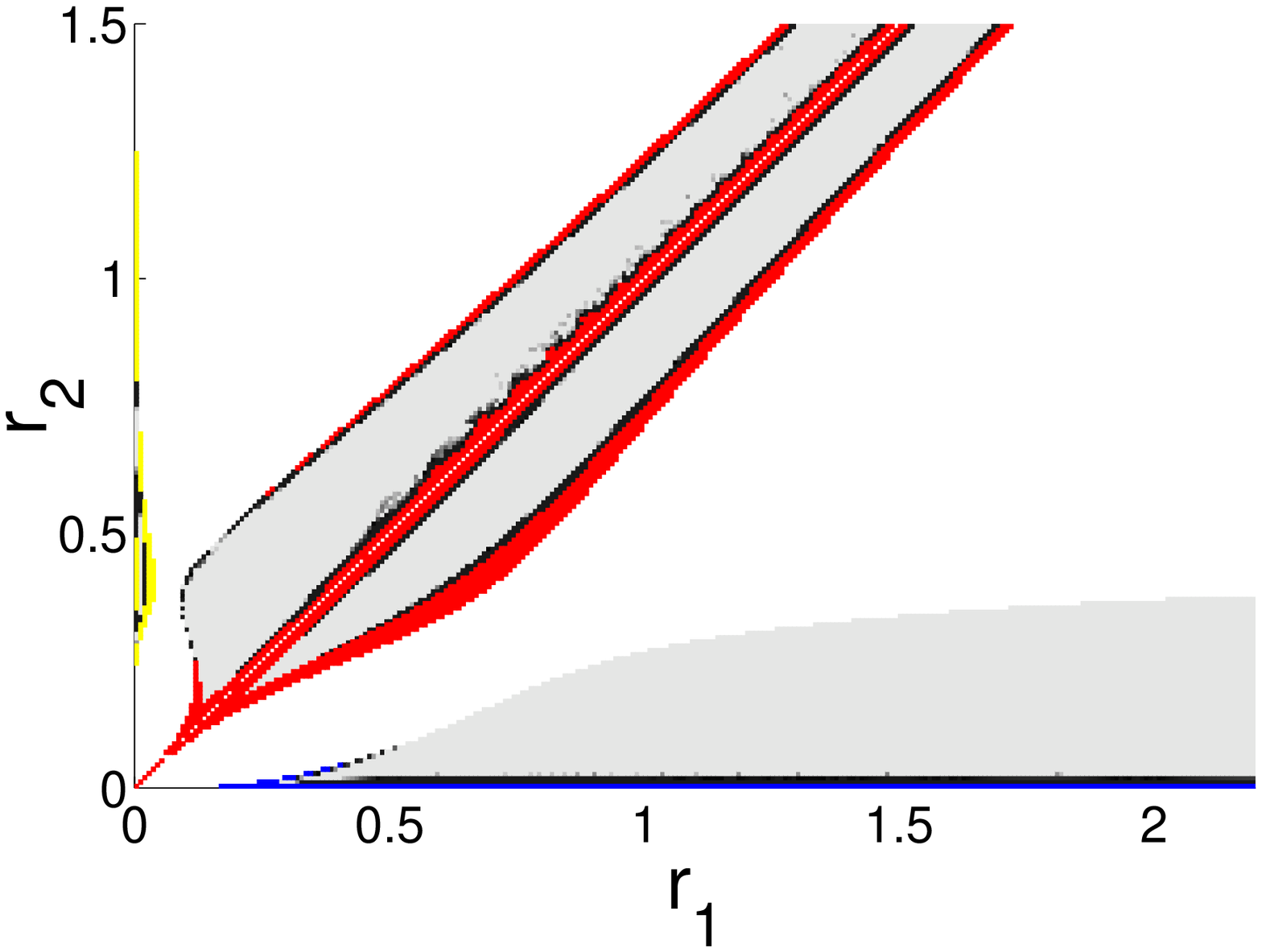}}
      \caption{$\mu = 0.1$, $E_0 = 1.2$. Graphs of the $r_1$, $r_2$ initial conditions
      for $C_0 = 0.4$, $0.8$, $0.9$ from left to right. The top
      line of graphs show the RLI categorizations and the bottom
      the SALI. The color coding is as before in Figure \ref{figmu1}.
              }
         \label{figmu01}
   \end{figure*}

\begin{description}
  \item[$C_0 = 0.4$] (Figure \ref{figmu01}, left). In this case $C_0$ is well below the critical
  value $C_{crit1}$.

The SB2 region is dominated by the  "$14$" (yellow) type of
collision orbits. Some chaotic (black)  orbits can be seen at
around $r_2  \approx 0.9$.

The DB2 region is very chaotic as suggested by the presence of
large number of collisions. There a large number of "$12$" (red)
and "$13$" (green) type of binary-binary  collisions which form a
regular pattern. Some "$14$" (yellow)  collisions can also be
found between them. The DB1 region has very similar behavior to
the $\mu = 1$ case. It contains a massive regular region with a
chaotic and collision region  at the boundaries of real motion.

The SB1 region has a very large number of collision orbits, where
"$13$" (green) and "$12$" (red) collision orbits are dominant.
Clear big  non-collision regular islands (light grey) can be found
between them.  It can be concluded that the phase space is quite
chaotic with  some regular islands.

\item[$C_0 = 0.8$] (Figure \ref{figmu01}, middle). In this case
$C_{crit1} < C_0 < C_{crit2}$.

The "$23$" hierarchy region is now totally disconnected from the
rest. There is a dramatic change in the SB1 region. It is  now
predominantly regular. Chaotic regions can be seen only at the
boundaries. The DB1 region is still the most chaotic.  Some
important changes can be found in the DB2 region. The
non-collision area has now grown to the full length of the DB2
region and is more regular as is  indicated by the white spots in
the RLI graph. It can be concluded that the phase space shows more
regular behavior.

\item[$C_0 = 0.9$] (Figure \ref{figmu01}, right). In this case
$C_0$ is  greater than $C_{crit2}$ and all hierarchy regions are
disconnected from  each other which means complete hierarchical
stability. The collision orbits have almost disappeared and most
of the orbits are regular.
\end{description}

Overall, there is a connection between the global hierarchical
stability criterion and the chaotic behavior of the phase space,
namely as $C_0$ increases and the system becomes more
hierarchically stable, the phase space becomes more regular.

\subsection{A comparison of the Sz\'ell et.al (2004) results with our CSFBP results}

We now compare the results of Sz\'ell et.al (2004) based on
symmetric initial conditions and symmetric equations of motion
with our results given in section 5.8 based on symmetric initial
conditions and the general four body equations of motion. The
following are the important and notable similarities:

\begin{enumerate}
\item The stability of the phase space of the CSFBP depends on the
$C_0$ value. The number of collision orbits decreases as the value
of $C_0$ increases. \item The SB1 region is the most stable as it
has very few collision orbits. \item  For $C_0<C_{crit}$ there is
always a big island of collision orbits at the junction of the
single binary and double binary regions.
\end{enumerate}

From Sz\'ell et.al (2004)at $C_0=0.9$, (Figure 5.12, right) and
our case for the integration time of $10^4$ (Figure 5.17) we have
almost identical results in terms of the collision orbits. In the
SB2 region \citeasnoun{AndrasMNRAS} have some chaotic orbits which
are also non-symmetric orbits in our case. The SB1 regions are
identical except for the existence of  a few collision orbits
along $r_2\approx 0$ in the \citeasnoun{AndrasMNRAS} case, which
are symmetric in our case.


\renewcommand{\baselinestretch}{1.5}
\begin{figure}[hbtp]
  \centerline{
    \epsfxsize=7.0cm
    \epsffile{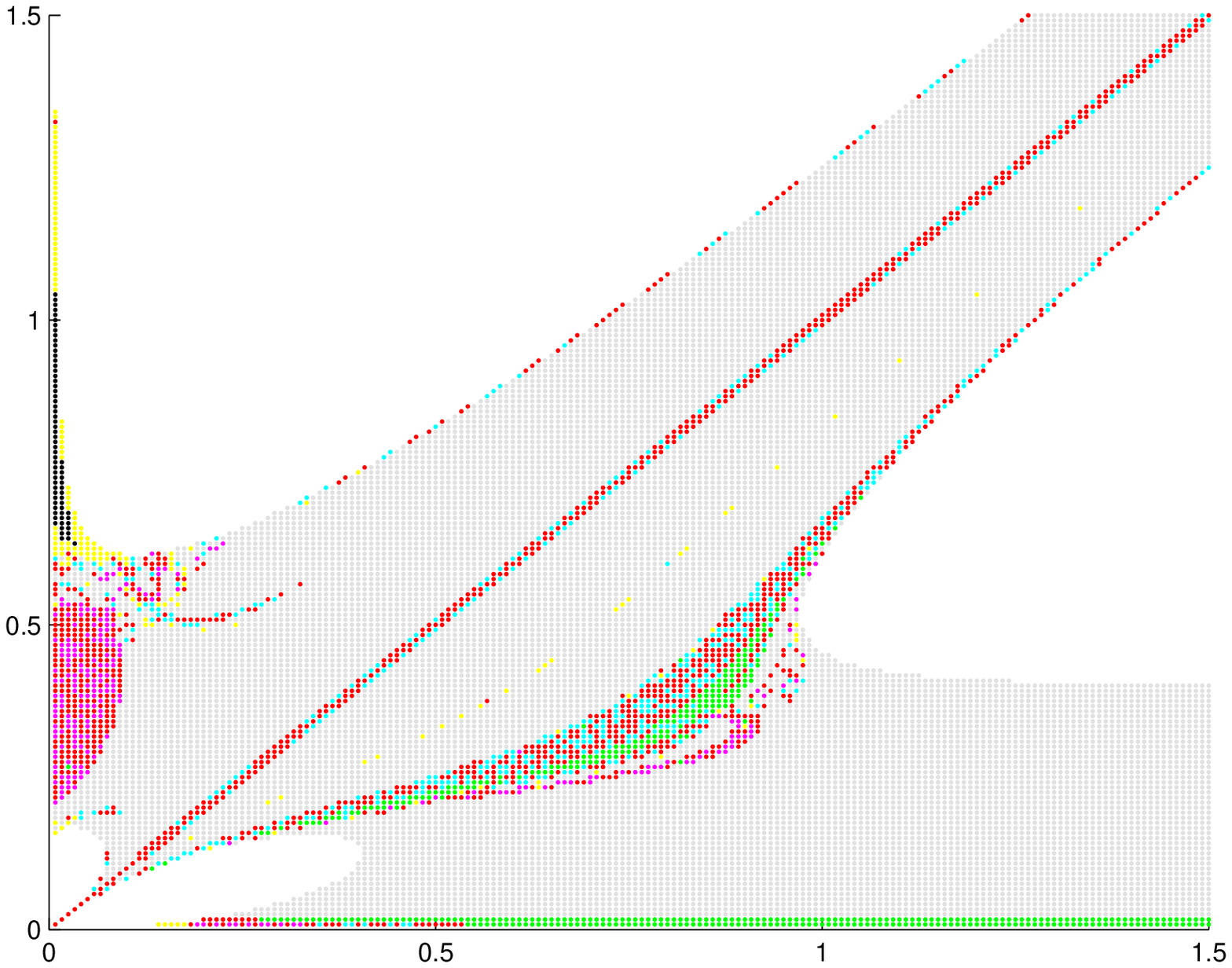}
  }
  \vspace{2pt}
   \centerline{(a)}
  \vspace{2pt}

  \centerline{
    \epsfxsize=7.0cm
    \epsffile{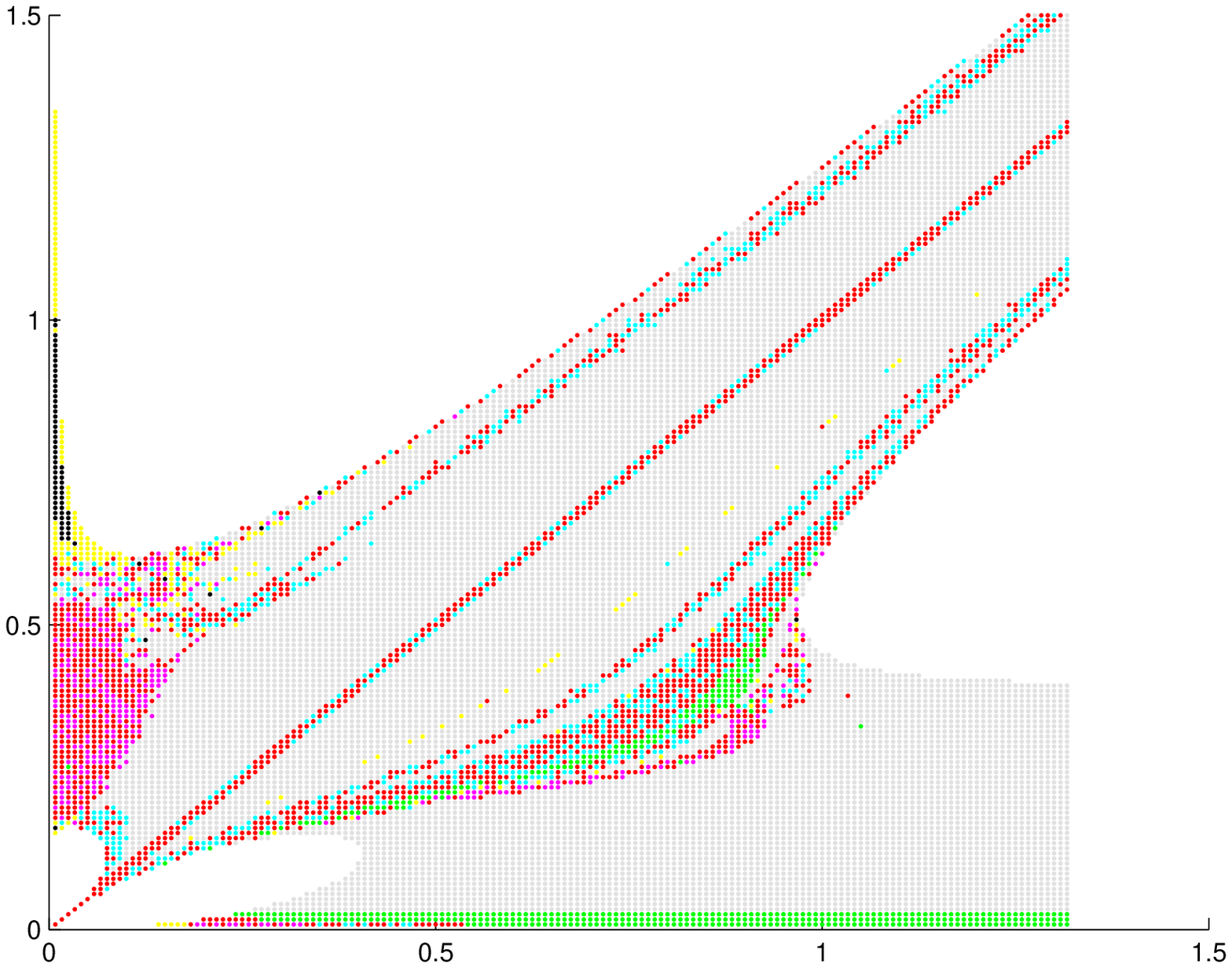}
  }
 \vspace{2pt}
  \centerline{(b)}
  \vspace{2pt}

  \centerline{
    \epsfxsize=7.0cm
    \epsffile{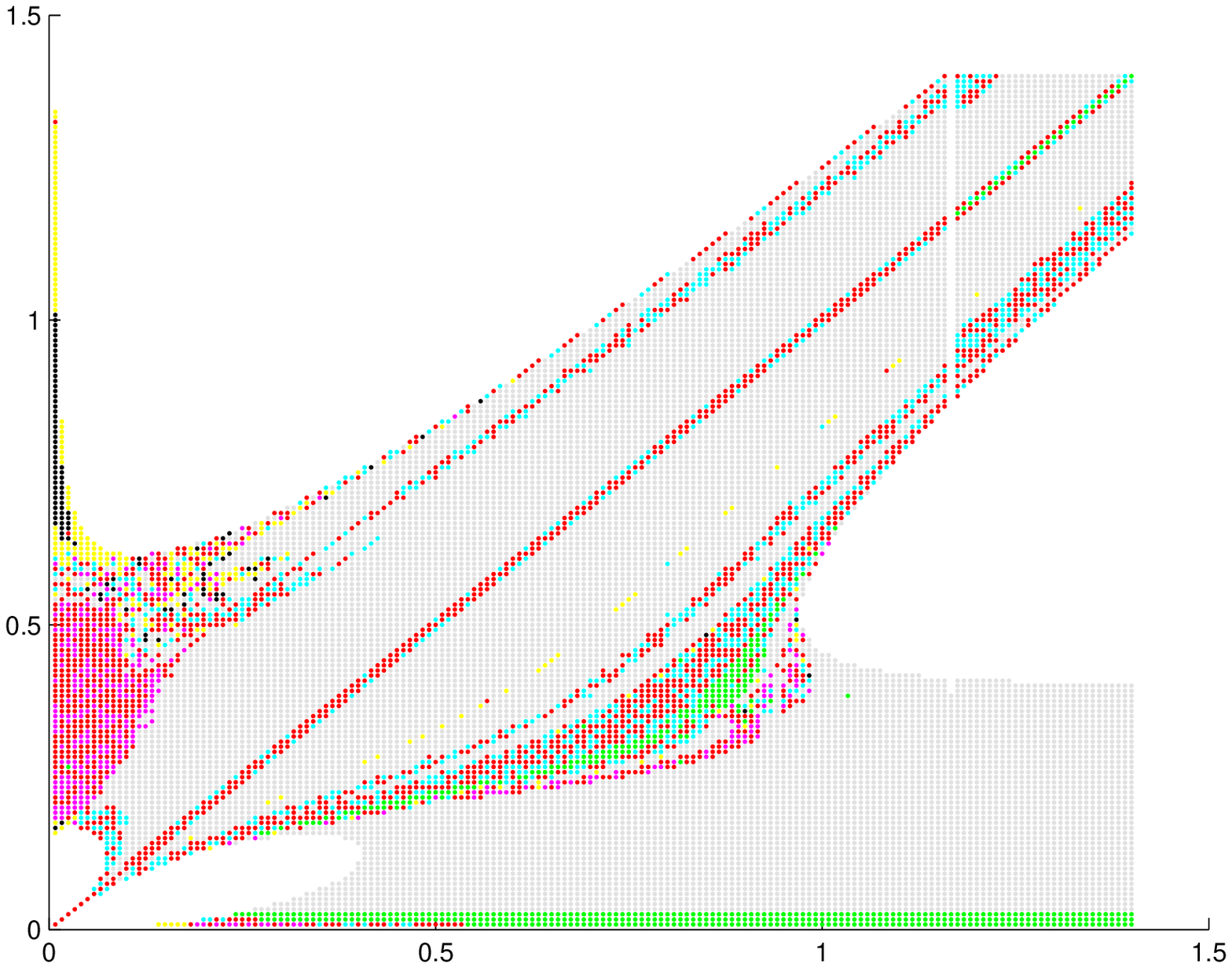}
  }
  \vspace{2pt}
  \centerline{(c)}
  \vspace{2pt}

  \caption{$\mu = 0.1, C_0=0.6, E_0= -1.2$ with no perturbation. Integration time a. $10^4$ time steps
   b. $10^5$ c. $10^6$. The colors indicate categories of orbits: red -12 type, yellow -13 type, magenta -14 type,
   blue -23 type, green -24 type, cyan -34 type, black -symmetry breaking and grey stable.}
  \label{fig4.11}
\end{figure}

\begin{figure}[hbtp]
  \centerline{
    \epsfxsize=7.0cm
    \epsffile{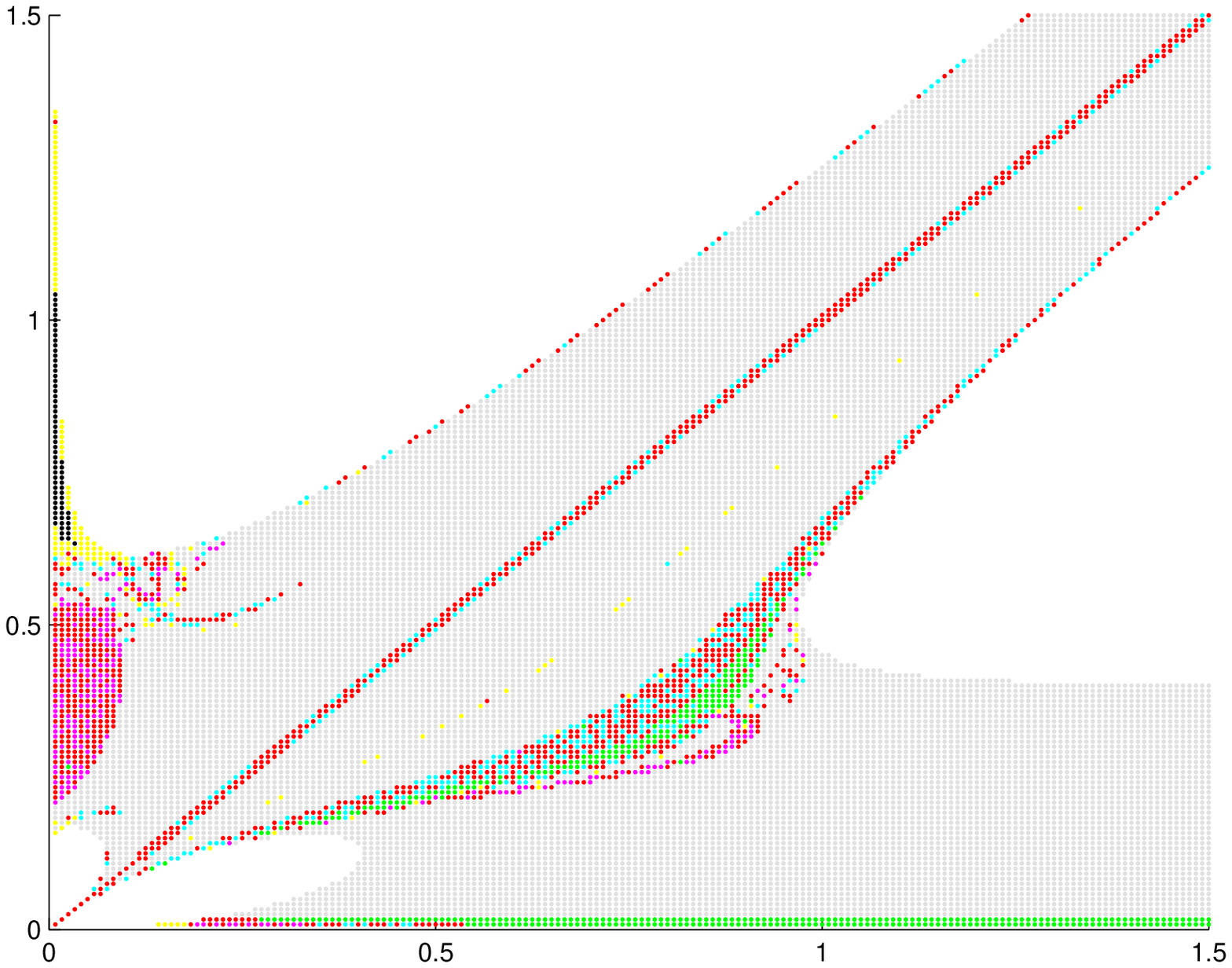}
  }
  \vspace{2pt}
   \centerline{(a)}
  \vspace{2pt}

  \centerline{
    \epsfxsize=7.0cm
    \epsffile{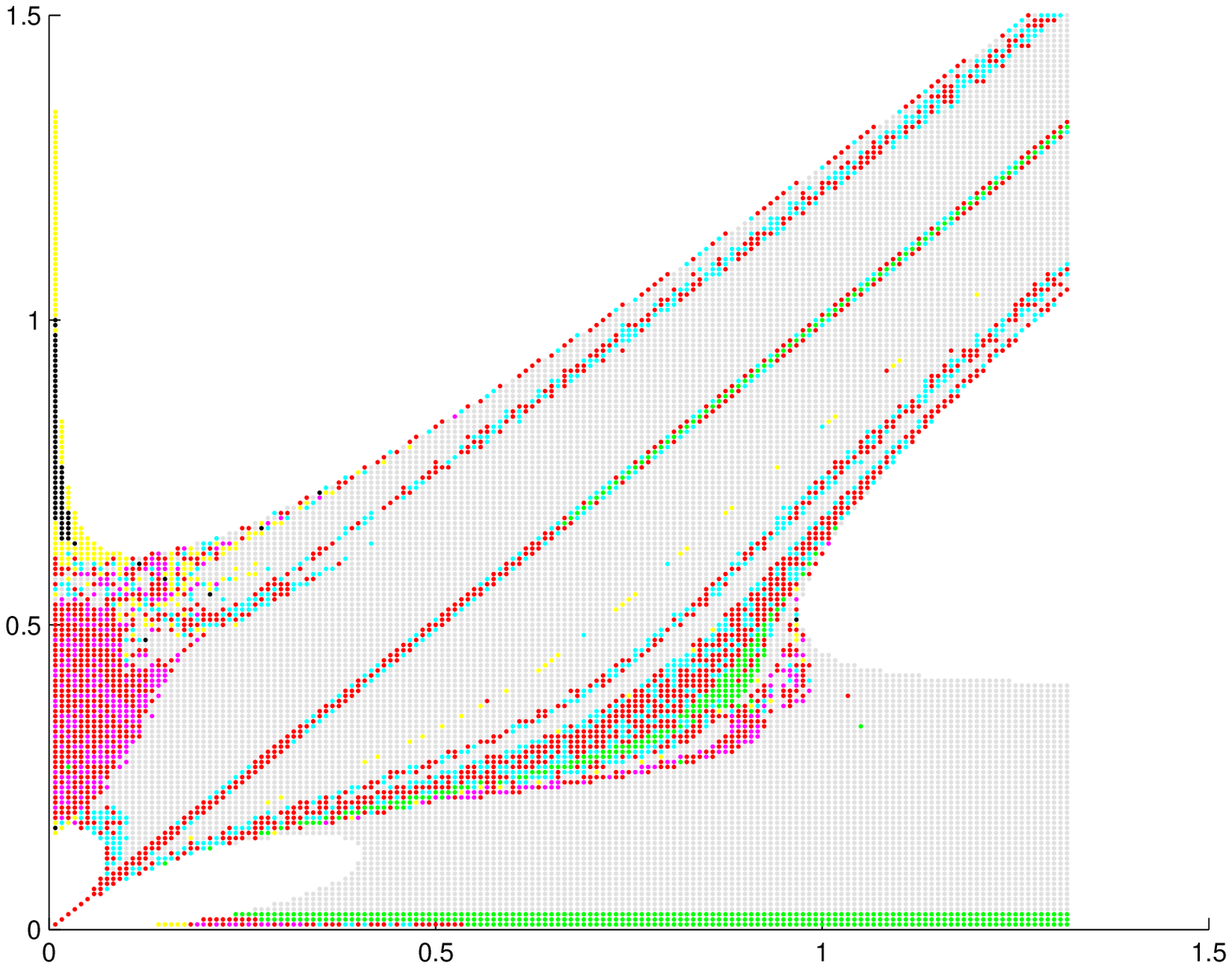}
  }
 \vspace{2pt}
  \centerline{(b)}
  \vspace{2pt}

  \centerline{
    \epsfxsize=7.0cm
    \epsffile{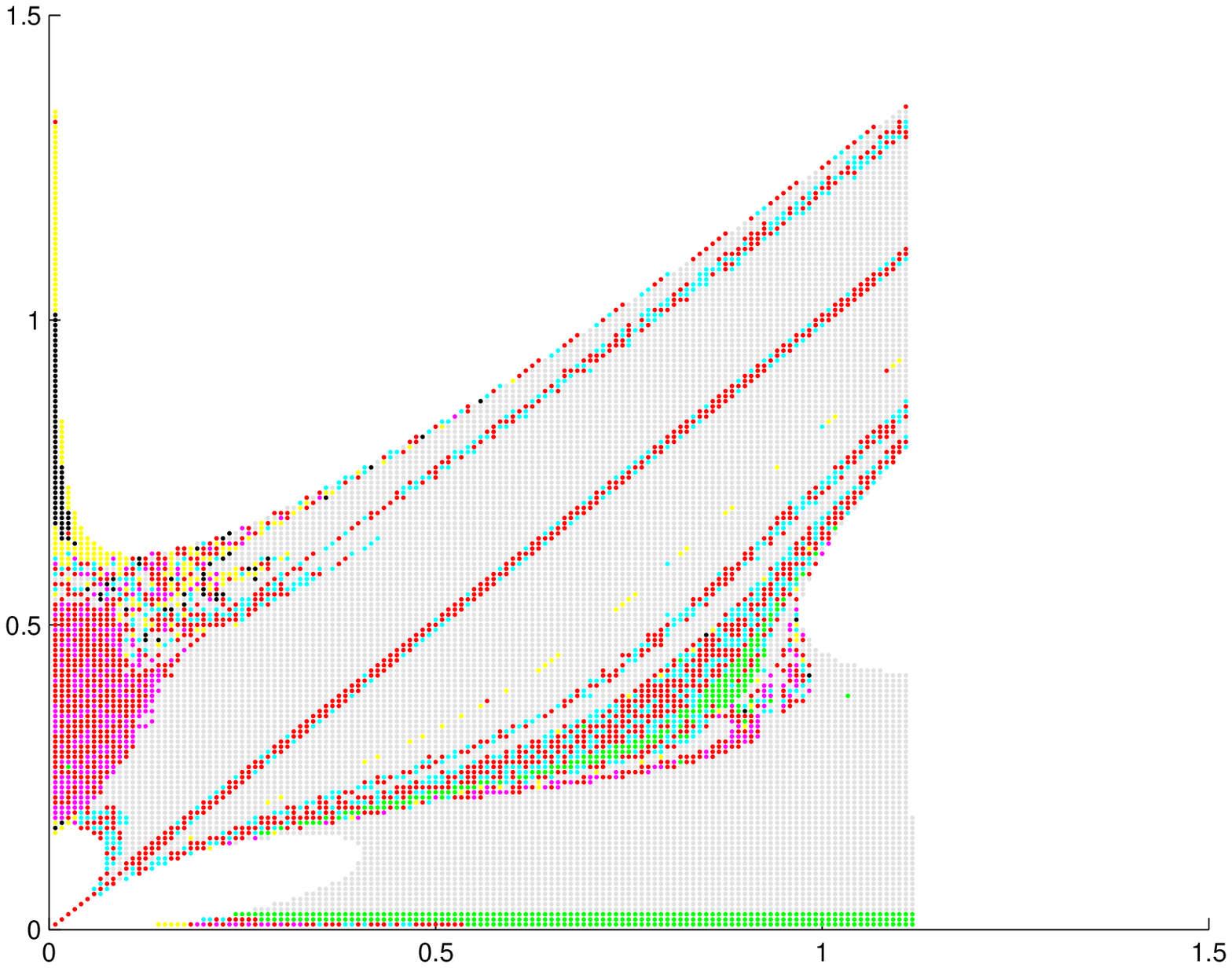}
  }
  \vspace{2pt}
  \centerline{(c)}
  \vspace{2pt}

  \caption{$\mu = 0.1, C_0=0.6, E_0= -1.2$ with perturbation of $10^{-5}$. Integration time a. $10^4$ time steps
   b. $10^5$ c. $10^6$. The colors indicate categories of orbits: red -12 type, yellow -13 type, magenta -14 type,
   blue -23 type, green -24 type, cyan -34 type, black -symmetry breaking and grey stable.}
  \label{fig4.12}
\end{figure}

\begin{figure}[hbtp]
  \centerline{
    \epsfxsize=7.0cm
    \epsffile{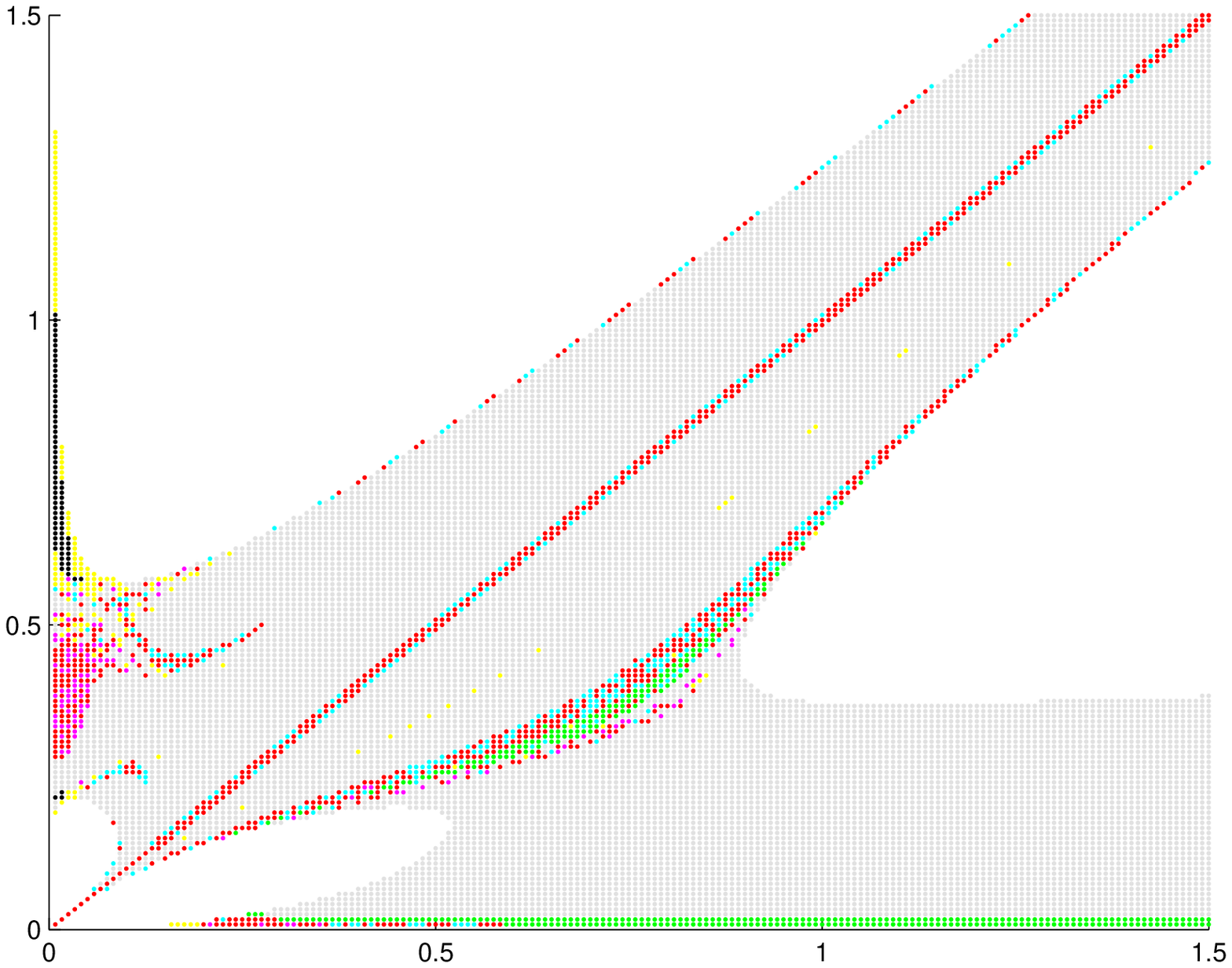}
  }
  \vspace{2pt}
   \centerline{(a)}
  \vspace{2pt}

  \centerline{
    \epsfxsize=7.0cm
    \epsffile{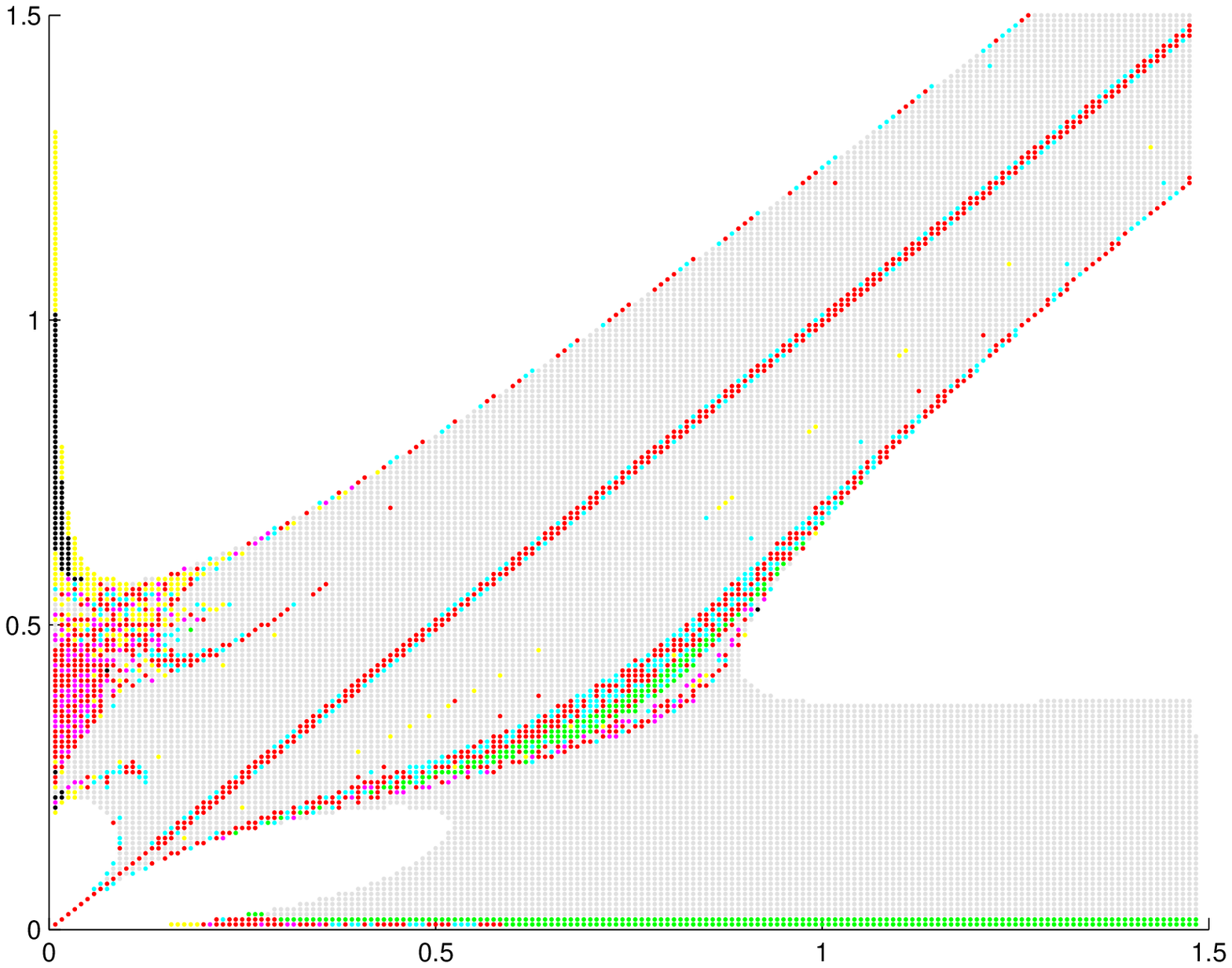}
  }
 \vspace{2pt}
  \centerline{(b)}
  \vspace{2pt}

  \centerline{
    \epsfxsize=7.0cm
    \epsffile{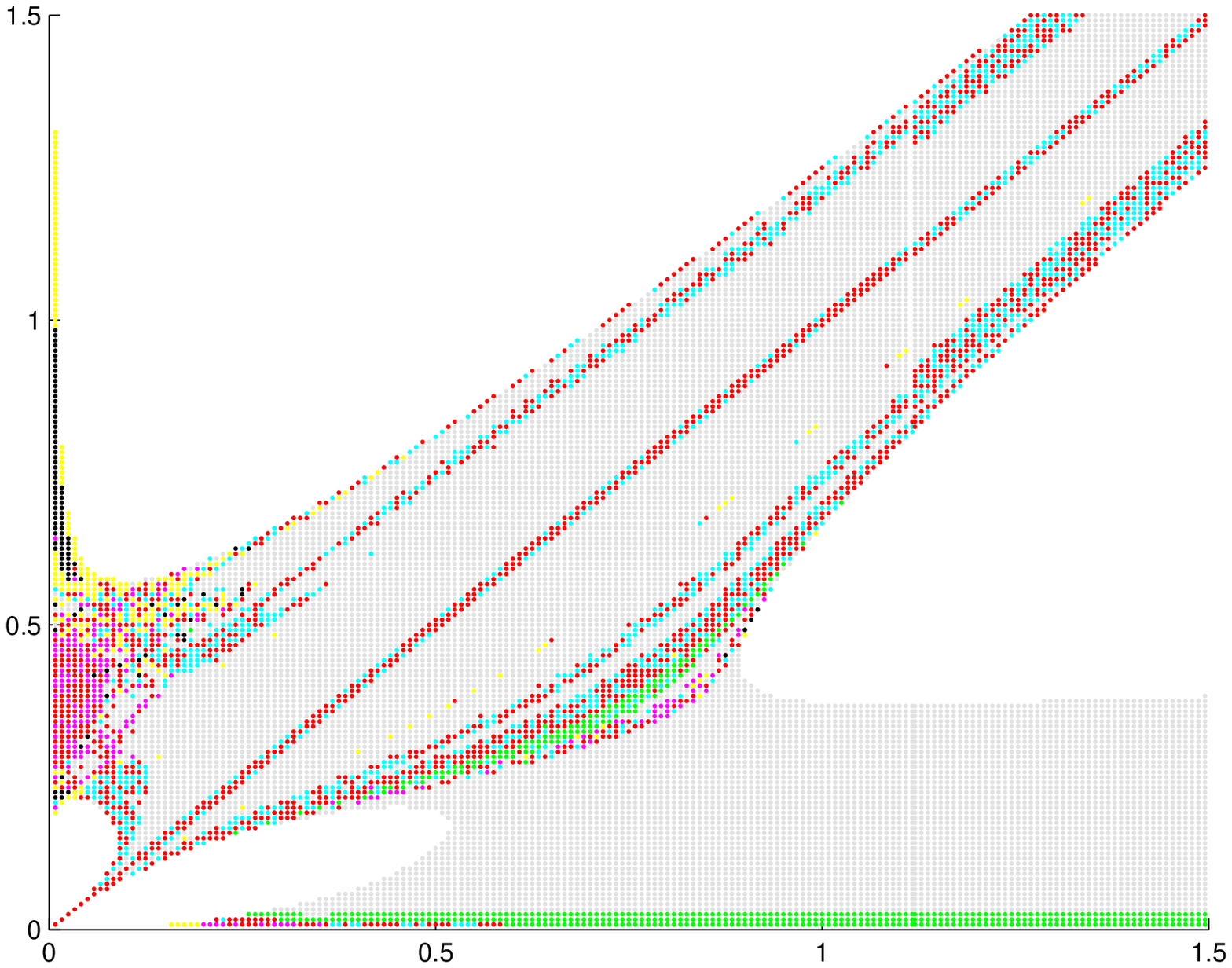}
  }
  \vspace{2pt}
  \centerline{(c)}
  \vspace{2pt}

  \caption{$\mu = 0.1, C_0=0.7, E_0= -1.2$ with no perturbation. Integration time a. $10^4$ time steps
   b. $10^5$ c. $10^6$. The colors indicate categories of orbits: red -12 type, yellow -13 type, magenta -14 type,
   blue -23 type, green -24 type, cyan -34 type, black -symmetry breaking and grey stable.}
  \label{fig4.13}
\end{figure}

\begin{figure}[hbtp]
  \centerline{
    \epsfxsize=7.0cm
    \epsffile{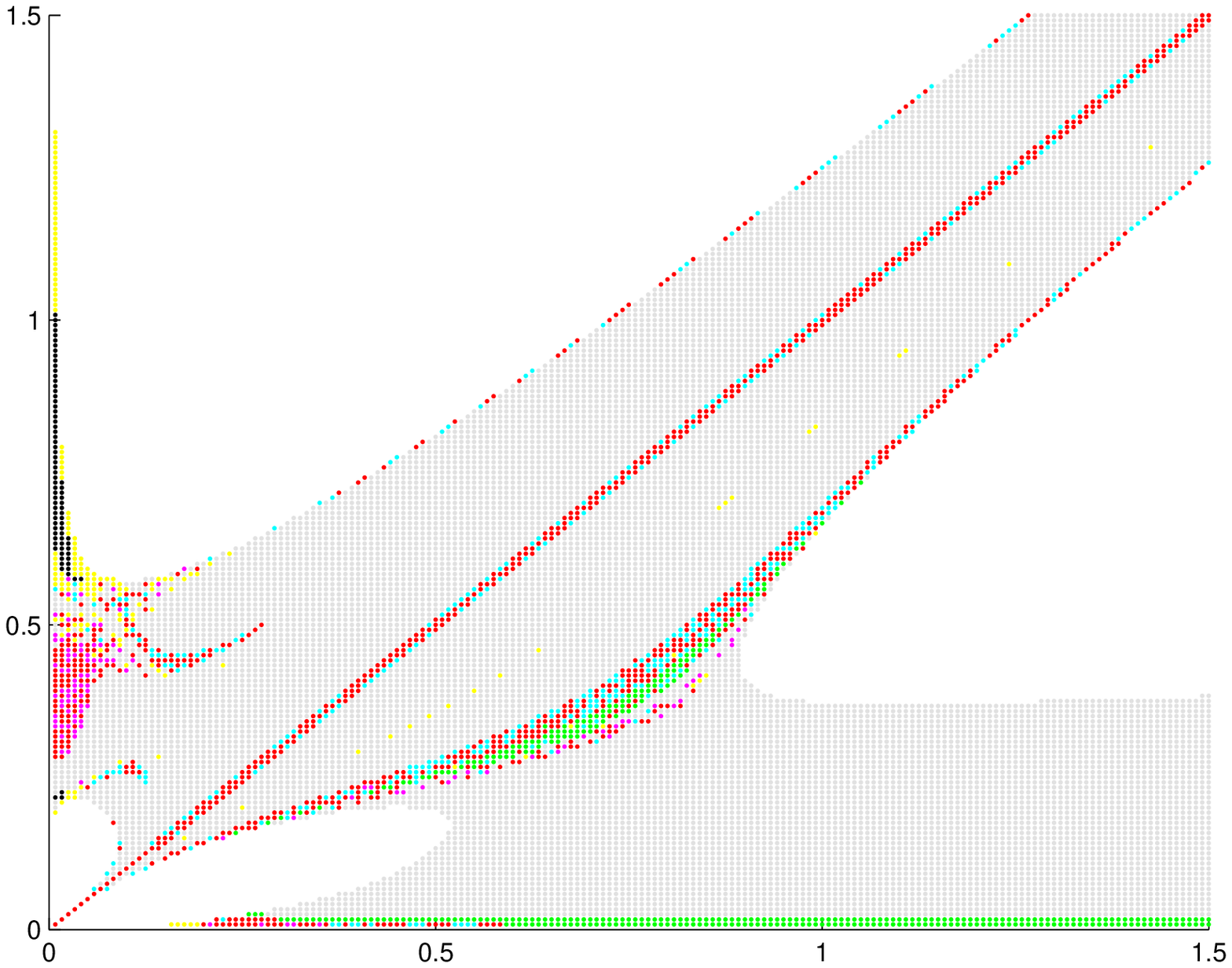}
  }
  \vspace{2pt}
   \centerline{(a)}
  \vspace{2pt}

  \centerline{
    \epsfxsize=7.0cm
    \epsffile{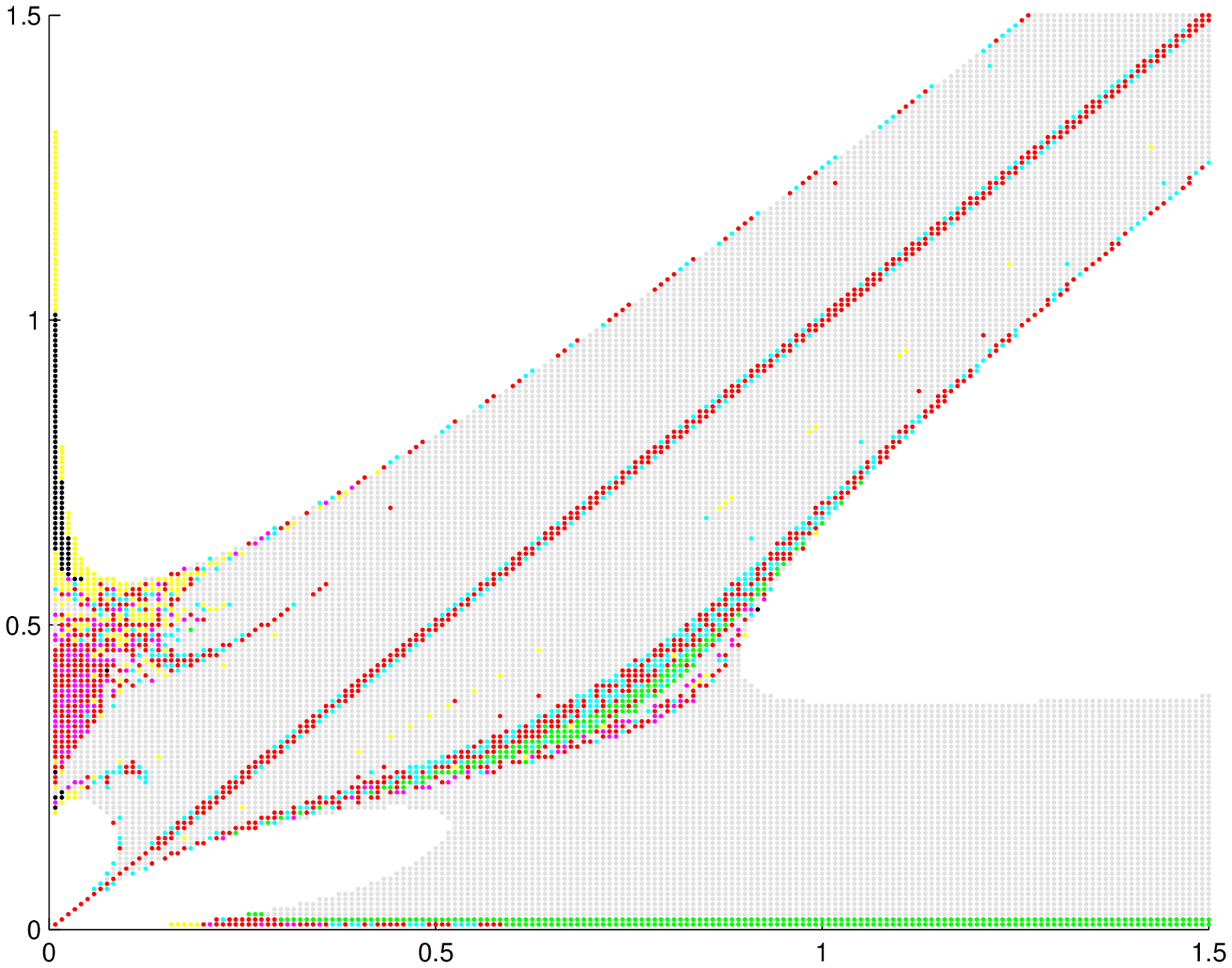}
  }
 \vspace{2pt}
  \centerline{(b)}
  \vspace{2pt}

  \centerline{
    \epsfxsize=7.0cm
    \epsffile{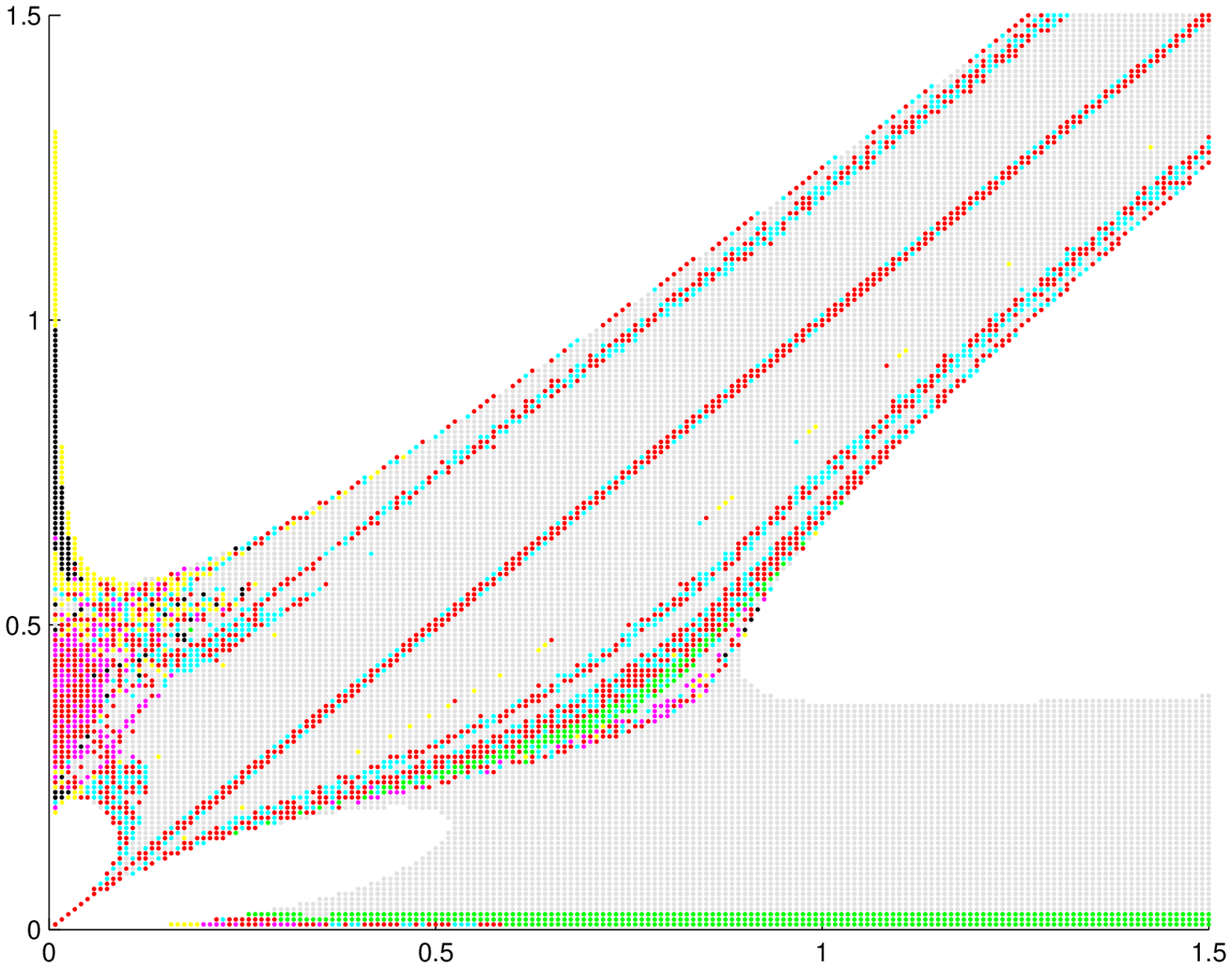}
  }
  \vspace{2pt}
  \centerline{(c)}
  \vspace{2pt}

  \caption{$\mu = 0.1, C_0=0.7, E_0= -1.2$ with perturbation of $10^{-5}$. Integration time a. $10^4$ time steps
   b. $10^5$ c. $10^6$. The colors indicate categories of orbits: red -12 type, yellow -13 type, magenta -14 type,
   blue -23 type, green -24 type, cyan -34 type, black -symmetry breaking and grey stable.}
  \label{fig4.14}
\end{figure}

\begin{figure}[hbtp]
  \centerline{
    \epsfxsize=7.0cm
    \epsffile{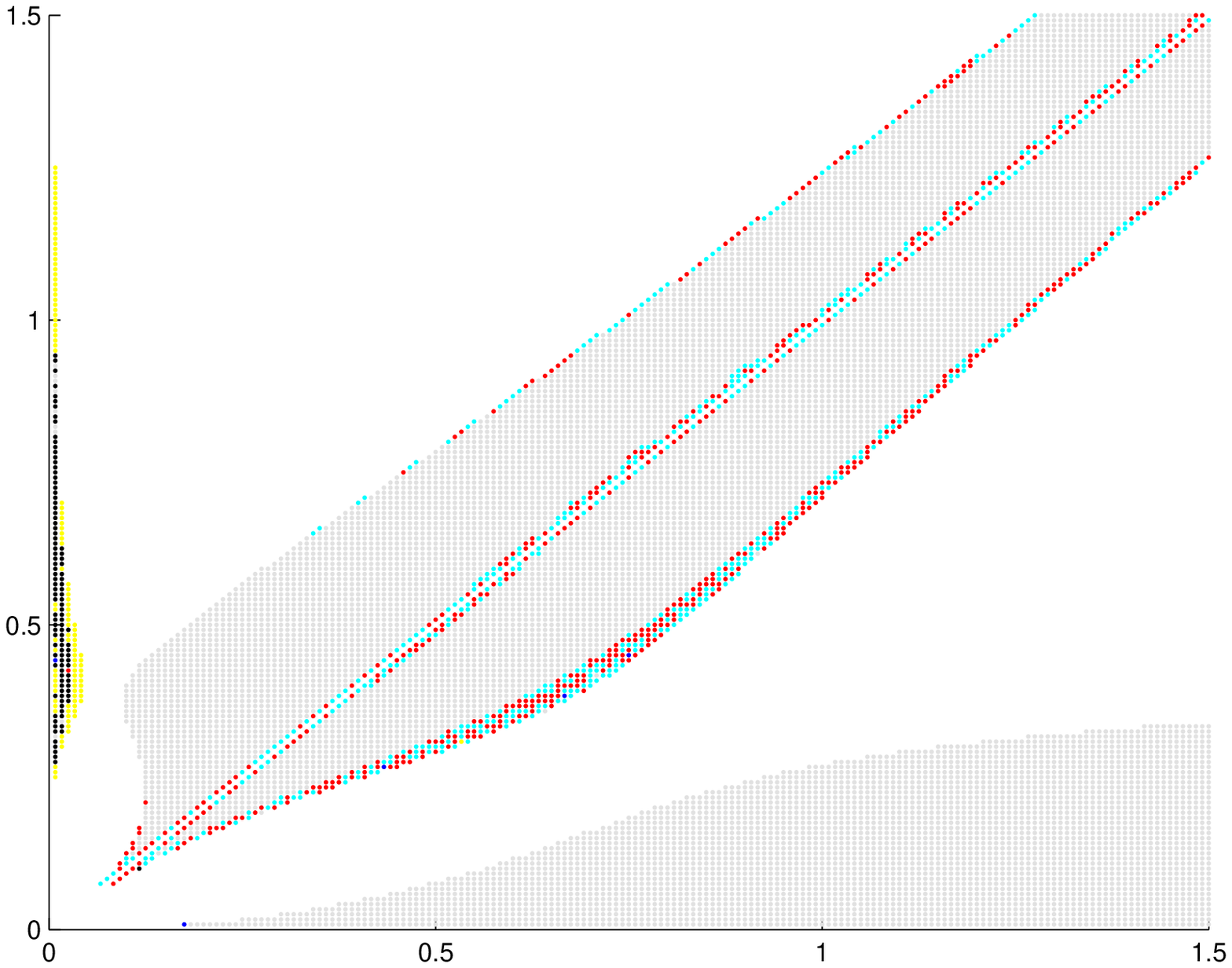}
  }
  \vspace{2pt}
   \centerline{(a)}
  \vspace{2pt}

  \centerline{
    \epsfxsize=7.0cm
    \epsffile{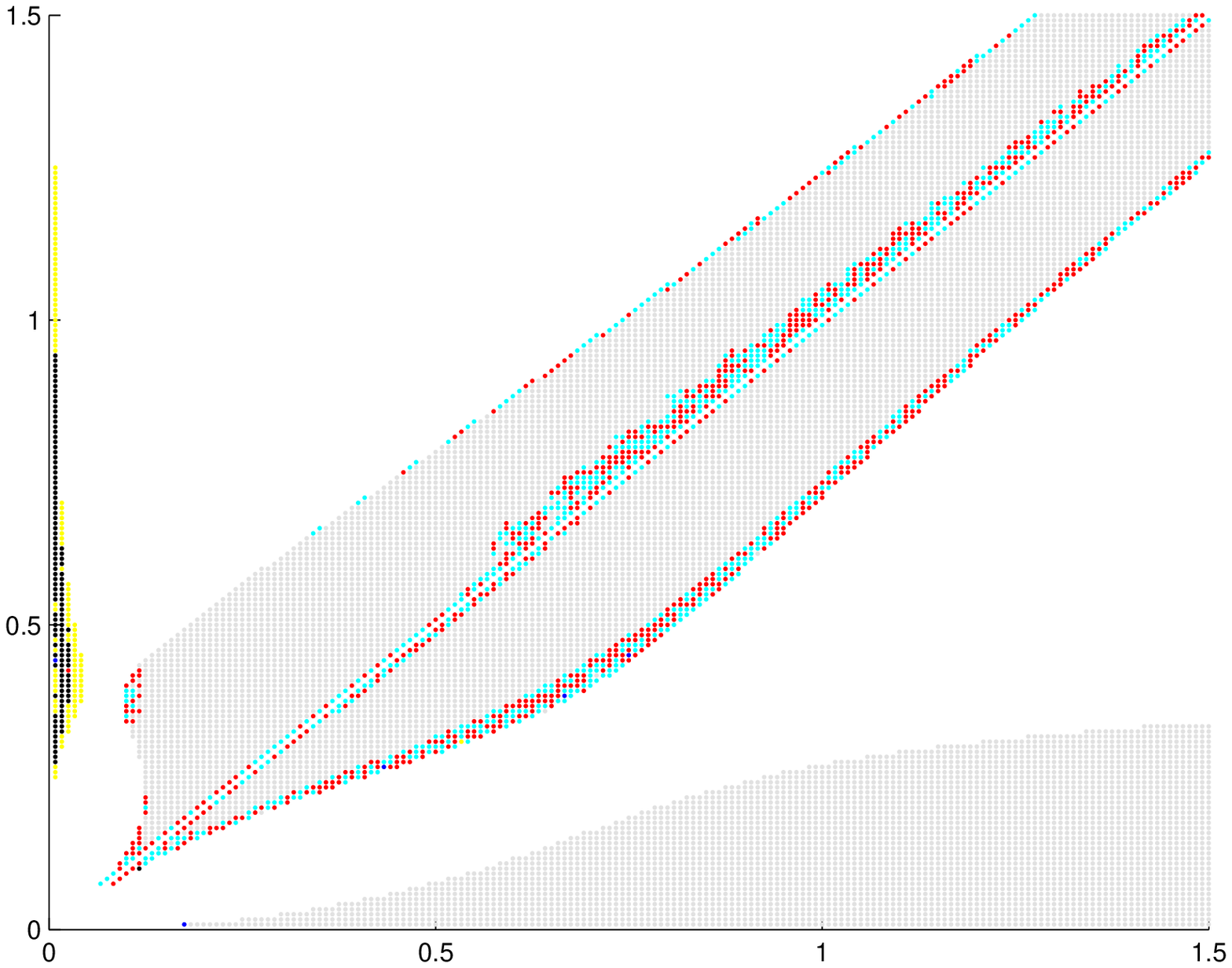}
  }
 \vspace{2pt}
  \centerline{(b)}
  \vspace{2pt}

  \centerline{
    \epsfxsize=7.0cm
    \epsffile{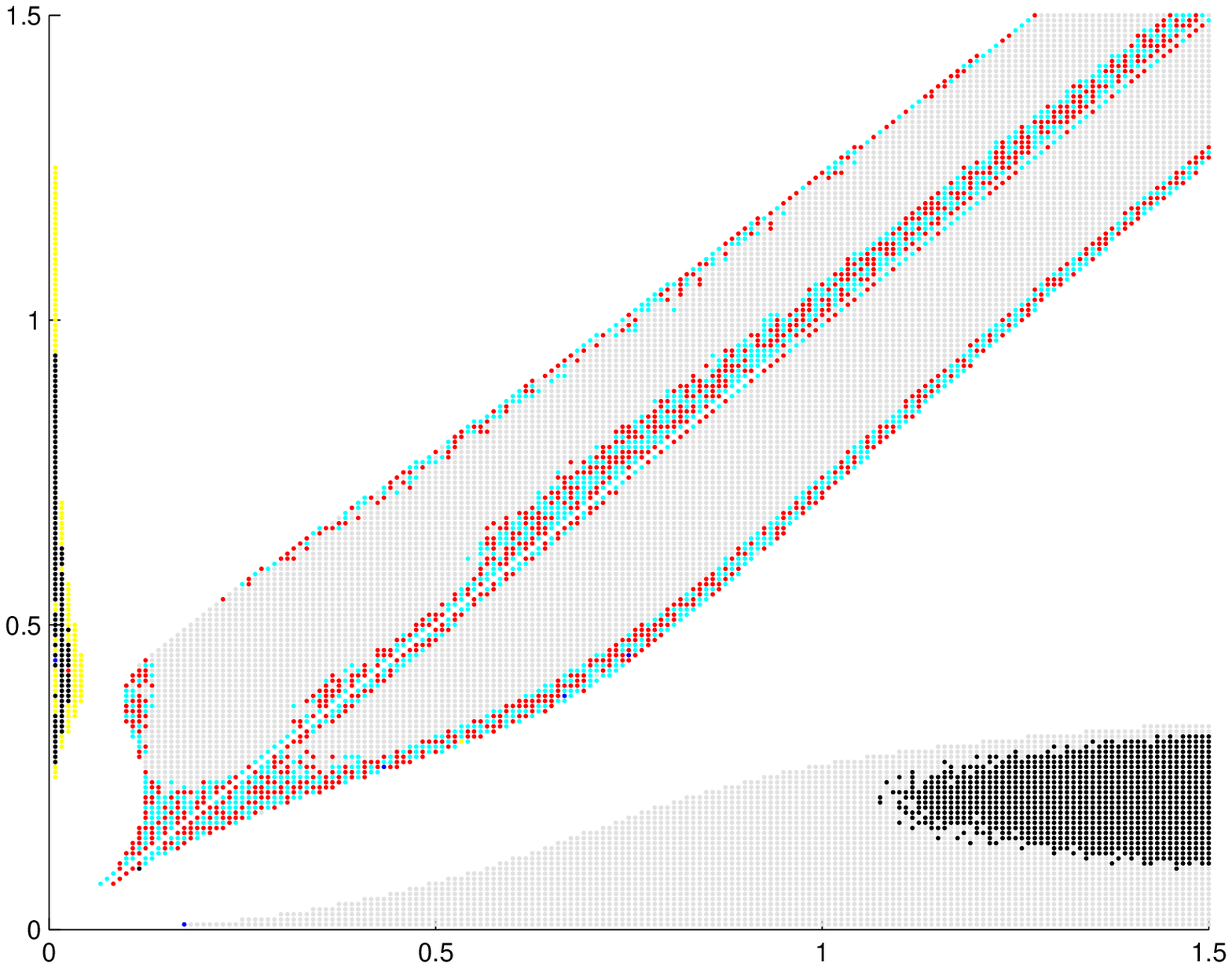}
  }
  \vspace{2pt}
  \centerline{(c)}
  \vspace{2pt}

  \caption{$\mu = 0.1, C_0=0.9, E_0= -1.2$ with no perturbation. Integration time a. $10^4$ time steps
   b. $10^5$ c. $10^6$. The colors indicate categories of orbits: red -12 type, yellow -13 type, magenta -14 type,
   blue -23 type, green -24 type, cyan -34 type, black -symmetry breaking and grey stable.}
  \label{fig4.15}
\end{figure}

\begin{figure}[hbtp]
  \centerline{
    \epsfxsize=7.0cm
    \epsffile{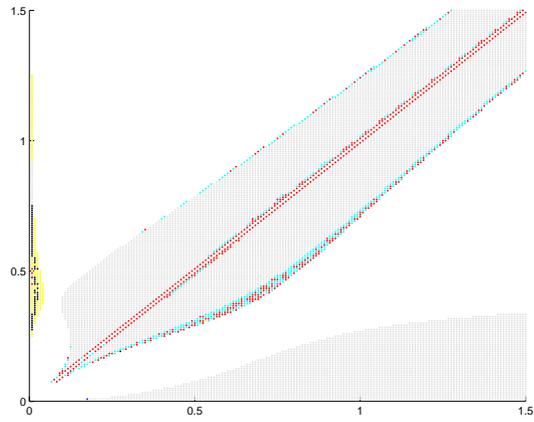}
  }
  \vspace{2pt}
   \centerline{(a)}
  \vspace{2pt}

  \centerline{
    \epsfxsize=7.0cm
    \epsffile{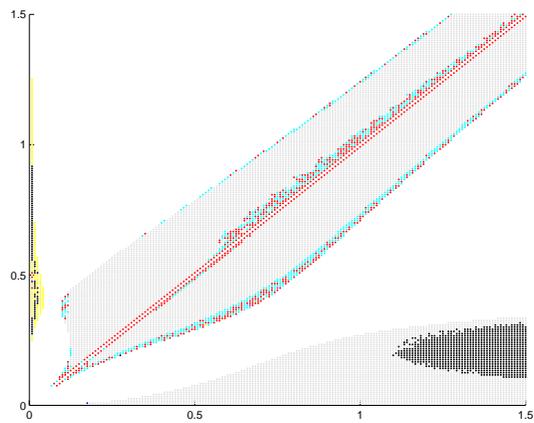}
  }
 \vspace{2pt}
  \centerline{(b)}
  \vspace{2pt}

  \centerline{
    \epsfxsize=7.0cm
    \epsffile{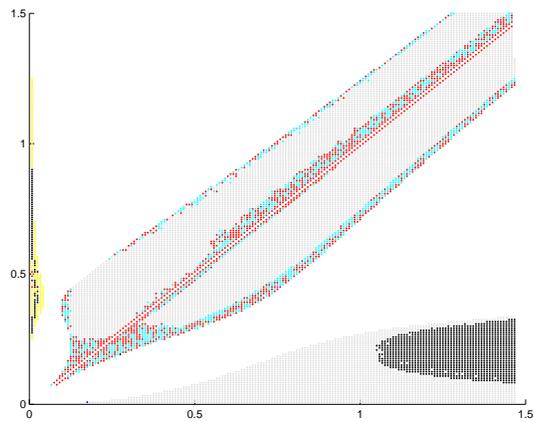}
  }
  \vspace{2pt}
  \centerline{(c)}
  \vspace{2pt}
  \caption{$\mu = 0.1, C_0=0.9, E_0= -1.2$ with perturbation of $10^{-5}$. Integration time a. $10^4$ time steps
   b. $10^5$ c. $10^6$. The colors indicate categories of orbits: red -12 type, yellow -13 type, magenta -14 type,
   blue -23 type, green -24 type, cyan -34 type, black -symmetry breaking and grey stable.}
  \label{fig4.16}
\end{figure}
\pagebreak
\renewcommand{\baselinestretch}{2}
\section{Summary and Conclusions}

In this chapter we investigated the stability of the symmetric
nature of the Caledonian Symmetric Four Body Problem (CSFBP) by
using nearly symmetric, slightly perturbed, initial conditions and
the general four body equations to see if the CSFBP system remain
symmetric.

We analyzed the phase space in detail in the $\mu=1$ and $\mu=0.1$
cases. For the $\mu$ values we drew graphs of the phase space of
the CSFBP with different $C_0$ values. Each point on the graphs,
Figures (\ref{fig4.2}) to (\ref{fig4.16}), describes a different
set of initial conditions of the CSFBP. In each graph we denote
different kinds of collisions with different colors. The orbits
which break the symmetry criterion are colored black and the
orbits which maintain the symmetry are colored grey. The main
behavior of the graphs were discussed in detail.

We found that the stability of the CSFBP orbits is dependent on
the value of the Szebehely constant. The larger the value of the
Szebehely constant the more stable is the CSFBP system. For the
equal mass case of the CSFBP, the initial conditions in the
$r_1-r_2$ space were very chaotic as most of the orbits ended in
collision orbits. The single binary regions were the most chaotic
as almost all of the orbits ended in collision. In the $\mu=0.1$
case, the SB2 region is the most chaotic and was comparatively
very small. The SB1 region is the most stable as overall there
were few collisions. To compare both the cases, $\mu=1$ and
$\mu=0.1$, the $\mu=0.1$ case is the most stable. In the $\mu=1$
case a large number of orbits fail the symmetry breaking
criterion. These orbits are not necessarily unstable therefore it
will be interesting to investigate, in future, the equal mass case
of the CSFBP without the symmetry restrictions. The effect of the
symmetry breaking is considerable in only one case when for
$\mu=0.1$ a large island of non-symmetric orbits appear in Figure
(\ref{fig4.16}b) which occurs after $10^5$ integration time steps
in the SB1 region. In all other cases of both $\mu=1$ and
$\mu=0.1$, there are not many breakings of the symmetry and
therefore it can be concluded that the CSFBP is stable to small
perturbations.

In both $\mu=1$ and $\mu=0.1$ case, the results were compared with
\citeasnoun{AndrasMNRAS}. It is found that the main features of
the CSFBP remain the same in both the analysis.

\begin{enumerate}
\item The stability of the CSFBP system increases as the value of
the Szebehely constant increases. \item The regions of real motion
are always surrounded by collision orbits. \item At the junction
of the single binary and double binary regions almost all of the
orbits are collision orbits \item The SB1 region is the most
stable in $\mu=0.1$ case as it has very few collision orbits,
while in $\mu=1$ case the double binary regions are the most
stable orbits. \item in all the comparisons made, the orbits which
maintained the dynamical symmetry over long times were recognized
as regular orbits in the Sz\'ell et.al (2004) investigations
\end{enumerate}

 In the next chapter we introduce a stationary mass to the
centre of mass of the CSFBP system to derive an analytical
stability criterion for the new system and also to see the effect
of the central stationary mass on the stability of the CSFBP.
Later in chapter 6 we will numerically investigate its
hierarchical stability.

%% file: ChapCS5BP.tex
\chapter{Caledonian Symmetric Five Body Problem (CS5BP)}

Steves and Roy (1998, 2000, 2001) have recently developed a
symmetrically restricted four body problem called the Caledonian
Symmetric Four Body Problem (CSFBP), for which they derive an
analytical stability criterion valid for all time. They show that
the hierarchical stability of the CSFBP depends solely on a
parameter they call the Szebehely Constant, $C_0$, which is a
function of the total energy and angular momentum of the system.
This stability criterion has been verified numerically by Sz\'ell,
Steves and \'erdi (2003a, 2003b), while the relationship between
the chaotic behavior of the phase space of the CSFBP and its
global stability as given by the Szebehely constant is analyzed by
Sz\'ell, \'erdi, S\'andor and Steves (2003).

Our aim in this chapter is to introduce a stationary mass to the
centre of mass of the CSFBP, to derive analytical stability
criterion for this five body system and to use it to discover the
effect on the stability of the whole system by adding a central
body. To do so we define a five body system in a similar fashion
to the CSFBP which we call the Caledonian Symmetric Five body
Problem (CS5BP).

Our motivation for studying this restricted or symmetric five body
problem comes from the three and four body problems, where
restriction methods utilizing assumptions of neglecting the masses
of some of the bodies or assumptions which require specific
conditions of symmetry have been very successful in reducing the
dimensions of the phase space to manageable levels while still
producing results which are meaningful to real physical systems.
For example a four body model requiring symmetrical restrictions
was used by Mikola, Saarinen and Valtonen (1984) as a means of
understanding multi-star formation in which symmetries produced in
the initial formation of the star system were maintained under the
evolution of the system. It is hoped that a restricted five body
model of this type which can easily be studied, may shed light on
the general five and four body problems in the same way that the
restricted three body problem has helped to deepen our
understanding of the general three body problem.

The Caledonian Symmetric Five Body Problem has direct applications
in dynamical systems where a very large mass exists at the centre
of mass with four smaller masses moving in dynamical symmetry
about it. The four small masses are affected by the central mass
but are small enough that they do not dynamically affect the
central body. Hence the central body remains stationary and
dynamical symmetry is maintained. This could occur, for example,
in exoplanetary systems of a star with four planets or a planetary
system with four satellites.

For completion we analyze the full range of mass ratios of central
body to the other four bodies: from the large central body system
described above to a five body system of equal masses to a small
central body with larger bodies surrounding it. In the case of
five equal masses or a small central body surrounded by larger
masses, the central body is unlikely to remain stationary as is
required by the CS5BP. The model may still be applicable to real
systems in which the outer bodies are well spaced and stationed
far away from the central body so that they have minimal effect on
the central body. The CS5BP model with $\mu_0=0$ i.e. the special
case of four bodies called the CSFBP, however, has no central body
required to remain stationary. It is therefore a realistic model
of symmetric four body systems: either four equal masses stellar
systems or two binary stars and two planets systems.

In section 5.1 we define the CS5BP in such a way that the CSFBP
becomes a special case of the CS5BP. The equations of motion and
Sundman's inequality, the key to the derivation of a stability
criterion, are given in section 5.2. We then derive in sections
5.3 to 5.5, the analytical stability criterion for the CS5BP. Here
we show that the hierarchical stability  depends solely on the
Szebehely constant which is a function of the total energy and
angular momentum. In section 5.5 we explain the role played by the
Szebehely constant $C_0$ in determining the topological stability
of the phase space. The topological stability of the phase space
for a wide range of mass ratios is discussed in section 5.6. The
conclusions are given in the final section 5.7.

This work constitutes a team effort of Steves, Shoaib and Roy. My
contribution to this work was to adopt the four body model of the
CSFBP by adding a central mass to it, derive the subsequent
stability criterion, and explore the stability of the CS5BP for
wide range of mass ratios. I also wrote the first draft of the
paper in the accompanying material to this thesis. I improved this
paper according to the suggestions of Steves and Roy and it has
subsequently formed this chapter.

\section{Definition of the Caledonian Symmetric Five Body Problem
(CS5BP)}

\begin{figure}
\centerline{\epsfig{file=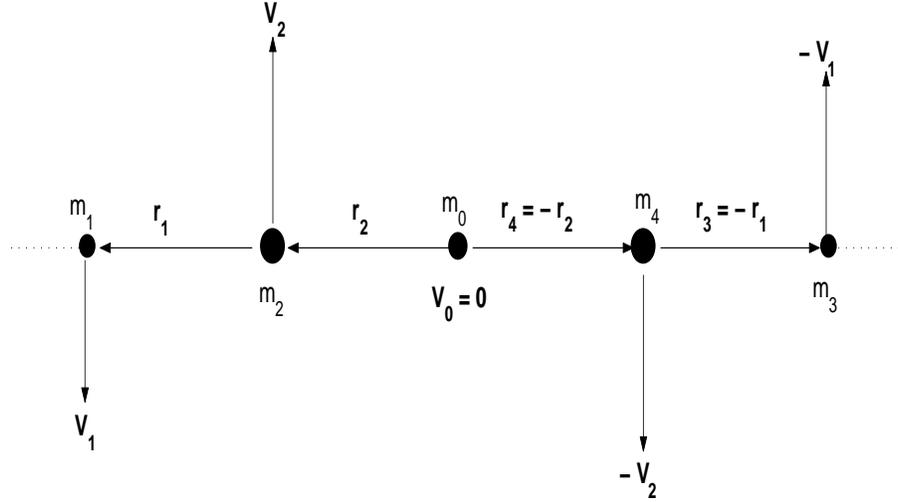,height=7cm,width=12cm}}

 \caption{The initial configuration of the CS5BP}
 \label{Col}
\end{figure}

The formulation of the CS5BP follows very closely from the
formulation of the CSFBP. The CSFBP is defined by Sz\'ell et.al
(2004) in the following manner. \begin{description}\item ''The
main feature of the model is its use of two types of symmetries.
1. past-future symmetry and 2. dynamical symmetry. Past future
symmetry exists in an n-body system when the dynamical evolution
of the system after $t=0$ is a mirror image of the dynamical
evolution of the system before $t=0$. It occurs whenever the
system passes through a mirror configuration, $i.e.$ a
configuration in which the velocity vectors of all the bodies are
perpendicular to all the position vectors from the centre of mass
of the system (Roy and Ovenden, 1955).
Dynamical symmetry exists when the dynamical evolution of two
bodies on one side of the centre of mass of the system is
paralleled by the dynamical evolution of the two bodies on the
other side of the centre of mass of the system. The resulting
configuration is always a parallelogram, but of varying length,
width and orientation.''
\end{description}
The above is a description of the CSFBP. However, it can be easily
seen that all symmetries are maintained if a fifth body of any
mass is placed at the centre of mass of the system and required to
be stationary there for all time. This constitutes the CS5BP.

Let us consider five bodies
$P_{0}$,$P_{1}$,$P_{2}$,$P_{3}$,$P_{4}$ of masses $
m_{0}$,$m_{1}$, $m_{2}$,$m_{3}$,$m_{4}$ respectively existing in
three dimensional Euclidean space. The radius and velocity vectors
of the bodies with respect to the centre of mass of the five body
system are given by $ \mathbf{r}_{i}$ and $\dot{\mathbf{r}}_{i}$
respectively, $i=0,1,2,3,4$. Let the centre of mass of the system
be denoted by $O$. The CS5BP has the following conditions:
\begin{enumerate}
\item All five bodies are finite point masses with:
\begin{equation}\label{1}
m_{1}=m_{3}, \qquad m_{2}=m_{4}
\end{equation}
\item $P_{0}$ is stationary at $O$, the centre of mass of the
system. $P_{1}$ and $P_{3}$ are moving symmetrically to each other
with respect to the centre of mass of the system. Likewise $P_{2}$
and $P_{4}$ are moving symmetrically to each other. Thus

\begin{eqnarray}\label{x6.2}
 \mathbf{r}_{1}=-\mathbf{r}_{3},\qquad
\mathbf{r}_{2}=-\mathbf{r}_{4},\qquad \mathbf{r}_{0}=0, \nonumber
\\
\mathbf{V}_{1}=\dot{\mathbf{r}}_{1}=-\dot{\mathbf{r}}_{3},\,\qquad
\mathbf{V}_{2}=\dot{\mathbf{r}}_{2}=-\dot{\mathbf{r}}_{4}, \qquad
\mathbf{V}_{0}=\dot{\mathbf{r}}_{0}=0,
\end{eqnarray}
This dynamical symmetry is maintained for all time $t$. \item At
time $t = 0$ the bodies are collinear with their velocity vectors
perpendicular to their line of position. This ensures past-future
symmetry and is described by:
\begin{eqnarray}
\mathbf{r}_{1}\times \mathbf{r}_{2}=0, \qquad \mathbf{r}_{1}\cdot
\dot{\mathbf{r}}_{1}=0, \qquad \mathbf{r}_{2}\cdot
\dot{\mathbf{r}}_{2}=0
\end{eqnarray}

\end{enumerate}
Figure (\ref{Col}) gives the initial configuration of the CS5BP.

It is useful to define the masses as ratios to the total mass. Let
the total mass $M$ of the system be
\begin{equation}\label{Tmass}
M=2\left( m_{1}+m_{2}\right) +m_{0}
\end{equation}
Let $\mu_{i}$ be the mass ratios defined as $\mu _{i}=
\frac{m_{i}}{M}$ for $i=0,1,2,3,4$. Equation (\ref{Tmass}) then
becomes
\begin{equation}
2\left( \mu _{1}+\mu _{2}\right) +\mu _{0}=1  \label{xrat}
\end{equation}
and
\begin{equation}
 0\leq\mu_0\leq1,\quad 0\leq\mu_1\leq0.5,\quad 0\leq\mu_2\leq0.5 \label{}
\end{equation}

\begin{figure}
\centerline{\epsfig{file=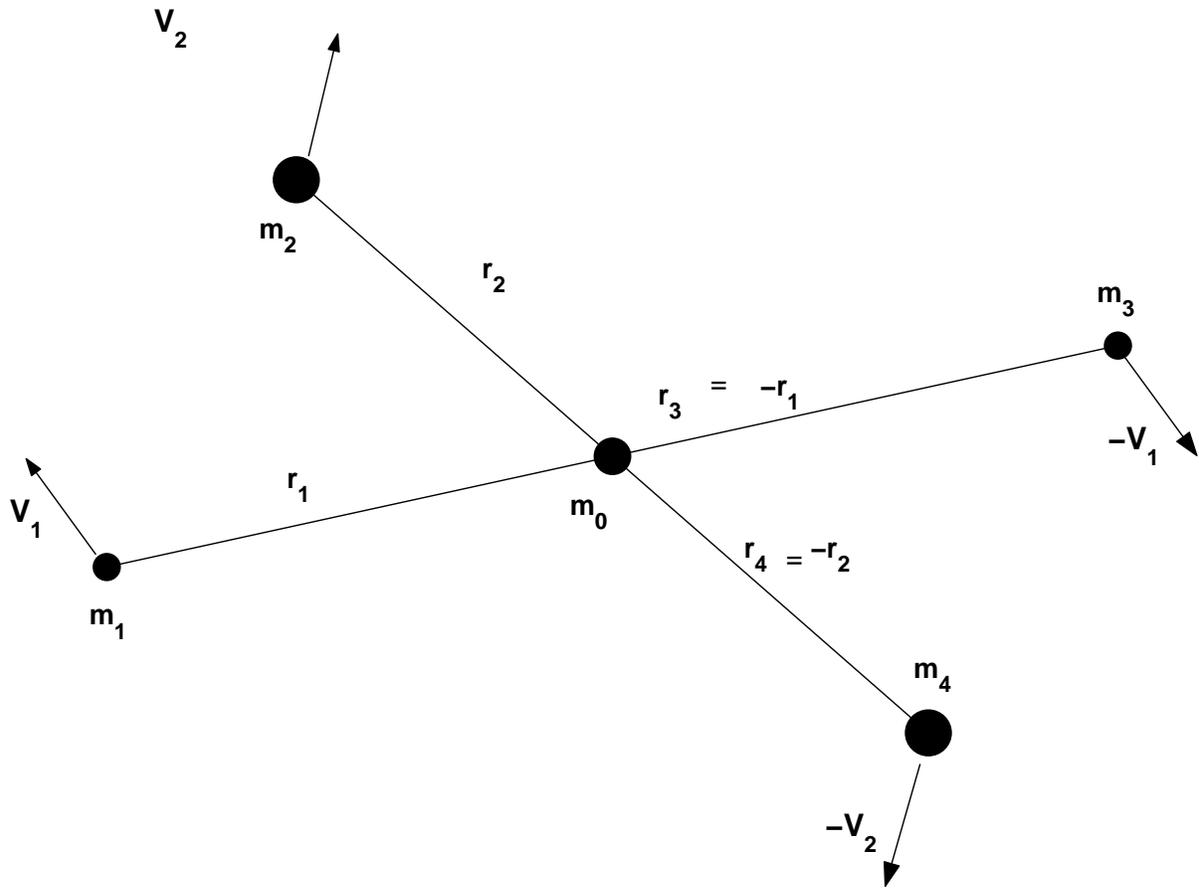,height=12cm,angle = 0}}

 \caption{The configuration of the coplanar CS5BP for $t > 0$}
 \label{5BP2}
\end{figure}

We simplify the problem yet further by studying solely the
coplanar CS5BP,where the radius and velocity vectors are coplanar.
Figure (\ref{5BP2}) gives the dynamical configuration of the
coplanar CS5BP at some time $t$. In the next sections we will
derive an analytical stability criterion for the CS5BP in such a
way that the CSFBP becomes a special case of the CS5BP. This
criterion will be applied to the CS5BP for a wide range of mass
ratios.

\section{The Equations of Motion and Sundman's Inequality for the CS5BP}
Taking the centre of mass of the system to be at rest and located
at the origin, the equations of motion of the general five body
system may be written as
\begin{equation}
m_{i}\ddot\mathbf{r}_{i}=\nabla _{i}U,\qquad i=0,1,2,3,4
\end{equation}
where $\mathbf{\nabla }_{i}=\mathbf{i}\frac{\partial }{\partial
x_{i}}+ \mathbf{j}\frac{\partial }{\partial
y_{i}}+\mathbf{k}\frac{\partial }{
\partial z_{i}}.\textrm { }\mathbf{i,j,k}$ are unit vectors, along the rectangular
axes $O_{x},O_{y},O_{z}$ respectively. $x_{i},y_{i},z_{i}$ are the
rectangular coordinates of body $P_{i}$ and $O$ is the centre of
mass of the system.

Given (\ref{x6.2}), the differential equations for the CS5BP
reduce to
\begin{equation}
m_{i}\ddot\mathbf{r}_{i}=\nabla _{i}U,\qquad i=0,1,2
\end{equation}

For the CS5BP the force function $U$ can be written as
$$
U=G\left[ \frac{1}{2}\left(
\frac{m_{1}^{2}}{r_{1}}+\frac{m_{2}^{2}}{r_{2}} \right)
+2m_{1}m_{2}\left( \frac{1}{r_{12}}+\frac{1}{\sqrt{2\left(
r_{1}^{2}+r_{2}^{2}\right) -r_{12}^{2}}}\right) \right .
$$
\begin{equation}\label{3}
\left . +2m_{0}\left( \frac{m_{1}}{
r_{1}}+\frac{m_{2}}{r_{2}}\right) \right ].
\end{equation}
where ${\mathbf{r}}_{12}={\mathbf{r}}_{1}-{\mathbf{r}}_{2}.$ Since
$\ m_{0}$ is stationary at the centre of mass, it does not
contribute to the kinetic energy T, the angular momentum
\textbf{c} or the moment of inertia I. Equations for these
quantities are given below
\begin{equation}
T = m_{1}\dot{r}_{1}^{2}+m_{2}\dot{r}_{2}^{2}, \label{4}
\end{equation}
\begin{equation}
\ \mathbf{c} = 2\left( m_{1}\mathbf{r}_{1}\times
\dot{\mathbf{r}_{1}}+m_{2}\mathbf{r}_{2}\times \dot{\mathbf{r}
_{2}}\right) ,  \label{5}
\end{equation}
\begin{equation}
I = 2\left( m_{1}r_{1}^{2}+m_{2}r_{2}^{2}\right) . \label{6}
\end{equation}

Sundman's inequality (Roy, 2004) may be written as
\begin{equation}
U+E\geq \frac{c^{2}}{2I}.
\end{equation}
For a given energy and angular momentum, Sundman's inequality
gives boundary surfaces in $\mathbf{r}_{i}$ space which define
regions of real motion. For the CS5BP, these boundary surfaces can
be written solely in terms of $r_{1}$$r_{2}$$r_{12}$ space as
follows.
 Let $E_{0}=-E.$ Then Sundman's inequality for the CS5BP becomes
$$
G\left [ \frac{1}{2}\left(
\frac{m_{1}^{2}}{r_{1}}+\frac{m_{2}^{2}}{r_{2}} \right)
+2m_{1}m_{2}\left( \frac{1}{r_{12}}+\frac{1}{\sqrt{2\left(
r_{1}^{2}+r_{2}^{2}\right) -r_{12}^{2}}}\right) \right .
$$
\begin{equation}\label{8}
\left . +2m_{0}\left(\frac{m_{1}}{
r_{1}}+\frac{m_{2}}{r_{2}}\right) \right ]\geq \frac{c^{2}}{4(
m_{1}r_{1}^{2}+m_{2}r_{2}^{2})}+E_{0}.
\end{equation}
Introducing the mass ratios $\mu_{i}$, (\ref{8}) becomes
$$
GM^{2}\left[ \frac{1}{2}\left( \frac{\mu
_{1}^{2}}{r_{1}}+\frac{\mu _{2}^{2}}{r_{2}}\right) +2\mu _{1}\mu
_{2}\left( \frac{1}{r_{12}}+\frac{1}{ \sqrt{2\left(
r_{1}^{2}+r_{2}^{2}\right) -r_{12}^{2}}}\right) \right .
$$
\begin{equation}\label{9}
\left . +2\mu _{0}\left(
\frac{\mu_{1}}{r_{1}}+\frac{\mu_{2}}{r_{2}}\right) \right ] \geq
\frac{c^{2}}{4M\left( \mu _{1}r_{1}^{2}+\mu _{2}r_{2}^{2}\right)}
+E_{0}.
\end{equation}
Let us now introduce dimensionless variables $\rho _{1}$, $\rho
_{2}$ and $\rho _{12}$, and a new dimensionless constant $C_{0},$
called the Szebehely constant (Steves and Roy, 2001) as follows
$$
\rho _{1}=\frac{E_{0}}{GM^{2}}r_{1},\qquad \rho _{2}=\frac{E_{0}}{
GM^{2}}r_{2},  \nonumber
$$
\begin{equation}\label{10}
\rho _{12} =\frac{E_{0}}{GM^{2}}r_{12} \qquad
C_{0}=\frac{c^{2}E_{0}}{G^{2}M^{5}},
\end{equation}
where $E_{0}\neq 0$. Then Sundman's inequality takes the following
form
$$
\frac{1}{2}\left( \frac{\mu _{1}^{2}}{\rho _{1}}+\frac{\mu
_{2}^{2}}{\rho _{2}}\right) +2\mu _{1}\mu _{2}\left( \frac{1}{\rho
_{12}}+\frac{1}{\sqrt{2\left( \rho _{1}^{2}+\rho _{2}^{2}\right)
-\rho _{12}^{2}}}\right)
$$
\begin{equation}\label{11}
+2\mu _{0}\left( \frac{\mu _{1}}{\rho _{1}}+\frac{\mu _{2}}{\rho
_{2}}\right)
\\
\geq \frac{C_{0}}{4\left( \mu _{1}\rho _{1}^{2}+\mu _{2}\rho
_{2}^{2}\right) }+1.
\end{equation}
Additionally we have the following kinematic constraints on the
problem
\begin{equation}
|r_{1}-r_{2}|\leq r_{12}\leq r_{1}+r_{2}
\end{equation}
which in the dimensionless variables becomes
\begin{equation}\label{eq18}
|\rho _{1}-\rho _{2}|\leq \rho _{12}\leq \rho _{1}+\rho _{2}
\end{equation}
Recall that given $\mu_{1}$ and $\mu_{0}$, $\mu_{2}$ can be always
determined using (\ref{xrat}). Thus for a CS5BP system with given
mass ratios of $\mu_{1}$ and $\mu_{0}$ and Szebehely constant
(i.e. angular momentum and energy combination) $C_0$, the
equalities of (\ref{11}) and (\ref{eq18}) define a surface in
dimensionless coordinate space $\rho_1$$\rho_2$$\rho_{12}$ which
confine the possible motions.

If any region of the possible space $\rho_1$$\rho_2$$\rho_{12}$ is
totally disconnected from any other, then the hierarchical
arrangement of the system in that region cannot physically evolve
into the hierarchical arrangements possible in the other regions
of real motion. Thus a CS5BP system existing in that hierarchical
arrangement would be hierarchically stable for all time. The
topology or disconnectedness of the boundary surface given by
(\ref{11}) and (\ref{eq18}) can therefore provide a stability
criterion for the system.

\section{Regions of motion in the CS5BP}

In this section, we construct explicit formulae for the boundary
surface of real motion, enabling us to draw it in section 4, and
in section 5 to identify the critical points for which the
topology and therefore the stability of the system changes.

It is useful to parameterize the surface in terms of variables
$y_{i}$ $(i=1,2)$ and $x_{12}$,
\begin{equation}
y_{i}=\frac{\rho _{i}}{\rho _{n}}\textrm{ where }\rho _{n}=\max
(\rho _{1},\rho _{2}) \textrm{ and }x_{12}=\frac{\rho _{12}}{\rho
_{n}}.
\end{equation}
This allows us to study two halves of the phase space separately,
depending on the relative magnitudes of $\rho_1$ and $\rho_2$.
\begin{enumerate}
\item if $\rho_1 > \rho_2$, then $\rho_n = \rho_1$; $y_1=1$;
$y_{2}=\frac{\rho _{2}}{\rho _{1}}$ and $x_{12}=\frac{\rho
_{12}}{\rho _{1}}$.

\item if $\rho_2 > \rho_1$, then $\rho_n = \rho_2$; $y_2=1$;
$y_{1}=\frac{\rho _{1}}{\rho _{2}}$ and $x_{12}=\frac{\rho
_{12}}{\rho _{2}}$.
\end{enumerate}

 In the new variables Sundman's inequality takes the following form
$$
\frac{1}{\rho _{n}}\left[ \frac{1}{2}\left( \frac{\mu
_{1}^{2}}{y_{1}}+ \frac{\mu _{2}^{2}}{y_{2}}\right) +2\mu _{1}\mu
_{2}\left( \frac{1}{x_{12}}+ \frac{1}{\sqrt{2\left(
y_{1}^{2}+y_{2}^{2}\right) -x_{12}^{2}}}\right) \right .
$$
\begin{equation}\label{12}
\left . +2\mu _{0}\left( \frac{\mu _{1}}{y_{1}}+\frac{\mu
_{2}}{y_{2}}\right) \right ] \geq \frac{1}{\rho
_{n}^{2}}\frac{C_{0}}{4\left( \mu _{1}y_{1}^{2}+\mu
_{2}y_{2}^{2}\right) }+1.
\end{equation}\label{20a}
In the new variables the kinematic constraint of (\ref{eq18})
becomes
\begin{equation}
|y_{1}-y_{2}|\leq x_{12}\leq y_{1}+y_{2}
\end{equation}
 Taking the equality sign in (\ref{12}), we obtain the
following quadratic equation which defines the boundary surface
between real and imaginary motion,
\begin{equation}
\rho _{n}^{2}-A\rho _{n}+B=0,
\end{equation}
where
\begin{eqnarray}
A\left( y_{1},y_{2},x_{12}\right)  &=&\frac{1}{2}\left( \frac{\mu
_{1}^{2}}{ y_{1}}+\frac{\mu _{2}^{2}}{y_{2}}\right) +2\mu _{1}\mu
_{2}\left( \frac{1}{ x_{12}}+\frac{1}{\sqrt{2\left(
y_{1}^{2}+y_{2}^{2}\right) -x_{12}^{2}}}
\right)   \nonumber \\
&&+2\mu _{0}\left( \frac{\mu _{1}}{y_{1}}+\frac{\mu
_{2}}{y_{2}}\right) ,
\end{eqnarray}

and
\begin{equation}
B=\frac{C_{0}}{4\left( \mu _{1}y_{1}^{2}+\mu _{2}y_{2}^{2}\right)
}. \label{B}
\end{equation}
The solution of the above quadratic equation is
\begin{equation}\label{rowm}
\rho _{n}=\frac{1}{2}\sqrt{\frac{C(y_{1},y_{2},x_{12})}{\mu
_{1}y_{1}^{2}+\mu _{2}y_{2}^{2}}}\left( 1\pm \sqrt{1-\frac{C_{0}}{
C(y_{1},y_{2,}x_{12})}}\right)
\end{equation}
where
\begin{equation}
C(y_{1},y_{2,}x_{12})=A^{2}\left( \mu _{1}y_{1}^{2}+\mu
_{2}y_{2}^{2}\right) . \label{C1}
\end{equation}

Thus for case 1. $\rho_1 > \rho_2$

\begin{equation}
\rho
_{1}(y_{2},x_{12})=\frac{1}{2}\sqrt{\frac{C(y_{2},x_{12})}{\mu
_{1}+\mu _{2}y_{2}^{2}}}\left( 1\pm
\sqrt{1-\frac{C_{0}}{C(y_{2},x_{12})}}\right)
\end{equation}
where
\begin{equation}
C(y_{2},x_{12})=\left( \mu _{1}+\mu _{2}y_{2}^{2}\right) \left[
\begin{array}{c}
\frac{1}{2}\left( \mu _{1}^{2}+\frac{\mu _{2}^{2}}{y_{2}^{2}}\right) + \\
2\mu _{1}\mu _{2}\left( \frac{1}{x_{12}}+\frac{1}{\sqrt{2\left(
1+y_{2}^{2}\right) -x_{12}^{2}}}\right) +2\mu _{0}\left( \mu
_{1}+\frac{\mu _{2}}{y_{2}}\right)
\end{array}
\right]^{2}
\end{equation}

with the constraints $0\leq y_{2}\leq 1$ and from (\ref{eq18})
\begin{equation}
1-y_{2}\leq x_{12}\leq 1+y_{2}  \label{19a}
\end{equation}
For a given $\rho _{1}$, the values of $\rho _{2}$ and $\rho
_{12}$ can be reconstructed by
\begin{equation}
\rho _{2}=y_{2}\rho _{1}\qquad \rho _{12}=x_{12}\rho _{1}.
\label{19bee}
\end{equation}

For case (2) $\rho _{2}>\rho _{1}$
\begin{equation}
\rho
_{2}(y_{1},x_{12})=\frac{1}{2}\sqrt{\frac{C(y_{1},x_{12})}{\mu
_{1}y_{1}^{2}+\mu _{2}}}\left( 1\pm
\sqrt{1-\frac{C_{0}}{C(y_{1},x_{12})} }\right)
\end{equation}
\begin{equation}
C(y_{1},x_{12})=\left( \mu _{1}y_{1}^{2}+\mu _{2}\right) \left[
\begin{array}{c}
\frac{1}{2}\left( \frac{\mu _{1}^{2}}{y_{1}^{2}}+\mu _{2}^{2}\right) + \\
2\mu _{1}\mu _{2}\left( \frac{1}{x_{12}}+\frac{1}{\sqrt{2\left(
1+y_{1}^{2}\right) -x_{12}^{2}}}\right) +2\mu _{0}\left( \frac{\mu
_{1}}{ y_{1}}+\mu _{2}\right)
\end{array}
\right]^{2}
\end{equation}
with the constraints $0\leq y_{1}\leq 1$ and from (\ref{eq18})
\begin{equation}
1-y_{1}\leq x_{12}\leq 1+y_{1}  \label{19b}
\end{equation}
For a given $\rho _{2}$, the value of $\rho _{1}$ and $\rho _{12}$
can be reconstructed by
\begin{equation}
\rho _{1}=y_{1}\rho _{2},\qquad \rho _{12}=x_{12}\rho _{2}.
\end{equation}

The cases 1 and 2 , given above, provide an explicit set of
formulae for determining points $(\rho_1,\rho_2,\rho_{12})$ on the
boundary surface. The parameters $(y_1,x_{12})$ or $(y_2,x_{12})$
are simply varied from 0 to 1. Here $y_1$ and $y_2$ are the
gradients of straight lines through the origin $O$ in the
$\rho_1$$O$$\rho_2$ plane. $x_{12}$ is the gradient of a straight
line through the origin $O$ in either the $\rho_1$$O$$\rho_{12}$
or the $\rho_2$$O$$\rho_{12}$ plane.

\begin{figure}
\centerline{\epsfig{file=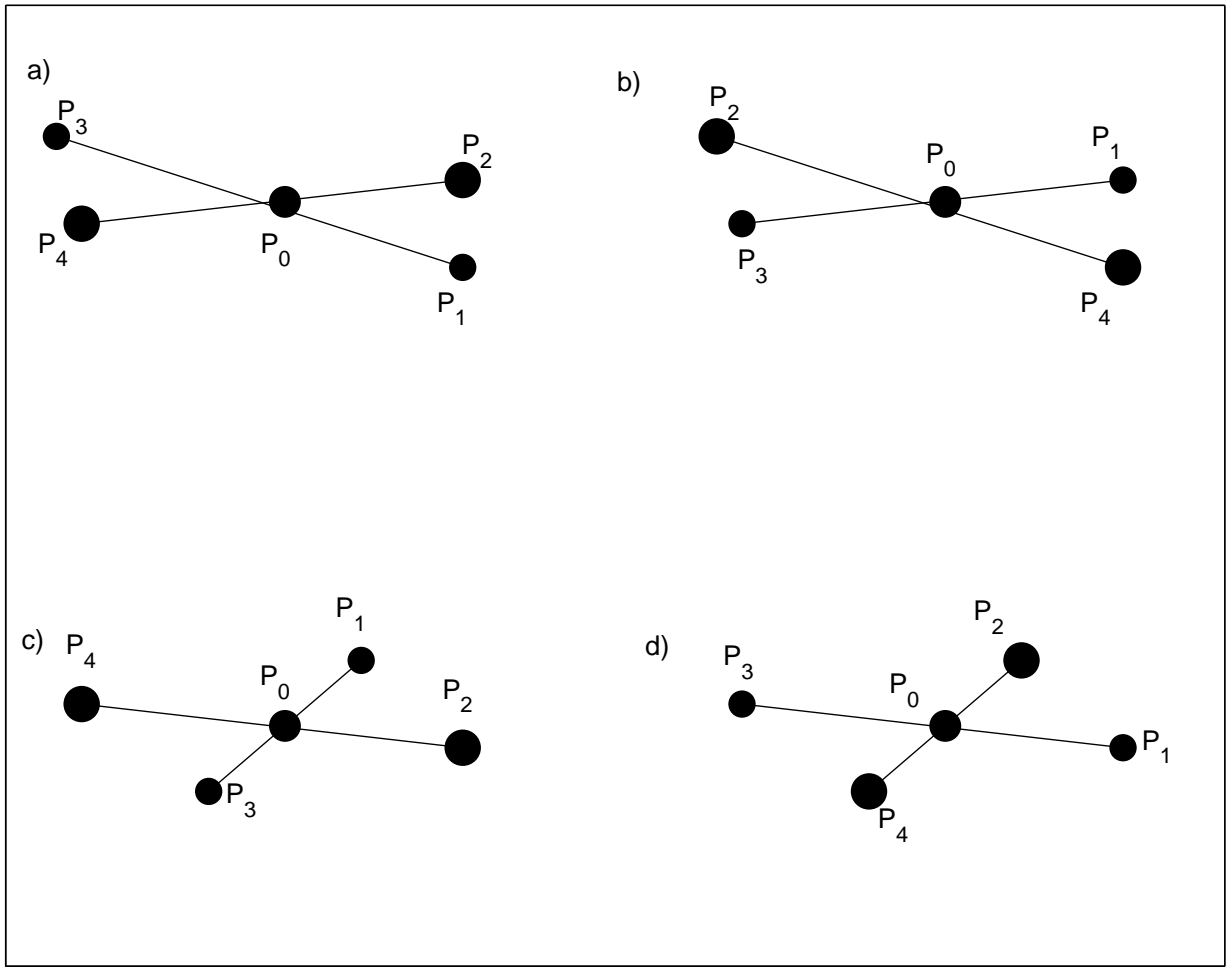,height=13cm,angle = 0}}

 \caption{The four possible hierarchies in the CS5BP}
 \label{hierarchies}
\end{figure}

Steves and Roy (2000,2001) showed that for the CSFBP, real motion
takes place in four tube-like regions of the
$\rho_1$$\rho_2$$\rho_{12}$ space that are connected to each other
near the origin for $C_0=0$. Each tube represents a particular
hierarchical arrangement. This is also the case in the CS5BP. See
figure (\ref{hierarchies}).

The four type of hierarchies possible in the CS5BP are as follows:
\begin{enumerate}
\item  $'12'$ type hierarchy, figure (\ref{hierarchies}a). A
double binary hierarchy where $P_{1}$ and $ P_{2}$ orbit their
centre of mass $C_{12}$. $P_{3}$ and $P_{4}$ orbit their centre of
mass $C_{34}$. The centre of masses $C_{12}$ and $C_{34}$ orbit
each other about the centre of mass of the five-body system $C$.

\item  $'14'$ type hierarchy, figure (\ref{hierarchies}b). A
double binary hierarchy where $P_{1}$ and $ P_{4}$ orbit their
centre of mass $C_{14}$. $P_{2}$ and $P_{3}$ orbit their centre of
mass $C_{23}$. The centre of masses $C_{14}$ and $C_{23}$ orbit
each other about the centre of mass of the five-body system $C$.
 \item  $'13'$ type hierarchy, figure (\ref{hierarchies}c).
A single binary hierarchy where $P_{1}$ and $ P_{3}$ orbit their
centre of mass $C$ in a small central binary. The $P_{2}$ and
$P_{4}$ orbit around the central binary.

\item  $'24'$ type hierarchy, figure (\ref{hierarchies}d). A
single binary hierarchy where $P_{2}$ and $ P_{4}$ orbit their
centre of mass $C$ in a small central binary. The $P_{1}$ and
$P_{3}$ orbit around the central binary.
\end{enumerate}

\section{Projections in the $\rho_1$-$\rho_2$ plane of real motion in the $\rho_1$$\rho_2$$\rho_{12}$
space}

In the CSFBP, Steves and Roy (2001) found that as $C_0$ is
increased, forbidden regions near the origin grow to the point
where they meet the boundary walls of the tubes, resulting in
disconnected regions. They found that it was possible to study the
connectivity of the regions of motion by projecting the
intersection of the boundary surface with the extreme values of
$\rho_{12}$ on to the $\rho_1$$O$$\rho_2$ plane. This is also the
case for the CS5BP.
\subsection{Maximum extension of the real motion projected in $\rho_1$$\rho_2$ space}
Equation (\ref{eq18}) and its equivalent in the new variables
(6.22) give the extreme values of $\rho_{12}$ and $x_{12}$
respectively. Intersection of the kinematic constraints with the
boundary surface (6.17) or (6.26)  give curves projected on the
$\rho_1$$\rho_2$ plane which show the maximum widths of the four
tubes as three arms. The two tubes located near
$\rho_1$$\approx$$\rho_2$ lay one on top of the other in the
projection, giving one arm near $\rho_1$$\approx$$\rho_2$.

The projection curves of the maximum widths of the tubes can be
found by substituting the equality of (6.22) into the boundary
surface (6.26). Both limits $x_{{12}_+}=y_1+y_2$ and
$x_{{12}_-}=|y_1-y_2|$ give the same equations, indicating that
the maximum widths at the $x_{{12}_+}$ upper location and the
lower location $x_{{12}_-}$ are identical.

The equations giving the maximum projections are:
\begin{equation}\label{34a}
\rho _{n}\left( y_{1},y_{2}\right)
=\frac{1}{2}\sqrt{\frac{C_{e}(y_{1},y_{2}) }{\mu _{1}y_{1}^{2}+\mu
_{2}y_{2}^{2}}}\left( 1\pm \sqrt{1-\frac{C_{0}}{
C_{e}(y_{1},y_{2})}}\right)
\end{equation}
where
\begin{eqnarray}\label{34b}
C_{e }(y_1,y_2)=\left( \mu _{1}y_{1}^{2}+\mu _{2}y_{2}^{2}\right)
\times \left[ \frac{1}{2}\left( \frac{\mu
_{1}^{2}}{y_{1}}+\frac{\mu _{2}^{2}}{y_{2}} \right)\right
.\nonumber
\\
\left . +2\mu _{0}\left( \frac{\mu _{1}}{y_{1}}+\frac{\mu
_{2}}{y_{2}} \right) +4\left( \mu _{1}\mu _{2}\frac{1}{
|y_{1}^{2}-y_{2}^{2}|}\right) \right ] ^{2}.
\end{eqnarray}
Recall that there are only two independent mass ratios required
since $2(\mu_1+\mu_2)+\mu_0=1$.

For case 1, $\rho_1>\rho_2$, (\ref{34a}) becomes
\begin{equation}\label{34c}
\rho _{1}(y_{2})=\frac{1}{2}\sqrt{\frac{C_{e}(y_{2})}{\mu _{1}+\mu
_{2}y_{2}^{2}}}\left( 1\pm
\sqrt{1-\frac{C_{0}}{C_{e}(y_{2})}}\right)
\end{equation}
where
\begin{eqnarray}\label{34d}
C_{e }(y_2)=\left( \mu _{1}+\mu _{2}y_{2}^{2}\right)\left[
\frac{1}{2} \left( \mu _{1}^{2}+\frac{\mu _{2}^{2}}{y_{2}}\right)
\right . \nonumber
\\
\left . +2\mu _{0}\left( \mu _{1}+\frac{\mu _{2}}{y_{2}}\right)
+4\mu _{1}\mu _{2}\frac{1}{1-y_{2}^{2}} \right] ^{2}.
\end{eqnarray}
The corresponding variable $\rho_2$ is given by
$$
\rho_2=y_2\rho_1.
$$
For case 2, $\rho_2>\rho_1$ (6.36) becomes
\begin{equation}\label{row2y1}
\rho _{2}(y_{1})=\frac{1}{2}\sqrt{\frac{C_{e}^{\prime
}(y_{1})}{\mu _{1}y_{1}^{2}+\mu _{2}}}\left( 1\pm
\sqrt{1-\frac{C_{0}}{C_{e}^{\prime }(y_{1})}}\right) ,
\end{equation}
where
\begin{eqnarray}\label{18}
C_{e }^{\prime }(y_1)=\left( \mu _{1}y_{1}^{2}+\mu _{2}\right)
\left[ \frac{1}{2}\left( \frac{\mu _{1}^{2}}{y_{1}}+\mu
_{2}^{2}\right) \right . \nonumber
\\
\left . +2\mu _{0}\left( \frac{\mu _{1}}{y_{1}}+\mu _{2}\right)
+4\mu _{1}\mu _{2}\frac{1}{ 1-y_{1}^{2}}\right] ^{2}.
\end{eqnarray}
The corresponding variable $\rho_1$ is given by
\begin{equation}
\rho_1=y_1\rho_2
\end{equation}
 Figure (\ref{fig4}) shows two typical examples of such
 projections for a given $\mu_1$,$\mu_0$ and $C_0$.

 Real motion is possible only in the white regions. For
 $C_0$$\neq$0,a forbidden region (grey) exists at the origin,
 which grows as $C_0$ is increased to meet the forbidden region
 surrounding the exterior of the arms. The $\rho_2<<\rho_1$ arm
 represents the `24` type of hierarchy, the $\rho_2 \approx\rho_1$
 arm represents both the `12` and `14` type of hierarchies and  the
 $\rho_1<<\rho_2$ arm represents the `13` hierarchy. The connected
 projection in figure (5a) for $C_0=0.039$ indicates that change from one
 hierarchical arrangement to another is possible. Figure
 (5b) gives the projection for $C_0=C_{crit}=0.055$, a critical value, at which the
 regions of real motion become disconnected. For $C_0>C_{crit}$
 all allowed regions are totally disconnected and evolution from
 one hierarchy to another is impossible. The system is therefore
 hierarchically stable.
 \subsection{The minima of the boundary surface of real motion projected in $\rho_1$$\rho_2$ space}
 The minima of the boundary surface give information on the three
 dimensional shape of the surface. Projection of the curves indicating where
 the minima are located in $\rho_1$$\rho_2$ space are useful in
 identifying when, as $C_0$ is increased, holes first appear within
 the boundary surface. Motion is still possible from one tube to another
  by moving around the central holes. See Steves and Roy (2001) for further details.

 For a given $y_1,y_2$, the minima of $\rho_n$ with respect to
 $x_{12}$ occur at $x_{12}=\sqrt{y_1^{2}+y_2^{2}}$. The projection
 of the minima onto the $\rho_1$$\rho_2$ plane are given by
 \begin{equation}
\rho _{n}\left( y_{1},y_{2}\right)
=\frac{1}{2}\sqrt{\frac{C_{m}(y_{1},y_{2}) }{\mu _{1}y_{1}^{2}+\mu
_{2}y_{2}^{2}}}\left( 1\pm \sqrt{1-\frac{C_{0}}{
C_{m}(y_{1},y_{2})}}\right),
\end{equation}
where
\begin{eqnarray}\label{34e}
C_{m}(y_1,y_2)=\left( \mu _{1}y_{1}^{2}+\mu _{2}y_{2}^{2}\right)
\left[ \frac{1}{2}\left( \frac{\mu _{1}^{2}}{y_{1}}+\frac{\mu
_{2}^{2}}{y_{2}} \right) \right .\nonumber
\\
\left .+2\mu _{0}\left( \frac{\mu _{1}}{y_{1}}+\frac{\mu _{2}}{y_{2}}%
\right) +4\left( \frac{\mu _{1}\mu
_{2}}{\sqrt{y_{1}^{2}+y_{2}^{2}}}\right) \right ]^{2}.
\end{eqnarray}
For case 1 $\rho_1>\rho_2$ (6.43) becomes
\begin{equation}
\rho _{1}(y_{2})=\frac{1}{2}\sqrt{\frac{C_{m}(y_{2})}{\mu _{1}+\mu
_{2}y_{2}^{2}}}\left( 1\pm
\sqrt{1-\frac{C_{0}}{C_{m}(y_{2})}}\right),
\end{equation}
where
\begin{equation}
C_{m}(y_{2})=\left( \mu _{1}+\mu _{2}y_{2}^{2}\right) \left[
\begin{array}{c}
\frac{1}{2}\left( \mu _{1}^{2}+\frac{\mu _{2}^{2}}{y_{2}}\right)
+\left(
\frac{4\mu _{1}\mu _{2}}{\sqrt{\left( 1+y_{2}^{2}\right) }}\right)  \\
+2\mu _{0}\left( \mu _{1}+\frac{\mu _{2}}{y_{2}}\right)
\end{array}
\right] ^{2}.
\end{equation}
For case 2 $\rho_2>\rho_1$ (6.43) becomes
\begin{equation}
\rho _{2}(y_{1})=\frac{1}{2}\sqrt{\frac{C_{m}^{\prime
}(y_{1})}{\mu _{1}y_{1}^{2}+\mu _{2}}}\left( 1\pm
\sqrt{1-\frac{C_{0}}{C_{m}^{\prime }(y_{1})}}\right) ,
\end{equation}
where
\begin{eqnarray}\label{17}
C_{m }^{\prime }(y_1)=\left( \mu _{1}y_{1}^{2}+\mu
_{2}\right)\left[ \frac{1}{2}\left( \frac{\mu _{1}^{2}}{y_{1}}+\mu
_{2}^{2}\right) \right . \nonumber
\\
\left . +2\mu _{0}\left( \frac{\mu _{1}}{y_{1}}+\mu _{2}\right)
+4\frac{\mu _{1}\mu _{2}}{ \sqrt{y_{1}^{2}+1}}\right] ^{2},
\end{eqnarray}

\section{The Szebehely ladder and Szebehely Constant}
Through the projections in the $\rho_1\rho_2$ plane given by the
maximum extensions and of the minima of the boundaries of real
motion in the $\rho_1\rho_2\rho_{12}$ space, we can study the
topology of the boundary surfaces and thus gain knowledge on the
hierarchical stability of the system. The topology changes as
$C_0$ increases. The critical values of $C_0$ at which the space
becomes disconnected therefore provide stability criterion.

The value of $\rho_n(y_1,y_2)$, for the maximum extensions and the
minima projections, explicitly depends on the value of
$C(y_1,y_2)$ i.e. it has two roots, a single double real root or
imaginary roots, if $C$ is greater than, equal to or less than
$C_0$ respectively.

The Equations $C_e(y_2)$, $C_e'(y_1)$,$C_m(y_2)$, $C_m'(y_1)$
therefore give information on the point at which the topology of
the projections changes. The critical changes occur at
$C(y_1,y_2)=C_0$, the single real root solution. For example $C_e$
can be evaluated for the range of $y_2$ from 0 to 1. Recall that
$y_2$ is the gradient of a straight line through the origin $O$ in
the $\rho_1O\rho_2$ plane.

The minimum value of $C_e(y_2)$, $C_e$(min), is the first value of
$C_0$, as it is increased, where there is only one solution
($\rho_1,\rho_2$) to the maximum projection curve. For
$C_0>C_e$(min), there are no solutions ($\rho_1,\rho_2$) at $y_2
=$ min and the projection becomes disconnected indicating a stable
system.

The minima of the four C-functions indicating the point of change
in the topology can be thought of as the rungs of a ladder, which
Steves and Roy (2001) call the Szebehely ladder. The rungs of the
ladder $R_1=C_m$(min); $R_2=C_m'$(min); $R_3=C_e$(min) and
$R_4=C_e'$(min) are dependent only on the masses of the system and
are invariant to the initial conditions or the $c$ and $E$ of the
system.

\begin{figure}
\centerline{\epsfig{file=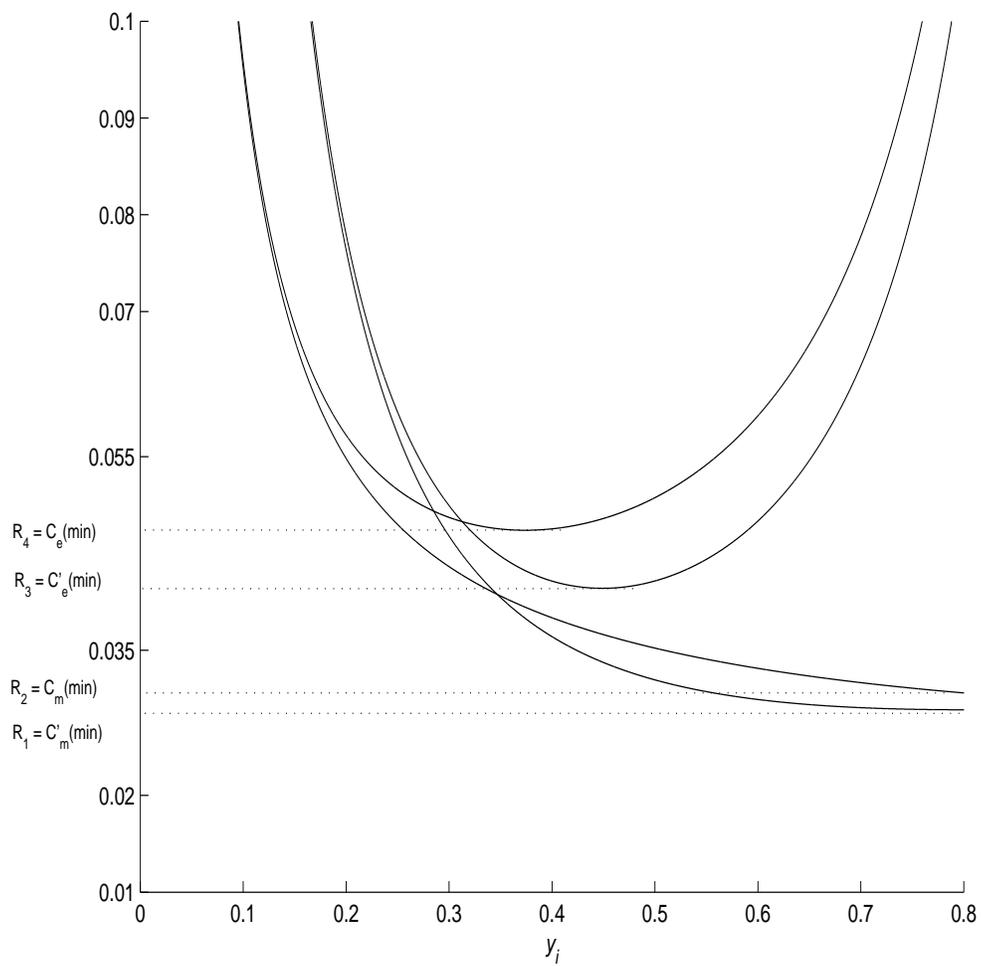 ,height=13cm, width=13cm}}
\caption{The Szebehely Ladder for $\mu_1=0.15, \mu_2=0.35$ and
$\mu_0=0.1$} \label{ladder}
\end{figure}

Both $ y_{1}$ and $y_{2}$ lie in the range $0$ to $1$. Hence we
can combine relations $C_e(y_2)$, $C_e'(y_1)$, $C_m(y_2)$,
$C_m'(y_1)$ in the same figure plotting $C$ against $ y $. See
figure (5.4), where the minima of the curves of the four $C-$
functions form the four rungs of the Szebehely ladder. The
stability of the system depends on the location of the Szebehely
constant $C_0$ for the system with respect to these rungs. Thus
for
\begin{enumerate}
\item   $C_{0}>R_{1}$, there is a hole of forbidden motion near
the central region connecting the four tubes within the boundary
surface for $\rho _{1}>\rho _{2}.$

\item  $C_{0}>R_{2}$, there is a hole of forbidden motion near the
central region connecting the four tubes within the boundary
surface for $\rho _{2}>\rho _{1}$.

\item  $C_{0}>R_{3}$, the arms in the projection of the maximum
extensions for $\rho _{1}>\rho _{2}$ are disconnected and the
$^{\prime }24^{\prime }$ hierarchy is stable.

\item  $C_{0}>R_{4}$, the arms in the projection of the maximum
extensions for $\rho _{2}>\rho _{1}$ are disconnected and the
$^{\prime }13^{\prime }$ hierarchy is stable.
\end{enumerate}
When $C_0>$ max$(R_3,R_4)$, all arms are disconnected and the
system is hierarchically stable.
\begin{enumerate}
\item If $\mu_2>\mu_1$, then $R_4>R_3>R_2>R_1$. \item If
$\mu_1>\mu_2$, then $R_3>R_4>R_1>R_2$. \item If $\mu_1=\mu_2$,
then $R_2=R_1$ and $R_3=R_4$.
\end{enumerate}
Thus the critical value of $C_0$ at which the whole system becomes
hierarchically stable for all time is
\begin{equation}
C_{crit}=\max (R_{3},R_{4})=\left\{
\begin{array}{c}
R_{3}=C_{e}(\min )\qquad \textrm{if }\mu _{1}>\mu _{2} \\
R_{4}=C_{e}^{\prime }(\min )\qquad \textrm{if }\mu _{2}>\mu _{1}
\end{array}
\right.   \label{46a}
\end{equation}
$\mu_1=\mu_2$ is the special case of equal masses where $C_0>
(C_{crit}=R_3$=$R_4)$ gives total hierarchical stability at one
critical point. Otherwise, hierarchical stability occurs in two
stages $C_{0}>(C_{crit_1}=R_{3})$ and $C_{0}>(C_{crit_2}=R_{4})$.
If $\mu_0=0$, then we have the special case of the CSFBP discussed
by Steves and Roy (2001). The critical value of $C_0$ at which the
system becomes hierarchically stable for all time is given by
(\ref{46a}). $R_3$ and $R_4$ are purely functions of $\mu_0$ and
$\mu_1$. For $\mu_1=\mu_2$, they are functions only of $\mu_0$.
Figure (6.8) plots these critical values as a function of $\mu_0$.
For $C_0>C_{crit}(\mu_0)$, hierarchical stability is guaranteed.
Figure (6.8) shows that $C_{crit}(\mu_0)$ has a maximum of
0.065667 at $\mu_0=0.183$. Thus if $C_0>0.065667$, all CS5BP's
with $\mu_1=\mu_2$ will be hierarchically stable.

We now show several examples for a range of mass ratios of how
rungs of the Szebehely ladder can be computed solely from $\mu_1$
and $\mu_0$. Then using the value of the Szebehely constant $C_0$
for the system, which depends on the initial conditions, the
hierarchical stability of the system can be determined.

\section{The Stability of the CS5BP systems with a range of different mass ratios}
\subsection{The Equal Mass CS5BP}
The equal mass CS5BP has $\mu_1=\mu_2=\mu_0=1/5$. In this case
there exists only two rungs of the Szebehely ladder; since
$C_m=C_m'$ and $C_e=C_e'$, viz.

\begin{equation}\label{19}
C_{m }(y)=\frac{1}{5}\left( 1+y^{2}\right) \left(
\frac{1}{10}\left( 1+ \frac{1}{y}\right)
+\frac{4}{25\sqrt{1+y^{2}}}\right) ^{2},
\end{equation}
\begin{equation}\label{20}
C_{e }(y)=\frac{1}{5}\left( 1+y^{2}\right) \left(
\frac{1}{10}\left( 1+ \frac{1}{y}\right) +\frac{4}{25\left(
1-y^{2}\right) }\right) ^{2} ,
\end{equation}
where $0\leq y\leq 1$.

The minimum values of $C_{m }(y)$ and $C_{e }(y)$ form the two
rungs of the Szebehely Ladder. The minimum value of $C_{m }$ is $
0.039222=R_{1} $ and the minimum value of $C_{e }$ is
$0.065551=R_{4}.$ $R_{1}$ and $R_{4}$ occur at $y=1$ and $y=0.472$
respectively.

\begin{figure}
\centerline{ \epsfig{file=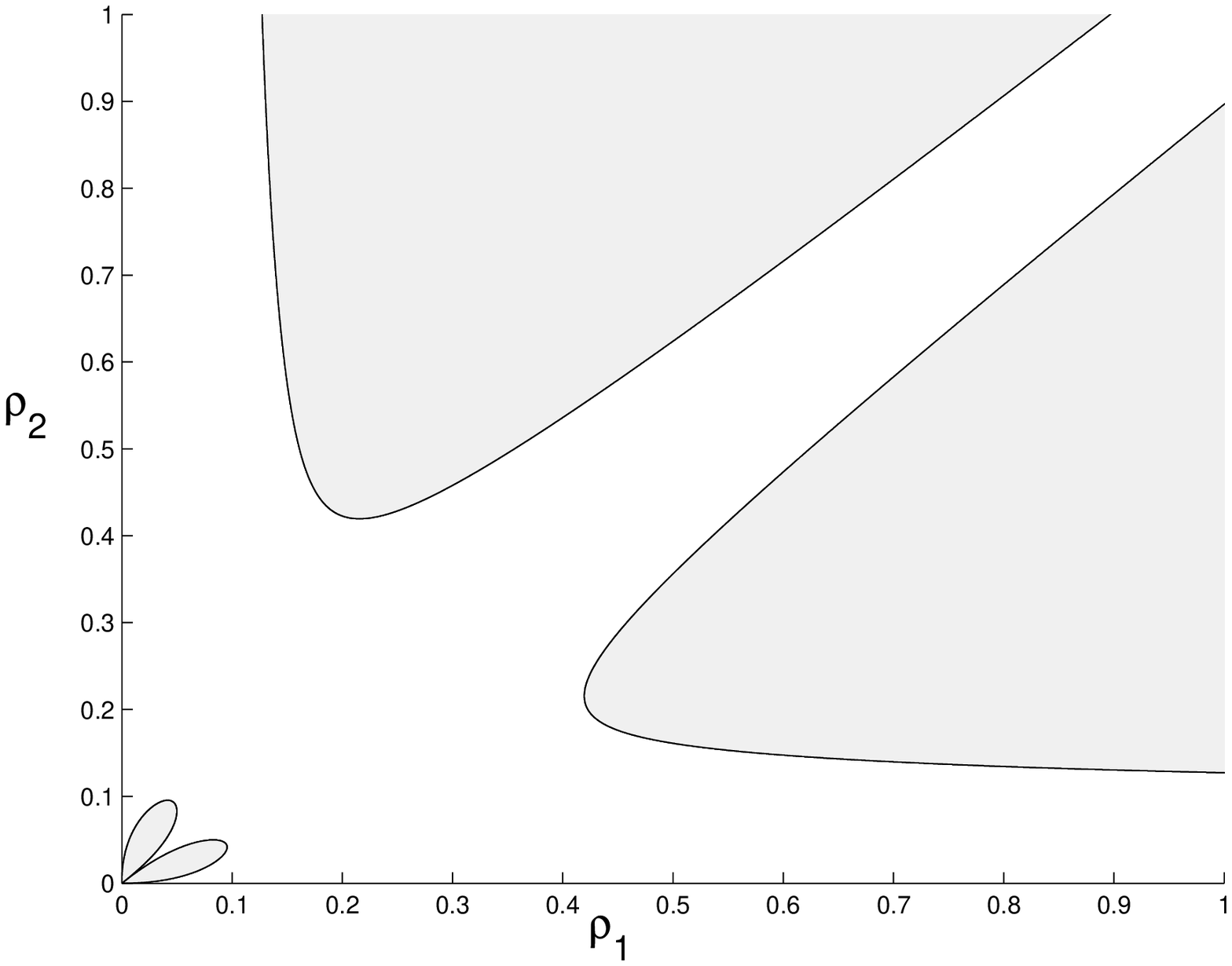 ,height=6cm}
\epsfig{file=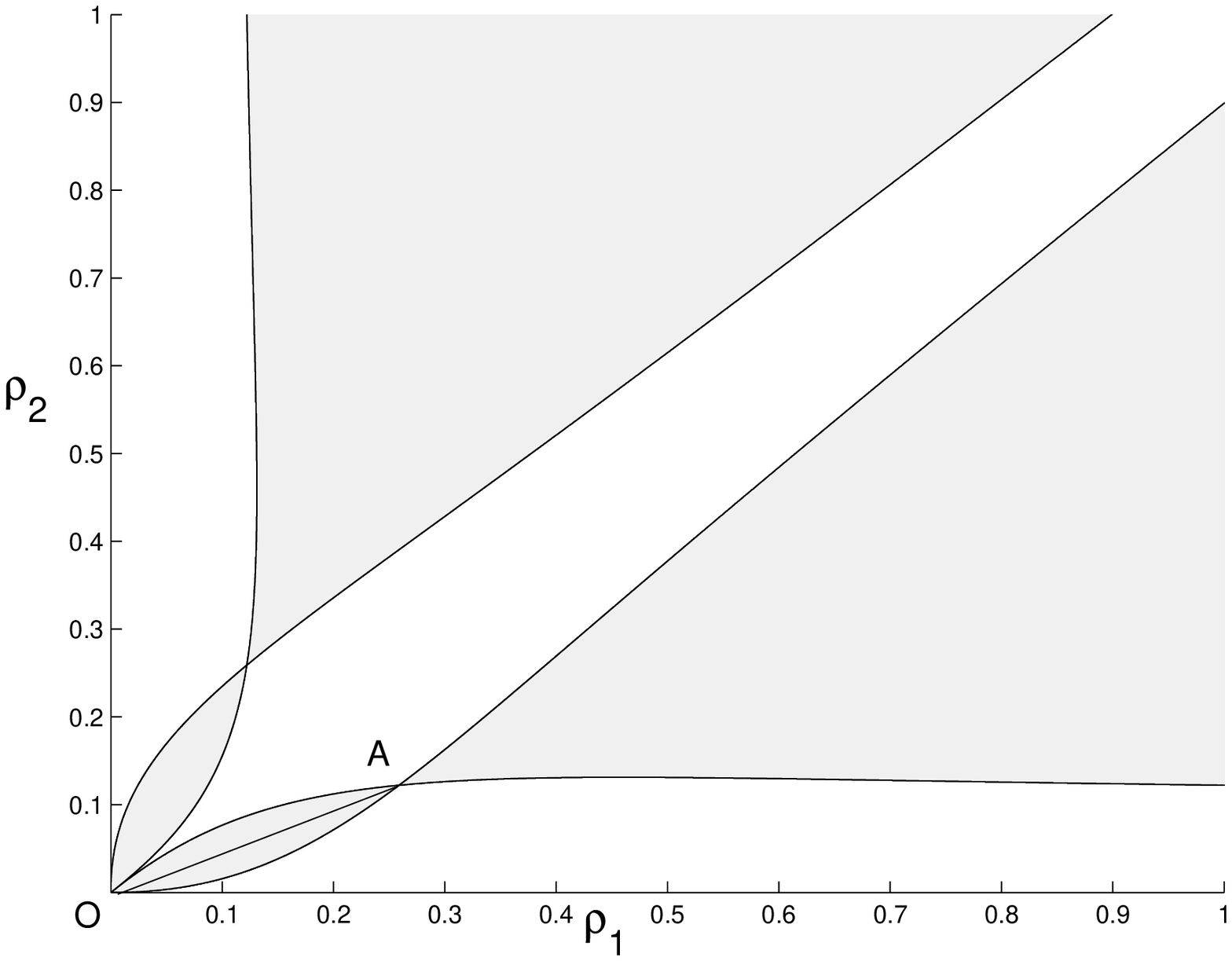 ,height=6cm}} \caption{$\mu _{1}=\mu
_{2}=\mu _{0}$: The projection of the boundary surface onto the
$\rho _{1}-\rho _{2}$ plane a. $C_{0}=R_{1}=0.0392219$\qquad b.
$C_{0}=R_{4}=0.0655514$} \label{fig4}
\end{figure}

Figure (\ref{fig4}a) shows the projection of the maximum
extensions for $C_{0}=R_{1}$. The phase space remains connected
but a small forbidden region exists near the origin. This
forbidden region grows as $C_0$ is increased until at $C_0=R_4$ at
the highest rung of the ladder, the phase space becomes
disconnected. See figure (\ref{fig4}b). Note that the gradient $y$
of the line $\bar{OA}$ (passing through the point of single
solution) is 0.472. The five body equal mass CS5BP is
hierarchically stable for values of $C_0$ greater than $R_4=$
0.06555.

\begin{figure}
\centerline{ \epsfig{file=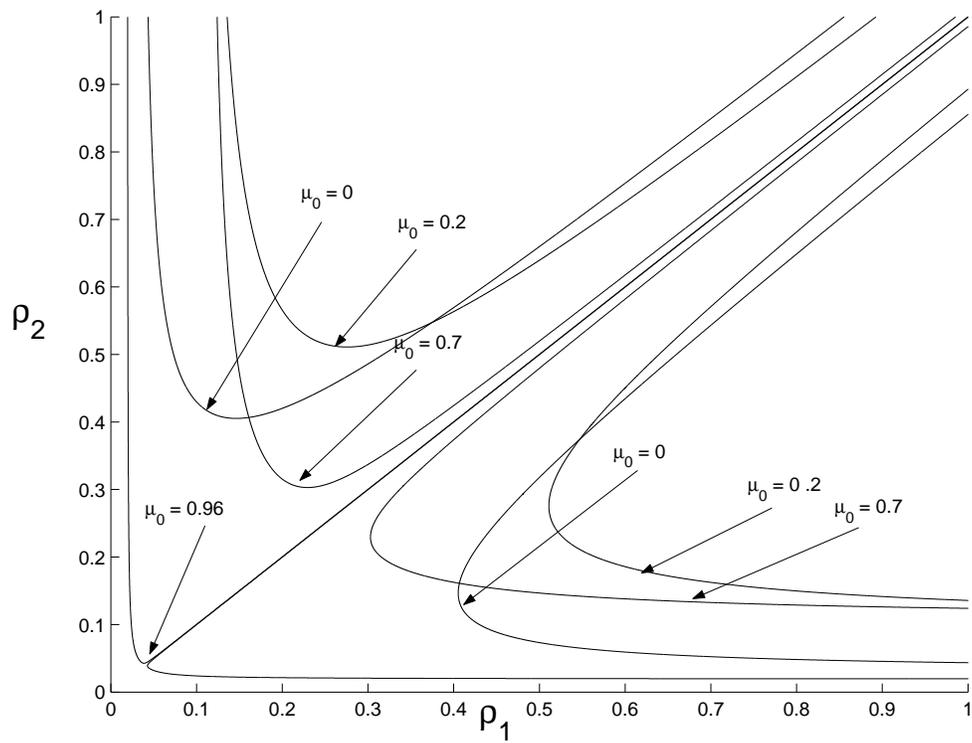 ,height=10cm}} \caption{The
projections of the boundary surfaces onto the $\rho _{1}-\rho
_{2}$ plane for $C_{0}=0$, $\mu _1=\mu _2$ and a range of $\mu_0$
from 0 to 0.96} \label{fig3}
\end{figure}

\subsection{Four equal masses with  a varying central mass $\mu_0$}

\begin{figure}
\centerline{ \epsfig{file=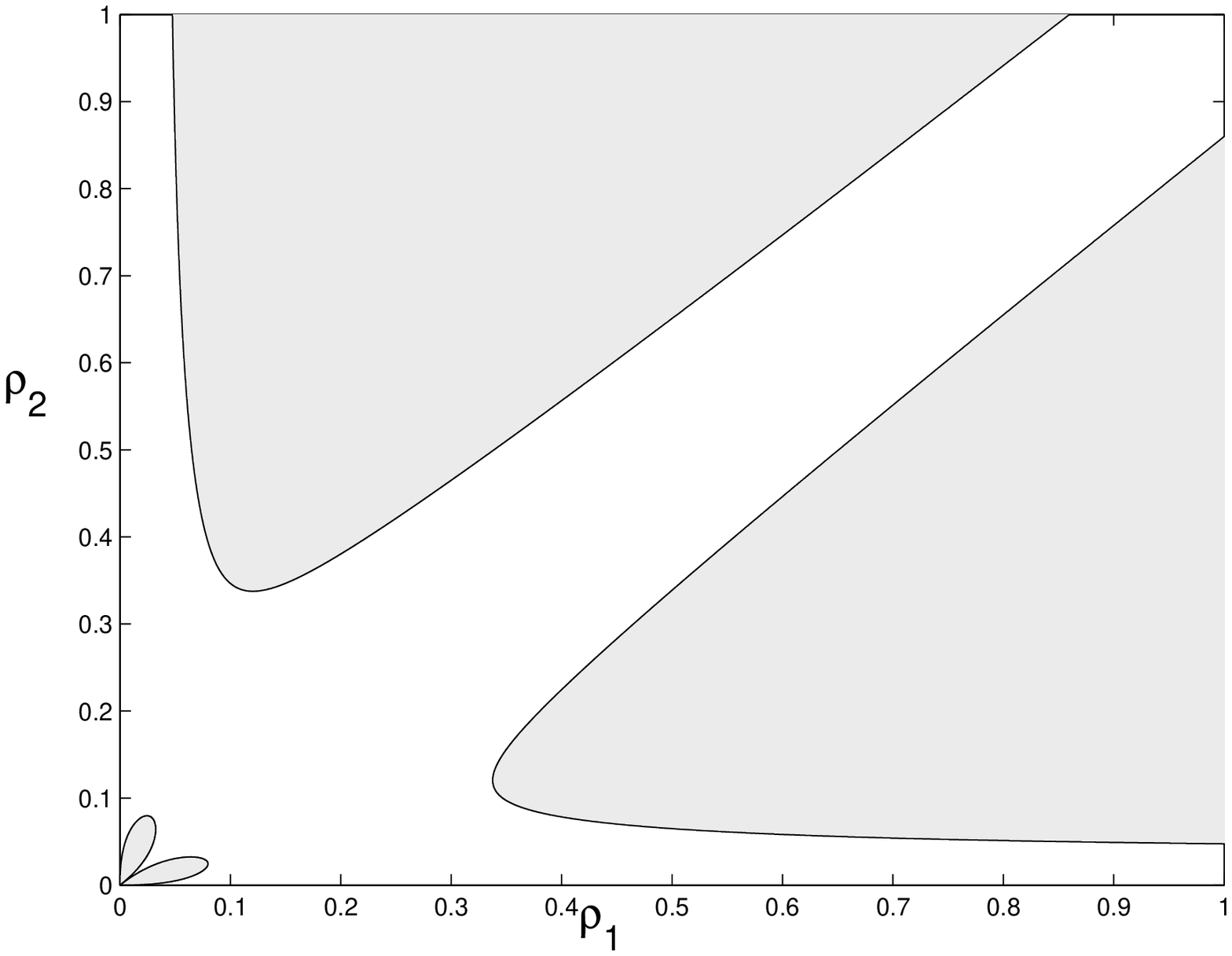 ,height=6cm}
\epsfig{file=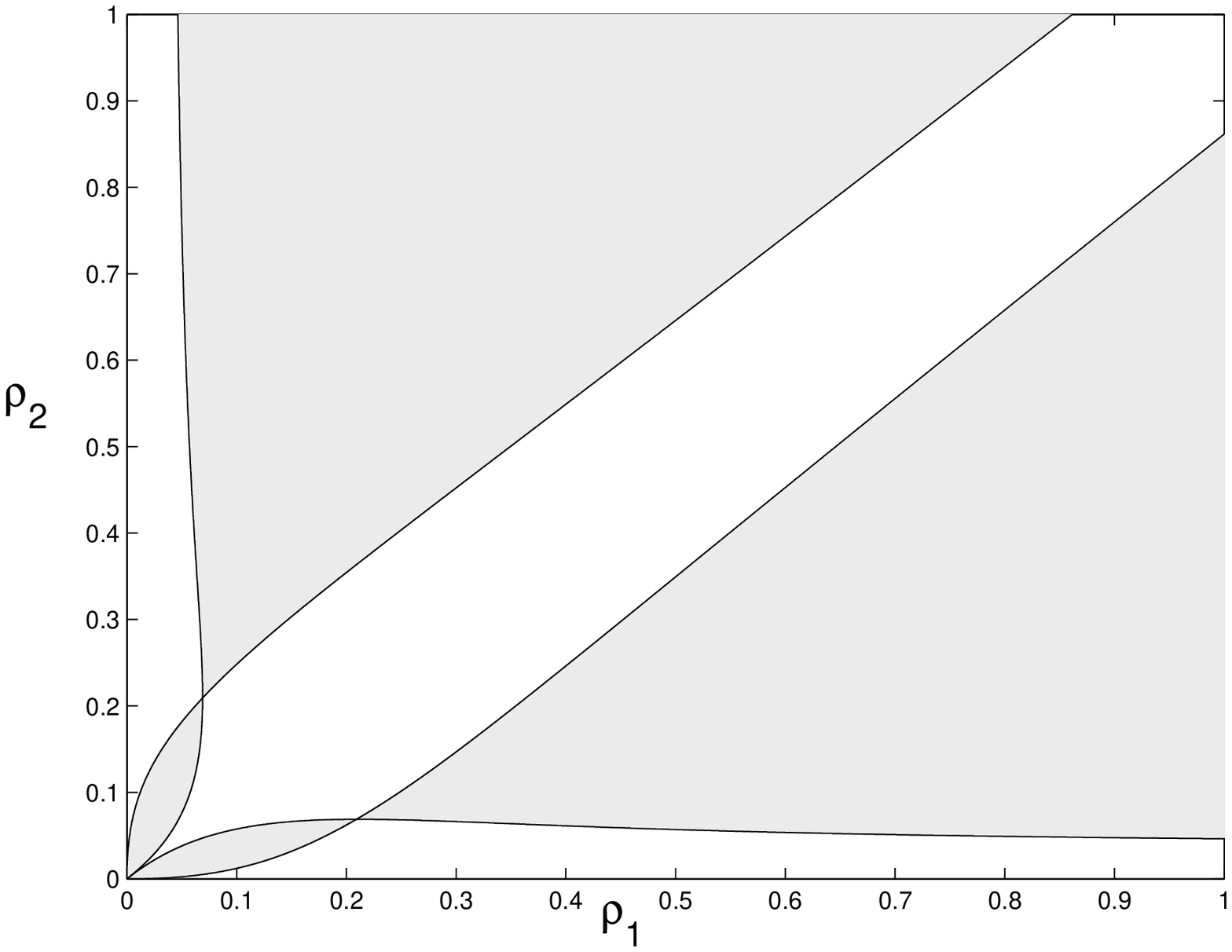 ,height=6cm}} \caption{$\mu _{1}=\mu
_{2}=\frac{22.475}{100},\mu _{0}=0.01$: The projection of the
boundary surface onto the $\rho _{1}-\rho _{2}$ plane at a.
$C_{0}=R_{1}=0.0295707$\qquad b. $C_{0}=R_{4}=0.048036$}
\label{fig6}
\end{figure}

In this case there exists only one independent mass ratio since
given $\mu_0$, $\mu_1$=1/4(1-$\mu_0$) from (6.5) and
$\mu_2=\mu_1$. Since $ \mu _{1}=\mu _{2}$, only two rungs of the
Szebehely ladder exist i.e. $C_{m }=C_{m }^{\prime }$ and $C_{e
}=C_{e}^{\prime }$
\begin{equation}
C_{m }=\mu _{1}\left( 1+y^{2}\right) \left( \left( \frac{1}{2}\mu
_{1}^{2}+2\mu _{0}\mu _{1}\right) \left( 1+\frac{1}{y}\right)
+\frac{4\mu _{1}^{2}}{\sqrt{1+y^{2}}}\right) ^{2}
\end{equation}
\begin{equation}
C_{e }=\mu _{1}\left( 1+y^{2}\right) \left( \left( \frac{1}{2}\mu
_{1}^{2}+2\mu _{0}\mu _{1}\right) \left( 1+\frac{1}{y}\right)
+\frac{4\mu _{1}^{2}}{\left( 1-y^{2}\right) }\right) ^{2}.
\end{equation}
Figures (\ref{fig6}) and (\ref{fig8}) show the projections of the
maximum extensions onto the $\rho_1\rho_2$ plane for two typical
cases:
\begin{enumerate}
\item a small central mass; $\mu _{1}=\mu
_{2}=\frac{22.475}{100},\mu _{0}=0.01$ and \item a large central
mass; $\mu _{1}=\mu _{2}=0.01,\mu _{0}=0.96$. In each figure, two
values of $C_0$, $C_0=R_1$, and $C_0=R_4$, have been selected.
\end{enumerate} For CS5BP's with a small central mass, the largest
region of real motion is found to be the central arms where
$\rho_1\approx\rho_2$. This indicates that such systems are most
likely to be moving in double binary hierarchies of type '12' and
'14', figure (\ref{fig6}).

\begin{figure}
\centerline{ \epsfig{file=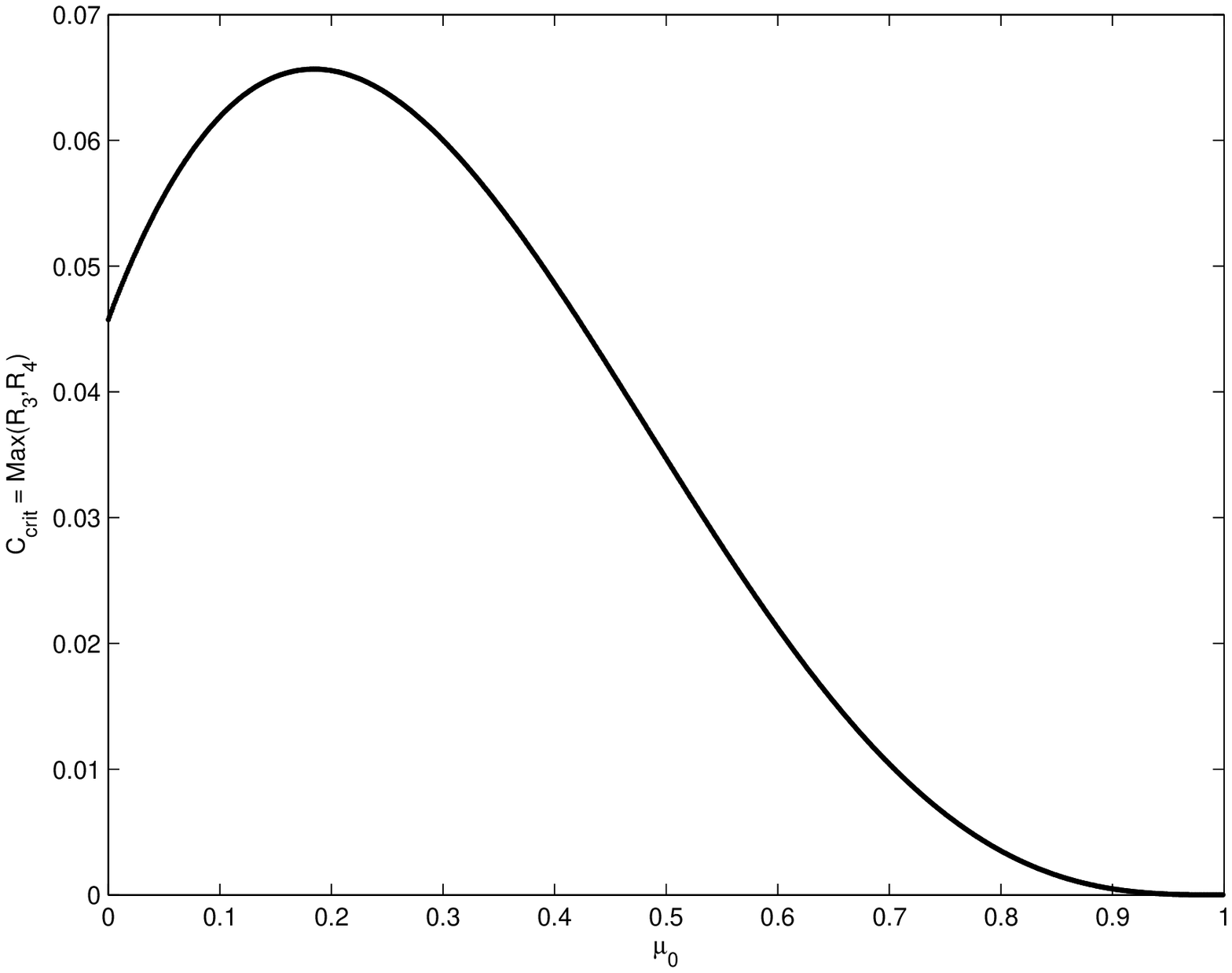 ,height=10cm}}
\caption{The critical values of $C_0$, $C_{crit}$, at which the
CS5BP becomes hierarchically stable as a function of $\mu_0$,
where $\mu_1=\mu_2$} \label{maximum1}
\end{figure}

\begin{figure}
\centerline{ \epsfig{file=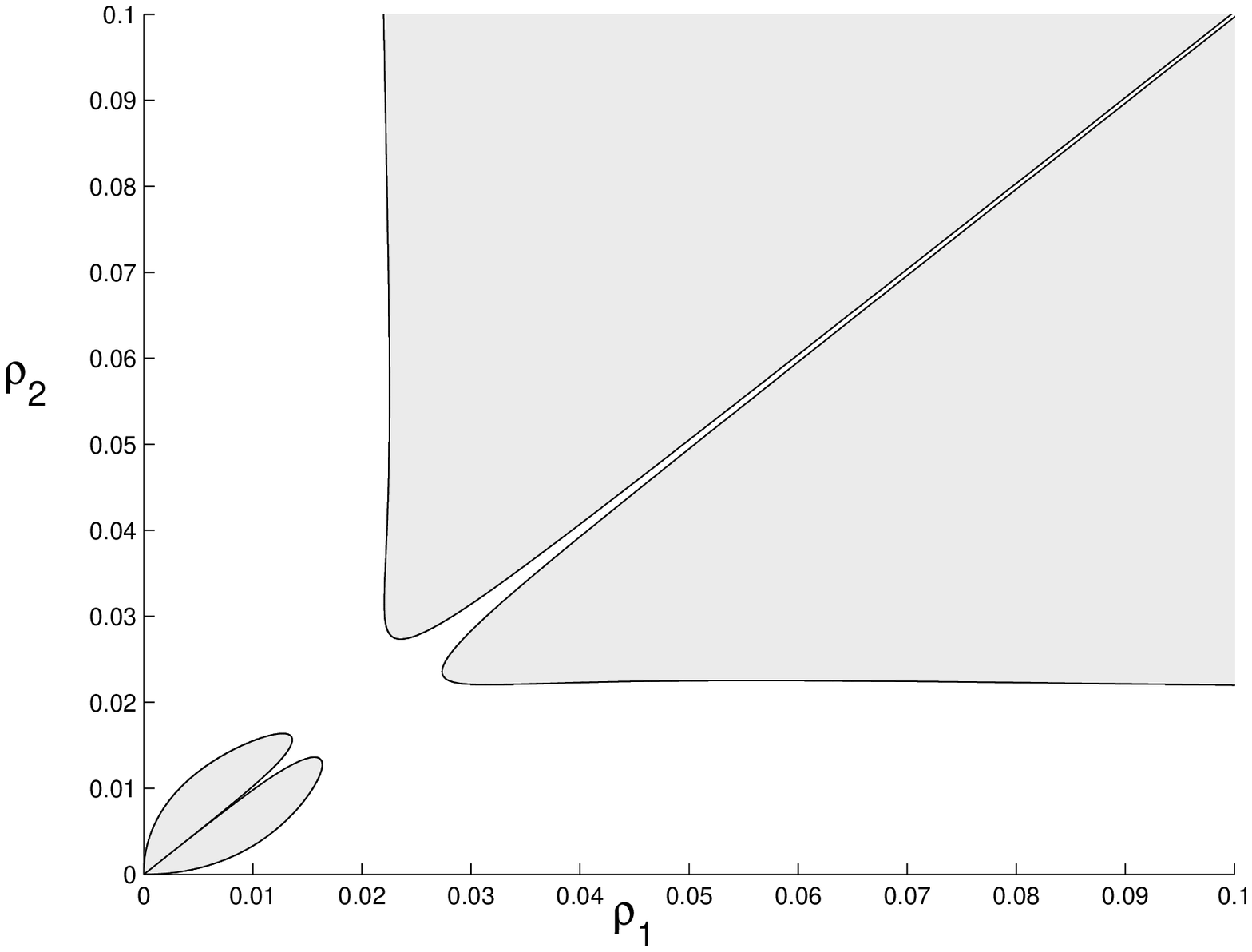 ,height=6cm}
\epsfig{file=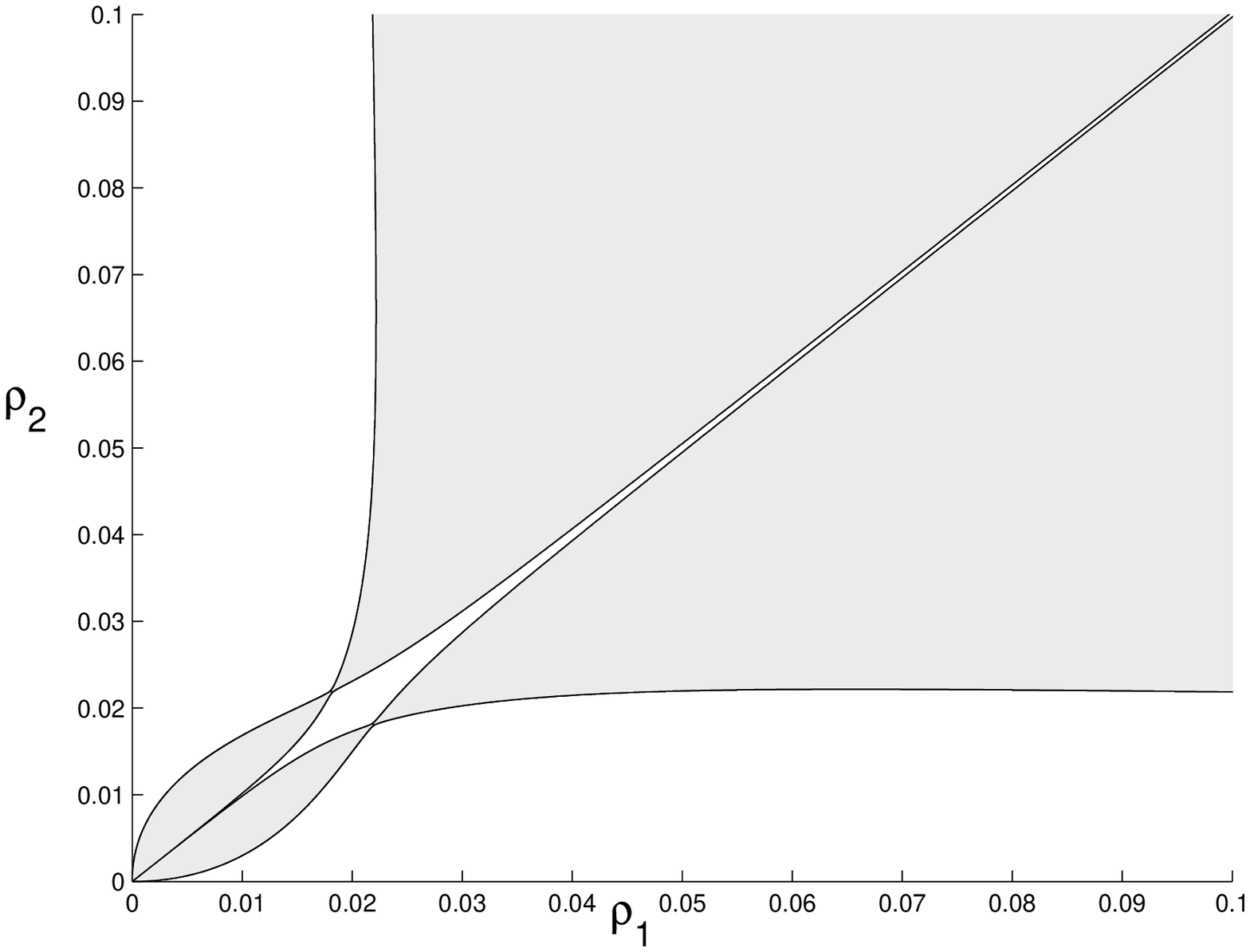 ,height=6cm}} \caption{$\mu _{1}=\mu
_{2}=0.01,\mu _{0}=0.96$: The projection of the boundary surface
onto the $\rho _{1}-\rho _{2}$ plane at a.
$C_{0}=R_{1}=0.0000301$\qquad b. $C_{0}=R_{4}=0.0000323$}
\label{fig8}
\end{figure}

CS5BP's with large central bodies, however, have their largest
region of real motion in the arms $\rho_1\approx0$ and
$\rho_2\approx0$, indicating that the single binary hierarchies
'13' and '24' will be the dominant systems.

It is interesting to study the effect of placing a small mass at
the centre of mass and increasing its mass from 0 to 1. Figure
(\ref{fig3}) shows the projections for $C_0= 0$ and for a range of
$\mu_0$.

For $\mu_0=0$ to 0.2, i.e a small central mass, the double binary
hierarchies dominate, with single binary hierarchies more
prevalent as $\mu_0$ increases. At $\mu_0$=0.2, i.e. the five body
equal mass case, the areas of real motion are of relatively equal
sizes for the double binary and single binary hierarchies,
suggesting neither is dominant. When comparing the area of real
motion available for $\mu_0=0$ (the four body equal mass case),
with that of $\mu_0=0.2$ (the five body equal mass case), we see
that the addition of a fifth body of equal mass at the centre
increases the area of real motion in both single binary and double
binary hierarchies. It thus most likely increases the chances of
moving from one hierarchy to another. Thus it is likely to be more
hierarchically unstable; as would be expected.

For $\mu_0=0.2$ to 1, i.e a larger central mass, single binary
hierarchies dominate, with double binary hierarchies becoming
virtually  non-existent for $\mu_0$ close to 1. $\mu_0$ close to 1
represents a star with four planets or a planet with four
satellites. In such situations, it is highly unlikely that the
four small bodies will form two binary pairs orbiting the central
body.

\subsection{Non-equal masses i.e. $\mu_{1}\neq \protect\mu _{2}\neq \protect\mu _{0}$}

\begin{figure}
\centerline{ \epsfig{file=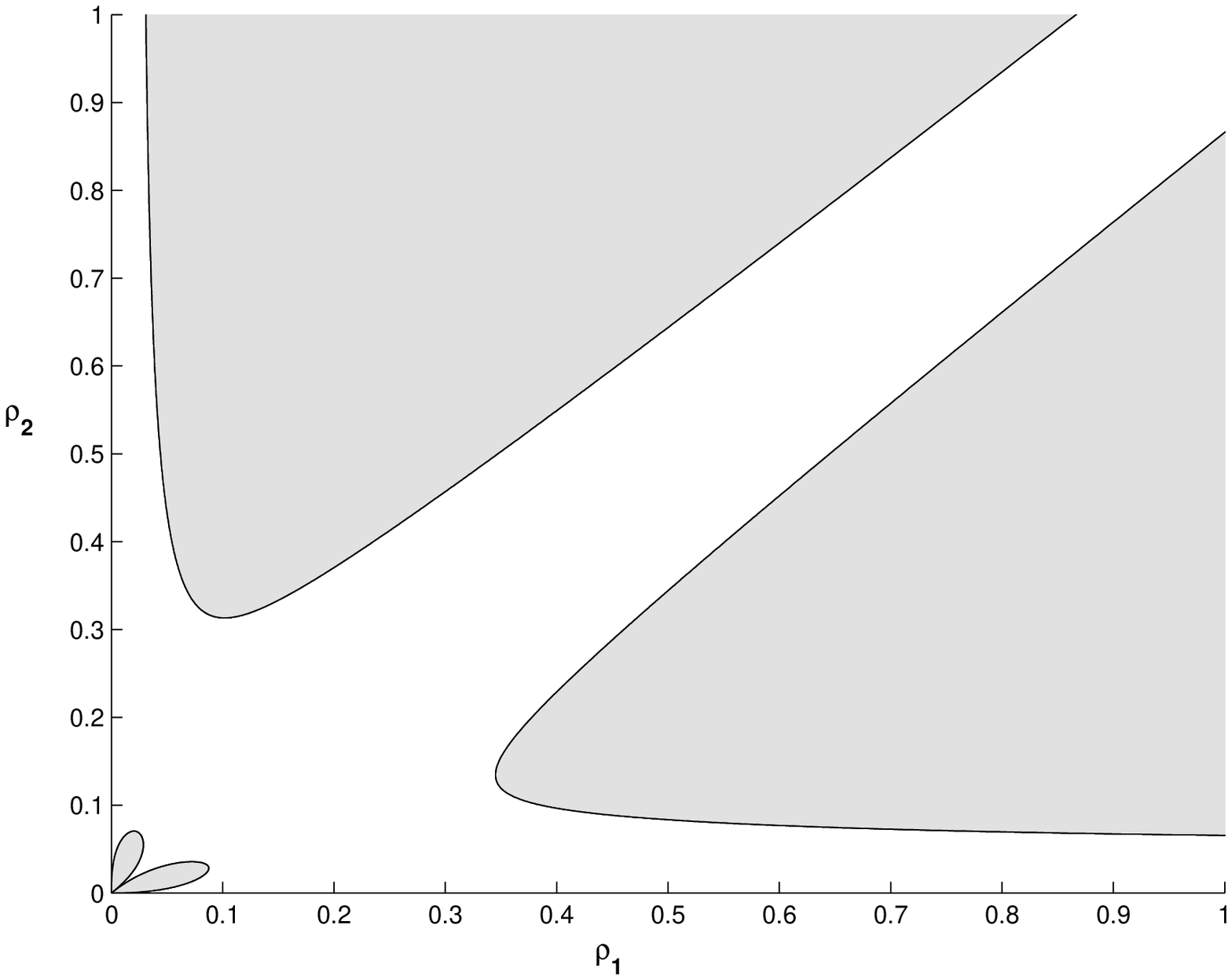 ,height=6cm}
\epsfig{file=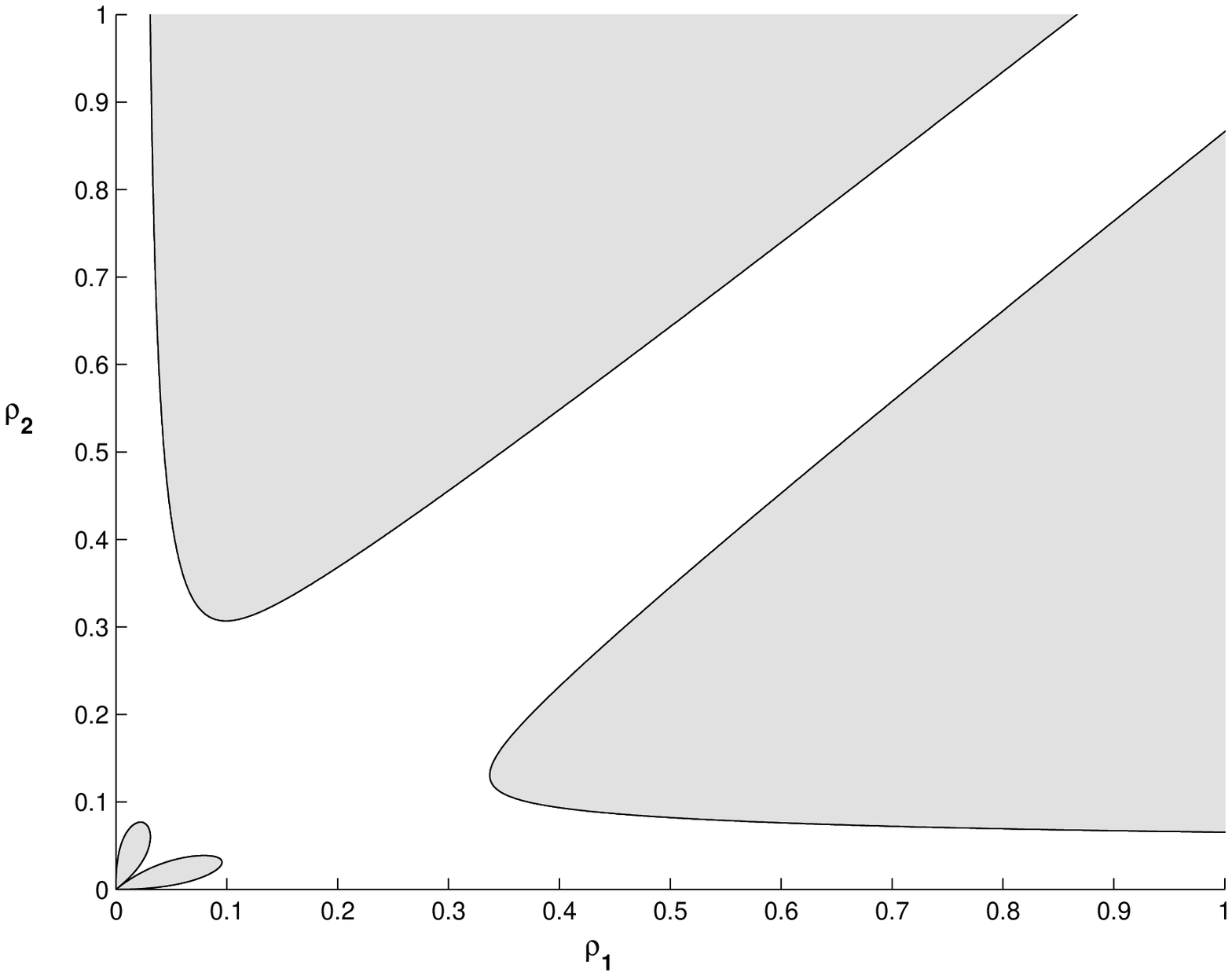 ,height=6cm}} \caption{$\mu _{1}=0.195,$
$\mu _{2}=0.3$ and $\mu _{0}=0.01.$ The projection of the boundary
surface onto the $\rho _{1}-\rho _{2}$ plane at a. $C_{0}=R_{1}=
0.0281$\qquad b. $C_{0}=R_{4}= 0.0270$} \label{fig10b}
\end{figure}

\begin{figure}
\centerline{ \epsfig{file=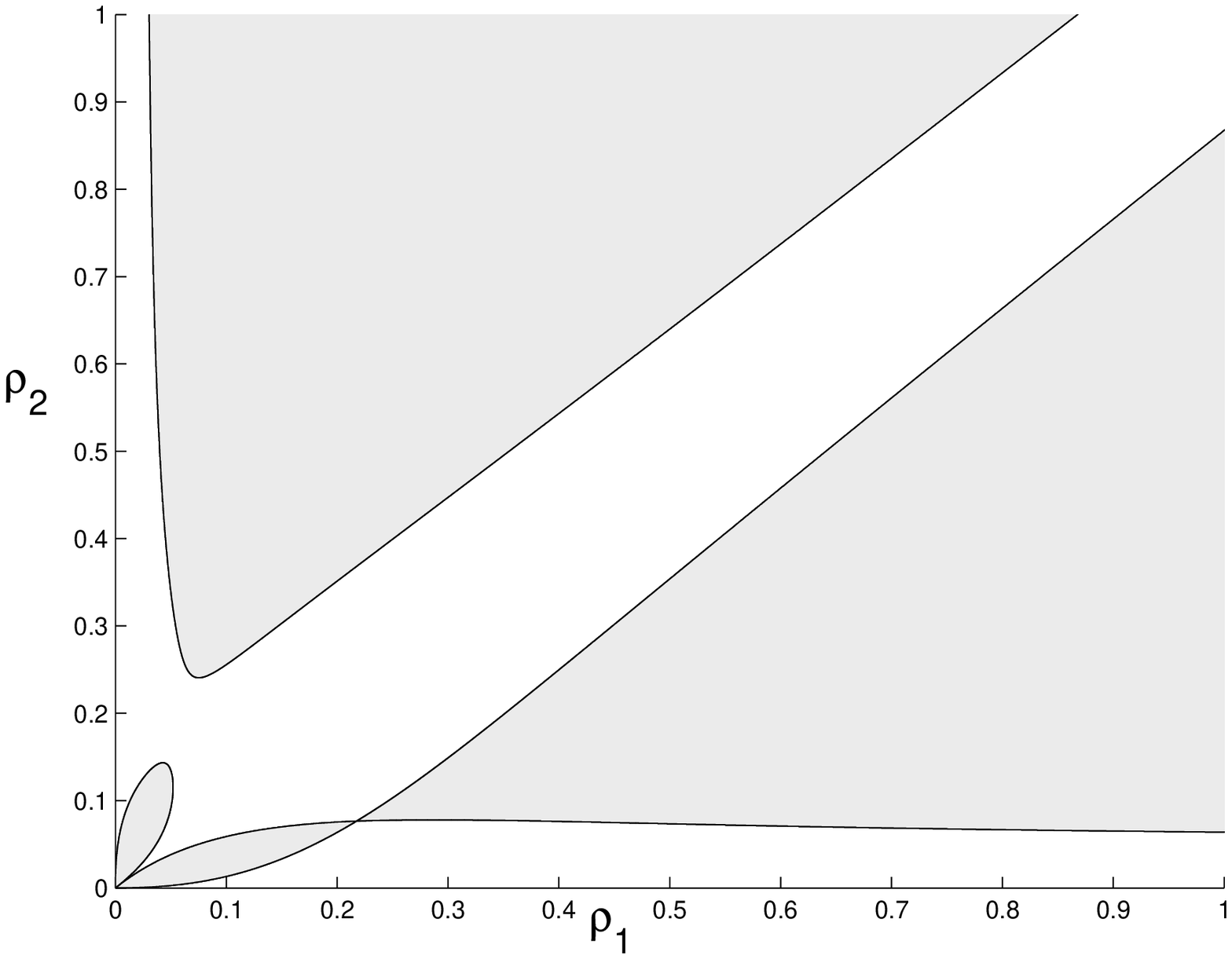 ,height=6cm}
\epsfig{file=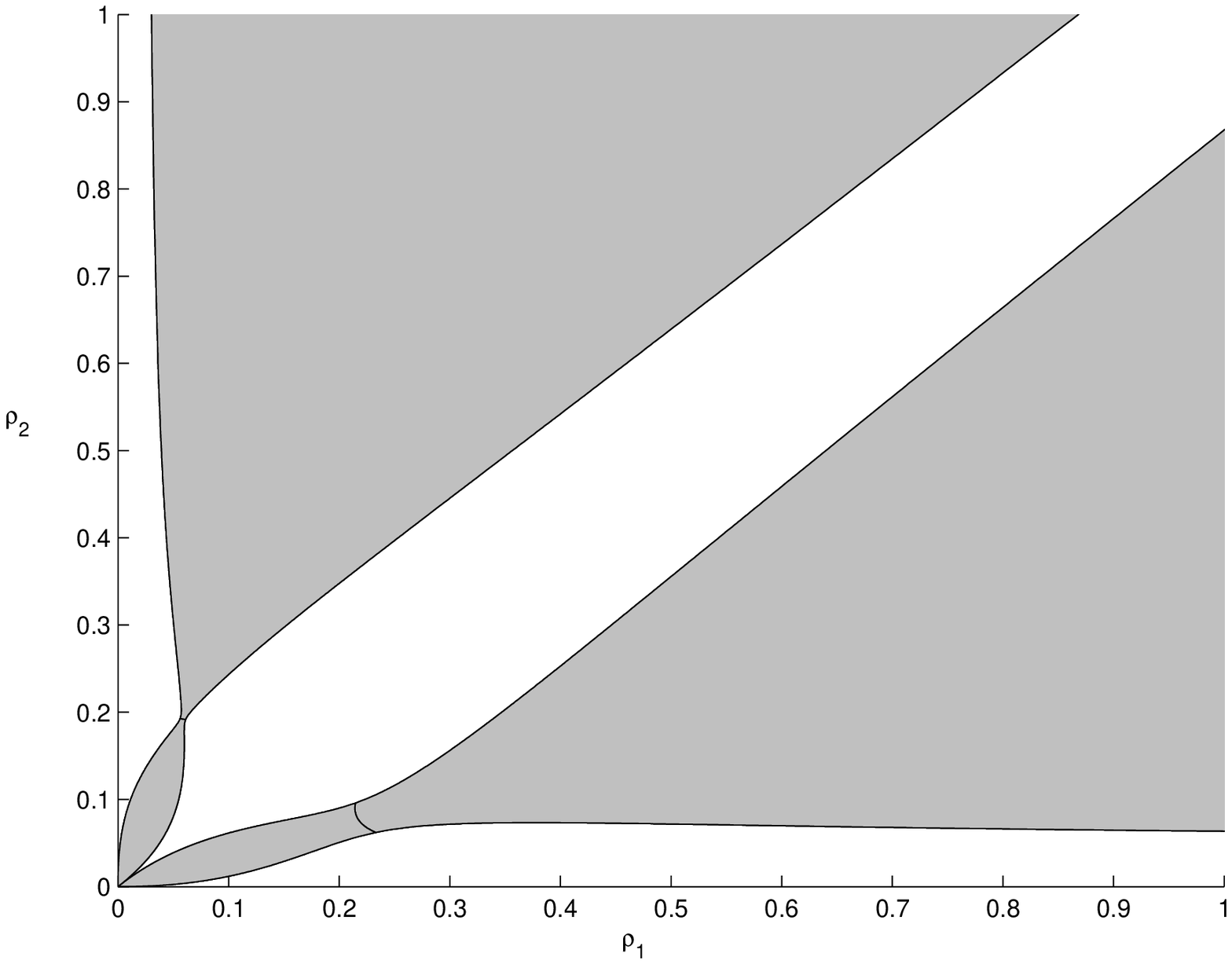 ,height=6cm}} \caption{$\mu _{1}=0.195,$
$\mu _{2}=0.3$ and $\mu_{0}=0.01.$ The projection of the boundary
surface onto the $\rho _{1}-\rho _{2}$ plane at a. $C_{0}=R_{3}=
0.0439$\qquad b. $C_{0}=R_{4}= 0.0470$} \label{fig10}
\end{figure}

\begin{figure}
\centerline{ \epsfig{file=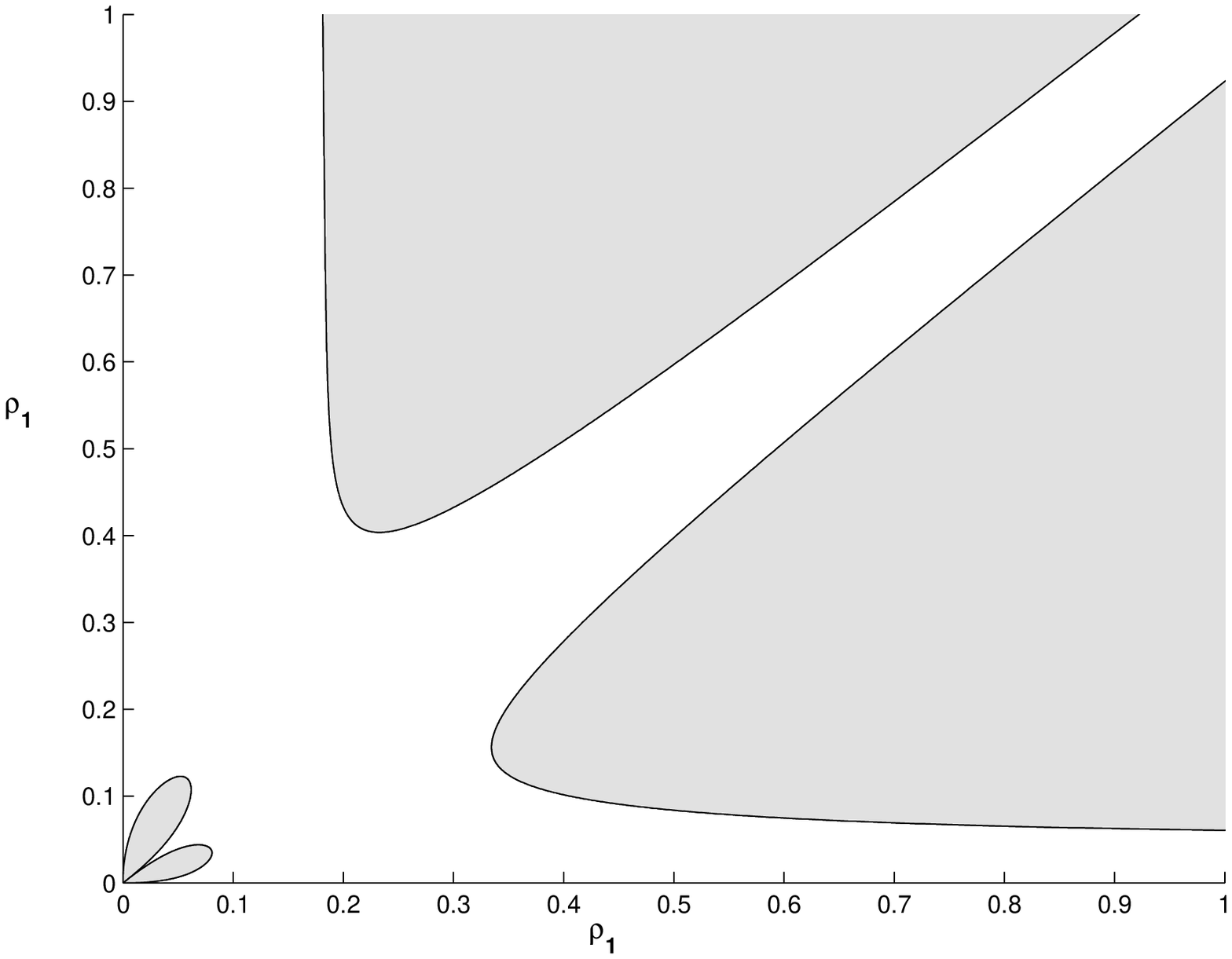 ,height=6cm}
\epsfig{file=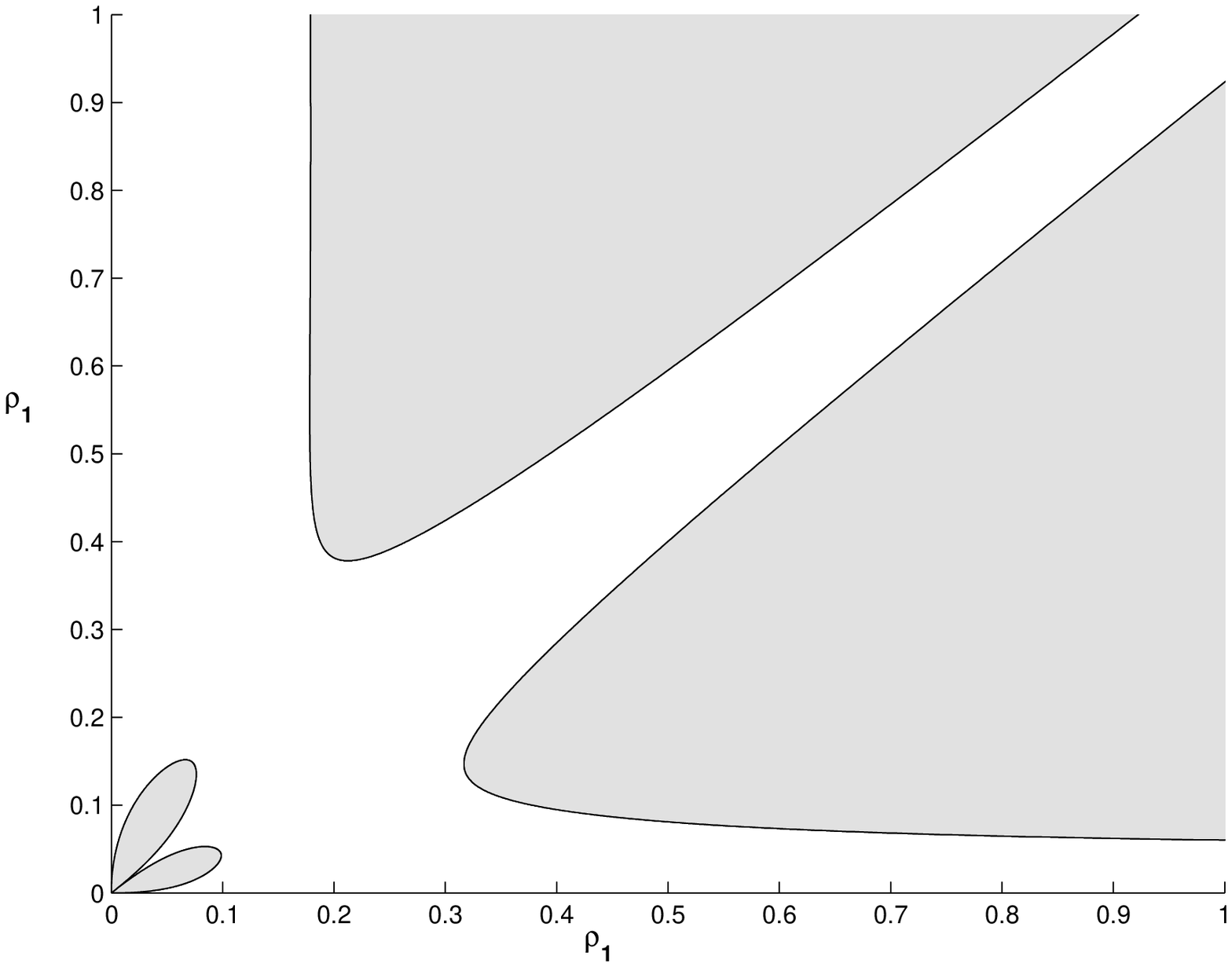 ,height=6cm}} \caption{$\mu _{1}=0.3,$
$\mu _{2}=0.1$ and $\mu _{0}=0.2.$ The projection of the boundary
surface onto the $\rho _{1}-\rho _{2}$ plane at a. $C_{0}=R_{1}=
0.0345$\qquad b. $C_{0}=R_{2}= 0.0348$} \label{fig12b}
\end{figure}

\begin{figure}
\centerline{ \epsfig{file=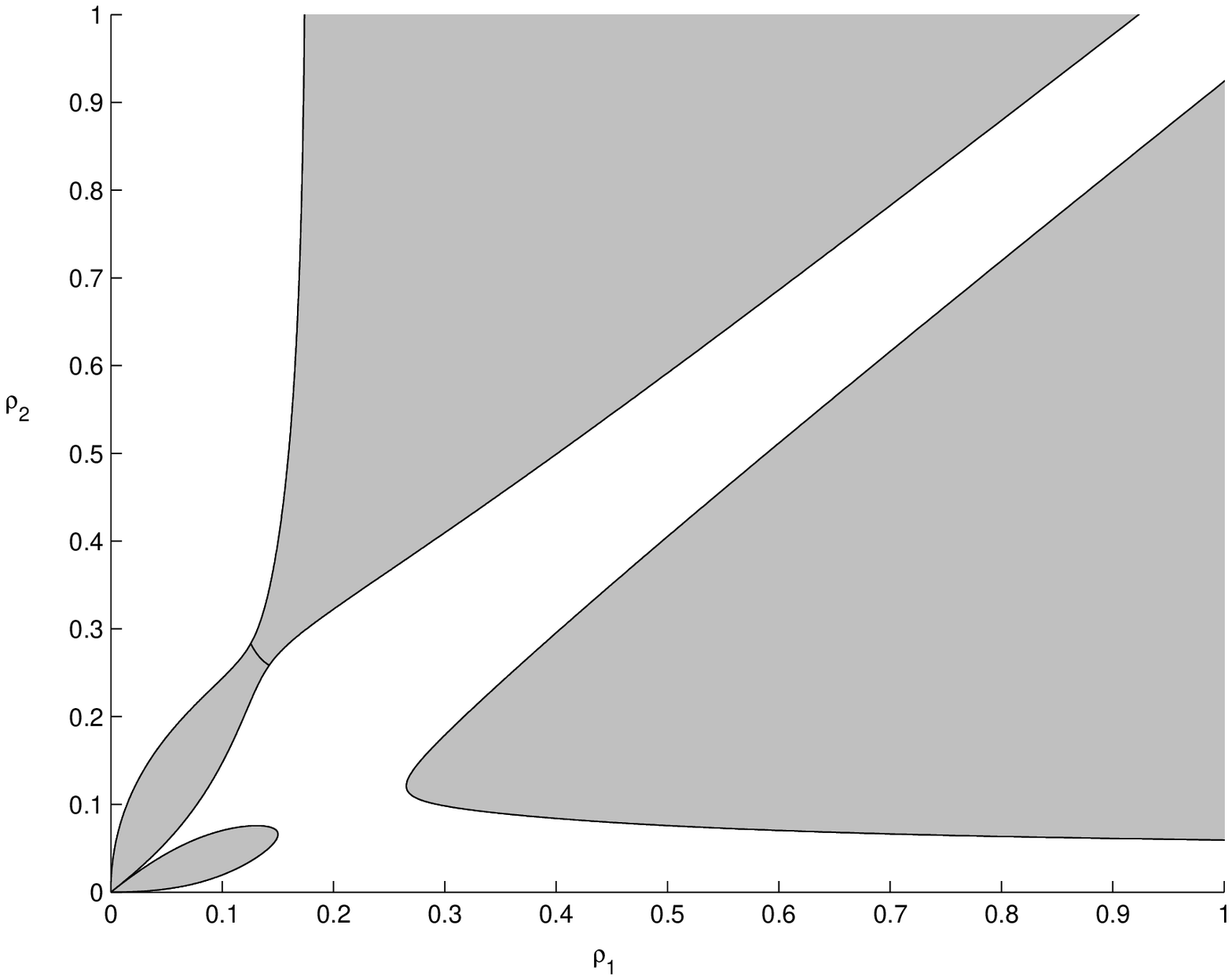 ,height=6cm}
\epsfig{file=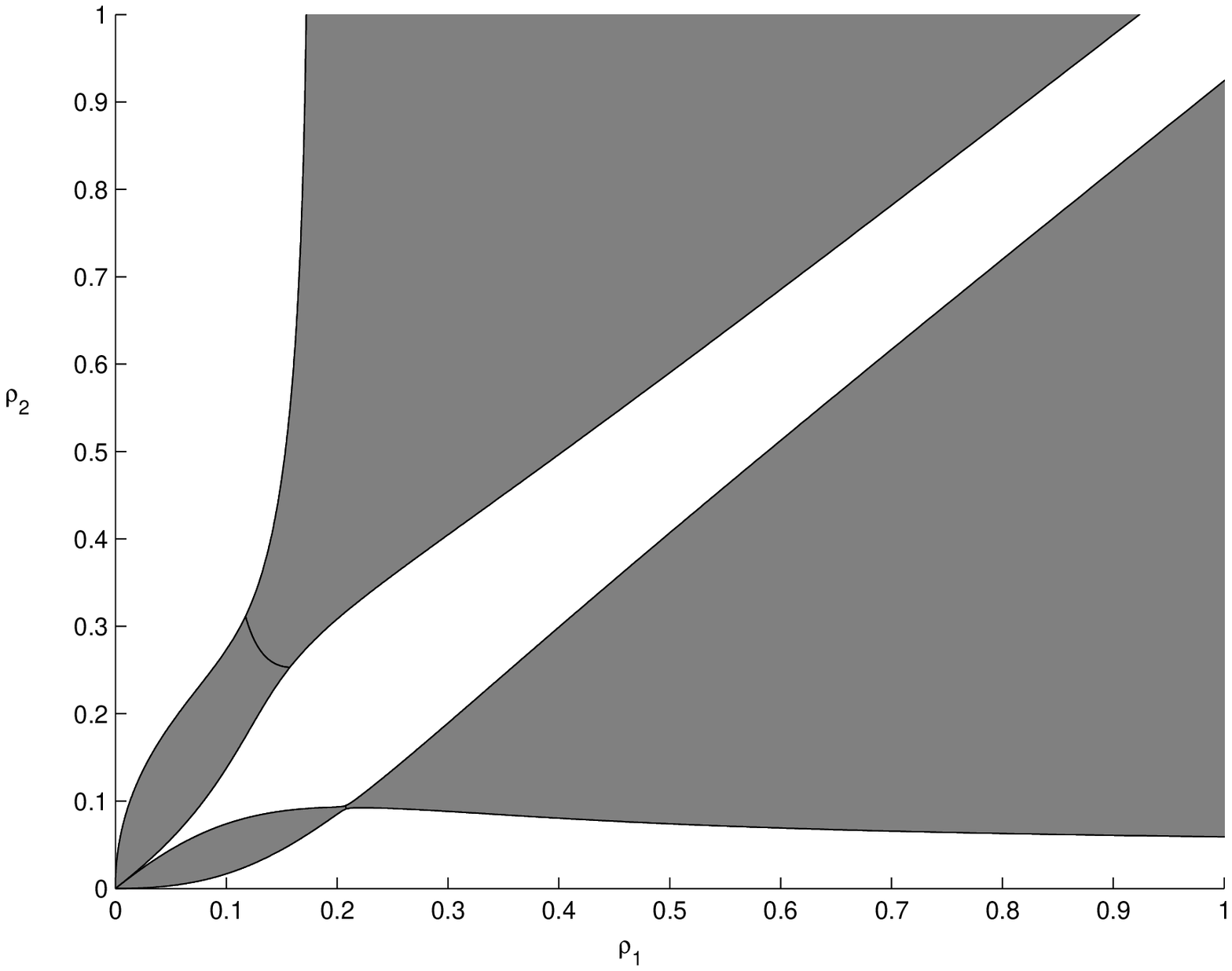 ,height=6cm}} \caption{$\mu _{1}=0.3,$
$\mu _{2}=0.1$ and $\mu _{0}=0.2.$ The projection of the boundary
surface onto the $\rho _{1}-\rho _{2}$ plane at a. $C_{0}=R_{3}=
0.051$\qquad b. $C_{0}=R_{4}= 0.0553$} \label{fig12}
\end{figure}

 With $\mu_{1}\neq \mu _{2}\neq \mu _{0}$ we now have two independent mass
ratios $\mu_0,\mu_1$, since $\mu_2=1/2(1-\mu _{0})-\mu _{1}$, by
(6.5). We also have four rungs of the Szebehely ladder i.e. $C_{m
}\neq C_{m }^{\prime }$ and $C_{e}\neq C_{e}^{\prime }$

Figure (\ref{fig10}) and (\ref{fig12}) give two typical examples
of projections for 1) $\mu_0<\mu_1<\mu_2$; and 2)
$\mu_2<\mu_0<\mu_1$. In each figure, two values of $C_0$,
$C_0=R_3$ and $C_0=R_4$ have been selected to show the two stages
of increasing hierarchical stability. For example, in figure
(\ref{fig10}) $\mu_1<\mu_2$, therefore when $R_3<C_0<R_4$, the arm
$\rho_2\approx0$ becomes disconnected first and any system in a
'24' hierarchy will be stable. If the system is in any other
hierarchy it is still free to change to all hierarchies, but the
'24' hierarchy. See Figure (\ref{fig10}a). Once $C_0>R_4$, all
arms become disconnected and the system is hierarchically stable
for all hierarchical arrangements. See figure (\ref{fig10}b).

Note that in figure (\ref{fig12}), where $\mu_1>\mu_2$ the arm
$\rho_1\approx0$ becomes disconnected first. Thus for
$R_3<C_0<R_4$, any system in a '13' hierarchy will be stable.

The critical value of $C_0$ at which the whole system becomes
stable is given by (\ref{46a}) and is only a function of $\mu_0$
and $\mu_1$.

Figure (5.14) plots these critical values as a function of $\mu_0$
and $\mu_1$. For $C_0>C_{crit}(\mu_0,\mu_1)$, hierarchical
stability is guaranteed. Figure (5.15) shows a cross-section of
Figure (5.14) for the value $\mu_0=0.2$, describing the two
curves, the critical values $R_3$ and $R_4$ as functions of
$\mu_1$. Figures (5.14) and (5.15) show that
$C_{crit}(\mu_0,\mu_1)$ has a maximum of $0.0659$ at
$(\mu_0,\mu_1) =( 0.2,0.223) $ and $(\mu_0,\mu_1)= (0.2, 0.185)$.
Thus if $C_0>0.0659$, all CS5BP's regardless of their mass ratios,
will be hierarchically stable.

\begin{figure}
\centerline{ \epsfig{file=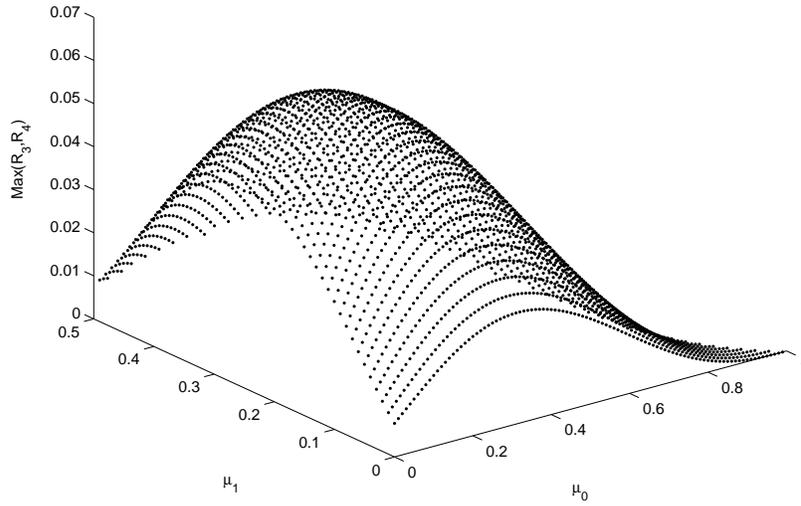 ,height=8cm}} \caption{The
critical values of $C_0$,$C_{crit}$, at which the CS5BP becomes
hierarchically stable as a function of $\mu_0, \mu_1$}
\label{maximum}
\end{figure}
\begin{figure}
\centerline{ \epsfig{file=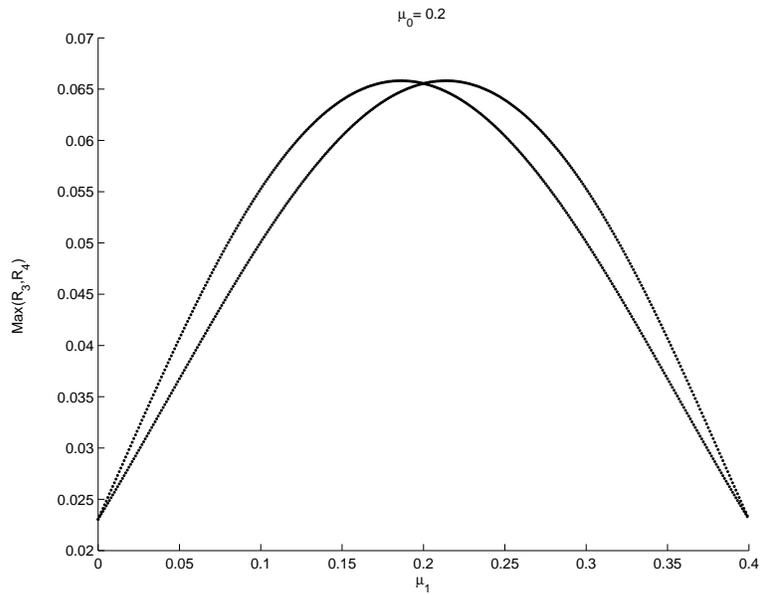 ,height=8cm} }
\caption{The critical values of $C_0$, $C_{crit}$, at which the
CS5BP becomes hierarchically stable as a function of $\mu_1$,
where $\mu_0=0.2$ } \label{maximum2}
\end{figure}
\section{Difference of Notation with Steves and Roy (2000, 2001) explained}
\begin{enumerate}
\item Steves and Roy (1998, 2000) began numbering their CSFBP in
numerical order $(1 2 3 4)$, thus the symmetric pairs were 1)
$P_1$ and $P_4$ and 2) $P_2$ and $P_3$.

In this current work it was realized that when the four body
problem was generalized to higher number of bodies, mathematically
it would be advantageous to number the bodies so that if the first
body in the pair was the j$th$, the second body in the pair would
be numbered $n+j$, where $2n$ is the total number of bodies in the
system. For the four and five body problems this meant the
symmetric pairs became: 1) $P_1$ and $P_3$ and 2) $P_2$ and $P_4$.
Thus when comparing Steves , Roy and Szell's original CSFBP
results with the results of this thesis, the following are
equivalent labels given in table (\ref{complabels}):

\begin{table*}
  \begin{center}
\caption{Difference in labelling of hierarchies with Steves and
Roy (1998, 2000)}\label{complabels}
\bigskip
  \begin{tabular}{c||c} \hline
Original Notation & Current notation\\ \hline \hline

12 hierarchy (double binary (DB))& 12 hierarchy (double binary
(DB))\\

13 hierarchy (double binary (DB))& 14 hierarchy (double binary
(DB))\\

14 hierarchy (single binary (SB))& 13 hierarchy (single binary
(SB))\\
23 hierarchy (single binary (SB))& 24 hierarchy (single binary
(SB))\\
  \end{tabular}
\end{center}
\end{table*}

\item  For the CSFBP, Steves and Roy (2001), need to define only
two different masses $m$ and $M$. They, therefore need only one
mass ratio to describe the system. Thus in the four body
symmetrical system, $\mu$ is defined as $\mu=m/M$. when more than
four bodies are included in the problem it is easier to use a more
general system of mass ratios. Thus $\mu_i$ is chosen to eb the
ratio of the ith body to the total mass of the system. Therefore
we have a scaling difference between the Steves and Roy original
notation and our notation of

\begin{equation}
\mu _{1}=\frac{\mu }{\left( 2\mu +2\right) } \qquad \mu
_{2}=\frac{1}{\left( 2\mu +2\right)},
\end{equation}

where $\mu =m/M$ as defined by Steves and Roy (2001). Hence for
$\mu =1$ i.e. the equal mass four body problem we have $\mu
_{1}=0.25$, $\mu _{2}=0.25$ and $\mu _{0}=0.$

\item  For the same reasons as above, we have a scaling difference
for the Szebehely Constant $C_{0}$ for which we give the following
conversion formula
\begin{equation}
C_{A}=\left( 2\mu +2\right) C_{S},
\end{equation}
where $C_{A}$ is the Szebehely Constant given by Steves and Roy
and $C_{S}$ is the Szebehely Constant given by ourselves.
\end{enumerate}

\section{Conclusions}
Steves and Roy (1998, 2000, 2001) have recently developed a
symmetrically restricted four body problem called the Caledonian
Symmetric Four Body Problem (CSFBP), for which they derive an
analytical stability criterion valid for all time. We introduced a
stationary mass to the centre of mass of the CSFBP and derived
analytical stability criterion for the resulting five body system
(CS5BP). The stability criterion was then used to determine the
effect of adding a central body on the stability of the whole
system.

The critical value of $C_0$ at which the whole system becomes
hierarchically stable for all time is
\begin{equation}
C_{crit}=\max (R_{3},R_{4})=\left\{
\begin{array}{c}
R_{3}=C_{e}(\min )\qquad \textrm{if }\mu _{1}>\mu _{2} \\
R_{4}=C_{e}^{\prime }(\min )\qquad \textrm{if }\mu _{2}>\mu _{1}
\end{array}
\right.
\end{equation}
$\mu_1=\mu_2$ is the special case of equal masses where $C_0>
(C_{crit}=R_3$=$R_4)$ gives total hierarchical stability at one
critical point. Otherwise, hierarchical stability occurs in two
stages $C_{crit1}=R_3<C_{0}>C_{crit_2}=R_{4}$ and
$C_{0}>(C_{crit_2}=R_{4})$. $R_3$ and $R_4$ are purely functions
of $\mu_0$ and $\mu_1$. For $\mu_1=\mu_2$, they are functions only
of $\mu_0$. Figure (6.8) plots these critical values as a function
of $\mu_0$. For $C_0>C_{crit}(\mu_0)$, hierarchical stability is
guaranteed. Figure (6.8) shows that $C_{crit}(\mu_0)$ has a
maximum of 0.065667 at $\mu_0=0.183$. Thus if $C_0>0.065667$, all
CS5BP's with $\mu_1=\mu_2$ will be hierarchically stable.

For the cases of non-equal masses and $C_{crit1}<C_0<C_{crit2}$,
one hierarchy type depending on the relative size of the masses
becomes disconnected and is therefore stable. For the cases of
either $\mu_0>\mu_2$ or $\mu _{0}>\mu _{1}$, the 13 hierarchy
state is hierarchically stable while in all other cases the 24
hierarchy state is hierarchically stable.

For $\mu_0=0$ to 0.2, i.e a small central mass surrounded by
larger masses, the double binary hierarchies dominate, with single
binary hierarchies becoming more prevalent as $\mu_0$ increases.
At $\mu_0=0.2$, i.e. the five body equal mass case, the areas of
real motion are of relatively equal sizes for the double binary
and single binary hierarchies, suggesting neither is dominant.
When comparing the area of real motion available for $\mu_0=0$
(the four body equal mass case), with that of $\mu_0=0.2$ (the
five body equal mass case), we see that the addition of a fifth
body of equal mass at the centre increases the area of real motion
in both single binary and double binary hierarchies.

In the CS5BP, Figures (5.14) and (5.15) show that
$C_{crit}(\mu_0,\mu_1)$ has a maximum of $0.0659$ at
$(\mu_0,\mu_1) =( 0.2,0.223) $ and $(\mu_0,\mu_1)= (0.2, 0.185)$.
Thus if $C_0>0.0659$, all CS5BP's regardless of their mass ratios,
will be hierarchically stable.

We also show in section 6.4 that when $\mu _{0}>\mu _{1}$ or
$\mu_0>\mu_{2}$, the 13 hierarchy state is hierarchically stable
for $R_3<C_{0}<R_{4}$ while in all other cases the 24 hierarchy
state is hierarchically stable for $R_3<C_{0}<R_{4}$.

The analytical stability criterion derived here tells us about the
complete hierarchical stability at the critical value of $C_0$. It
does not tell much about what happens in between $C_0=0$ and
$C_{crit}$. To find out this answer in chapter 6 we will
numerically integrate many CS5BP systems by using the analytical
stability criterion as a guide. We will then analyze the stability
of the different hierarchy states by studying the frequency of
changes from  each specified hierarchy to another specified
hierarchy.

%% file: ChapNumInvist.tex
\chapter{Numerical investigation of Hierarchical Stability of the Caledonian Symmetric Five Body Problem (CS5BP)}

In the last chapter we derived an analytical criterion for the
topological stability of the CS5BP. In this chapter we investigate
the hierarchical stability of the CS5BP numerically, and compare
the results with that of the analytical stability criterion. We
verify that for $C_0>R_4$, the CS5BP system is hierarchically
stable and for $R_3<C_0<R_4$ the CS5BP system is partially stable.
It is also shown that with increasing value of $C_0$, the system
becomes more stable.

In section 6.1 we provide a brief review of the numerical
investigation of the hierarchical stability of the Caledonian
Symmetric Four-Body Problem \cite{Andras1}. In section 6.2 we
determine the equations of motion of the coplanar case of the
CS5BP and describe our criterion for detecting the hierarchy
changes. In section 6.3 we briefly describe the integrator
software designed by ourselves to solve the differential equations
given in section 6.2. The equations for the initial conditions are
derived in section 6.4. In section 6.5 we give a detailed analysis
of the hierarchical stability of the CS5BP in which we also
discuss the CSFBP as a special case of the CS5BP. Here we are
interested in understanding the relationship between the Szebehely
constant $C_0$ and hierarchical stability. Section 6.6 contains
the conclusions.
\section{Review of the Hierarchical Stability of the Caledonian Symmetric Four Body Problem (CSFBP)}

\begin{figure}[tbp]
\begin{center}
\mbox{\epsfig{file=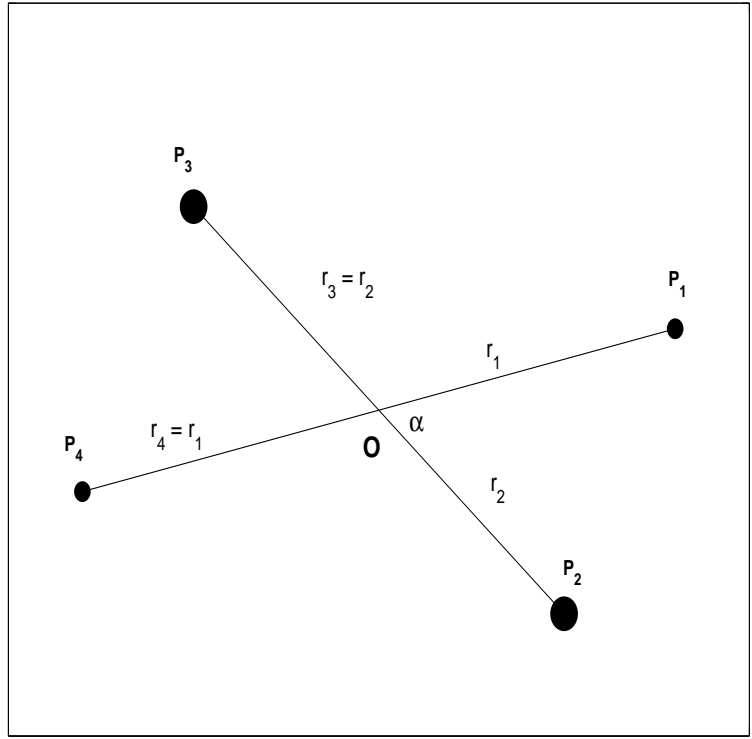,height=10cm,width=10cm}}
\end{center}
\caption{The Caledonian Symmetric Four Body (CSFBP) model}
\label{CSFBP}
\end{figure}

Let $\alpha $ be the angle between $P_{1}$ and $P_{2}$ in figure
(\ref{CSFBP}). With the help of $\alpha $ we can determine the
hierarchical evolution of the system \cite{Andras1}. The CSFBP has
four different hierarchy states which are defined by Sz\'ell as
follows. Please note that the definition of hierarchy states for
the CS5BP is slightly different from those of the CSFBP, see
section 5.7 for details.

\begin{enumerate}
\item  $12$ type hierarchy. A double binary (DB) hierarchy where $P_{1}$ and $%
P_{2}$ orbit their centre of mass $C_{12}$. $P_{3}$ and $P_{4}$ orbit their
centre of mass $C_{34}$. The centre of masses $C_{12}$ and $C_{34}$ orbit
each other about the centre of mass of the four-body system $C$.

\item  $13$ type hierarchy (14 in the case of the CS5BP). A double
binary (DB) hierarchy where $P_{1}$ and $ P_{3}$ orbit their
centre of mass $C_{13}$. $P_{2}$ and $P_{4}$ orbit their centre of
mass $C_{24}$. The centre of masses $C_{13}$ and $C_{24}$ orbit
each other about the centre of mass of the four-body system $C$.

\item  $14$ type hierarchy (13 in the case of the CS5BP). A single
binary (SB) hierarchy where $P_{1}$ and $ P_{4}$ orbit their
centre of mass $C$ in a small central binary. The $P_{2}$ and
$P_{3}$ orbit around the central binary.

\item  $23$ type hierarchy (24 in the case of the CS5BP). A single
binary (SB) hierarchy where $P_{2}$ and $ P_{3}$ orbit their
centre of mass $C$ in a small central binary. The $P_{1}$ and
$P_{4}$ orbit around the central binary.
\end{enumerate}

Sz\'ell (2003) uses the following criterion to determine the
hierarchical position of the four body system under discussion.

Let $k$ be an integer. When $\alpha $ oscillates inside the
interval $(2k\pi -\frac{\pi }{2},2k\pi +\frac{\pi }{2})$, then we
have a $12$ type hierarchy. When it oscillates inside the interval
$(2k\pi +\pi -\frac{\pi }{2 },2k\pi +\pi +\frac{\pi }{2})$, $k\in
{\Bbb Z}$, interval then we have a $ 13 $ type hierarchy.

In the case when $\alpha $ is a monotonically increasing function
and there exists $t_{1}$ and $t_{2}>t_{1}$ for which $\alpha
(t_{1})=(4k\pm 1)\frac{\pi }{2}$ and $\alpha (t_{2})=(4k\pm
3)\frac{\pi }{2}$ we have either a ''$
23 $'' type hierarchy or ''$14$'' type hierarchy. If $r_{1}>r_{2}$ it is a ''%
$23 $'' type hierarchy and if $r_{1}<r_{2}$ it is a ''$14$'' type hierarchy

Now we will briefly discuss the results of  Sz\'ell's numerical
integration into the frequency and type of hierarchy changes for
different mass ratios. We will particularly use later the results
for $\mu=1$, when comparisons are made with the CS5BP numerical
investigations.


\subsection{Hierarchy changes for different mass ratios of the CSFBP}
In this section we give a short summary of the results of Sz\'ell
(2003) for $\mu=1$, $\mu=0.1$, $\mu=0.01$ and $\mu=0.001$ of the
CSFBP.

${\bf \mu =1}$ {\bf Case} : There are no hierarchy changes for
largest critical value of $C_{0}$ i.e. $R_{3}=R_{4}$. There are
numerous hierarchy changes for all smaller values of $C_{0}$. The
number of hierarchy changes reduces as $C_{0}$ increases. Thus it
can be said that the system is more unstable for small $C_{0}$
values. The most unlikely changes are 12 $\rightarrow$ 13 and 13
$\rightarrow$ 12, which is a change from one double binary
hierarchy to another double binary hierarchy. The most likely
changes for $C_{0}$ less than $R_{1}$ are 23 $\rightarrow$ 14 and
14 $\rightarrow$ 23 ( single binary to single binary hierarchy
changes and vice versa). For bigger $C_{0}$ these hierarchy
changes decrease and the 23 $\rightarrow$ 12 and 23 $\rightarrow$
13 (SB to DB and vice versa) hierarchy changes become dominant.
Near the critical value $R_{2}$ the number of hierarchy changes
decreases as one would expect. Only the 23 $\rightarrow$ 12 and 23
$\rightarrow$ 13 type of hierarchy changes remain considerable.

${\bf \mu =0.1}$ {\bf Case : }There are no hierarchy changes to
and from the 23 hierarchy state near $R_{4}$. With small $C_{0}$
values 14 $\rightarrow$ 23 and 23 $\rightarrow$ 14 type of
hierarchy changes are dominant. As in the previous case 12
$\rightarrow$ 13 and 13 $\rightarrow$ 12 are the most unlikely
hierarchy changes.

${\bf \mu =0.01}$ {\bf Case : }There are no 12 $\rightarrow$ 23,
13 $\rightarrow$ 23, 23  $\rightarrow$ 13 and 23  $\rightarrow$ 14
hierarchy changes for $C_{0}$ near $R_{4}.$ With small $C_{0}$,
the 14 $\rightarrow$ 23 and 23 $\rightarrow$ 14 hierarchy changes
are dominant.

${\bf \mu =0.001}$ {\bf Case :} There are very few hierarchy
changes for this mass ratio which indicates that the system is
very nearly hierarchically stable for all $C_0$ values chosen.

The overall behavior of the system for different mass ratios
indicate that as the value of $\mu $ decreases the system becomes
hierarchically more stable.

We will now derive the equations of motion and initial conditions
in order to perform a similar numerical investigation of the CS5BP
problem.

\section{The Equations of Motion of the CS5BP and Hierarchy Changing Criterion}

Note that from now we will use the CS5BP notation for numbering
the five bodies. See section 5.7. The classical equations of
motion for the general n-body problem are given by
\begin{equation}\label{6.1}
m_{i}\stackrel{..}{\mathbf{r}_{i}}=\sum_{i\neq
j}\frac{m_{i}m_{j}}{r_{ij}^{3} }\mathbf{r}_{ij},\qquad
i=0,1,2,3...
\end{equation}
where $\mathbf{r}_i=(x_i,y_i)$ and
$\mathbf{r}_{ij}=\mathbf{r}_j-\mathbf{r}_i$. For a five-body
problem we will get the following equations of motion from
(\ref{6.1}) above
\begin{equation}
\stackrel{..}{\mathbf{r}}_{0}=\frac{m_{1}\mathbf{r}_{01}}{r_{01}^{3}}+\frac{
m_{2}\mathbf{r}_{02}}{r_{02}^{3}}+\frac{m_{3}\mathbf{r}_{03}}{r_{03}^{3}}+
\frac{m_{4}\mathbf{r}_{04}}{r_{04}^{3}}
\end{equation}
\begin{equation}
\stackrel{..}{\mathbf{r}}_{1}=\frac{m_{0}\mathbf{r}_{10}}{r_{10}^{3}}+\frac{
m_{2}\mathbf{r}_{12}}{r_{12}^{3}}+\frac{m_{3}\mathbf{r}_{13}}{r_{13}^{3}}+
\frac{m_{4}\mathbf{r}_{14}}{r_{14}^{3}}
\end{equation}
\begin{equation}
\stackrel{..}{\mathbf{r}}_{2}=\frac{m_{0}\mathbf{r}_{20}}{r_{20}^{3}}+\frac{
m_{1}\mathbf{r}_{21}}{r_{21}^{3}}+\frac{m_{3}\mathbf{r}_{23}}{r_{23}^{3}}+
\frac{m_{4}\mathbf{r}_{24}}{r_{24}^{3}}
\end{equation}
\begin{equation}
\stackrel{..}{\mathbf{r}}_{3}=\frac{m_{0}\mathbf{r}_{30}}{r_{30}^{3}}+\frac{
m_{1}\mathbf{r}_{31}}{r_{31}^{3}}+\frac{m_{2}\mathbf{r}_{32}}{r_{32}^{3}}+
\frac{m_{4}\mathbf{r}_{34}}{r_{34}^{3}}
\end{equation}
\begin{equation}
\stackrel{..}{\mathbf{r}}_{4}=\frac{m_{0}\mathbf{r}_{40}}{r_{40}^{3}}+\frac{
m_{1}\mathbf{r}_{41}}{r_{41}^{3}}+\frac{m_{2}\mathbf{r}_{42}}{r_{42}^{3}}+
\frac{m_{3}\mathbf{r}_{43}}{r_{43}^{3}}
\end{equation}
By utilizing all the symmetries of the coplanar CS5BP, see chapter
5 for details, we reduce the number of equations from fifteen to
four. As $m_0$ is stationary at the origin for all time, therefore
we have
$\mathbf{r}_0=\dot\mathbf{r}_0=\ddot\mathbf{r}_0=\mathbf{0}$.
Also, as $\mathbf{r}_1=-\mathbf{r}_3$ and
$\mathbf{r}_2=-\mathbf{r}_4$ therefore we need to solve for
$\mathbf{r}_1$ and $\mathbf{r}_2$ or $\mathbf{r}_3$ and
$\mathbf{r}_4$ only. The equation of motion for the coplanar CS5BP
in the simplified form are the following
\begin{equation}
\stackrel{..}{\mathbf{r}}_{1}=-\frac{1}{r_{1}^{3}}\left(
m_{0}+\frac{m_{1}}{4 }\right) \mathbf{r}_{1}-m_{2}\left(
\frac{\mathbf{r}_{1}-\mathbf{r}_{2}}{
r_{12}^{3}}+\frac{\mathbf{r}_{1}+\mathbf{r}_{2}}{r_{14}^{3}}\right)
\label{6.2}
\end{equation}

\begin{equation}
\stackrel{..}{\mathbf{r}}_{2}=-\frac{1}{r_{2}^{3}}\left(
m_{0}+\frac{m_{2}}{4 }\right) \mathbf{r}_{2}-m_{2}\left(
\frac{\mathbf{r}_{2}-\mathbf{r}_{1}}{
r_{12}^{3}}+\frac{\mathbf{r}_{1}+\mathbf{r}_{2}}{r_{23}^{3}}\right)
\label{6.3}
\end{equation}
As $\ddot\mathbf{r}_1=(\ddot{x}_1,\ddot{y}_1)$ and
$\ddot\mathbf{r}_2=(\ddot{x}_2,\ddot{y}_2)$, equations (\ref{6.2})
and (\ref{6.3}) can be rewritten as
\begin{eqnarray}
\stackrel{..}{x}_{1} &=&-\frac{1}{\left(
x_{1}^{2}+y_{1}^{2}\right) ^{\frac{3
}{2}}}\left( m_{0}+\frac{m_{1}}{4}\right) x_{1}-  \nonumber \\
&&m_{2}\left( \frac{x_{1}-x_{2}}{\left(
(x_{1}-x_{2})^{2}+(y_{1}-y_{2}\right)
^{2})^{\frac{3}{2}}}+\frac{x_{1}+x_{2} }{\left(
(x_{1}+x_{2})^{2}+(y_{1}+y_{2}\right) ^{2})^{\frac{3}{2}}}\right)
\label{6.4}
\end{eqnarray}
\begin{eqnarray}
\stackrel{..}{y}_{1} &=&-\frac{1}{\left(
x_{1}^{2}+y_{1}^{2}\right) ^{\frac{3
}{2}}}\left( m_{0}+\frac{m_{1}}{4}\right) y_{1}-  \nonumber \\
&&m_{2}\left( \frac{y_{1}-y_{2}}{\left(
(x_{1}-x_{2})^{2}+(y_{1}-y_{2}\right)
^{2})^{\frac{3}{2}}}+\frac{y_{1}+y_{2} }{\left(
(x_{1}+x_{2})^{2}+(y_{1}+y_{2}\right) ^{2})^{\frac{3}{2}}}\right)
\label{6.5}
\end{eqnarray}
\begin{eqnarray}
\stackrel{..}{x}_{2} &=&-\frac{1}{\left(
x_{2}^{2}+y_{2}^{2}\right) ^{\frac{3
}{2}}}\left( m_{0}+\frac{m_{2}}{4}\right) x_{2}-  \nonumber \\
&&m_{1}\left( \frac{x_{2}-x_{1}}{\left(
(x_{1}-x_{2})^{2}+(y_{1}-y_{2}\right)
^{2})^{\frac{3}{2}}}+\frac{x_{1}+x_{2} }{\left(
(x_{1}+x_{2})^{2}+(y_{1}+y_{2}\right) ^{2})^{\frac{3}{2}}}\right)
\label{6.6}
\end{eqnarray}
\begin{eqnarray}
\stackrel{..}{y}_{2} &=&-\frac{1}{\left(
x_{2}^{2}+y_{2}^{2}\right) ^{\frac{3
}{2}}}\left( m_{0}+\frac{m_{2}}{4}\right) y_{2}-  \nonumber \\
&&m_{1}\left( \frac{y_{2}-y_{1}}{\left(
(x_{1}-x_{2})^{2}+(y_{1}-y_{2}\right)
^{2})^{\frac{3}{2}}}+\frac{y_{1}+y_{2} }{\left(
(x_{1}+x_{2})^{2}+(y_{1}+y_{2}\right) ^{2})^{\frac{3}{2}}}\right)
\label{6.7}
\end{eqnarray}
\begin{figure}\label{example}
\centerline{ \epsfig{file=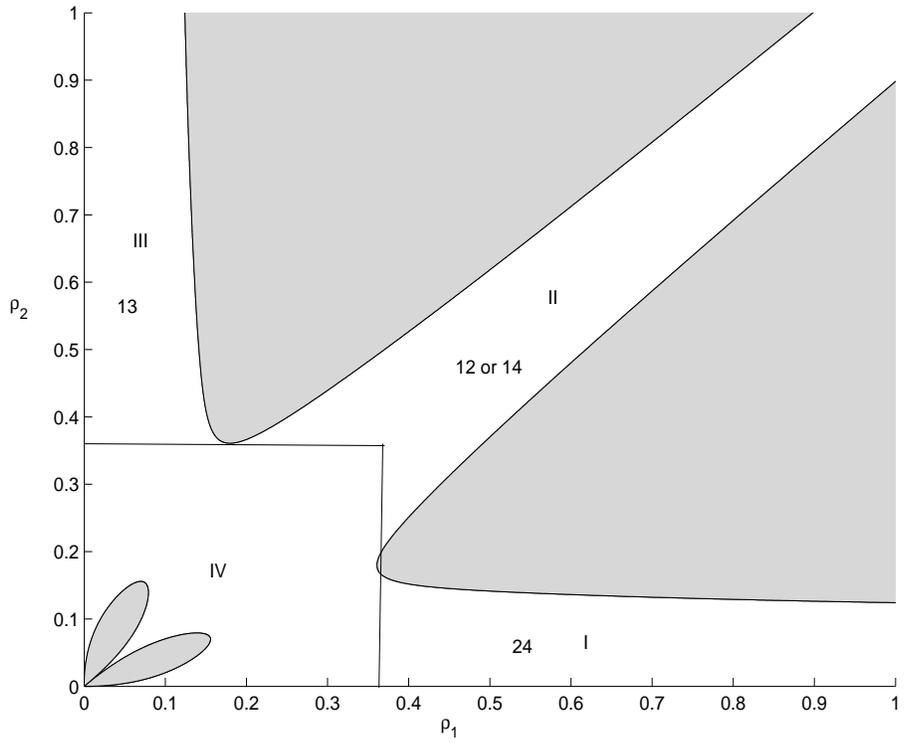,height=10cm,width=12cm}}
\caption{Regions of allowed real motion (white) in the
$\rho_1-\rho_2-\rho_{12}$ space projected onto the $\rho_1-\rho_2$
plane}
\end{figure}

To monitor the hierarchies and its changes from one state to
another we  solve numerically equations (\ref{6.4}) to (\ref{6.7})
at any time $t$ to give the locations of the four bodies in the
phase space with respect to each other. We will explain section
6.3 how we solve these equations. Once we find the position
co-ordinates of all the four bodies, then it is not difficult to
determine what hierarchy state the CS5BP system is in. See section
6.7 for the definition of hierarchy states for the CS5BP.  To
monitor the hierarchy changes during the integration time we make
use of figure (\ref{example}). Please note that figure
(\ref{example}) is different for each set of initial conditions.
Region I and III represent single binary areas while region II
represent double binary areas. If the system is in region I we say
that it is in the 24 type of hierarchy state, if it is in region
II we say that it is in the 12 or 14 hierarchy state depending on
the distances $|P_1P_2|$ and $|P_1P_4|$. If $|P_1P_2|$ $<$
$|P_1P_4|$, then the system is in the 12 type of hierarchy state.
Otherwise it is in the 14 type of hierarchy state. If the system
is in region III then we say that it is in the 13 type of
hierarchy state. But if the system is in region IV then we do not
recognize any hierarchy state until it enters one of the other
three regions. When the system transfers from one region to
another, we record a hierarchy change. For example if the system
goes out of region I and enters region IV and then region III we
record this as a 24 $\rightarrow$ 13 (2313) hierarchy change and
vice versa.

Note that this criterion for recognizing a hierarchy change is
different from that of \citeasnoun{Andras1}. Our criterion is
based on the sizes of $r_1$ and $r_2$ relative to each other while
Sz\'ell's criterion is also based on changes from libration to
rotation and vice versa of the angle $\alpha$ between $P_1$ and
$P_2$. In practical terms, both methods produce similar
frequencies of hierarchy type changes.

\section{The Integrator}
The equations of motion for the CS5BP, equations (\ref{6.4}) to
(\ref{6.7}), are highly non-linear second order coupled
differential equations. It is not possible to find their
analytical solution. Therefore we have to use some numerical
technique for integrating this system of the differential
equation. The software must take as input data, the initial
parameters of the differential equations and give as output data,
the solution ( position and velocity coordinates) of differential
equations belonging to the input data. To develop the integrator
we first needed to find an accurate and fast numerical method and
then develop an environment where the input and output data is
efficiently handled.

The numerical method, we chose, for this integration project is a
15th order method with a adaptive step size control called the
Radau method of Everhart  \cite{everhart}. This method makes use
of Gauss-Radau spacing. We have adopted the FORTRAN code from the
example of Everhart's three body problem. We used Microsoft Visual
C++ 6.0 to construct the environment of the integrator.
\section{Initial Conditions}

The numerical integrations of (\ref{6.4}) to (\ref{6.7}) require
initial values for $\mathbf{r_1}$ and $\mathbf{r_2}$. In order to
satisfy the initial conditions of the CS5BP we immediately have
\begin{equation}
\mathbf{r_1} = (x_1,0) \qquad \textrm{ and } \qquad \mathbf{r_2} =
(x_2,0)
\end{equation}

We choose a grid of  values of $x_1$ and $x_2$ so that they range
within the intervals (0, 3] and (0, 2.5], separated by an
increment size 0.05. The initial velocities $\mathbf{V_1}$ of
$P_1$ and $\mathbf{V_2}$ of $P_2$ are calculated using the
following relations.

\begin{equation}\label{velocity1}
{V_{1y}} =\frac
{x_1cm_1-x_2\sqrt{m_1m_2(-c^2+4(x_1^2m_1+x_2^2m_2)(E_0+U))}}{2m_1(m_1x_1^2+m_2x_2^2)}
\end{equation}

\begin{equation}\label{velocity2}
{V_{2y}} =\frac
{x_2cm_2+x_1\sqrt{m_1m_2(-c^2+4(x_1^2m_1+x_2^2m_2)(E_0+U))}}{2m_2(m_1x_1^2+m_2x_2^2)}
\end{equation}

Where $V_{1y}$ and $V_{2y}$ are the $y$ components of
$\mathbf{V_1}$ and $\mathbf{V_2}$ respectively, $c =
\sqrt{\frac{C_0}{-E_0}}$ is the angular momentum of the system,
$E_0$ is negative of the energy $E$ and $U$ is the potential given
in chapter 4. The $x$ components of $\mathbf{V_1}$ and
$\mathbf{V_2}$ are set to be zero for $t=0$.

Our aim in this chapter is to examine the motion and stability of
a variety of the CS5BP systems each with a different Szebehely
constant $C_0$. We also wish to compare systems with the same
Szebehely constant. Thus we select the initial conditions so that
they result in the same $C_0$. For a given $C_0$, the variables
$V_{1y}$ and $V_{2y}$ can be calculated from equations
(\ref{velocity1}) and (\ref{velocity2}).

We investigate the following sets of mass ratios of the CS5BP with
several values of $C_0$ for each set of mass ratios. Diagram of
the initial configuration for each set of mass ratios are given in
table (6.1), along with section in which each system is discussed.

\begin{enumerate}
\item The Four Body Cases i.e the CSFBP
\begin{enumerate}
\item $\mu_1 = \mu_2=0.25$ and $\mu_0=0$, in CSFBP notation
$\mu=0$.

$C_0=\{ 0.01,0.028,0.031,0.038,0.042,0.046\}$
 \item $\mu_1=0.05$, $\mu_2=0.45$ and $\mu_0=0$, in CSFBP notation
$\mu=0.1$.

$C_0=\{0.012,0.013,0.015,0.018,0.019\}$ \item $\mu_1=0.00495$,
$\mu_2=0.49505$ and $\mu_0=0$, in CSFBP notation $\mu=0.01$.

$C_0=\{0.008,0.0082,0.0085,0.0086,0.0087\}$ \item
$\mu_1=0.0004995$, $\mu_2=0.4995095$ and $\mu_0=0$, in CSFBP
notation $\mu=0.001$.

$C_0=\{0.0078,0.00786,0.00788,0.00789,0.0079\}$
\end{enumerate}

\item Equal mass case of the CS5BP i.e. $\mu_1=\mu_2=\mu_0=0.2$

$C_0=\{0.03,0.06,0.07\}$ \item Four equal masses with a varying
central mass $\mu_0$
\begin{enumerate}
\item $\mu_1 = \mu_2 = \frac{22.475}{100}$ and $\mu_0 =
\frac{1}{100}$; Four equal masses with a very small central mass

$C_0=\{0.026,0.04,0.05\}$ \item $\mu_1 = \mu_2 = \frac{2}{9}$ and
$\mu_0 = \frac{1}{9}$; Four equal masses with a comparatively
larger central mass but still smaller than the outer bodies

$C_0=\{0.036,0.055,0.063\}$ \item $\mu_1 = \mu_2 = \frac{1}{100}$
and $\mu_0 = \frac{96}{100}$; Four equal masses with a large
central mass and smaller outer bodies.

$C_0=\{0.00003,0.000031,0.000033\}$
\end{enumerate}
\item Three equal masses and two increasing symmetrically

\begin{enumerate}
\item $\mu_1 = \mu_0 = 0.326$ and $\mu_2 = 0.11$

$C_0=\{0.02,0.027,0.028,0.029,0.03\}$

\item $\mu_1 = \mu_0 = 0.15$ and $\mu_2 = 0.275$

$C_0=\{0.03,0.036,0.05,0.06,0.063\}$

 \item $\mu_1 = \mu_0 = 0.01$ and $\mu_2 = 0.485$
\end{enumerate}

\item Non-equal masses
\begin{enumerate}
 \item $\mu_1 = 0.195$, $\mu_2= 0.3$ and $\mu_0 =0.01$

 $C_0=\{0.02,0.0281,0.04,0.044,0.055\}$

 \item $\mu_1 = 0.3$, $\mu_2= 0.1$ and $\mu_0 =0.2$

 $C_0=\{0.03,0.0346,0.05,0.055,0.06\}$
 \item $\mu_1 = 0.35$, $\mu_2= 0.01$ and $\mu_0= 0.28$

 $C_0=\{0.02,0.026,0.028,0.029,0.03\}$
 \end{enumerate}

\end{enumerate}
 The values of $C_0$ were chosen so as to represent each important
 region of the Szebehely ladder. Recall that $R_1$, $R_2$, $R_3$ and
 $R_4$, the rungs of the Szebehely ladder, represent boundaries
 for $C_0$ at which topology of the phase space changes.

 In order to analyze the behavior of the motion, for each $(\mu_0,\mu_1,\mu_2,
 C_0)$ case, approximately 3000 integrations were performed for over 1
 million time-steps each. A total of 93$\times$3000 = 279000 orbits were
 investigated numerically. Therefore the total integration time
 was well over 34 million time steps.

Initially the system is in the 12 (DB) or 24 (SB) hierarchy state
when $r_1>r_2$, see figure (7.3a), and it is in the 12 (DB) or 13
(SB) hierarchy state when $r_1<r_2$, see figure (7.3b). Overall
the CSFBP system is initially either in the 12 (DB), 13 (SB) or 24
(SB) hierarchy states. We do not initially place the system in a
14 hierarchy state, since it is equivalent to the 12
 hierarchy state where only the numbering of the bodies differ.
Both the cases 1) initially two larger masses in the centre and
two smaller masses located outside ( $r_1>r_2$), and 2)two smaller
masses in the centre and two larger masses located outside
($r_1<r_2$) are dealt with.

\renewcommand{\baselinestretch}{1.5}
\begin{table*}
\begin{center}
\caption{Different Four and Five body systems
investigated}\label{mussebat}
\begin{tabular}{||c||c||c||ccc|} \hline

Diagram                   & mass ratios       & Section number \\
\hline\hline
            &  CSFBP&\\\cline {2-2}
\large {\textbf{O}}--\large {\textbf{O}}--$\cdot$--\large{\textbf{O}}--\large {\textbf{O}}&$\mu_1=\mu_2=0.25,\mu_0=0$& Section 7.5.1 \\
\textbf{O}--{\Large{\textbf{O}}}--$\cdot$--{\Large{\textbf{O}}}--\textbf{O} & $\mu_1=0.05,\mu_2=0.45,\mu_0=0$& \\
{\textbf{o}}--{\Large{\textbf{O}}--$\cdot$--\Large{\textbf{O}}}--\textbf{o}& $\mu_1=0.00495,\mu_2=0.49505,\mu_0=0$& \\
o--{\LARGE{\textbf{O}}}--$\cdot$--{\LARGE{\textbf{O}}}--o&
$\mu_1=0.0004995,\mu_2=0.4995095,\mu_0=0$&
  \\\hline
  &Equal mass case of the CS5BP&\\ \cline{2-2}
\large {\textbf{O}}--\large
{\textbf{O}}--{\textbf{O}}--\large{\textbf{O}}--\large
{\textbf{O}}&$\mu_1=\mu_2=\mu_0=0.2$& Section 7.5.2 \\ \hline
 &Four equal masses with a varying central mass&\\ \cline{2-2}
{\Large
{\textbf{O}}--\textbf{O}}--o--{\Large\textbf{O}--\textbf{O}}&
$\mu_1=\mu_2=\frac{22.475}{100},\mu_0=\frac{1}{100}$& Section 7.5.3 \\
{\large{\textbf{O}}--\textbf{O}}--\textbf{O}--{\large\textbf{O}--\textbf{O}}&
$\mu_1=\mu_2=\frac{2}{9},\mu_0=\frac{1}{9}$&  \\

o--o--{\LARGE\textbf{O}}--o--o&
$\mu_1=\mu_2=\frac{1}{100},\mu_0=\frac{96}{100}$&  \\\hline

    &Three equal masses and two increasing& \\&symmetrically&\\ \cline{2-2}
    {\large\textbf{O}}--\textbf{O}--{\large\textbf{O}}--\textbf{O}--{\large\textbf{O}}&$\mu_1=\mu_0=0.326,\mu_2=0.11$&
    Section 7.5.4\\
    {\textbf{O}}--{\large\textbf{O}}--{\textbf{O}}--{\large\textbf{O}}--{\textbf{O}}&$\mu_1=\mu_0=0.326,\mu_2=0.11$& \\
{{O}}--{\Huge\textbf{O}}--{{O}}--{\Huge\textbf{O}}--{{O}}&$\mu_1=\mu_0=0.01,\mu_2=0.485$&
\\\hline

&Non-equal masses& \\ \cline{2-2}

\textbf{O}--{\large\textbf{O}}--o--{\large\textbf{O}}--\textbf{O}&$\mu_1=0.195,\mu_2=0.3,\mu_0=0.01$&
Section 7.5.5\\

{\large\textbf{O}}--{\textbf{o}}--{\large\textbf{o}}--{\textbf{o}}--{\large\textbf{O}}&$\mu_1=0.3,\mu_2=0.1,\mu_0=0.2$&
 \\
 {\Large\textbf{O}}--{{o}}--{\large\textbf{O}}--{{o}}--{\Large\textbf{O}}&$\mu_1=0.35,\mu_2=0.01,\mu_0=0.28$&
 \\\hline\hline

\end{tabular}
\end{center}
\end{table*}
\renewcommand{\baselinestretch}{2}
\section{Hierarchical Stability of the CS5BP}
We have given an analytical stability criterion for the CS5BP, in
chapter 5, and we have shown that the CS5BP is hierarchically
stable for $C_0\geq$$C_{crit}$. We have also shown that it is
possible for the CS5BP to move from one hierarchy state to another
for $C_0<C_{crit}$. We aim to understand the behavior of the CS5BP
during the transition from the phase space being connected to it
being disconnected.

To discuss the hierarchical stability of the CS5BP, we selected
five $C_0$ values i.e. $C_0<R_1$, $R_1<C_0<R_2$, $R_2<C_0<R_3$,
$R_3<C_0<R_4$ and $C_0>R_4$ where $R_4$ is the critical value of
$C_0$ denoted by $C_{crit}$ for each mass ratio. The main reason
for integrating for $C_0\geq$$C_{crit}$ was to validate that the
integrator was working properly and the numerical investigation
agreed with the analytical prediction; i.e that the phase space is
disconnected for $C_0\geq$$C_{crit}$ and therefore the system is
hierarchically stable.

For each $(\mu_0,\mu_1,\mu_2,C_0)$ set we integrated numerically
approximately 3000 different orbits for about 1 million time steps
of integration time. In order to study the hierarchical stability
for each $(\mu_0,\mu_1,\mu_2,C_0)$ set we gathered data from the
numerical integrations outputs on the number of hierarchy changes.

We have constructed two kinds of tables for each
$(\mu_0,\mu_1,\mu_2,C_0)$ set, namely
\begin{enumerate}
\item The frequency distribution of hierarchy changes: Columns
represent fixed $C_0$ values while rows represent the number of
each type of hierarchy changes. The number of hierarchy changes
were calculated by determining the total number of hierarchy
changes of each type which occurred throughout the many different
integrations with the same $(\mu_0,\mu_1,\mu_2,C_0)$ set. \item
Percentage of hierarchy changes: Columns represent fixed $C_0$
values while rows represent percentage of hierarchy changes.
\end{enumerate}
Now we will discuss the hierarchical stability of the CS5BP for
different mass ratios. First we will discuss the $\mu_0=0$ case
which is a special case of the CS5BP called the CSFBP in the
following section.

\subsection{Hierarchical stability of the CSFBP, ($\mathbf{\mu_1=\mu_2}$
and $\mathbf{\mu_0=0}$)}

\begin{figure}
\centerline{
\epsfig{file=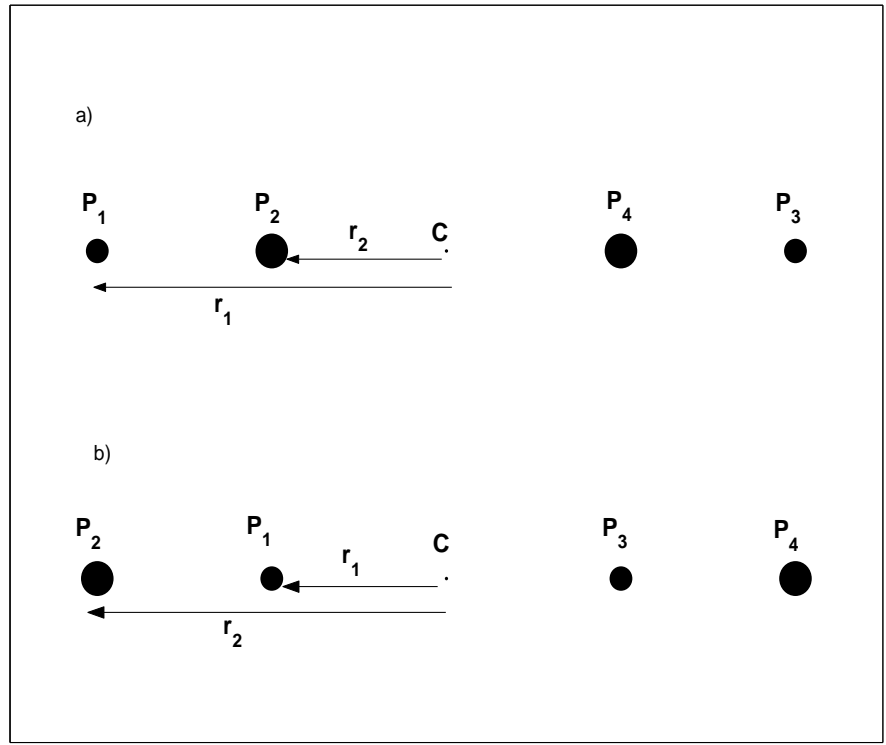,height=10cm,width=12cm }}
\caption{The two possible initial states in the CSFBP a) $r_1>r_2$
b) $r_2>r_1$} \label{CollinearCSFBP}
\end{figure}

In this section we will discuss the $\mu_0=0$ case of the CS5BP
which is the CSFBP (Sz\'ell, 2003). Sz\'ell (2003) discussed a
special case of the CSFBP for hierarchical stability. He discussed
the case when $r_1>r_2$ or in other words when the larger masses
were nearer to the centre of mass than the smaller masses, see
figure (\ref{CollinearCSFBP}a). Therefore he covered half of the
phase space. He also always started in the 12, double binary, or
the 24, single binary, hierarchy state or using the CSFBP notation
the 12 or 23 hierarchy states. Recall that we have a difference of
notation with \citeasnoun{Andras1}, the CSFBP 13 (DB) hierarchy
state is 14 (DB) and its 23 (SB) is 24 (SB) in our case and vice
versa, for more details see section 6.1.  We will use our notation
in the whole thesis and whenever a comparison is necessary we will
give the CSFBP notation in brackets as against to ours.

Our study of the $\mu_0=0$ case will complete Sz\'ell's analysis
of hierarchical stability of the CSFBP, as we will consider both
$r_1>r_2$ and $r_1<r_2$ cases, see figure (\ref{CollinearCSFBP}).

\subsubsection{Equal mass case of the CSFBP $\mu_1=\mu_2=0.25,\mu_0=0$ (or in the CSFBP notation $\mu=1$)}

The critical values in this case are $R_1=R_2=0.0286267$ and
$R_3=R_4=0.0457437$. The results of integration are contained in
Table (\ref{tab1}).

There are no hierarchy changes for $C_0\geq C_{crit}$ which
confirm the analytical results we obtained in chapter 6. The most
likely hierarchy changes for $C_0=0.01$ and $C_0 = 0.028$ are 12
$\rightarrow$ 24 and 12 $\rightarrow$ 13 and vice versa. The 12
$\rightarrow$ 24 hierarchy changes die out for higher $C_0$
values, while the 12 $\rightarrow$ 13 remain dominant. For higher
$C_0$ values, the 13 $\rightarrow$ 12 hierarchy changes joins the
race for the most likely hierarchy changes and remains
considerably close to the 12 $\rightarrow$ 13 hierarchy changes.
All of the above changes are from SB to DB and vice versa.

The most unlikely hierarchy changes are from double binary to
double binary. These kind of hierarchy changes usually do not
happen directly. There is usually a third kind of hierarchy change
of intermediate hierarchy state which takes the system in to a new
configuration. If we look at the average number of hierarchy
changes from different hierarchy states we come to the conclusion
that the 12 is the most unstable while 14 is comparatively more
stable than all others. For example there are nearly twelve
hundred hierarchy changes from the 12 hierarchy state to another,
which is the most, and there are a little above five hundred from
the 14 hierarchy state to another.

\subsubsection{$\mathbf{\mu_1= 0.05}$ and $\mathbf{\mu_2 = 0.45},\mu_0=0$ or in CSFBP notation $\mu=0.1$}

The critical values in this case are $R_1=0.012952, R_2=0.0138015,
R_3=0.0159447, R_4=0.0182837$. The total number of hierarchy
changes of each for different values of $C_0$ are listed in Table
(\ref{tab2}). The corresponding percentages of different hierarchy
changes are given in Table (\ref{tab2b}).

For $C_0=0.019>C_{crit}$ the phase space is disconnected and
therefore there are no hierarchy changes. At
$R_3<C_0=0.018<C_{crit}=R_4$ there are no hierarchy changes to or
from the 24 hierarchy state which is exactly what the analytical
stability criterion predicts i.e. there are no 12 $\rightarrow$
24, 14$\rightarrow$24, 13 $\rightarrow$ 24, 24 $\rightarrow$ 12,
24 $\rightarrow$ 14 and 24 $\rightarrow$ 13 hierarchy changes.

The most likely hierarchy changes for small $C_0$ values are the
12 $\rightarrow$ 24. And the most unlikely hierarchy changes are
the 24 $\rightarrow$ 14 and 14 $\rightarrow$ 24. Overall hierarchy
changes from 14 to any other hierarchy state for all $C_0$ values
are very small except for small $C_0$ values eg. $C_0 = 0.018$
where the 14 $\rightarrow$ 12 and 12 $\rightarrow$ 14 are the
highest hierarchy changes. Hierarchy changes from double binary to
double binary are very small as one would expect.

The main characteristics of this case is similar to those of the
equal mass case. With small $C_0$, we have more hierarchy changes
and as $C_0$ increases the number of hierarchy changes decreases.
The most unlikely hierarchy changes are still from double binary
to double binary.

The main difference in this case and the equal mass case is that
we have fewer hierarchy changes. Therefore the system is
hierarchically more stable. Moreover in the equal mass case we
were not able to identify any, comparatively, stable hierarchy
states as all hierarchy states were equally unstable, but in this
case we have some hierarchy states which can be considered
hierarchically stable. For example Hierarchy changes from the 14
(DB) hierarchy state is approximately 1 percent, which is very
nearly stable. As we decrease the value of $\mu_1$ we will be
getting closer to a perturbed two+two body system and therefore we
should expect a comparatively more stable situation.
\subsubsection{$\mathbf{\mu_1= 0.00495}$ and $\mathbf{\mu_2 = 0.49505},\mu_0=0$ (or in CSFBP notation $\mu=0.01$)}

The critical values in this case are $R_1=0.00828195,
R_2=0.00838264, R_3=0.00850907, R_4=0.00869902$. The total number
of hierarchy changes of each for different values of $C_0$ are
listed in Table (\ref{tab3}). The corresponding percentages of
different hierarchy changes are given in Table (\ref{tab3b}).

There are no hierarchy changes with $C_0=0.0087$, since this value
is greater than $C_{crit}$ and therefore the phase space is
disconnected.

At $C_0=0.0086$ the Szebehely constant is a little less than
$C_{crit}$ and the phase space is partially disconnected.
Therefore the hierarchy changes to or from the 24 hierarchy state
are forbidden. In table (\ref{tab3}) it can be seen that there are
no 12 $\rightarrow$ 24, 14 $\rightarrow$ 24, 13 $\rightarrow$ 24,
24 $\rightarrow$ 12, 24 $\rightarrow$ 14 and 24 $\rightarrow$ 13
hierarchy changes, confirming numerically the analytical stability
criterion.

\begin{figure}[tbp]
\begin{center}
\epsfig{file=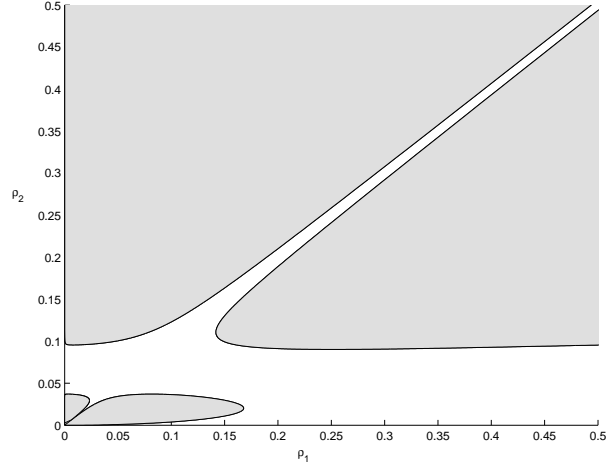,width=8cm}
\end{center}
\caption{Projection found in the same manner as those in chapter 6
but for the case when $\mu_1=0.00495$ and $\mu_2=0.49505$,
$\mu_0=0$} \label{aakhireee}
\end{figure}

As we were expecting from our experience of the previous
$\mu_1=0.05$, case the number of hierarchy changes decrease
further as we decrease the value of $\mu_1$. But the main
characteristics remain the same. With small $C_0$, 12
$\rightarrow$ 24, 24 $\rightarrow$ 12, and 13 $\rightarrow$ 24
hierarchy changes are dominant. This makes sense physically since
as $\mu_2$ is increased, the situation of two large masses in the
centre forming a single binary with two smaller masses orbiting
their centre of mass (eg the 24 hierarchy) will begin to dominate
the system. We can see this growing dominance of 24 hierarchies
visually in the projection of allowed motion for this case figure
(\ref{aakhireee}) which shows an increased region of allowed
motion. When $C_0$ is increased the number of hierarchy changes
drop dramatically and the system is comparatively much more stable
than the previous cases discussed so far. Again the most unlikely
hierarchy changes are from double binary to double binary.

The main difference between the previous  case and this case is
that the system is hierarchically more stable as the system is now
closer to the perturbed two+two body system.

\subsubsection{$\mathbf{\mu_1= 0.0004995}$ and $\mathbf{\mu_2 = 0.4995095,\mu_0=0}$ (or in CSFBP notation $\mu=0.001$)}

The critical values in this case are $R_1=0.00785938,
R_2=0.00786963, R_3=0.00788154, R_4=0.00789903$. The total number
of hierarchy changes of each for different values of $C_0$ are
listed in Table (\ref{tab4}). The corresponding percentages of
different hierarchy changes are given in Table (\ref{tab4b}).

There are only five hierarchy changes for the smallest $C_0$
value, which are the only hierarchy changes for the 15,000 orbits
integrated for one million time steps of integration time. These
few hierarchy changes are from 13 $\rightarrow$ 24, 24
$\rightarrow$ 13 and 24 $\rightarrow$ 12. The very small number of
hierarchy changes indicate that the system is generally
hierarchically stable. This is so because the system is close to a
slightly perturbed two body system.
\subsubsection{Comparison with Sz\'ell's Analysis }

To compare our analysis of the Caledonian Symmetric Problem
(CSFBP) with Sz\'ell's we have the following similarities and
differences. Recall the difference in method of research with
\citeasnoun{Andras1} which might produce some differences.
\begin{enumerate}
\item \citeasnoun{Andras1} counts for only half of the phase space
($r_1>r_2$). This might produce a bias in favor of some hierarchy
states. \item Our hierarchy changing criterion is different from
\citeasnoun{Andras1} as our criterion is based on the size of
$r_1$ and $r_2$ relative to each other, while Sz\'ell's criterion
is also based on change from libration to rotation or vice versa
of the angle $\alpha$ between $P_1$ and $P_2$.
\end{enumerate}

\citeasnoun{Andras1} number and percent of hierarchy changes are
tabulated in Appendix 1. Recall that when making comparisons you
must take account of Sz\'ell's different numbering of the four
bodies from our numbering.

 \textbf{Differences}
\begin{enumerate}
\item For $\mu_1=0.25$ (or CSFBP notation $\mu=1$), the 12
$\rightarrow$ 24 and 24 $\rightarrow$ 12, (in CSFBP notation 12
$\rightarrow 23$ and 23 $\rightarrow$ 12), hierarchy changes are
much higher in our case than in Sz\'ell's case. They are twenty
percent to fourteen percent in our case and four percent to two
percent in Sz\'ell's case. See tables (\ref{tab1}) and
(\ref{tab1b}) and table 1 and 2 in appendix 1. These differences
are most likely because of the inclusion of the second half of the
phase space into our research which almost doubles the number of
starting conditions for 12 hierarchy state. See appendix for
\citeasnoun{Andras1} results. When the value of $C_0$ increases,
the hierarchy changes discussed  above decrease in our case while
it increases in Sz\'ell's case.

\item For all values of the Szebehely constant $C_0$ in the $\mu_1
= 0.05$ (or CSFBP notation $\mu = 0.1$) case, the 12 $\rightarrow$
24 (CSFBP notation 12 $\rightarrow$ 23)hierarchy changes are very
high in our case while it is very small in Sz\'ell's case. The 24
$\rightarrow$ 13 (CSFBP notation 23 $\rightarrow$ 14) hierarchy
changes are very small in our case and stand at a maximum of nine
percent. In Sz\'ell's case they are very high and stands at thirty
six percent maximum and ten percent minimum. See tables
(\ref{tab2}) and (\ref{tab2b}) and table 3 and 4 in appendix 1.
These differences are also most likely because of the inclusion of
the second half of the phase space. The size of the region of real
motion in the non e-qual mass cases for the 13 (CSFBP notation 14)
hierarchy state is very small therefore we have fewer starting
conditions for this hierarchy state and thus we will have fewer
hierarchy changes to or from this hierarchy state.

\item For $\mu_1 = 0.00495$ (or CSFBP notation $\mu=0.01$), the 13
$\rightarrow$ 24 hierarchy changes are very small (maximum fifteen
percent) in our case and becomes zero near the critical value of
the Szebehely constant $C_0$ which is the prediction of the
analytical stability criterion. In Sz\'ell's case these hierarchy
changes are comparatively very high (maximum forty percent) and do
not become zero near the critical value of the Szebehely constant
$C_0$ which is against the prediction of the analytical stability
criterion. See tables (\ref{tab3}) and (\ref{tab3b}).
 Sz\'ell (2003) do not give any reason for this, see
appendix for their results.

\item For $\mu_1 = 0.00495$ (0r CSFBP notation $\mu=0.01$), the 24
$\rightarrow$ 14 (CSFBP notation 23 $\rightarrow$ 13) hierarchy
changes are zero in our case while it goes up to fifteen percent
in Sz\'ell's case.

\item For $\mu_1 = 0.0004995$ (0r $\mu=0.001$), the 12
$\rightarrow$ 13 (CSFBP notation 12 $\rightarrow$ 14) and 12
$\rightarrow$ 24 (CSFBP notation 12 $\rightarrow$ 23) hierarchy
changes are zero in our case while they are non-zero in Sz\'ell's
case and 12 $\rightarrow$ 13 (CSFBP notation 12 $\rightarrow$ 14)
change increases from zero to thirty one percent when $C_0$
increases.
\end{enumerate}

\textbf{Similarities}

\begin{enumerate}
\item the main characteristics of both the analysis remain the
same. The number of hierarchy changes decrease as the value of
$C_0$ increases and the number of hierarchy changes are zero for
$C_0 \geq C_{crit}$

\item For $\mu_1=\mu_2=0.25$, (in CSFBP notation $\mu=1$), and
$\mu_1=0.05$, (in CSFBP notation $\mu=0.1$), the number of
hierarchy changes from double binary to double binary are very
small. Recall that the double binaries in our case are 12 and 14
while they are 12 and 13 in Sz\'ell's case. See tables
(\ref{tab1}) and (\ref{tab2}).

\item For $\mu_1=0.05$, in CSFBP notation $\mu=0.1$, the 24
$\rightarrow$ 12 (CSFBP: 23 $\rightarrow$ 12) hierarchy changes
are very small for all $C_0$. Also the 24 $\rightarrow$ 14 (CSFBP:
23 $\rightarrow$ 13) and 14 $\rightarrow$ 24 (CSFBP: 13
$\rightarrow$ 23) are the most unlikely hierarchy changes in both
the cases. See tables (\ref{tab2}) and (\ref{tab2b})

\item For $\mu_1=0.00495$, in CSFBP notation $\mu=0.01$,the 14
$\rightarrow$ 24 (CSFBP: 13 $\rightarrow$ 23) hierarchy changes
are the most unlikely in both cases.

\item For $\mu_1=0.0004995$, in CSFBP notation $\mu=0.001$, the 13
$\rightarrow$ 24 (CSFBP: 14 $\rightarrow$ 23) and 24 $\rightarrow$
13 (CSFBP: 23 $\rightarrow$ 14) hierarchy changes are the most
likely in both the cases. Also the number of hierarchy changes are
very small with 12 $\rightarrow$ 13, 14 $\rightarrow$ 12, 14
$\rightarrow$ 13 and 13 $\rightarrow$ 14 (CSFBP: 12 $\rightarrow$
14, 13 $\rightarrow$ 12, 13 $\rightarrow$ 14, and 14 $\rightarrow$
13) being zero.

\end{enumerate}
\renewcommand{\baselinestretch}{1}
\begin{table*}
  \begin{center}
  \caption{Total number of hierarchical changes for the equal mass
   case of the CSFBP, $\mu_1=\mu_2=0.25,\mu_0=0$; or in CSFBP notation $\mu=1$}\label{tab1}
\bigskip
  \begin{tabular}{|c|c|c|c|c|c|c|} \hline
   & $C_0=0.01$ & $C_0=0.028$
& $0.031$ & $0.038$ & $0.042$ & $C_0 = 0.046$
\\\hline \hline
  12 13 & 179 & 143 & 150 & 125 & 104  & 0    \\
  12 14  &  5 &  0  & 10 & 22 & 12   & 0\\
  12 24  &  204 & 162 & 24 & 33 & 11   & 0\\
  13 12  &  91 & 9 &  141& 119& 90    & 0 \\
  13 14  &  66 & 10&  21&  20&  28    & 0  \\
  13 24  &  56 & 18 &  28&  27&  20  & 0 \\
  14 12  &  5 &  0  & 11 & 13 & 19   & 0  \\
  14 13  &  60 & 32&  24  &41 & 38   & 0 \\
  14 24   & 80  &84 & 36  &37 & 26    & 0\\
  24 12 &   146 &130& 32  &33&  21    & 0 \\
  24 13  &  130 &90 & 32 & 33 & 41     & 0 \\
  24 14  &  85  &74 & 26 &25  &26 & 0\\ \hline
 \hline
  Total & 1022   & 678 &509 &503 & 410  & 0  \\ \hline
  \end{tabular}
\end{center}
\end{table*}
\begin{table*}
  \begin{center}
\caption{Percentage of hierarchical changes for the equal mass
case of the CSFBP, $\mu_1=\mu_2=0.25,\mu_0=0$; or in CSFBP
notation $\mu=1$}\label{tab1b}
\bigskip
  \begin{tabular}{|c|c|c|c|c|c|c|} \hline
   & $C_0=0.01$& $C_0=0.028$
& $0.031$ & $0.038$ & $0.042$ & $C_0 = 0.046$\\
\hline\hline
  12 13 & 17.51  & 21.09  & 22.12   &18.44   &15.34
  & 0    \\
  12 14  & 0.49  &  0   &1.47  &  3.24    &1.77
   & 0\\
  12 24  &  19.96  & 24.89   &3.54  &  4.87   & 1.62
  & 0\\
  13 12  &  8.9 &1.33    &20.8   & 17.55   &13.27
    & 0 \\
  13 14  &  6.46  &  1.47    &3.1& 2.95  &  4.13
    & 0  \\
  13 24  & 5.48    &2.65   & 4.13  &  3.98 &   2.95
  & 0 \\
  14 12  &  0.49   & 0  & 1.62  &  1.92 &   2.8
& 0  \\
  14 13  & 5.87   & 4.72   & 3.54  &  6.05   & 5.6
   & 0 \\
  14 24   & 7.83   & 12.39  & 5.31   & 5.46   & 3.83
    & 0\\
  24 12 &  14.29  & 19.17&   4.72 &   4.87 &   3.1
   & 0 \\
  24 13  & 12.72 &  13.27  & 4.72   & 4.87  &  6.05
     & 0 \\
  24 14  &  8.32   &  10.91   & 3.83   &  3.69    & 3.83
 & 0\\ \hline

  \end{tabular}
\end{center}
\end{table*}

\begin{table*}
  \begin{center}
\caption{Total number of hierarchical changes for $\mu_1=0.05$,
$\mu_2=0.45$ and $\mu_0=0$, in CSFBP notation
$\mu=0.1$}\label{tab2}
\bigskip
  \begin{tabular}{|c||c|c|c|c|c|c|cc|} \hline
 & $C_0=0.012$ & $C_0=0.013$ & $C_0 = 0.015$  &
$C_0 = 0.018$ & $C_0 = 0.019$     \\\hline \hline
  12 13  &  132   &   98   &   121 &  9   & 0    \\
  12 14  &  1     &   5    &   5   &  16  & 0\\
  12 24  &  360   &   348  &   90  &  0   & 0\\
  13 12  &  55    &   61   &   73  &  2   & 0 \\
  13 14  &  2     &   5    &   3   &  0   & 0  \\
  13 24  &  136   &   138  &   46  &  0   & 0 \\
  14 12  &  5     &   11   &   3   &  16  & 0  \\
  14 13  &  5     &    8   &   6   &  0   & 0 \\
  14 24  &  0     &   2    &   2   &  0   & 0\\
  24 12  &  78    &   57   &   22  &  0   & 0 \\
  24 13  &  77    &   56   &   22  &  0   & 0 \\
  24 14  &  0     &   3    &   1   &  0   &  0\\ \hline
 \hline
  Total  & 851    & 792     &394   & 43   & 0  \\ \hline
  \end{tabular}
\end{center}
\end{table*}

\begin{table*}
  \begin{center}
\caption{Percentage of hierarchical changes for $\mu_1=0.05$,
$\mu_2=0.45$ and $\mu_0=0$,  in CSFBP notation
$\mu=0.1$}\label{tab2b}
\bigskip
  \begin{tabular}{|c||c|c|c|c|c|c|c|c|} \hline
& $C_0=0.012$ & $C_0=0.013$ & $C_0 = 0.015$  & $C_0 = 0.018$ &
$C_0 = 0.019$     \\\hline \hline
  12 13 & 16 & 12 & 31 & 21
  & 0    \\
  12 14  & 0   &1  & 1 &  37

   & 0\\
  12 24  &  42 & 44  &24 & 0

  & 0\\
  13 12  &  6 &  8  & 19 & 5

    & 0 \\
  13 14  &  0  & 1  & 1 &  0

    & 0  \\
  13 24  & 16  &17  &12 & 0

  & 0 \\
  14 12  &  1  & 1  & 1  & 37

& 0  \\
  14 13  & 1 &  1  & 2 &  0

   & 0 \\
  14 24   & 0  & 0  & 1  & 0
    & 0\\
  24 12 &  9  &  7   & 6  &  0

   & 0 \\
  24 13  & 9   &7  & 6 &  0

     & 0 \\
  24 14  &  0  & 0  & 0  & 0

 & 0\\ \hline

  \end{tabular}
\end{center}
\end{table*}

\begin{table*}
  \begin{center}
\caption{Total number of hierarchical changes for $\mu_1=0.00495$,
$\mu_2=0.49505$ and $\mu_0=0$, in CSFBP notation
$\mu=0.01$}\label{tab3}
\bigskip
  \begin{tabular}{|c||c|c|c|c|c|c|cc|} \hline
& $C_0=0.008$ & $C_0=0.0082$ & $C_0 = 0.0085$  & $C_0 = 0.0086$ &
$C_0 = 0.0087$     \\\hline \hline
  12 13  &  41   &   17   &   4    &     8   & 0    \\
  12 14  &  15   &   0    &   0    &  0  & 0\\
  12 24  &  78   &   3    &   2    &  0   & 0\\
  13 12  &  19   &   5    &   4    &  4   & 0 \\
  13 14  &  2     &   0    &   0   &  0   & 0  \\
  13 24  &  46   &   1  &   3  &  0   & 0 \\
  14 12  &  9     &   0   &   0   &  0  & 0  \\
  14 13  &  12     &    0   &   0   &  0   & 0 \\
  14 24  &  0     &   0   &   0   &  0   & 0\\
  24 12  &  47    &   0   &   2  &  0   & 0 \\
  24 13  &  37    &   0   &   4  &  0   & 0 \\
  24 14  &  0     &   0    &   0   &  0   &  0\\ \hline
 \hline
  Total  & 306    & 27     &19   & 12   & 0  \\ \hline
  \end{tabular}
\end{center}
\end{table*}

\begin{table*}
  \begin{center}
\caption{Percentage of hierarchical changes for $\mu_1=0.00495$,
$\mu_2=0.49505$ and $\mu_0=0$,  in CSFBP notation
$\mu=0.01$}\label{tab3b}
\bigskip
  \begin{tabular}{|c||c|c|c|c|c|c|c|c|} \hline
& $C_0=0.008$ & $C_0=0.0083$ & $C_0 = 0.0085$  & $C_0 = 0.0086$ &
$C_0 = 0.0087$     \\\hline \hline
  12 13 & 13  &63 & 21 & 67

  & 0    \\
  12 14  & 5  & 0  & 0  & 0
   & 0\\
  12 24  &  25  &11 & 11 & 0
  & 0\\
  13 12  &  6 &  19 & 21 & 33
    & 0 \\
  13 14  &  1  & 0  & 0 &  0
    & 0  \\
  13 24  & 15 & 4  & 16 & 0
  & 0 \\
  14 12  &  3  & 0 &  0 &  0
& 0  \\
  14 13  & 4  & 0 &  0  & 0
   & 0 \\
  14 24   & 0  & 0  & 0  & 0
    & 0\\
  24 12 &  15  &4  & 11 & 0
   & 0 \\
  24 13  & 12  &0  & 21 & 0
     & 0 \\
  24 14  &  0  & 0  & 0  & 0
 & 0\\ \hline

  \end{tabular}
\end{center}
\end{table*}

\begin{table*}
  \begin{center}
\caption{Total number of hierarchical changes for
$\mu_1=0.0004995$, $\mu_2=0.4995005$ and $\mu_0=0$, in CSFBP
notation $\mu=0.001$}\label{tab4}
\bigskip
  \begin{tabular}{|c||c|c|c|c|c|c|cc|} \hline
& $C_0=0.0078$ & $C_0=0.00786$ & $C_0 = 0.00788$ & $C_0 = 0.00789$
& $C_0 = 0.0079$     \\\hline \hline
  12 13  &  0   &   0   &   0 &  0   & 0    \\
  12 14  &  0   &   0   &   0 &  0   & 0    \\
  12 24  &  0   &   0   &   0 &  0   & 0    \\
  13 12  &  0   &   0   &   0 &  0   & 0    \\
  13 14  & 0   &   0   &   0 &  0   & 0     \\
  13 24  &  2   &   0   &   0 &  0   & 0    \\
  14 12  &  0   &   0   &   0 &  0   & 0    \\
  14 13  &  0   &   0   &   0 &  0   & 0 \\
  14 24  &  0   &   0   &   0 &  0   & 0\\
  24 12  &  1   &   0   &   0 &  0   & 0 \\
  24 13  &  2   &   0   &   0 &  0   & 0 \\
  24 14  &  0   &   0   &   0 &  0   & 0\\ \hline
 \hline
  Total  & 5   &   0   &   0 &  0   & 0 \\ \hline
  \end{tabular}
\end{center}
\end{table*}

\begin{table*}
  \begin{center}
\caption{Percentage of hierarchical changes for $\mu_1=0.0004995$,
$\mu_2=0.4995005$ and $\mu_0=0$,  in CSFBP notation
$\mu=0.001$}\label{tab4b}
\bigskip
  \begin{tabular}{|c||c|c|c|c|c|} \hline
& $C_0=0.0078$ & $C_0=0.00786$ & $C_0 = 0.00788$ & $C_0 = 0.00789$
& $C_0 = 0.0079$     \\\hline \hline
  12 13  &  0   &   0   &   0 &  0   & 0    \\
  12 14  &  0   &   0   &   0 &  0   & 0    \\
  12 24  &  0   &   0   &   0 &  0   & 0    \\
  13 12  &  0   &   0   &   0 &  0   & 0    \\
  13 14  & 0   &   0   &   0 &  0   & 0     \\
  13 24  &  40   &   0   &   0 &  0   & 0    \\
  14 12  &  0   &   0   &   0 &  0   & 0    \\
  14 13  &  0   &   0   &   0 &  0   & 0 \\
  14 24  &  0   &   0   &   0 &  0   & 0\\
  24 12  &  20   &   0   &   0 &  0   & 0 \\
  24 13  &  40   &   0   &   0 &  0   & 0 \\
  24 14  &  0   &   0   &   0 &  0   & 0\\ \hline

  \end{tabular}
\end{center}
\end{table*}
\pagebreak

\renewcommand{\baselinestretch}{2}
\subsection{Equal mass case of the CS5BP, $\mu_1=\mu_2=\mu_0=0.25$}\label{1.4.2}

The critical values in this case are $R_1 = R_2=0.039222$ and $R_3
= R_4=0.065551$. The results are contained in Table (\ref{tab5}).

There are no hierarchy changes for $C_0\geq C_{crit}$ which is
also the prediction of the analytical criterion. The most likely
hierarchy changes for $C_0=0.03$ are the 14 $\rightarrow$ 24 and
24 $\rightarrow$ 14. When we increase the value of $C_0$ to 0.06
the 12 $\rightarrow$ 14 changes become the most likely hierarchy
change which is from double binary to double
binary. 
The most unlikely hierarchy changes are for all the $C_0$ values
are those involving changes to and from 13.

The main characteristic of this case is similar to that of the
CSFBP i.e. the number of hierarchy changes decreases as the value
of $C_0$ increases; however there are some differences as follows.
\begin{enumerate}
\item There are fewer hierarchy changes overall than for the equal
mass case of the CSFBP which means that the extra central mass has
played a stabilizing role in terms of hierarchical stability and
therefore the equal mass CS5BP is more stable than the equal mass
CSFBP. \item The 13 $\rightarrow$ 12 and 12 $\rightarrow$ 13 were
among the most likely hierarchy changes in the case of the equal
mass CSFBP which clearly is the least in this case and the double
binary to double binary hierarchy changes are most likely in the
CS5BP case for large $C_0$. See tables (\ref{tab1}) and
(\ref{tab5})
\end{enumerate}
\renewcommand{\baselinestretch}{1}
\begin{table*}
  \begin{center}
\caption{Total number of hierarchical changes for equal mass case
of the CS5BP,$\mu_1=\mu_2=\mu_0=0.25$}\label{tab5}
\bigskip
  \begin{tabular}{|c||c|c|c|c|c|c|cc|} \hline
& $C_0=0.03$ & $C_0=0.06$ & $C_0 = 0.07$\\\hline \hline
  12 13  &  14   &   0   &   0    \\
  12 14  &  54   &   50   &   0     \\
  12 24  &  19   &   10   &   0     \\
  13 12  &  8  &   2   &   0     \\
  13 14  & 18   &   6  &   0      \\
  13 24  &  10   &   1   &   0    \\
  14 12  &  16   &   9  &   0    \\
  14 13  &  17   &   11   &   0 \\
  14 24  &  83   &   15   &   0\\
  24 12  &  18  &   20   &   0 \\
  24 13  &  15  &   7   &   0 \\
  24 14  &  67  &   10   &   0\\ \hline
 \hline
  Total  & 339 &  141   &   0  \\ \hline
  \end{tabular}
\end{center}
\end{table*}

\begin{table*}
  \begin{center}
\caption{Percentage of hierarchical changes for equal mass case of
the CS5BP,$\mu_1=\mu_2=\mu_0=0.25$}\label{tab5b}
\bigskip
  \begin{tabular}{|c||c|c|c|c|c|c|cc|} \hline
& $C_0=0.03$ & $C_0=0.06$ & $C_0 = 0.07$\\\hline \hline
  12 13  &  4   &   0   &   0    \\
  12 14  &  16   &   36   &   0     \\
  12 24  &  6   &   7   &   0     \\
  13 12  &  2  &   1   &   0     \\
  13 14  & 5   &   4  &   0      \\
  13 24  &  3  &   1   &   0    \\
  14 12  &  5   &   6  &   0    \\
  14 13  &  5   &   8   &   0 \\
  14 24  &  24   &   11   &   0\\
  24 12  &  5  &   14   &   0 \\
  24 13  &  4  &   5   &   0 \\
  24 14  &  20  &   7   &   0\\ \hline

  \end{tabular}
\end{center}
\end{table*}
\renewcommand{\baselinestretch}{2}
\subsection{Four equal masses with a varying central mass $\mu_0$}
In this section we introduce a stationary mass to the centre of
the equal mass CSFBP and will use it to discover the effect of
changing its mass on the hierarchical stability of the system.
First we place a small mass at the centre of mass and then we
increase it until the mass of the central body is much higher than
the outer bodies. We discuss the following three different sets of
mass ratios:
\begin{enumerate}
\item $\mu_1 = \mu_2 = \frac{22.475}{100}$ and $\mu_0 =
\frac{1}{100}$: Four equal masses with a very small central mass;
\item $\mu_1 = \mu_2 = \frac{2}{9}$ and $\mu_0 = \frac{1}{9}$:
Four equal masses with a comparatively larger central mass but
still smaller than the outer bodies; \item $\mu_1 = \mu_2 =
\frac{1}{100}$ and $\mu_0 = \frac{96}{100}$: Four equal masses
with a large central mass and small outer bodies.
\end{enumerate}

\subsubsection{Four equal masses with a very small central mass ($\mathbf{\mu_1=\mu_2= \frac{22.475}{100}}$ and $\mathbf{\mu_0 = \frac{1}{100}}$)}

The critical values in this case are $R_1=0.0292835, R_2=
0.0293423, R_3= 0.0468416$ and $R_4= 0.0482373$. Results are
contained in tables (\ref{tab6}) and  (\ref{tab6b}).

There are no hierarchy changes for $C_0\geq C_{crit}$ which is
also the prediction of the analytical stability criterion. The
most likely hierarchy changes for small $C_0$ are 12 $\rightarrow$
13 and 13 $\rightarrow$ 12. We have seen in the equal mass CS5BP
that the single binary to double binary hierarchy changes and vice
versa were small FOR large $C_0$. These have increased after
reducing the size of the central mass with in particular to and
from 13 changes become larger.

The main difference between this case and the equal mass CSFBP is
that we have fewer hierarchy changes. Compared to the equal mass
CS5BP, this case has more hierarchy changes. An increasing central
mass appears to stabilize the system. We shall see this confirmed
in the next two sections.

\subsubsection{Four equal masses with a slightly bigger central mass ($\mathbf{\mu_1=\mu_2= \frac{2}{9}}$ and $\mathbf{\mu_0 = \frac{1}{9}}$)}

The critical values in this case are $R_1=0.0368188=R_2, R_3=
0.062839=R_4$. Results are contained in tables (\ref{tab7}) and
(\ref{tab7b}).

We have increased the mass of the central body further but is
still smaller than the four outer bodies. The total number of
hierarchy changes decreased further but are still more than we had
in the equal mass case of the CS5BP. The introduction of the
central mass is indeed playing a stabilizing role. The most likely
hierarchy changes in this case are 14 $\rightarrow$ 24 which were
the most unlikely in the previous case with a smaller central
mass. The most unlikely hierarchy changes are from 13 to any other
hierarchy state for $C_0=0.036$.

The main characteristic of both the cases remain the same. There
are no hierarchy changes for $C_0>C_{crit}$ and the number of
hierarchies decreases as the value of the Szebehely constant
increases.
\subsubsection{Four equal masses with a very large central mass ($\mathbf{\mu_1=\mu_2= \frac{1}{100}}$ and $\mathbf{\mu_0 = \frac{96}{100}}$)}

The critical values in this case are $R_1=0.0000300822=R_2,
R_3=0.000032294 =R_4$. Results are contained in tables
(\ref{tab8}) and (\ref{tab8b}).

We have further increased the mass of the central body so that the
central body is 96 times bigger than the outer bodies i.e.
$\mu_0=96\mu_1=96\mu_2$. This system is close to a planetary
system with a star in the centre with four planets.

The total number of hierarchy changes does not decrease much and
is almost the same as we had in the previous case suggesting a
similar level of instability. This is  deceiving if we do not look
at the individual members of the table. Most of the hierarchy
changes are coming from 14 $\rightarrow$ 24 and vice versa. For
small $C_0$ ($C_0= 0.00003$), 14 $\rightarrow$ 24 and 24
$\rightarrow$ 14 changes contribute 74 percent to the total number
of hierarchy changes and for large $C_0$ ($C_0= 0.000033$), they
contribute more than 99.9 percent of the total hierarchy changes.
See tables (\ref{tab8}) and (\ref{tab8b}).

There are no hierarchy changes from 12 $\rightarrow$ 13, 12
$\rightarrow$ 14, 13 $\rightarrow$ 12, 14 $\rightarrow$ 12 and 14
$\rightarrow$ 13 throughout the integrations carried out for this
case. This system is comparatively much more stable than any other
case of four equal masses with or without a stationary central
mass.

As we have seen from all other cases of four equal masses with a
stationary central mass  and without a stationary central mass,
the number of hierarchy changes decreases as the central mass
increases. In the equal mass case of the CSFBP with no central
mass the number of hierarchy changes were highest then after the
introduction of the central mass the number of hierarchy changes
decreased as we increased the central mass. Therefore now we can
conjecture that the larger the central mass the more stable the
system becomes.

\renewcommand{\baselinestretch}{1}

\begin{table*}
  \begin{center}
\caption{Total number of hierarchical changes for
$\mu_1=\mu_2=\frac{22.475}{100}$ and $\mu_0=0.01$. Four equal
masses with a very small central mass}\label{tab6}
 \bigskip
  \begin{tabular}{|c||c|c|c|c|c|c|cc|} \hline
& $C_0=0.026$ & $C_0=0.04$ & $C_0 = 0.05$\\\hline \hline
  12 13  &  96   &   60   &   0    \\
  12 14  &  30   &   12   &   0     \\
  12 24  &  22   &   20   &   0     \\
  13 12  &  83  &   65   &   0     \\
  13 14  & 31  &   25 &   0      \\
  13 24  &  26   &   10   &   0    \\
  14 12  &  44   &   31  &   0    \\
  14 13  &  63   &   51   &   0 \\
  14 24  &  26   &   5   &   0\\
  24 12  &  28  &   15   &   0 \\
  24 13  &  39  &   40   &   0 \\
  24 14  &  19  &   8   &   0\\ \hline
 \hline
  Total  & 507 &  342   &   0  \\ \hline
  \end{tabular}
\end{center}
\end{table*}

\begin{table*}
  \begin{center}
\caption  {Percentage of hierarchical changes for
$\mu_1=\mu_2=\frac{22.475}{100}$ and $\mu_0=0.01$. Four equal
masses with a very small central mass}\label{tab6b}
\bigskip
  \begin{tabular}{|c||c|c|c|c|c|c|cc|} \hline
& $C_0=0.026$ & $C_0=0.04$ & $C_0 = 0.05$\\\hline \hline
  12 13  &  19   &   18   &   0    \\
  12 14  &  6   &   4   &   0     \\
  12 24  &  4   &   6   &   0     \\
  13 12  &  16  &   19   &   0     \\
  13 14  & 6  &   7 &   0      \\
  13 24  &  5   &   3   &   0    \\
  14 12  &  9   &   9  &   0    \\
  14 13  &  12   &   15   &   0 \\
  14 24  &  5   &   1   &   0\\
  24 12  &  6  &   4   &   0 \\
  24 13  &  8  &   12   &   0 \\
  24 14  &  4  &   2   &   0\\ \hline

  \end{tabular}
\end{center}
\end{table*}

\begin{table*}
  \begin{center}
\caption{Total number of hierarchical changes for $\mu_1
=\mu_2=\frac{2}{9}$ and $\mu_0=\frac{1}{9}$. Four equal masses
with a slightly bigger central mass}\label{tab7}
\bigskip
  \begin{tabular}{|c||c|c|c|c|c|c|cc|} \hline
& $C_0=0.036$ & $C_0=0.055$ & $C_0 = 0.063$\\\hline \hline
  12 13  &  45   &   4   &   0    \\
  12 14  &  35   &   27   &   0     \\
  12 24  &  29   &   19   &   0     \\
  13 12  &  22  &   2   &   0     \\
  13 14  & 22 &   10 &   0      \\
  13 24  &  22   &   15   &   0    \\
  14 12  &  30   &   20  &   0    \\
  14 13  &  32   &   25   &   0 \\
  14 24  &  93   &   37   &   0\\
  24 12  &  34  &   14   &   0 \\
  24 13  &  24  &   12   &   0 \\
  24 14  &  75  &   29   &   0\\ \hline
 \hline
  Total  & 462 &  214   &   0  \\ \hline
  \end{tabular}
\end{center}
\end{table*}

\begin{table*}
  \begin{center}
\caption{Percentage of hierarchical changes for $\mu_1
=\mu_2=\frac{2}{9}$ and $\mu_0=\frac{1}{9}$. Four equal masses
with a slightly bigger central mass}\label{tab7b}
\bigskip
  \begin{tabular}{|c||c|c|c|c|c|c|cc|} \hline
& $C_0=0.036$ & $C_0=0.055$ & $C_0 = 0.063$\\\hline \hline
  12 13  &  10   &   2   &   0    \\
  12 14  &  8   &   13   &   0     \\
  12 24  &  6   &   9   &   0     \\
  13 12  &  5  &  1   &   0     \\
  13 14  & 5 &   5 &   0      \\
  13 24  &  5   &   7   &   0    \\
  14 12  &  6   &   9  &   0    \\
  14 13  &  7   &   12   &   0 \\
  14 24  &  20   &   17   &   0\\
  24 12  &  7  &   7   &   0 \\
  24 13  &  5  &   6   &   0 \\
  24 14  &  16  &   14   &   0\\ \hline

  \end{tabular}
\end{center}
\end{table*}

\begin{table*}
  \begin{center}
\caption {Total number of hierarchical changes for $\mu_1
=\mu_2=\frac{1}{100}$ and $\mu_0=\frac{96}{100}$. Four equal
masses with a very large central mass}\label{tab8}
\bigskip
  \begin{tabular}{|c||c|c|c|c|c|c|cc|} \hline
& $C_0=0.00003$ & $C_0=0.000031$ & $C_0 = 0.000033$\\\hline \hline
  12 13  &  0   &   0   &   0    \\
  12 14  &  0   &   0   &   0     \\
  12 24  &  32   &   0   &   0     \\
  13 12  &  0  &   0   &   0     \\
  13 14  & 1 &   0 &   0      \\
  13 24  &  26   &   0   &   0    \\
  14 12  &  0   &   0  &   0    \\
  14 13  &  0   &   1  &   0 \\
  14 24  &  136   &  52   &   0\\
  24 12  &  32  &   0   &   0 \\
  24 13  &  27  &   0  &   0 \\
  24 14  &  196  &   93   &   0\\ \hline
 \hline
  Total  & 450 &  146   &   0  \\ \hline
  \end{tabular}
\end{center}
\end{table*}

\begin{table*}
  \begin{center}
\caption{Percentage of hierarchical changes for $\mu_1
=\mu_2=\frac{1}{100}$ and $\mu_0=\frac{96}{100}$. Four equal
masses with a very large central mass}\label{tab8b}
\bigskip
  \begin{tabular}{|c||c|c|c|c|c|c|cc|} \hline
& $C_0=0.00003$ & $C_0=0.000031$ & $C_0 = 0.000033$\\\hline \hline
  12 13  &  0   &   0   &   0    \\
  12 14  &  0   &   0   &   0     \\
  12 24  &  7   &   0   &   0     \\
  13 12  &  0  &   0   &   0     \\
  13 14  & 0 &   0 &   0      \\
  13 24  &  6   &   0   &   0    \\
  14 12  &  0   &   0  &   0    \\
  14 13  &  0   &   1  &   0 \\
  14 24  &  30   &  36   &   0\\
  24 12  &  7  &   0   &   0 \\
  24 13  &  6  &   0  &   0 \\
  24 14  &  44  &   64   &   0\\ \hline

  \end{tabular}
\end{center}
\end{table*}

 \subsection{Three equal masses and two increasing symmetrically}

In this section we will discuss a new case of the Caledonian
Symmetric Five Body Problem. This is a completely different
situation from those discussed earlier and little comparison with
the previous cases will be made. Here we take the masses of three
of the bodies to be equal and two of them to be varying
symmetrically from being very small to very large and effectively
approaching a situation of a perturbed two body problem. The three
bodies with equal masses will comprise of two outer bodies and the
central body i.e. $\mu_1=\mu_0$. $\mu_2$ will increase to the
largest masses from being the smallest. We will discuss the
following three cases.
\begin{enumerate}
\item $\mu_1 = \mu_0 = 0.326$ and $\mu_2 = 0.11$

\item $\mu_1 = \mu_0 = 0.15$ and $\mu_2 = 0.275$

 \item $\mu_1 = \mu_0 = 0.01$ and $\mu_2 = 0.485$
\end{enumerate}

\subsubsection{$\mathbf{\mu_1=\mu_0=0.326}$ and $\mathbf{\mu_2=0.11}$}

The critical values in this case are $R_1=0.0269644,R_2=0.0270025,
R_3=0.0286002, R_4=0.0294447$. Results are contained in tables
(\ref{tab9}) and (\ref{tab9b}).

There are no hierarchy changes for $C_0=0.03$. Since this value is
greater than the critical value, this fact shows that the
numerical integrations follow well the analytical stability
criterion.

The most unlikely hierarchy changes are 14 $\rightarrow$ 24. The
12 $\rightarrow$ 13, 13 $\rightarrow$ 12, 13 $\rightarrow$ 24, 24
$\rightarrow$ 13 and 24 $\rightarrow$ 14 hierarchy changes are
also very small for small $C_0$ ($C_0=0.02$). The 14 $\rightarrow$
24 and 13 $\rightarrow$ 24 hierarchy changes remains the smallest
for all values of $C_0$.

For small $C_0 = 0.02$, 12 $\rightarrow$ 14 and 14 $\rightarrow$
12 hierarchy changes are the most likely and remain amongst the
highest throughout the integrations carried out for this case of
the CS5BP. These hierarchy changes remain the highest throughout
the integration.

At $C_0=0.029$ the Szebehely constant is near $C_{crit}$ and the
phase space is partially disconnected. Therefore the hierarchy
changes to or from the 13 hierarchy state are not allowed. This is
numerically confirmed in table (\ref{tab8}) where there are no 12
$\rightarrow$ 13, 13 $\rightarrow$ 12, 13 $\rightarrow$ 14, 13
$\rightarrow$ 24, 14 $\rightarrow$ 13 and 24 $\rightarrow$ 13
hierarchy changes.

The main characteristic of this case is similar to all other cases
of the CS5BP. With small $C_0$ we have more hierarchy changes and
as $C_0$ increases the number of hierarchy changes decrease.

\subsubsection{$\mathbf{\mu_1=\mu_0=0.15}$ and $\mathbf{\mu_2=0.275}$}

The critical values in this case are $R_1=0.0364159,R_2=0.0365447,
R_3=0.0578382, R_4=0.0622745$. Results are contained in tables
(\ref{tab10}) and (\ref{tab10b}).

In this case we increase the value of $\mu_2$ to 0.275 and
consequently the value of $\mu_1$ decreases and hence we have
$\mu_2 > \mu_1 = \mu_0$.

There are no hierarchy changes for $C_0 > C_{crit}$. For
$C_0=0.06$ which is near the critical value there are no hierarchy
changes from and to the 24 hierarchy state which is also the
prediction of analytical stability criterion. The most likely
hierarchy changes for small $C_0$ ($C_0 = 0.03$) are 14
$\rightarrow$ 24 and 12 $\rightarrow$ 24. When $C_0$ increases the
14 $\rightarrow$ 24 hierarchy changes decrease and the 12
$\rightarrow$ 24 changes remains the most likely until it dies out
for large $C_0$ ($C_0=0.06$). The most unlikely hierarchy changes
are 12 $\rightarrow$ 14 and 13 $\rightarrow$ 14.

The main characteristic of this case and all other cases remain
the same i.e. the number of hierarchy changes decreases as $C_0$
increases.

\subsubsection{$\mathbf{\mu_1=\mu_0=0.01}$ and $\mathbf{\mu_2=0.485}$}

The critical values in this case are
$R_1=0.00968676,R_2=0.00988033, R_3=0.0102131, R_4=0.0107386$.
Results are contained in tables (\ref{tab11}) and (7.23).

In this case we further decrease the value of $\mu_1$ and $\mu_0$
and consequently the value of $\mu_2$ increases. As $\mu_1 = \mu_0
\ll \mu_2$ therefore this can be considered a perturbed two body
problem.

There are no hierarchy changes for any value of $C_0$ we have
investigated.  See table (\ref{tab11}). We have investigated about
15,000 orbits each for 1 million time steps of integration time
and there are no hierarchy changes. Therefore this is the most
stable of the CS5BP systems we have discussed so far in this
chapter.
\begin{table*}
  \begin{center}
  \caption{Total number of hierarchical changes for $\mu_1=\mu_0=0.326$, $\mu_2=0.11$}\label{tab9}
\bigskip
  \begin{tabular}{|c||c|c|c|c|c|c|cc|} \hline
& $C_0=0.02$ & $C_0=0.027$ & $C_0 = 0.028$ & $C_0 = 0.029$ & $C_0
= 0.03$     \\\hline \hline
  12 13  &  15    &   7   &  11  &  0   & 0    \\
  12 14  &  108   &   24  &  21  &  15  & 0    \\
  12 24  &  45    &   26  &   24 &  15  & 0    \\
  13 12  &  22    &   6   &   6  &  0   & 0    \\
  13 14  & 78     &   7   &   10 &  0   & 0     \\
  13 24  &  11    &   3   &   3  &  0   & 0    \\
  14 12  &  103   &   24  &   28 &  21  & 0    \\
  14 13  &  84    &   7   &   3  &  0   & 0 \\
  14 24  &  6     &   2   &   1  &  1   & 0\\
  24 12  &  35    &   14  &   8  &  4   & 0 \\
  24 13  &  15    &   6   &   8  &  0   & 0 \\
  24 14  &  14    &   5   &   5  &  7   & 0\\ \hline
 \hline
  Total  & 536   &   130   &  128 & 63   & 0 \\ \hline
  \end{tabular}
\end{center}
\end{table*}

\input{ChapNumInvistTables}

%% file: ChapNumInvistTables.tex
\begin{table*}
  \begin{center}
  \caption
  {Percentage of  hierarchical changes for $\mu_1=\mu_0=0.326$, $\mu_2=0.11$}\label{tab9b}
\bigskip
  \begin{tabular}{|c||c|c|c|c|c|c|cc|} \hline
 & $C_0=0.02$ & $C_0=0.027$ & $C_0 = 0.028$
& $C_0 = 0.0029$ & $C_0 = 0.03$     \\\hline \hline
  12 13  &  3    &   5    &  9  &  0    & 0    \\
  12 14  &  20   &   18   &  16 &  24   & 0    \\
  12 24  &  8    &   20   &  19 &  24   & 0    \\
  13 12  &  4    &   5    &   5 &   0   & 0    \\
  13 14  & 15    &   5    &   8 &   0   & 0    \\
  13 24  &  2    &   2    &   2 &   0   & 0    \\
  14 12  &  19   &   18   &   22 &  33  & 0    \\
  14 13  &  16   &   5    &  2   &  0   & 0    \\
  14 24  &  1    &   2    &   1  &  2   & 0    \\
  24 12  &  7    &   11   &   6  &  6   & 0    \\
  24 13  &  3    &   5    &   6  &  0   & 0    \\
  24 14  &  3    &   4    &   4  &  11  & 0    \\ \hline

  \end{tabular}
\end{center}
\end{table*}

\begin{table*}
\begin{center}
  \caption{Total number of hierarchical changes for $\mu_1=\mu_0=0.15$, $\mu_2=0.275$}\label{tab10}
\bigskip
  \begin{tabular}{|c||c|c|c|c|c|} \hline
& $C_0=0.03$ & $C_0=0.036$ & $C_0 = 0.05$ & $C_0 = 0.06$ & $C_0 =
0.063$     \\\hline
 \hline
  12 13  &  38    &  24   &  16  &  0  & 0    \\
  12 14  &  26   &   13   &  74  &  14  & 0    \\
  12 24  &  67    &   70  &   99 &  0  & 0    \\
  13 12  &  20    &  20   &   8 &  0  & 0    \\
  13 14  & 14     &   7   &   38 &  3   & 0     \\
  13 24  &  47    &  26   &   27 &  0  & 0    \\
  14 12  &  24   &   17   &   72 &  22  & 0    \\
  14 13  &  30    &   18  &   36 &  4  & 0 \\
  14 24  &  86    &   42  &   16  &  0   & 0\\
  24 12  &  53    &   28  &   37 &  0   & 0 \\
  24 13  &  44    &   17  &   34 &  0  & 0 \\
  24 14  &  47    &   35   &  24  &  0   & 0\\ \hline
 \hline
  Total  & 496   &   317   &  480 & 43   & 0 \\ \hline
  \end{tabular}
\end{center}
\end{table*}

\begin{table*}
\begin{center}
  \caption{Total number of hierarchical changes for $\mu_1=\mu_0=0.15$, $\mu_2=0.275$}\label{tab10b}
\bigskip
  \begin{tabular}{|c||c|c|c|c|c|} \hline
& $C_0=0.03$ & $C_0=0.036$ & $C_0 = 0.05$ & $C_0 = 0.06$ & $C_0 =
0.063$     \\\hline
 \hline
  12 13  &  8    &  8   &  3  &  0  & 0    \\
  12 14  &  5   &   4   &  15  &  33  & 0    \\
  12 24  &  14   &   22  &   21 &  0  & 0    \\
  13 12  &  4    &  6   &   2 &  0  & 0    \\
  13 14  & 3     &   2   &   8 &  7   & 0     \\
  13 24  &  9    &  8   &   6 &  0  & 0    \\
  14 12  &  5   &   5   &   15 &  51  & 0    \\
  14 13  &  6    &   6  &   8 &  9  & 0 \\
  14 24  &  17    &   13  &  3  &  0   & 0\\
  24 12  &  11    &   9  &   8 &  0   & 0 \\
  24 13  &  9    &   5  &   7 &  0  & 0 \\
  24 14  &  9    &   11   &  5  &  0   & 0\\ \hline

  \end{tabular}
\end{center}
\end{table*}

\begin{table*}
\begin{center}
  \caption{Total number of hierarchical changes for $\mu_1=\mu_0=0.01$, $\mu_2=0.485$}\label{tab11}
\bigskip
  \begin{tabular}{|c||c|c|c|c|c|} \hline
& $C_0=0.009$ & $C_0=0.0098$ & $C_0 = 0.01$ & $C_0 = 0.0106$ &
$C_0 = 0.0108$     \\\hline
 \hline
  12 13  &  0   &   0   &   0 &   0   &   0   \\
  12 14  &  0   &   0   &   0    &   0   &   0 \\
  12 24  &  0   &   0   &   0 &   0   &   0    \\
  13 12  &  0  &   0   &   0    &   0   &   0 \\
  13 14  & 0 &   0 &   0     &   0   &   0 \\
  13 24  &  0  &   0   &   0   &   0   &   0 \\
  14 12  &  0   &   0  &   0   &   0   &   0 \\
  14 13  &  0   &   0  &   0&   0   &   0 \\
  14 24  &  0   &  0  &   0&   0   &   0\\
  24 12  &  0  &   0   &   0 &   0   &   0\\
  24 13  &  0  &   0  &   0 &   0   &   0\\
  24 14  &  0  &   0   &   0&   0   &   0\\ \hline
 \hline
  Total  & 0 &  0   &   0 &   0   &   0 \\ \hline

  \end{tabular}
\end{center}
\end{table*}

\begin{table*}
\begin{center}
  \caption{Percentage of hierarchical changes for $\mu_1=\mu_0=0.01$, $\mu_2=0.485$}\label{tab11b}
\bigskip
  \begin{tabular}{|c||c|c|c|c|c|} \hline
& $C_0=0.009$ & $C_0=0.0098$ & $C_0 = 0.01$ & $C_0 = 0.0106$ &
$C_0 = 0.0108$     \\\hline
 \hline
  12 13  &  0   &   0   &   0 &   0   &   0   \\
  12 14  &  0   &   0   &   0    &   0   &   0 \\
  12 24  &  0   &   0   &   0 &   0   &   0    \\
  13 12  &  0  &   0   &   0    &   0   &   0 \\
  13 14  & 0 &   0 &   0     &   0   &   0 \\
  13 24  &  0  &   0   &   0   &   0   &   0 \\
  14 12  &  0   &   0  &   0   &   0   &   0 \\
  14 13  &  0   &   0  &   0&   0   &   0 \\
  14 24  &  0   &  0  &   0&   0   &   0\\
  24 12  &  0  &   0   &   0 &   0   &   0\\
  24 13  &  0  &   0  &   0 &   0   &   0\\
  24 14  &  0  &   0   &   0&   0   &   0\\ \hline
  \end{tabular}
\end{center}
\end{table*}

\subsection{Non-Equal mass cases of the CS5BP}

 The cases of the CS5BP we have discussed so far had at least two
 or more of the mass ratios equal which included, $\mu_1=\mu_2$
 and $\mu_0=0$, $\mu_1=\mu_2=\mu_0$, $\mu_1=\mu_2$ and $\mu_0\neq
 0$, and $\mu_1=\mu_0$ and $\mu_0\neq 0$. In this section we discuss
 the cases of the CS5BP where none of the mass ratios are equal. We
 will discuss the following three sets of mass ratios:

 \begin{enumerate}
 \item $\mu_1 = 0.195$, $\mu_2= 0.3$ and $\mu_0 =0.01$
 \item $\mu_1 = 0.3$, $\mu_2= 0.1$ and $\mu_0 =0.2$
 \item $\mu_1 = 0.35$, $\mu_2= 0.01$ and $\mu_0= 0.28$
 \end{enumerate}

\subsubsection{$\mathbf{\mu_1 = 0.195}$, $\mathbf{\mu_2= 0.3}$ and $\mathbf{\mu_0= 0.01}$}

The critical values in this case are $R_1=0.0280812,R_2=0.0283641,
R_3=0.0439109, R_4=0.0469418$. Results are contained in tables
(\ref{tab12}) and (\ref{tab12b}).

There are no hierarchy changes for $C_0 = 0.055 > C_{crit}$ as was
predicted by the analytical stability criterion. For $C_0 = 0.044$
there are no hierarchy changes to or from  the 24 hierarchy state
which is exactly what the analytical stability criterion predicts.

The most likely hierarchy changes are always to or from the 24
hierarchy state until it becomes disconnected from all other
hierarchy states. For small $C_0$ ($C_0=0.02$), the 24
$\rightarrow$ 13 hierarchy changes are the highest, for
$C_0=0.0281$, the 13 $\rightarrow$ 24 hierarchy changes is the
highest and for large $C_0$ ($C_0=0.04$), the 24 $\rightarrow$ 13
hierarchy changes are again the highest which means that the 24
hierarchy state is the most unstable for small values of the
Szebehely constant, $C_0$. Also the most number of hierarchy
changes are from the 24 hierarchy state which is another reason
for branding the 24 hierarchy state the most unstable for small
$C_0$.

The most unlikely hierarchy changes are from the 14 hierarchy
state for $C_0=0.02,0.028$ and 0.04. This suggests that  the 14
hierarchy state is the most stable hierarchy state but perhaps it
is the result of not starting in the 14 hierarchy state. Fewer
starts in the 14 hierarchy mean fewer changes from it are
possible. At $C_0$ large, $C_0=0.044$, the 14 hierarchy state
appears to be contributing more than 50 percent to the total
hierarchy changes and turns out to be unstable. For large values
of the Szebehely constant near the critical value the 24 hierarchy
state is the most stable opposite to the fact that it is the most
unstable situation for small $C_0$.

The main characteristic of this case and all other cases of the
CS5BP discussed so far remain the same. There are no hierarchy
changes for $C_0>C_{crit}$ and the number of hierarchy changes
decrease as the value of $C_0$ increases. There is a sudden
decrease in the total number of hierarchy changes near $C_{crit}$
but this is not unexpected as at this point the 24 hierarchy state
which was the main contributor disconnects and therefore the total
number of hierarchy changes drop.

 \subsubsection{$\mathbf{\mu_1 = 0.3}$, $\mathbf{\mu_2= 0.1}$ and $\mathbf{\mu_0= 0.2}$}

The critical values in this case are $R_1=0.0334268,R_2=0.0336685,
R_3=0.0526554, R_4=0.0577769$. Results are contained in tables
(\ref{tab13}) and (\ref{tab13b}).

We increase the value of $\mu_0$ from 0.01 to 0.2 from the last
case and the total number of hierarchy changes drop dramatically.
As we said earlier, the bigger the central mass, the more stable
the system would be and this behavior compliments our statement.

There are no hierarchy changes for $C_0>C_{crit}$. For $C_0$
large, $C_0 = 0.055 <  C_{crit}$ the number of hierarchy changes
are zero which is  another reason to believe that this system is
comparatively more  stable than the $\mu_1 = 0.195$, $\mu_2= 0.3$
and $\mu_0= 0.01$ case discussed earlier. The number of hierarchy
changes from the 14 hierarchy state are very small throughout the
integrations performed for this set of mass ratios and becomes
zero for $C_0=0.05$ and therefore 14 is  the most stable hierarchy
state. Recall that 14 was the most  stable hierarchy state in the
previous case too. The 24  $\rightarrow$ 12 and 24 $\rightarrow$
13 hierarchy changes are  also very small and is the most unlikely
for $C_0=0.03$ and  $C_0=0.0346$. See table (\ref{tab13}).

  The most likely hierarchy changes for $C_0=0.03$ and
  $C_0=0.0346$ are 12 $\rightarrow$ 13. For $C_0=0.05$ the highest
  number of hierarchy changes are from 13 to 24. See table
  (\ref{tab13}).

\subsubsection{$\mathbf{\mu_1 = 0.35}$, $\mathbf{\mu_2= 0.01}$ and $\mathbf{\mu_0= 0.28}$}

The critical values in this case are $R_1=0.0243191,R_2=0.0251055,
R_3=0.0361804, R_4=0.0408993$. Results are contained in tables
(\ref{tab14}) and (\ref{tab14b}).

The total number of hierarchy changes further dropped as the value
of $\mu_0$ increased and $\mu_2$ decreased. There are no hierarchy
changes for $C_0=0.029$ and $C_0=0.03>C_{crit}$. There are no
hierarchy changes from the 14 hierarchy state except one 14
$\rightarrow$ 24 hierarchy changes for $C_0=0.028$, see table
(\ref{tab14}). Therefore the 14 hierarchy state continue to be the
most stable for all the three sets of mass ratios discussed in
this section. There are also no 12 $\rightarrow$ 14 and 14
$\rightarrow$ 12 hierarchy changes which is the double binary to
double binary. For this particular set of mass ratios the double
binary hierarchy states i.e. 12 and 14 appear to be more stable as
there are very few hierarchy changes from 12 as well.

The most likely hierarchy changes are to or from the 24 hierarchy
state as the size of the region of real motion is larger than the
others for 24 hierarchy state. The 24 $\rightarrow$ 13 and the 13
$\rightarrow$ 24 hierarchy changes are the highest for $C_0=0.02$
and for $C_0=0.026$. For $C_0=0.028$ the 24 $\rightarrow$ 12 and
12 $\rightarrow$ 24 hierarchy changes are the highest.

It is very interesting to see that in all the three sets of mass
ratios discussed in this section, the 14 hierarchy state which is
a double binary was the most stable and the 24 hierarchy state,
which a single binary, was the most unstable.

The main characteristic of all the cases of the CS5BP discussed in
this chapter remains the same. There are no hierarchy changes for
the Szebehely constant greater than the critical value and the
number of hierarchy changes decreases as $C_0$ is increased.

\begin{table*}
\begin{center}
  \caption{Total number of hierarchical changes for $\mu_1=0.195,\mu_0=0.01$, and $\mu_2=0.3$}\label{tab12}
\bigskip
  \begin{tabular}{|c||c|c|c|c|c|} \hline
& $C_0=0.02$ & $C_0=0.0281$ & $C_0 = 0.04$ & $C_0 = 0.044$ & $C_0
= 0.055$     \\\hline
 \hline
  12 13  &  58  &    50   &   56    &   14   &   0   \\
  12 14  &  14  &    7    &   22    &   12   &   0 \\
  12 24  &  48  &   40    &   50   &   0   &   0    \\
  13 12  &  56  &   34   &   21   &   2   &   0 \\
  13 14  &  9   &   7     &   6   &   0   &   0 \\
  13 24  &  38  &   67    &   17   &   0   &   0 \\
  14 12  &  10  &   12    &   14   &   20   &   0 \\
  14 13  &  16  &   12   &   9    &   10   &   0 \\
  14 24  &  6   &   9     &   0   &   0   &   0\\
  24 12  & 85  &   53    &   33    &   0   &   0\\
  24 13  & 93  &   60   &   49    &   0   &   0\\
  24 14  &  9  &   5     &   0   &   0   &   0\\ \hline
 \hline
  Total  & 442 &  356  &   277 &   58   &   0 \\ \hline

  \end{tabular}
\end{center}
\end{table*}

\begin{table*}
\begin{center}
  \caption{Percentage of hierarchical changes for $\mu_1=0.195,\mu_0=0.01$, and $\mu_2=0.3$}\label{tab12b}
\bigskip
  \begin{tabular}{|c||c|c|c|c|c|} \hline
 & $C_0=0.02$ & $C_0=0.0281$ & $C_0 = 0.04$ &
$C_0 = 0.044$ & $C_0 = 0.055$     \\\hline
 \hline
  12 13  & 13   &   14   &  20   &  24  &   0   \\
  12 14  &  3   &   2    &   8   &  21   &   0  \\
  12 24  &  11   &  11   &  18  &   0   &   0    \\
  13 12  &  13    & 10   &   8  &   3   &   0 \\
  13 14  & 2    &   2    &   2  &   0   &   0 \\
  13 24  &  9  &   19    &   6  &   0   &   0 \\
  14 12  &  2   &   3    &   5   &  34   &   0 \\
  14 13  &  4   &   3    &   3   &  17  &   0 \\
  14 24  &  1   &   3    &   0   &   0   &   0\\
  24 12  &  19  &   15   &   12  &   0   &   0\\
  24 13  &  21  &   17   &  18  &    0   &   0\\
  24 14  &  2  &    1    &   0   &   0   &   0\\ \hline
  \end{tabular}
\end{center}
\end{table*}

\begin{table*}
\begin{center}
  \caption{Total number of hierarchical changes for $\mu_1=0.3,\mu_0=0.2$, and $\mu_2=0.1$}\label{tab13}
\bigskip
  \begin{tabular}{|c||c|c|c|c|c|} \hline
& $C_0=0.03$ & $C_0=0.0346$ & $C_0 = 0.05$ & $C_0 = 0.055$ & $C_0
= 0.06$     \\\hline
 \hline
  12 13  &  84  &   55   &   1    &   0   &   0   \\
  12 14  &  16  &   16    &   4    &   0   &   0 \\
  12 24  &  12  &   18    &   18   &   0   &   0    \\
  13 12  &  26  &   12    &   5   &   0   &   0 \\
  13 14  &  19   &  19     &  0   &   0   &   0 \\
  13 24  &  4   &   10    &  30   &   0   &   0 \\
  14 12  &  9   &   6    &   0   &   0   &   0 \\
  14 13  &  19  &   19    &  0    &  0   &   0 \\
  14 24  &  19  &   28   &   0   &   0   &   0\\
  24 12  & 8   &    5   &   6    &   0   &   0\\
  24 13  & 8   &   13   &   5    &   0   &   0\\
  24 14  & 26  &  35     &   0   &   0   &   0\\ \hline
 \hline
  Total  & 250 &  246  &   69 &   0   &   0 \\ \hline

  \end{tabular}
\end{center}
\end{table*}

\begin{table*}
\begin{center}
  \caption{Percentage of hierarchical changes for $\mu_1=0.3,\mu_0=0.2$, and $\mu_2=0.1$}\label{tab13b}
\bigskip
  \begin{tabular}{|c||c|c|c|c|c|} \hline
& $C_0=0.03$ & $C_0=0.0346$ & $C_0 = 0.05$ & $C_0 = 0.055$ & $C_0
= 0.06$     \\\hline
 \hline
  12 13  &  34  &   24   &   1    &   0   &   0   \\
  12 14  &  6  &   7    &   6    &   0   &   0 \\
  12 24  &  5   &   8    &   26   &   0   &   0    \\
  13 12  &  10  &   5    &   7   &   0   &   0 \\
  13 14  &  8    &  8     &  0   &   0   &   0 \\
  13 24  &  2   &   4     &  43   &   0   &   0 \\
  14 12  &  4   &   3    &   0   &   0   &   0 \\
  14 13  &  8  &   8    &  0    &  0   &   0 \\
  14 24  &  8   &  12    &   0   &   0   &   0\\
  24 12  & 3   &    2   &   9    &   0   &   0\\
  24 13  & 3   &   6    &   7    &   0   &   0\\
  24 14  & 10  &  15     &   0   &   0   &   0\\ \hline
 \hline
  \end{tabular}
\end{center}
\end{table*}

\begin{table*}
\begin{center}
  \caption{Total number of hierarchical changes for $\mu_1=0.35,\mu_0=0.28$, and $\mu_2=0.01$}\label{tab14}
\bigskip
  \begin{tabular}{|c||c|c|c|c|c|} \hline
& $C_0=0.02$ & $C_0=0.026$ & $C_0 = 0.028$ & $C_0 = 0.029$ & $C_0
= 0.03$     \\\hline
 \hline
  12 13  &  0   &   3    &   0    &   0   &   0   \\
  12 14  &  0   &   0    &   0    &   0   &   0 \\
  12 24  &  8   &   6    &   2    &   0   &   0    \\
  13 12  &  0  &   3    &   0   &   0   &   0 \\
  13 14  &  0   &  0    &  0   &   0   &   0 \\
  13 24  &  17   &   3  &  0   &   0   &   0 \\
  14 12  &  0   &   0   &   0   &   0   &   0 \\
  14 13  &  0  &   0    &  0    &  0   &   0 \\
  14 24  &  0  &   0    &  1    &   0   &   0\\
  24 12  & 14  &    6   &   1    &   0   &   0\\
  24 13  & 17  &   4    &   0    &   0   &   0\\
  24 14  & 2   &  0      &   1   &   0   &   0\\ \hline
 \hline
  Total  & 58 &  25  &   5 &   0   &   0 \\ \hline

  \end{tabular}
\end{center}
\end{table*}
\begin{table*}
\begin{center}
  \caption{Percentage of hierarchical changes for $\mu_1=0.35,\mu_0=0.28$, and $\mu_2=0.01$}\label{tab14b}
\bigskip
  \begin{tabular}{|c||c|c|c|c|c|} \hline
& $C_0=0.02$ & $C_0=0.026$ & $C_0 = 0.028$ & $C_0 = 0.029$ & $C_0
= 0.03$     \\\hline
 \hline
  12 13  &  0   &   12    &   0    &   0   &   0   \\
  12 14  &  0   &   0    &   0    &   0   &   0 \\
  12 24  &  14  &   24    &  40    &   0   &   0    \\
  13 12  &  0  &   12    &   0   &   0   &   0 \\
  13 14  &  0   &  0    &  0   &   0   &   0 \\
  13 24  &  29   &   12  &  0   &   0   &   0 \\
  14 12  &  0   &   0   &   0   &   0   &   0 \\
  14 13  &  0  &   0    &  0    &  0   &   0 \\
  14 24  &  0  &   0    &  20    &   0   &   0\\
  24 12  & 24  &    24   &   20    &   0   &   0\\
  24 13  & 29  &   16    &   0    &   0   &   0\\
  24 14  & 3   &  0      &   20   &   0   &   0\\ \hline
  \end{tabular}
\end{center}
\end{table*}
\section{Conclusions}

The numerical investigations shows that the introduction of a
small stationary mass to the centre of mass of the Caledonian
Symmetric Four Body Problem (CSFBP) plays a stabilizing role. The
number of hierarchy changes decreases as we increase the value of
$\mu_0$ i.e. the central mass. This can be applied to a system of
four planets and one star at the centre of mass of the system. We
have seen that such system is hierarchically stable, see for
example the analysis and tables provided for $\mu_1 = \mu_2 =
\frac{1}{100}$ and $\mu_0=\frac{96}{100}$.

The numerical investigations support the analytical predictions
i.e. that the phase space is disconnected for $C_0 \geq C_{crit}$.
In some cases the analytical curves predicted that the phase space
is partially disconnected eg. for $R_3< C_0<R_4$. This is also
supported by the numerical investigations.

The stability of the four and five body systems increases as the
value of $\mu_1$ or $\mu_2$ decreases. With small $\mu_1$ or
$\mu_2$ the system is close to a three body system perturbed by
two bodies of negligible mass. In the case of $\mu_1 = \mu_0$ or
$\mu_2 = \mu_0$ very small as compared to the other mass ratio
then, the system is close to a perturbed two body system.

Numerical investigations of  all the CS5BP systems discussed in
this chapter reveal that as the value of the Szebehely constant
increases the number of hierarchy changes decreases and hence
their hierarchical stability improves. Therefore it can be stated
for the Szebehely constant near zero the system will be very
unstable, for increasing $C_0$ it will become more stable and for
greater than the critical value it will be hierarchically stable
for all time.

We have seen that the studies of the CSFBP and the CS5BP have
given valuable information to understand better the general
problems. Therefore it is of interest to generalize the CS5BP
system in the next chapter for any number of bodies.

%% file: ChapCSNBP.tex
\chapter{The Caledonian Symmetric N Body Problem}

In chapter 5 we investigated the Caledonian Symmetric five Body
Problem (CS5BP), which is a symmetrically reduced five body
problem containing all possible symmetries in the phase space. We
have found an analytical stability criterion for the CS5BP. In
chapter 5 we numerically verified this stability criterion and
have shown that $C_0$, the Szebehely constant, plays a very
important role in determining the hierarchical stability criterion
of the CS5BP.

In this chapter we generalize the CS5BP to the CSNBP i.e the
Caledonian Symmetric N Body Problem in such a way that it can be
easily used for any number of bodies such as three, four (CSFBP)
and five (CS5BP) etc. In section 7.1 we define the CSNBP.  The
equations of motion and Sundman's inequality, the key to the
stability criterion, are given in sections 7.2 and 7.3
respectively. We then derive in sections 7.4 and 7.5 the
analytical stability criterion for the CSNBP. The conclusions are
given in section 7.6.
\section{Definition of the Caledonian Symmetric N Body Problem (CSNBP)}

The CSNBP is formulated by using all possible symmetries. The main
feature of the model is its use of two types of symmetries. 1.
past-future symmetry and 2. dynamical symmetry. Past future
symmetry exists in an n-body system when the dynamical evolution
of the system after $t=0$ is a mirror image of the dynamical
evolution of the system before $t=0$. It occurs whenever the
system passes through a mirror configuration, $i.e.$ a
configuration in which the velocity vectors of all the bodies are
perpendicular to all the position vectors from the centre of mass
of the system (Roy and Ovenden, 1955).

Let us consider $2n+1$ bodies $P_{0}$, $P_{1}$, $P_{2}$, $\cdots$,
$P_{2n}$, of masses $ m_{0}$, $m_{1}$, $m_{2}$, $\cdots$, $m_{2n}$
respectively existing in three dimensional Euclidean space. The
radius and velocity vectors of the bodies with respect to the
centre of mass of the five body system are given by $
\mathbf{r}_{i}$ and $\dot{\mathbf{r}}_{i}$ respectively,
$i=0,1,2,\cdots,n$. Let the centre of mass of the system be
denoted by $O$. The CSNBP has the following conditions:
\begin{enumerate}
\item All $2n+1$ bodies are finite point masses with:
\begin{equation}\label{1}
m_{i}=m_{n+i}, \qquad i=0,1,2,...,n.
\end{equation}

\item $P_{0}$ is stationary at $O$, the centre of mass of the
system. $P_{i}$ and $P_{i+n}$ are moving symmetrically to each
other with respect to the centre of mass of the system. Likewise
$P_{j}$ and $P_{n+j}$ are moving symmetrically to each other, see
figure (\ref{CSNBP}). Thus

\begin{eqnarray}\label{2}
 \mathbf{r}_{i}=-\mathbf{r}_{i+n},\qquad
\mathbf{r}_{j}=-\mathbf{r}_{j+n},\qquad \mathbf{r}_{0}=0,
\nonumber
\\
\mathbf{V}_{i}=\dot{\mathbf{r}}_{i}=-\dot{\mathbf{r}}_{i+n},\,\qquad
\mathbf{V}_{j}=\dot{\mathbf{r}}_{j}=-\dot{\mathbf{r}}_{j+n},
\qquad \mathbf{V}_{0}=\dot{\mathbf{r}}_{0}=0,
\end{eqnarray}
This dynamical symmetry is maintained for all time $t$. \item At
time $t = 0$ the bodies are collinear with their velocity vectors
perpendicular to their line of position. This ensures past-future
symmetry and is described by:
\begin{eqnarray}
\mathbf{r}_{i}\times \mathbf{r}_{j}=0, \qquad \mathbf{r}_{i}\cdot
\dot{\mathbf{r}}_{i}=0, \qquad \mathbf{r}_{j}\cdot
\dot{\mathbf{r}}_{j}=0
\end{eqnarray}

\end{enumerate}
Figure (\ref{CSNBP}) gives the dynamical configuration of the
CSNBP.

It is useful to define the masses as ratios to the total mass. Let
the total mass $M$ of the system be

\begin{equation}\label{Tmass2}
M=2\sum\limits_{i=1}^{n}m_{i}+m_{0}
\end{equation}
Let $\mu_{i}$ be the mass ratios defined as $\mu _{i}=
\frac{m_{i}}{M}$ for $i=0,1,2,\cdots,2n$. Equation (\ref{Tmass2})
then becomes
\begin{equation}
M=(2\sum\limits_{i=1}^{n}\mu_{i}+\mu_{0})M  \label{rat}
\end{equation}
Thus
\begin{equation}
\sum\limits_{i=1}^{n}\mu _{i}+\frac{\mu _{0}}{2}=\frac{1}{2}.
\label{massratios}
\end{equation}
 and
\begin{equation}
 0\leq\mu_0\leq1,\quad0\leq\mu_i\leq0.5,\qquad i = 1,2,\cdots,n \label{}
\end{equation}

We simplify the problem yet further by studying solely the
coplanar CSNBP, where the radius and velocity vectors are
coplanar. Figure (\ref{CSNBP}) gives the dynamical configuration
of the coplanar CSNBP at some time $t$. In the next sections we
will derive an analytical stability criterion for the CSNBP. This
criterion was applied to the CS5BP for a wide range of mass ratios
in chapter 5.

\section{The Equations of Motions}

Let there be $2n+1$ bodies of masses $m_{0},m_{1},m_{2},...,m_{2n-1},m_{2n}.$
Then their equations of motion may be written as
\begin{equation}
m_{i}\ddot{r}_{i}=\nabla _{i}U,\qquad i=0,1,2,...k\textrm{ \ where
}k=2n
\end{equation}
where ${\bf \nabla }_{i}={\bf i}\frac{\partial }{\partial x_{i}}+{\bf j}%
\frac{\partial }{\partial y_{i}}+{\bf k}\frac{\partial }{\partial z_{i}},%
{\bf i,j,k,}$ being unit vectors, along the rectangular axes $%
O_{x},O_{y},O_{z}$ respectively, $x_{i},y_{i},z_{i}$ being the rectangular
coordinates of body $P_{i}$ and $O$ being the center of mass of the system.

The force function $U$ is given by
\begin{equation}
U=G\sum\limits_{i=1}^{k}\sum\limits_{j=1}^{k}\frac{m_{i}m_{j}}{r_{ij}}
+m_{0}\sum\limits_{i=1}^{k}\frac{m_{i}}{r_{i}},\textrm{ \ \ \ \
}i\neq j,j<i, \label{U}
\end{equation}
where
\[
{\bf r}_{ij}={\bf r}_{i}-{\bf r}_{j}.
\]

Then the energy $E$ of the system may be written as
\begin{equation}
E=T-U
\end{equation}
where $T$ is the kinetic energy given by
\begin{equation}
T=\frac{1}{2}\sum\limits_{i=1}^{N}m_{i}|\stackrel{.}{{\bf r}_{i}}|^{2},
\end{equation}
Note:  $m_{0}$ remains stationary at the centre of mass so that
there is no contribution to the kinetic energy, angular momentum
and moment of inertia from $m_{0}.$ The angular momentum ${\bf C}$
is given by

\begin{equation}
{\bf C=}\sum\limits_{i=1}^{N}m_{i}{\bf r}_{i}\times \stackrel{.}{{\bf r}}%
_{i}.
\end{equation}

We may also write the moment of inertia of the system $I$ as
\begin{equation}
I=\sum\limits_{i=1}^{N}m_{i}{\bf r}_{i}^{2}.
\end{equation}

Now we may introduce the following symmetries.

\begin{enumerate}
\item  By the Roy-Ovenden mirror theorem the orbital history of the system
after $t=0$ is a mirror image of the history after $t=0$, given that their
velocity vectors are perpendicular to their relative radius vectors at $t=0$.

\item  The dynamic symmetry at any time $t$. Divide the $2n$ bodies into two
sets of bodies $P_{\alpha },\alpha =1,2,...n,$ and $P_{\beta },\beta
=n+1,n+2,...2n-1,2n$ and let $P_{\alpha }$ in the $\alpha $ set have mass $%
m_{\alpha }$ and position and velocity vectors ${\bf r}_{\alpha }$ and $%
\stackrel{.}{{\bf r}_{\alpha }\textrm{ }}$at time $t$. Let the
body $P_{\beta } $ in the $\beta $ set have mass $m_{\alpha }$ and
position and velocity
vectors $-{\bf r}_{\alpha }$ and $\stackrel{.}{-{\bf r}_{\alpha }}$ at time $%
t.$ The $(2n+1)^{th}$ body has mass $m_{0}$ and position and velocity
vectors ${\bf r}_{0}=0$ and $\stackrel{.}{{\bf r}_{0}}=0$ at time $t.$
\end{enumerate}

\bigskip Then the kinetic energy $T$, the angular momentum $C$ and the
moment of inertia $I$ may be written as

\begin{equation}
T=\sum\limits_{i=1}^{n}m_{i}|\stackrel{.}{{\bf r}_{i}}|^{2},
\end{equation}
\begin{equation}
{\bf C=}2\sum\limits_{i=1}^{n}m_{i}{\bf r}_{i}\times
\stackrel{.}{{\bf r}} _{i}.
\end{equation}
\begin{equation}
I=2\sum\limits_{i=1}^{n}m_{i}{\bf r}_{i}^{2}.
\end{equation}

\begin{figure}[tbp]
\begin{center}
\epsfig{file=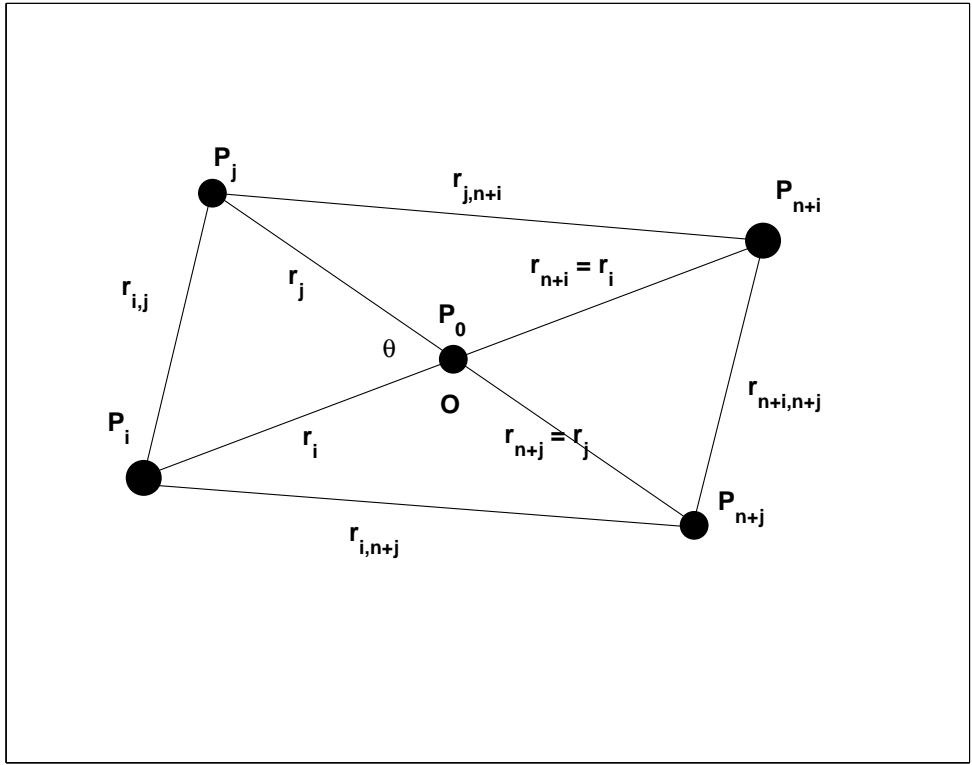, width=13cm}
\end{center}
\caption{Model of the Caledonian Symmetric N Body Problem (CSNBP)}
\label{CSNBP}
\end{figure}

Now consider the force function given by (\ref{U}). The figure
(\ref{CSNBP}) defined by any two symmetric pairs and a fifth body
at the centre of mass is always a parallelogram of changing shape
and orientation. Now

\begin{equation}
r_{ij}=r_{n+i,n+j};\textrm{ \ \ }r_{j,n+i}=r_{i,n+j}
\end{equation}
also
\begin{equation}
r_{i,n+i}=2r_{i};\textrm{ \ \ \ }r_{j,n+j}=2r_{j}.
\end{equation}
Consider $\Delta ^{\prime }s$ $P_{i}P_{j}O,$ $P_{j}OP_{n+{i}}.$
Let $\angle P_{i}OP_{j}=\theta .$

Then
\begin{equation}
r_{ij}^{2}=r_{i}^{2}+r_{j}^{2}-2r_{i}r_{j}\cos \theta .
\end{equation}
Also
\begin{equation}
r_{j,n+i}^2=r_{n+i}^{2}+r_{j}^{2}+2r_{n+i}r_{j}\cos \theta .
\end{equation}
But $r_{n+i}=r_{i},$ so that
\begin{equation}
r_{j,n+i}^{2}=2(r_{i}^{2}+r_{j}^{2})-r_{ij}^{2}.
\end{equation}

Therefore the six mutual radius vectors in the parallelogram may be written
as
\begin{equation}
P_{i}P_{n+i}=2r_{i};
\end{equation}
\begin{equation}
P_{j}P_{n+i}=P_{i}P_{n+j}=\sqrt{2\left( r_{i}^{2}+r_{j}^{2}\right)
-r_{ij}^{2}};
\end{equation}
\begin{equation}
P_{n+i}P_{n+j}=P_{i}P_{j}=r_{ij}.
\end{equation}
Now using the above symmetries the force function can be written as
\begin{equation}
U=G\left( \sum\limits_{i=1}^{n}\frac{m_{i}^{2}}{2r_{i}}+2\sum%
\limits_{i=1}^{n}\sum\limits_{j=1}^{n}m_{i}m_{j}\left( \frac{1}{r_{ij}}+%
\frac{1}{\sqrt{2\left( r_{i}^{2}+r_{j}^{2}\right) -r_{ij}^{2}}}\right)
+2m_{0}\sum\limits_{i=1}^{n}\frac{m_{i}}{r_{i}}\right) ,  \label{U2}
\end{equation}
\[
\textrm{ }i\neq j,\textrm{ \ \ }j<i.
\]

\section{Sundman's Inequality}

Sundman's inequality may be written as (see Orbital Motion by
Archie Roy (2004))
\begin{equation}
T\geq \frac{c^{2}}{2I}.
\end{equation}
Now
\begin{equation}
E=T-U,
\end{equation}
so that
\begin{equation}
U+E\geq \frac{c^{2}}{2I}.
\end{equation}
Let $E_{0}=-E.$ Then for real motion, we must have
\begin{equation}
U\geq \frac{C^{2}}{2I}+E_{0}.  \label{sundman1}
\end{equation}
Using (\ref{U2}), (\ref{sundman1}) becomes
\begin{eqnarray}
&&G\left( \sum\limits_{i=1}^{n}\frac{m_{i}^{2}}{2r_{i}}+2\sum
\limits_{i=1}^{n}\sum\limits_{j=1}^{n}m_{i}m_{j}\left(
\frac{1}{r_{ij}}+ \frac{1}{\sqrt{2\left(
r_{i}^{2}+r_{j}^{2}\right) -r_{ij}^{2}}}\right)
+2m_{0}\sum\limits_{i=1}^{n}\frac{m_{i}}{r_{i}}\right)  \nonumber \\
&\geq &\frac{C^{2}}{4\sum_{i=1}^{n}m_{i}r_{i}^{2}}+E_{0},\textrm{
where }i\neq j\textrm{ and }j<i.  \label{sundman2}
\end{eqnarray}

Let $M$ be the total mass of the system, so that
\begin{equation}
M=2\sum\limits_{i=1}^{n}m_{i}+m_{0}.  \label{totalmass}
\end{equation}
Let
\begin{equation}
\mu _{i}=\frac{m_{i}}{M},\textrm{ so that }0<\mu _{i}<1.
\end{equation}
Then (\ref{totalmass}) becomes
\begin{equation}
M=\left( 2\sum\limits_{i=1}^{n}\mu _{i}+\mu _{0}\right) M,
\end{equation}
thus
\begin{equation}
\sum\limits_{i=1}^{n}\mu _{i}+\frac{\mu _{0}}{2}=\frac{1}{2}.
\label{massratios}
\end{equation}
Thus the Sundman's inequality becomes
\begin{eqnarray}
&&GM^{2}\left( \sum\limits_{i=1}^{n}\frac{\mu _{i}^{2}}{2r_{i}}%
+2\sum\limits_{i=1}^{n}\sum\limits_{j=1}^{n}\mu _{i}\mu _{j}\left(
\frac{1}{ r_{ij}}+\frac{1}{\sqrt{2\left(
r_{i}^{2}+r_{j}^{2}\right) -r_{ij}^{2}}} \right) +2\mu
_{0}\sum\limits_{i=1}^{n}\frac{\mu _{i}}{r_{i}}\right)
 \nonumber\\
&\geq &\frac{C^{2}}{4M^{2}\sum_{i=1}^{n}\mu
_{i}r_{i}^{2}}+E_{0},\textrm{ where }i\neq j\textrm{ and }j<i.
\label{sundman3}
\end{eqnarray}
Let us now introduce dimensionless variables $\rho _{i},\rho
_{ij}$ and a new dimensionless constant $C_{0}$, called the
Szebehely constant Steves and Roy (2001) defined by the following
equations
\begin{eqnarray}
\rho _{i}=\frac{E_{0}}{GM^{2}}r_{i},\qquad \rho
_{ij}=\frac{E_{0}}{GM^{2}}r_{ij} \\
C_{0}=\frac{C^{2}E_{0}}{G^{2}M^{5}}, \label{newvariables}
\end{eqnarray}
where $E_0 \neq 0$. Then Sundman inequality takes the following
form
\begin{eqnarray}
&&\sum\limits_{i=1}^{n}\frac{\mu _{i}^{2}}{2\rho _{i}}+ 2\sum
\limits_{i=1}^{n}\sum\limits_{j=1}^{n}\mu _{i}\mu _{j}\left(
\frac{1}{\rho _{ij}}+\frac{1}{\sqrt{2\left( \rho _{i}^{2}+\rho
_{j}^{2}\right) -\rho _{ij}^{2}}}\right)+
2\mu_{0}\sum\limits_{i=1}^{n}\frac{\mu _{i}}{\rho
_{i}}\nonumber\\&\geq &\frac{C_{0}}{4M^{2}\sum_{i=1}^{n}\mu
_{i}\rho _{i}^{2}}+1,\qquad \textrm{where}i\neq j\textrm{ and
}j<i. \label{sundman4}
\end{eqnarray}
The geometry of triangles places the following constraint on the
radius vectors.
\begin{equation}
|r_{i}-r_{j}|\leq r_{ij}\leq r_{i}+r_{j}.  \label{rij}
\end{equation}
Then from (\ref{newvariables}) and (\ref{rij}) we get the following relation
between the new variables,
\begin{equation}
|\rho _{i}-\rho _{j}|\leq \rho _{ij}\leq \rho _{i}+\rho _{j}.  \label{rowij}
\end{equation}

\section{Regions of motion in the CSNBP}

In this section we construct explicit formulae for the boundary
surface of real motion.

It is useful to parameterize the surface in terms of variables
$y_i$ $(i=1,2,\cdots,n)$ and $x_ij$. We introduce two more
variables namely \ $y_{i}$ and $x_{ij}$ which are defined as
follows
\begin{equation}
y_{i}=\frac{\rho _{i}}{\rho _{n}}\textrm{, \ here }\rho
_{n}\textrm{ is the largest value of }\rho _{i}\textrm{, for all
}i.
\end{equation}
\begin{equation}
x_{ij}=\frac{\rho _{ij}}{\rho _{n}}.
\end{equation}
In the new variables, the kinematic constraints of  (\ref{rowij})
become
\begin{equation}
|y_{i}-y_{j}|\leq x_{ij}\leq y_{i}+y_{j}.
\end{equation}
Then Sundman's inequality takes the following form after introducing the
above new variables,

\begin{eqnarray}
&&\frac{1}{\rho _{n}}\left(
\frac{1}{2}\sum\limits_{i=1}^{n}\frac{\mu
_{i}^{2}}{y_{i}}+2\sum\limits_{i=1}^{n}\sum\limits_{j=1}^{n}\mu
_{i}\mu _{j}\left( \frac{1}{x_{ij}}+\frac{1}{\sqrt{2\left(
y_{i}^{2}+y_{j}^{2}\right) -x_{ij}^{2}}}\right) +2\mu
_{0}\sum\limits_{i=1}^{n}\frac{\mu _{i}}{y_{i}}\right)  \nonumber \\
&\geq &\frac{C_{0}}{4\rho _{n}^{2}\sum_{i=1}^{n}\mu
_{i}y_{i}^{2}}+1,\textrm{ where }i\neq j\textrm{ and }j<i.
\label{sundman5}
\end{eqnarray}
Hence
\begin{equation}
\frac{1}{\rho _{n}}A\geq \frac{1}{\rho _{n}^{2}}B+1,
\label{sundman6}
\end{equation}
where
\begin{eqnarray}
A(y_1,y_2,\cdots,y_n,x_{ij})&=&\frac{1}{2}\sum\limits_{i=1}^{n}\frac{\mu
_{i}^{2}}{y_{i}} +2\sum\limits_{i=1}^{n}\sum\limits_{j=1}^{n}\mu
_{i}\mu _{j}\left( \frac{1}{ x_{ij}}+\frac{1}{\sqrt{2\left(
y_{i}^{2}+y_{j}^{2}\right) -x_{ij}^{2}}} \right) \nonumber \\
&+&2\mu_{0}\sum\limits_{i=1}^{n}\frac{\mu _{i}}{y_{i}},i\neq
j\textrm{ and }j<i  \label{A1}
\end{eqnarray}
and
\begin{equation}
B(y_1,y_2,\cdots,y_n)=\frac{C_{0}}{4\sum_{i=1}^{n}\mu
_{i}y_{i}^{2}}.  \label{B}
\end{equation}
Taking the equality sign in (\ref{sundman6}) which defines a boundary
between real and imaginary motion.
\begin{equation}
\frac{1}{\rho _{n}}A=\frac{1}{\rho _{n}^{2}}B+1,
\end{equation}
this implies that
\begin{equation}
\rho _{n}^{2}-A\rho _{n}+B=0.
\end{equation}
After solving this quadratic equation and then simplifying we get the
following
\begin{equation}
\rho _{n}(y_1,y_2,\cdots,y_n,x_{12})=\frac{A}{2}\left( 1\pm
\sqrt{1-\frac{C_{0}}{A^{2}\sum_{i=1}^{n}\mu _{i}y_{i}^{2}}}\right)
.  \label{rowm}
\end{equation}
Define $C(y_1,y_2,\cdots,y_n,x_{12})$ by
\begin{equation}
C(y_1,y_2,\cdots,y_n,x_{12})=A^{2}\sum_{i=1}^{n}\mu _{i}y_{i}^{2},
\label{c1}
\end{equation}
so that
\begin{equation}
\rho _{n}(y_1,y_2,\cdots,y_n,x_{12})=\frac{A}{2}\left( 1\pm
\sqrt{1-\frac{C_{0}}{C}}\right) . \label{rowmfinal2}
\end{equation}
or
\begin{eqnarray}
\rho
_{n}(y_1,y_2,\cdots,y_n,x_{12})=\frac{1}{2}(\sqrt{\frac{C(y_1,y_2,\cdots,y_n,x_{12})}{\sum_{i=1}^{n}\mu
_{i}y_{i}^{2}}})\times\nonumber\\\left(1\pm
\sqrt{1-\frac{C_{0}}{C(y_1,y_2,\cdots,y_n,x_{12})}}\right) .
\label{rowmfinal}
\end{eqnarray}
 The value of $\rho _{n}$ depends explicitly on $C(y_1,y_2,\cdots,y_n,x_{12})$, if

\begin{enumerate}
\item  $C(y_1,y_2,\cdots,y_n,x_{12})>C_{0},$ there are two real
roots for $\rho _{n}$ \item $C(y_1,y_2,\cdots,y_n,x_{12})=C_{0},$
there is a double real root for $\rho _{n}$ \item
$C(y_1,y_2,\cdots,y_n,x_{12})<C_{0},$ there are two imaginary
roots for $\rho _{n}$
\end{enumerate}

\section{Maximum and Minimum extensions of the real motion projected in the phase space}

We define a new function $K_{ij}$, given by
\begin{equation}
K_{ij}=\mu _{i}\mu _{j}\left( \frac{1}{x_{ij}}+\frac{1}{\sqrt{2\left(
y_{i}^{2}+y_{j}^{2}\right) -x_{ij}^{2}}}\right) .  \label{kij}
\end{equation}
We may write
\begin{equation}
K_{ij}=\mu _{i}\mu _{j}\frac{1}{\sqrt{2\left( y_{i}^{2}+y_{j}^{2}\right) }}%
W_{ij},  \label{kij2}
\end{equation}
where
\begin{equation}
W_{ij}=\frac{1}{\omega _{ij}}+\frac{1}{\sqrt{1-\omega _{ij}^{2}}}
\label{Wij}
\end{equation}
and
\begin{equation}
\omega _{ij}=\frac{x_{ij}}{\sqrt{2\left( y_{i}^{2}+y_{j}^{2}\right) }}.
\end{equation}
Therefore (\ref{c1}) becomes

\begin{eqnarray}
C(y_1,y_2,\cdots,y_n,x_{12})=\left( \sum_{i=1}^{n}\mu
_{i}y_{i}^{2}\right)\times  \nonumber \\
\left( \frac{1}{2} \sum\limits_{i=1}^{n}\frac{\mu
_{i}^{2}}{y_{i}}+2\sum\limits_{i=1}^{n}\sum
\limits_{j=1}^{n}K_{ij}+2\mu _{0}\sum\limits_{i=1}^{n}\frac{\mu
_{i}}{y_{i}} \right) ^{2}\label{c1final}\\\textrm {where } i\neq
j\textrm{ and }j<i. \nonumber
\end{eqnarray}

By the close inspection of $W_{ij}$ and $\omega _{ij}$ we find the
following extreme values of $W_{ij}.$

\begin{equation}
\left.
\begin{array}{c}
W_{{ij}_{\min }}=2\sqrt{2}\textrm{ at }\omega _{ij}=\frac{1}{\sqrt{2}} \\
W_{{ij}_{\max }}=\frac{2\sqrt{2}max\{y_{i},y_j\}\left(
y_{i}^{2}+y_{j}^{2}\right) ^{1/2}}{
|y_{i}^{2}-y_{j}^{2}|}, \\
\textrm{when }\omega _{ij}\textrm{\ \ is either
}\frac{|y_{i}-y_{j}|}{\sqrt{ 2\left( y_{i}^{2}+y_{j}^{2}\right)
}}\textrm{ or }\frac{y_{i}+y_{j}}{\sqrt{ 2\left(
y_{i}^{2}+y_{j}^{2}\right) }}.
\end{array}
\right\}
\end{equation}

We know from (\ref{kij2}) that
\begin{equation}
K_{ij_{\min }}=\mu _{i}\mu _{j}\frac{1}{\sqrt{2\left(
y_{i}^{2}+y_{j}^{2}\right) }}W_{ij_{\min }}.
\end{equation}
As $W_{ij_{\min }}=2\sqrt{2}$ therefore
\begin{equation}
K_{ij_{\min }}=2\mu _{i}\mu _{j}\frac{1}{\sqrt{y_{i}^{2}+y_{j}^{2}}}.
\end{equation}
And hence
\begin{eqnarray}
C_{\min }=C_m=\left( \sum_{i=1}^{n}\mu _{i}y_{i}^{2}\right) \times
\left( \frac{1}{2}\sum\limits_{i=1}^{n}\frac{\mu _{i}^{2}}{y_{i}}
+4\sum\limits_{i=1}^{n}\sum\limits_{j=1}^{n}\frac{\mu _{i}\mu
_{j}}{\sqrt{ y_{i}^{2}+y_{j}^{2}}}+2\mu
_{0}\sum\limits_{i=1}^{n}\frac{\mu _{i}}{y_{i}} \right)^{2}\label{c1min}\\
\textrm{where  }i\neq j\textrm{ and }j<i.  \nonumber
\end{eqnarray}

Therefore the equations giving minimum projections are

\begin{eqnarray}
\rho
_{n}(y_1,y_2,\cdots,y_n,x_{12})=\frac{1}{2}\left(\sqrt{\frac{C_m(y_1,y_2,\cdots,y_n,x_{12})}{\sum_{i=1}^{n}\mu
_{i}y_{i}^{2}}}\right)\times \nonumber \\ \left( 1\pm
\sqrt{1-\frac{C_{0}}{C_m(y_1,y_2,\cdots,y_n,x_{12})}}\right) .
\label{rownfinal}
\end{eqnarray}

Proceeding on the same lines as for $K_{{ij}_{\min }}$ we get
$K_{{ij}_{\max }}$
\begin{equation}
K_{{ij}_{\max }}=2\mu_{i}\mu_{j}y_{i}.
\end{equation}
And thus
\begin{eqnarray}
C_{\max }=C_e=\left( \sum_{i=1}^{n}\mu_{i}y_{i}^{2}\right) \times
\left( \frac{1}{2}\sum\limits_{i=1}^{n}\frac{\mu_{i}^{2}}{y_{i}}
+4\sum\limits_{i=1}^{n}\sum\limits_{j=1}^{n}2\mu
_{i}\mu_{j}\frac{y_{i}}{|y_{i}^{2}-y_{j}^{2}|}+2\mu
_{0}\sum\limits_{i=1}^{n}\frac{\mu_{i}}{y_{i}} \right)
^{2}\label{c1max}\\ \textrm{where }i\neq j\textrm{ and }j<i.
\nonumber
\end{eqnarray}

The equations giving the maximum projections are:
\begin{eqnarray}
\rho
_{n}(y_1,y_2,\cdots,y_n,x_{12})=\frac{1}{2}\left(\sqrt{\frac{C_e(y_1,y_2,\cdots,y_n,x_{12})}{\sum_{i=1}^{n}\mu
_{i}y_{i}^{2}}}\right)\times \nonumber \\ \left( 1\pm
\sqrt{1-\frac{C_{0}}{C_e(y_1,y_2,\cdots,y_n,x_{12})}}\right) .
\label{rownfinal}
\end{eqnarray}
\section{Summary and Conclusions}

We have generalized the Caledonian Symmetric Five Body Problem
(CS5BP) to the Caledonian Symmetric N Body Problem (CSNBP) and
derived a general form of the stability criterion involving the
Szebehely constant $C_0$. Research in chapters 6 and 7 show that
the general form of the stability criterion can be used for four
and five body symmetrical systems. \citeasnoun{Andras1} has shown
that a guarantee of complete hierarchical stability for all time
is not possible for the six or more body problems because, unlike
the phase space of the CS5BP and CSFBP, the phase space of the
CSNBP is a connected manifold and transition from one hierarchy to
another is always possible. However it is hoped that the general
stability criterion can still be useful in determining
restrictions placed on the ability of the system to change from
one specific hierarchy to another in cases of more than five body
configurations. It may also be possible to find a hierarchical
stability criterion that gives partial stability for some
hierarchies in a similar manner as the criterion $R_3<C_0<R_4$
found for the four and five body problem. Also, although there may
be no guarantee of hierarchical stability for all time for
$C_0>C_{crit}$, it may be that numerically and effectively there
is hierarchical stability, since the likelihood of changing from
one hierarchy to another becomes very small as the connections
between the hierarchies in the phase space narrow. In other words,
a measure of the size of $C_0$ relative to $C_{crit}$ may well
give an indication of when a system is very unlikely to change its
hierarchy, even though theoretically it is still possible.

%% file: SummAndConc.tex
\chapter{Summary and Conclusions}

We have investigated the Caledonian Symmetric $N$ Body Problem
with particular focus on the Caledonian Symmetric Four Body
Problem (CSFBP) and Caledonian Symmetric Five Body Problem
(CS5BP). Both the CSFBP and the CS5BP are symmetrically restricted
problems with all possible symmetries utilized. We have also
studied, numerically, a slightly perturbed non-symmetric extension
of the CSFBP by using the general four body equations.

In this chapter we provide a brief summary of the results obtained
in this thesis. For this purpose we divide this chapter into four
sections. In Section 8.1 we summarize the results obtained in
Chapter 3 which includes equilibrium solutions of four body
problem and its linear stability analysis. In Section 8.2 we
summarize the results of Chapter 4 dealing with the slightly
perturbed Caledonian Symmetric Four Body Problem (CSFBP). In
Section 8.3 we summarize the results of both analytical and
numerical work undertaken to understand the global characteristics
of the Caledonian Symmetric Five Body Problem (CS5BP). In Section
8.4 we summarize the results of Chapter 7 dealing with the
Caledonian Symmetric N Body Problem (CSNBP).

\section{Equilibrium configurations of four body problem and its linear stability analysis}

The main goal of the first part of the thesis was to review the
literature on analytical solutions to four and five body problems
(chapter 2), to find equilibrium solutions for the four body
problem (Chapter 3) and to discuss their linear stability (Chapter
4).

In Section 3.1 of chapter 3 we gave a detailed review of the
equilibrium configurations of equal mass cases of the four body
problem \cite{BonnieRoy1998}. In Section 3.2 we discussed the non
equal mass cases of the four body problem. The non equal mass
cases have two pairs of equal masses of masses $m$ and $M$ with
$m$ being the smaller one. We defined the ratio between $m$ and
$M$ as $\mu =m/M \leq 1$. We allowed $\mu$ to reduce from 1 to 0
to obtain the Lagrange five equilibrium points $L_1, L_2, L_3,L_4$
and $L_5$ of the Copenhagen problem \cite{BonnieRoy1998}. We
completed the analysis of \citeasnoun{BonnieRoy1998} to include
two more examples of equilibrium solutions of four body problem:
\begin{enumerate}
\item The Triangular equilibrium configuration of four body
problem with the two bigger masses making the base of the
triangle. \item The Triangular equilibrium configuration of four
body problem with the two smaller masses making the base of the
triangle.
\end{enumerate}
We allowed the two smaller masses to reduce from 1 to 0. Figures
(3.9) and (3.11) showed the evolution of all the four masses when
$\mu$ goes to zero from 1 in the above two cases of the triangular
equilibrium configurations. In the Triangular Case-I for $\mu=1$
we get the two well known equal mass solutions i.e. the
equilateral triangle solution and the isosceles triangle solution.
For the triangular case-I as $\mu$ is reduced from 1 to 0 there
always exists two continuous family of solutions: 1) Solution 1
starting at $\mu=1$ with the isosceles triangle solution and
ending as $\mu \rightarrow 0$ with $P_2 \rightarrow L_4$; and 2)
solution 2 starting at $\mu=1$ with the equilateral triangle
solution and ending as $\mu \rightarrow 0$ with $P_4 \rightarrow
L_1$. In the Triangular Case-II for $\mu=1$ the equilateral
triangle solution is obtained. In this case there is a continuous
family of solutions for $\mu$ between 1 and 0.9972.

\section{Stability analysis of the CSFBP}

In chapter 4 we analyzed the evolution of the nearly symmetric
Caledonian Symmetric Four Body systems over a range of initial
conditions. We identified the stable and unstable regions of
orbits by integrating over 1 million time steps. The possible
endpoints of the numerical integration were: 1. close encounter
between any two of the bodies occurs, 2. the dynamical Symmetry is
broken or 3.  1 million time steps of integration. Two types of
systems were analyzed, the symmetric and nearly symmetric CSFBP by
using a general four body integrator. The general four body
integrator was developed using the Microsoft Visual C++ software.
The results obtained were processed using Matlab 6.5 to produce
the graphs given in Figures (\ref{fig4.2}) to (\ref{fig4.16})

We analyzed the phase space of the symmetric and nearly symmetric
CSFBP's in detail for the $\mu=1$ and $\mu=0.1$ cases. Graphs of
the phase space of the CSFBP's with different $C_0$ values were
determined. Each grid point $r_1,r_2$ on the  graphs denoting
initial conditions of an orbit, Figures (\ref{fig4.2}) to
(\ref{fig4.16}), was color coded according to the different final
outcomes of the evolution of the orbits.

We found that the size of the stable regions in the phase space of
the  CSFBP is dependent on the value of the Szebehely constant.
The larger the value of the Szebehely constant the more stable
overall  is the CSFBP system across the phase space. For the equal
mass case of the CSFBP the phase space was very chaotic as most of
the orbits were collision orbits with the single binary regions
being the most chaotic. In the single binary cases, almost all of
the orbits were collision orbits. In the $\mu=0.1$ case, the SB2
region is the most chaotic and was comparatively very very small.
The SB1 region is the most stable as overall there were few
collisions. When comparing the two cases, $\mu=1$ and $\mu=0.1$
with each other, the $\mu=0.1$ case is the most stable. In the
$\mu=1$ case a large number of orbits fail the symmetry breaking
criterion. We know that those orbits are not necessarily unstable.
Therefore it will be interesting to investigate, in future, the
long term stability of the equal mass case of the symmetric and
nearly symmetric CSFBP past the symmetry breaking point.

\section{Topological stability of the CS5BP and numerical
verification of the analytical results}

The main goal of this part of the thesis (Chapter 5, 6) was to
derive an analytical stability criterion to discuss the global
hierarchical stability of the CS5BP and then to confirm the
prediction of these theoretical results using numerical
integrations. In order to do the first part we introduced a
stationary mass to the centre of mass of the CSFBP. The resulting
system was named The Caledonian Symmetric Five Body problem
(CS5BP).

In Section 1 of Chapter 5 we defined the CS5BP in such a way that
the CSFBP became a special case of the CS5BP. We derived the
Sundman's inequality in its simplest possible form which is the
key to the derivation of the stability criterion. In Section 3 to
5 we derived the analytical stability criterion. We showed that
the hierarchical stability of the CS5BP for a range of different
mass ratios solely depends on the Szebehely constant, $C_0$ which
is a function of the total energy and angular momentum. We also
showed that $C_0> 0.0659$ will guarantee that the CS5BP for all
values of the mass ratios $\mu_i,i=0,1,2$  will be hierarchically
stable. It is also shown that when $\mu_0>\mu_1$ or $\mu_2$, the
14 hierarchy state is hierarchically stable for $R_3<C_0<R_4$
while in all other cases the 24 hierarchy state is hierarchically
stable at $R_3<C_0<R_4$. In all cases hierarchical stability for
all states occurred when $C_0>R_4$. Of course when $\mu_1=\mu_2$,
hierarchical stability for all time occurs when $C_0>R_3=R_4$

In Chapter 6 we confirmed the theoretical results of Chapter 5
with the help of numerical integrations, i.e. we showed that the
Szebehely constant can be used to predict hierarchical stability.
We also confirmed that when the value of $C_0$ was increased, the
system became more hierarchically stable. In order to do this an
appropriate form of equations of motion were derived and a
numerical integrator was developed using Fortran and Microsoft
Visual C++ software. Initial conditions were set in such a way
that the whole phase space was covered. We used the Matlab 6.5
Software package to process the results. The results of thousands
of integrations were analyzed in Section 5 and 6 and the results
were collected in different kinds of tables.

The table of hierarchy changes contained the number of hierarchy
changes throughout the whole range of different integrations. It
can easily be seen from these tables that as we increase the value
of the Szebehely constant, the number of hierarchy changes drop
until it becomes zero for $C_0>C_{crit}=R_4$. Therefore it
confirms the analytical predictions of Chapter 6. For values of
$C_0$ , $R_3<C_0<R_4$, the analytically predicted hierarchically
stable states were verified as well. These tables also showed that
the introduction of a stationary mass into the centre of mass of
the CSFBP played a stabilizing role. The number of hierarchy
changes decreased as we increased the value of the central mass.
Also when one of the $\mu_1$ or $\mu_2$ was very small we had
fewer number of hierarchy changes.

Our analysis in Chapter 6, also extended and  completed the
analysis of \citeasnoun{Andras1} for the CSFBP, since he checked
the hierarchical stability of the CSFBP for only half of the phase
space. We have discussed the CSFBP as a special case of the CS5BP,
where $\mu_0=0$, and covered the whole phase space.

\section{The Caledonian Symmetric N Body Problem (CSNBP)}

In Chapter 7 we generalized the Caledonian Symmetric Five Body
Problem (CS5BP) to the Caledonian Symmetric N Body Problem (CSNBP)
which is the most general form of the Caledonian problem.
\citeasnoun{Andras1} showed that there cannot be any hierarchical
stability criterion for the six or more body problem which is
valid for all time. Our generalization showed that the Szebehely
constant still exists and still defines the topological behavior
of the phase space. We have shown in Chapter 5 and 6 that there is
a relation between the Szebehely constant and the hierarchical
stability. Therefore it is very likely that this constant will
also play some role in determining the hierarchical stability of
the $N$ Body Problem where $n \geq 6$. It has been proven by
\citeasnoun{Andras1} that it cannot give complete  hierarchical
stability criterion for all time, but perhaps it can give a
criterion that stipulates the hierarchical stability of certain
states of hierarchy as $R_3$ did in the CSFBP and CS5BP. Or
perhaps even though there is no guarantee of hierarchical
stability for all time using $C_0$ as a stability criterion, it
may be that the criterion can provide a measure that in effect
provide stability. i.e. analytically it is possible for the system
to change from one hierarchy to another but effectively it does
not because the chances of this occurring are so slim. We will
address these question in the future.

%% file: futurework.tex
\chapter{Future Work}
We investigated Caledonian Symmetric N Body Problems. The main
focus was on the Caledonian Symmetric Five Body Problem (CS5BP)
with the Caledonian Symmetric Four Body Problem (CSFBP) studied as
a special case. Research on the CS5BP and hence the CSFBP is far
from complete. In every chapter we answered some very interesting
questions but with every answer more questions arose which we
would like to answer in the future. We will list some of these
questions in this chapter with some explanations on how these
questions might be addressed.

Section 9.1, 9.2 and 9.3 discusses the questions still unanswered,
with some proposed methods of tackling the solutions for the
following areas respectively, CSFBP, CS5BP and CSNBP.
\section{Future lines of research on the CSFBP}
In Chapter 4 we investigated the phase space of the CSFBP for
stable and unstable regions using the general four body
integrator. We investigated the four body problem with two types
of initial conditions i.e. symmetric initial conditions and nearly
symmetric initial condition where symmetry was broken slightly in
the $x$-component of $P_1$.

In the equal mass case there are a large number of orbits which
fails the symmetry criterion. We know that because an orbit fails
to maintain its symmetry does not necessarily mean that it is
unstable. Therefore it will be interesting to repeat the analysis
of Chapter 4 for those regions of the phase space of the CSFBP, in
the equal mass case , that failed the symmetry, without the
symmetry restrictions. This research can be done with our existing
numerical tools except for a slight modification in the
integrator. We could use the same equations of motion and initial
conditions given in Chapter 4.

The long-term integration of the general four body equations is a
very important exercise but at the same time very time consuming
too. For example only long-term integrations can tell you which of
the perturbed system will remain close to the symmetric CSFBP
systems. There is no alternate method for this particular exercise
except long-term integration.

There are, however some fast chaos detection methods which can be
used to study the chaotic behavior of the CSFBP. Sz\'ell, \'erdi,
S\'andor and Steves (2003) have investigated the chaotic behavior
of the CSFBP using the fast chaos detection methods of Relative
Lyapunov Indicators (RLI) and Smaller Alignment Indices (SALI). By
using the general four body equations, the CSFBP system becomes
closer to  the general four body system. It would be interesting
to investigate the phase space of the CSFBP by using the fast
chaos detection methods like RLI and SALI with the general four
body equations.

This exercise will provide us with the unique opportunity of
comparing these results with our results obtained in Chapter 4 and
Sz\'ell, \'erdi, S\'andor and Steves (2003) results. We expect to
get similar results to Sz\'ell, \'erdi, S\'andor and Steves (2003)
about the connection between the chaotic and regular behavior of
the phase space of the CSFBP with the Szebehely constant i.e. the
phase space of the CSFBP is more regular for $C_0$ near $C_{crit}$
and particularly for $C_0>C_{crit}$. We also think that by using
the general four body integrator the phase space of the CSFBP will
be more chaotic which is indicated by the results in Chapter 4. To
be able to do this exercise we would have to modify the integrator
or perhaps write a new integrator.

Another aspect of the CSFBP which is completely unexplored is an
investigation of the non-coplanar case. The degrees of freedom
will increase which will make it comparatively more challenging
than the coplanar case. The analytical results obtained by Steves
and Roy (2001) will not change, but we will have to modify the
equations of motion and hence the integrator to carry out this
research. The three dimensional CSFBP model will be a very good
approximation of the general four body model near a symmetrical
system. The results obtained in this case could be used to model
three dimensional symmetrical quadruple stellar and planetary
systems.

We believe that apart from being an approximation to the general
four body problem, the CSFBP has theoretical as well as practical
applications. The phase space of the CSFBP is a subspace of the
phase space of the general four body problem. Therefore some
typical behavior of the CSFBP may occur in the general four body
problem. We would like to examine how the disconnectedness of the
restricted phase space influences the global phase space nearby.

\section{Future lines of research on the CS5BP}\label{sec9.2}

In Chapter 5 we developed a symmetrically restricted five body
problem called the Caledonian Symmetric Five Body problem (CS5BP),
for which we derived an analytical stability criterion valid for
all time. Following in the footsteps of Steves and Roy (1998,
2000, 2001) we have shown that the hierarchical stability of the
CS5BP depends solely on a parameter called the Szebehely constant,
$C_0$. We verified this stability criterion numerically in Chapter
7. The CS5BP is modelled by introducing a stationary mass to the
centre of mass of the CSFBP and we have seen that the bigger the
central mass, the more stable the CS5BP system becomes. This
generates some very interesting questions.

The characteristics of the CS5BP studied in Chapter 5 and 6 are
its global characteristics. In future we would like to look at the
finer details of the CS5BP and study it locally. We will repeat
the exercise of Chapter 4 for the CS5BP i.e. we will perform long
term integrations for different mass ratios of the CS5BP with the
general five body integrator for both symmetric and nearly
symmetric cases. This will help us to discover the effect on the
stability of the general system near the symmetric system, of
adding a central body to the CSFBP. We will also be able to find
out if the system remains close to symmetry for a longer time than
it was found in the CSFBP.

To be able to complete this research we would need to extend the
general four body integrator to include five bodies. We would also
need to find suitable initial conditions and derive the equations
of motion for the general CS5BP. We should also be prepared to
dedicate a lot of CPU time.

Another method of studying the CS5BP locally is to use the fast
chaos detection methods like RLI and SALI. Proceeding along the
same lines as we suggested for the CSFBP, we plan to study the
chaotic and regular behavior of the CS5BP. In this research we
will have the flexibility to vary the value of the central mass.
Therefore we can approximate this CS5BP system by CSFBP by
approximating the central mass to zero. This exercise will provide
us with the unique opportunity of comparing these results with the
results gained by investigating the long-term evolution of CS5BP
orbits that are  both symmetric and nearly symmetric. We expect to
get results similar to the long term integrations.

Like the CSFBP the non-coplanar case of the CS5BP is completely
unexplored. Recall that in this case the degrees of freedom
increases in the non-coplanar case and therefore will be more
challenging to explore. The method of research will be the same as
was for the CSFBP. In fact we can analyze the CS5BP first and then
let the mass of the central body be equal to zero in order to
analyze the CSFBP. The three dimensional CS5BP model could be a
good approximation of the general five body model. These results
can be used to model planetary systems.

After completing the proposed research on the CS5BP in this
section we should be able to answer the following very important
question.

Does the central mass of the CS5BP forces the system to remain
close to symmetry for longer than the CSFBP and what role does the
central mass play in the overall stability/instability of the
CS5BP in the more general problem.

Our current analysis of the CS5BP suggests that if the mass of the
central body is bigger than the other bodies, then the system will
be more stable.

\section{Future lines of research on the CSNBP}

In chapter 7 we have generalized the CS5BP to any number of bodies
to give the CSNBP. In the CSNBP there exist critical values of the
Szebehely constant but we know that they do not predict
hierarchical stability for all time because unlike the phase space
of the CS5BP the phase space of the CSNBP is a connected manifold
and transition from one hierarchy state to another is always
possible. Although complete hierarchical stability is not
possible, we think that the Szebehely constant could still have
some role to play in the hierarchical stability of more than five
body systems. For example for a very large value of $C_0$, greater
than $C_{crit}$, the six or seven body system should be more
stable than it will be for a smaller $C_0$. We plan to investigate
this numerically and confirm or contradict our expectations.

A very interesting application of the CSNBP can be in Galactic
Dynamics or stellar cluster dynamics. For example by taking N to
be a very large number or in other words by taking $N\rightarrow
\infty$ it may be possible to model globular clusters or galaxies
and their stability at least in an empirical sense.

%% file: Appendix1.tex

\chapter*{Appendix A}

\section*{Results of \citeasnoun{anrashieStab} on the numerical investigation of hierarchical stability of the CSFBP}
In this appendix results on the numerical investigations of the
hierarchical stability of the CSFBP are given \cite{anrashieStab},
which include tables on total number of hierarchy changes and
percentage of hierarchy changes for different values of $\mu$.
\appendix
\renewcommand{\baselinestretch}{1}

\begin{table}
  \begin{center}
  \caption{Hierarchical changes for $\mu = 1$}\label{hiechange}
 \centerline{Total number of hierarchy changes}

  \begin{tabular}{c|cc|ccc||cc} \hline
& $C_0=10$ & $C_0=29$ & $C_0 = 30$  & $C_0 = 40$  &
$C_0 = 46$  &  $C_0 = 47$ & $C_0 = 60$  \\
\hline
  12 13 & 0    &  1   & 0    &  3   &  0   &  0    & 0   \\ \cline {1-6}
  12 14 & 230  &  136 & 120  &  79  &  7   &  0    & 0   \\
  12 23 & 143  &  111 & 74   &  87  &  10  &  0    & 0   \\ \cline {1-6}
  13 12 & 0    &  2   & 2    &  4   &  0   &  0    & 0   \\ \cline {1-6}
  13 14 & 133  &  79  & 88   &  68  &  0   &  0    & 0   \\
  13 23 & 64   &  97  & 75   &  69  &  9   &  0    & 0   \\
  14 12 & 137  &  106 & 97   &  56  &  2   &  0    & 0   \\
  14 13 & 95   &  112 & 108  &  65  &  4   &  0    & 0   \\
  14 23 & 1119 &  185 & 158  &  42  &  0   &  0    & 0   \\
  23 12 & 80   &  164 & 151  &  170 &  56  &  0    & 0   \\
  23 13 & 217  &  198 & 191  &  184 &  57  &  0    & 0   \\
  23 14 & 1316 &  267 & 236  &  42  &  1   &  0    & 0   \\
  \end{tabular}

\vspace{1cm}

\centerline{Percentage of hierarchical changes }
  \begin{tabular}{c|cc||ccc||cc} \hline
 & $C_0 = 10$  & $C_0 = 29$  & $C_0 = 30$  & $C_0 = 40$  & $C_0 = 46$ &  $C_0 = 47$& $C_0 = 60$  \\ \hline
    12 13 &  0.0  &  0.1  &  0.0  &  0.3  &  0.0  &  0.0  &  0.0 \\ \cline {1-6}
    12 14 &  6.5  &  9.3  &  9.2  &  9.1  &  4.8  &  0.0  &  0.0 \\
    12 23 &  4.0  &  7.6  &  5.7  &  10.0 &  6.8  &  0.0  &  0.0  \\ \cline {1-6}
    13 12 &  0.0  &  0.1  &  0.2  &  0.5  &  0.0  &  0.0  &  0.0 \\ \cline {1-6}
    13 14 &  3.8  &  5.4  &  6.8  &  7.8  &  0.0  &  0.0  &  0.0 \\
    13 23 &  1.8  &  6.7  &  5.8  &  7.9  &  6.2  &  0.0  &  0.0 \\
    14 12 &  3.9  &  7.3  &  7.5  &  6.4  &  1.4  &  0.0  &  0.0 \\
    14 13 &  2.7  &  7.7  &  8.3  &  7.5  &  2.7  &  0.0  &  0.0 \\
    14 23 &  31.7 &  12.7 &  12.2 &  4.8  &  0.0  &  0.0  &  0.0 \\
    23 12 &  2.3  &  11.2 &  11.6 &  19.6 &  38.4 &  0.0  &  0.0 \\
    23 13 &  6.1  &  13.6 &  14.7 &  21.2 &  39.0 &  0.0  &  0.0 \\
    23 14 &  37.2 &  18.3 &  18.2 &  4.8  &  0.7  &  0.0  &  0.0 \\
  \end{tabular}
\end{center}
\end{table}

\clearpage

\begin{table}
 \begin{center}
  \caption{Hierarchical changes for $\mu = 0.1$}\label{hiechange0.1}
  \centerline{Total number of hierarchical changes}
  \begin{tabular}{c|cc|c|c||c||c} \hline
 & $C_0 = 0.4$  & $C_0 = 0.6$  & $C_0 = 0.65$  & $C_0 = 0.7$  & $C_0 = 0.8$  &  $C_0 = 0.9$   \\ \hline
    12 13 &  0   & 1   &  3   & 2  & 0 &  0 \\ \cline {1-5}
    12 14 &  166 & 49  &  41  & 51 & 3 &  0 \\ \cline {6-6}
    12 23 &  64  & 49  &  32  & 38 & 0 &  0 \\ \cline {1-6}
    13 12 &  2   & 0   &  0   & 0  & 0 &  0 \\ \cline {1-5}
    13 14 &  28  & 30  &  37  & 50 & 1 &  0 \\ \cline {6-6}
    13 23 &  33  & 33  &  38  & 18 & 0 &  0 \\ \cline {6-6}
    14 12 &  38  & 27  &  31  & 37 & 1 &  0 \\
    14 13 &  31  & 27  &  32  & 30 & 1 &  0 \\ \cline {6-6}
    14 23 &  481 & 220 &  147 & 90 & 3 &  0 \\ \cline {6-6}
    23 12 &  69  & 65  &  57  & 62 & 0 &  0 \\
    23 13 &  47  & 62  &  64  & 61 & 0 &  0 \\ \cline {6-6}
    23 14 &  541 & 244 &  177 & 84 & 1 &  0 \\ \cline {6-6}
  \end{tabular}

\vspace{1cm}

\centerline{Percentage of hierarchical changes}
  \begin{tabular}{c|cc|c|c||c||c} \hline
 & $C_0 = 0.4$  & $C_0 = 0.6$  & $C_0 = 0.65$  & $C_0 = 0.7$  & $C_0 = 0.8$  &  $C_0 = 0.9$   \\ \hline
    12 13 &  0.0  &  0.1  &  0.5  &  0.4  &  0.0  &  0.0 \\ \cline {1-5}
    12 14 &  11.1 &  6.1  &  6.2  &  9.8  &  30.0 &  0.0 \\ \cline {6-6}
    12 23 &  4.3  &  6.1  &  4.9  &  7.3  &  0.0  &  0.0 \\ \cline {1-6}
    13 12 &  0.1  &  0.0  &  0.0  &  0.0  &  0.0  &  0.0 \\ \cline {1-5}
    13 14 &  1.9  &  3.7  &  5.6  &  9.6  &  10.0 &  0.0 \\ \cline {6-6}
    13 23 &  2.2  &  4.1  &  5.8  &  3.4  &  0.0  &  0.0 \\ \cline {6-6}
    14 12 &  2.5  &  3.3  &  4.7  &  7.1  &  10.0 &  0.0 \\
    14 13 &  2.1  &  3.3  &  4.9  &  5.7  &  10.0 &  0.0 \\ \cline {6-6}
    14 23 &  32.1 &  27.3 &  22.3 &  17.2 &  30.0 &  0.0 \\ \cline {6-6}
    23 12 &  4.6  &  8.1  &  8.6  &  11.9 &  0.0  &  0.0 \\
    23 13 &  3.1  &  7.7  &  9.7  &  11.7 &  0.0  &  0.0 \\ \cline {6-6}
    23 14 &  36.1 &  30.2 &  26.9 &  16.1 &  10.0 &  0.0 \\ \cline {6-6}
  \end{tabular}
  \end{center}
\end{table}

\clearpage

\begin{table}
\begin{center}
  \caption{Hierarchical changes for $\mu = 0.01$}\label{hiechange0.01}
  \centerline{Total number of hierarchical changes}
  \begin{tabular}{c|cc|c||c||c} \hline
 & $C_0 = 0.2$  & $C_0 = 0.27$  & $C_0 = 0.283$  & $C_0 = 0.29$  & $C_0 = 0.3$   \\ \hline
    12 13 &  0   & 0  & 0  & 1 &  0 \\ \cline {1-4}
    12 14 &  17  & 16 & 7  & 6 &  0 \\ \cline{5-5}
    12 23 &  14  & 7  & 0  & 0 &  0 \\ \cline{5-5}
    13 12 &  0   & 1  & 0  & 1 &  0 \\ \cline {1-4}
    13 14 &  3   & 1  & 6  & 5 &  0 \\ \cline{5-5}
    13 23 &  3   & 3  & 4  & 0 &  0 \\ \cline{5-5}
    14 12 &  3   & 4  & 1  & 6 &  0 \\
    14 13 &  1   & 0  & 3  & 7 &  0 \\ \cline{5-5}
    14 23 &  105 & 38 & 13 & 2 &  0 \\ \cline{5-5}
    23 12 &  2   & 4  & 4  & 0 &  0 \\
    23 13 &  5   & 6  & 8  & 0 &  0 \\ \cline{5-5}
    23 14 &  113 & 41 & 7  & 2 &  0 \\ \cline{5-5}
  \end{tabular}

\vspace{1cm}

\centerline{Percentage of hierarchical changes}
  \begin{tabular}{c|cc|c||c||c} \hline
  & $C_0 = 0.2$  & $C_0 = 0.27$  & $C_0 = 0.283$ & $C_0 = 0.29$  & $C_0 = 0.3$   \\ \hline
    12 13 &  0.0  &  0.0  &  0.0  &  3.3  &  0.0 \\ \cline{1-4}
    12 14 &  6.4  &  13.2 &  13.2 &  20.0 &  0.0 \\ \cline{5-5}
    12 23 &  5.3  &  5.8  &  0.0  &  0.0  &  0.0 \\ \cline{5-5}
    13 12 &  0.0  &  0.8  &  0.0  &  3.3  &  0.0 \\ \cline{1-4}
    13 14 &  1.1  &  0.8  &  11.3 &  16.7 &  0.0 \\ \cline{5-5}
    13 23 &  1.1  &  2.5  &  7.5  &  0.0  &  0.0 \\ \cline{5-5}
    14 12 &  1.1  &  3.3  &  1.9  &  20.0 &  0.0 \\
    14 13 &  0.4  &  0.0  &  5.7  &  23.3 &  0.0 \\ \cline{5-5}
    14 23 &  39.5 &  31.4 &  24.5 &  6.7  &  0.0 \\ \cline{5-5}
    23 12 &  0.8  &  3.3  &  7.5  &  0.0  &  0.0 \\
    23 13 &  1.9  &  5.0  &  15.1 &  0.0  &  0.0 \\ \cline{5-5}
    23 14 &  42.5 &  33.9 &  13.2 &  6.7  &  0.0 \\ \cline{5-5}
  \end{tabular}
  \end{center}
\end{table}

\clearpage

\begin{table}
\begin{center}
  \caption{Hierarchical changes for $\mu = 0.001$}\label{hiechange0_001}
  \centerline{Total number of hierarchical changes}
  \begin{tabular}{c|cc|c||c||c} \hline
 & $C_0 = 0.25$ & $C_0 = 0.2529$ & $C_0 = 0.2533$ & $C_0 = 0.2538$  & $C_0 = 0.2545$   \\ \hline
    12 13 &  0  & 0 &  0 &  0 &  0 \\ \cline {1-4}
    12 14 &  0  & 1 &  5 &  0 &  0 \\
    12 23 &  2  & 2 &  0 &  0 &  0 \\ \cline {1-4}
    13 12 &  0  & 0 &  0 &  0 &  0 \\ \cline {1-4}
    13 14 &  0  & 0 &  2 &  0 &  0 \\
    13 23 &  1  & 2 &  0 &  0 &  0 \\
    14 12 &  0  & 0 &  1 &  0 &  0 \\
    14 13 &  0  & 1 &  3 &  0 &  0 \\
    14 23 &  56 & 4 &  2 &  0 &  0 \\
    23 12 &  0  & 2 &  1 &  0 &  0 \\
    23 13 &  1  & 1 &  1 &  0 &  0 \\
    23 14 &  75 & 9 &  1 &  0 &  0 \\
  \end{tabular}

\vspace{1cm}

 \centerline{ Percentage of hierarchical changes}
  \begin{tabular}{c|cc|c||c||c} \hline
  & $C_0 = 0.25$ & $C_0 = 0.2529$  & $C_0 = 0.2533$  & $C_0 = 0.2538$  & $C_0 = 0.2545$   \\ \hline
    12 13 &  0.0  &  0.0  &  0.0  &  0.0  &  0.0 \\ \cline {1-4}
    12 14 &  0.0  &  4.5  &  31.3 &  0.0  &  0.0 \\
    12 23 &  1.5  &  9.1  &  0.0  &  0.0  &  0.0 \\ \cline {1-4}
    13 12 &  0.0  &  0.0  &  0.0  &  0.0  &  0.0 \\ \cline {1-4}
    13 14 &  0.0  &  0.0  &  12.5 &  0.0  &  0.0 \\
    13 23 &  0.7  &  9.1  &  0.0  &  0.0  &  0.0 \\
    14 12 &  0.0  &  0.0  &  6.3  &  0.0  &  0.0 \\
    14 13 &  0.0  &  4.5  &  18.8 &  0.0  &  0.0 \\
    14 23 &  41.5 &  18.2 &  12.5 &  0.0  &  0.0 \\
    23 12 &  0.0  &  9.1  &  6.3  &  0.0  &  0.0 \\
    23 13 &  0.7  &  4.5  &  6.3  &  0.0  &  0.0 \\
    23 14 &  55.6 &  40.9 &  6.3  &  0.0  &  0.0 \\
  \end{tabular}
  \end{center}
\end{table}